\documentclass[10pt,preprint]{aastex6}
\usepackage{natbib}
\usepackage{multirow}
\usepackage{upgreek}
\usepackage{wasysym}
\usepackage{txfonts}
\usepackage{gensymb}
\usepackage{morefloats}
\usepackage{threeparttable}
\usepackage{rotating}
\usepackage{afterpage}
\slugcomment{version \today}
\shorttitle{Constraining the Physical State of the Hot Gas Halos in NGC 4649 and NGC 5846}
\shortauthors{Paggi et al.}

\begin{document}

\title{Constraining the Physical State of the Hot Gas Halos in NGC 4649 and NGC 5846}

\author{Alessandro Paggi\altaffilmark{1}, Dong-Woo Kim\altaffilmark{1}, Craig Anderson\altaffilmark{1}, Doug Burke\altaffilmark{1}, Raffaele D'Abrusco\altaffilmark{1}, Giuseppina Fabbiano\altaffilmark{1}, Antonella Fruscione\altaffilmark{1}, Tara Gokas\altaffilmark{1}, Jen Lauer\altaffilmark{1}, Michael McCollough\altaffilmark{1}, Doug Morgan\altaffilmark{1}, Amy Mossman\altaffilmark{1}, Ewan O'Sullivan\altaffilmark{1}, Ginevra Trinchieri\altaffilmark{2}, Saeqa Vrtilek\altaffilmark{1}, Silvia Pellegrini\altaffilmark{3}, Aaron J. Romanowsky\altaffilmark{4,5} \& Jean Brodie\altaffilmark{5}}
\affil{\altaffilmark{1}Harvard-Smithsonian Center for Astrophysics, 60 Garden St, Cambridge, MA 02138, USA: 
\href{mailto:apaggi@cfa.harvard.edu}{apaggi@cfa.harvard.edu}\\
\altaffilmark{2}INAF-Osservatorio Astronomico di Brera, via Brera 28, 20121 Milano, Italy\\
\altaffilmark{3}Department of Physics and Astronomy, University of Bologna, viale Berti Pichat 6/2, 40127 Bologna, Italy\\
\altaffilmark{4}Department of Physics \& Astronomy, San Jos\'{e} State University, San Jos\'{e}, CA 95192, USA\\
\altaffilmark{5}University of California Observatories, 1156 High Street, Santa Cruz,CA 95064, USA
}

\begin{abstract}
We present results of a joint \textit{Chandra}/\textit{XMM-Newton} analysis of the early-type galaxies NGC 4649 and NGC 5846 aimed at investigating differences between mass profiles derived from X-ray data and those from optical data, to probe the state of the hot ISM in these galaxies. If the hot ISM is at a given radius in hydrostatic equilibrium (HE) the X-ray data can be used to measure the total enclosed mass of the galaxy. Differences from optically-derived mass distributions therefore yield information about departures from HE in the hot halos. The X-ray mass profiles in different angular sectors of NGC 4649 are generally smooth with no significant azimuthal asymmetries within \(12\) kpc. Extrapolation of these profiles beyond this scale yields results consistent with the optical estimate. However, in the central region (\(r < 3\) kpc) the X-ray data underpredict the enclosed mass, when compared with the optical mass profiles. Consistent with previous results we estimate a non-thermal pressure component accounting for \(30\%\) of the gas pressure, likely linked to nuclear activity. In NGC 5846 the X-ray mass profiles show significant azimuthal asymmetries, especially in the NE direction. Comparison with optical mass profiles in this direction suggests significant departures from HE, consistent with bulk gas compression and decompression due to sloshing on \(\sim 15\) kpc scales; this effect disappears in the NW direction where the emission is smooth and extended. In this sector we find consistent X-ray and optical mass profiles, suggesting that the hot halo is not responding to strong non-gravitational forces.
\end{abstract}


\section{Introduction}\label{sec:intro}

About 85\% of the mass of the Universe takes the form of invisible Dark Matter (DM). {DM} plays a key role in {the} formation and evolution of galaxies. {Numerical simulations based on} the cold dark matter (\(\Lambda\)CDM) model \citep[e.g.,][]{1997ApJ...490..493N} successfully reproduce observed large-scale structures of galaxies through the hierarchical mergers of DM halos with baryonic matter (galaxies) trapped in their gravitational potential. DM also influences the star formation efficiency \citep[e.g.,][]{2010ApJ...721L.163A}, the major star formation epoch \citep[e.g.,][]{2005ApJ...621..673T}, the growth of {super-massive black holes} \citep[e.g.,][]{2010MNRAS.405L...1B}, and the fate of hot gas \citep{1991ApJ...376..380C,2013ApJ...776..116K}.

A key test of the \(\Lambda\)CDM predictions is the radial profile of DM halos in galaxies. X-ray observations of the hot ISM can be used to measure the total mass within a given radius if the hot gas is hydrostatic equilibrium (HE), but departures from HE can seriously affect mass measurements (e.g., \citealt{1984ApJ...286..186F, 2012ASSL..378..235B}; see also \citealt{1985ApJ...296..447T, 1987ApJ...315...46F, 1987ApJ...312..503C, 1990ApJ...359...29D, 1991ApJ...369..121D, 1992ApJS...80..645K, 1992ApJ...393..134K, 2012ASSL..378..235B}). Recently, comparisons with mass profiles derived from optical indicators (GCs/PNe) have uncovered significant discrepancies (e.g., see Figure 8 in \citealt{2015MNRAS.450.1962P} for NGC 1407; Figure 7 in \citealt{2014MNRAS.439..659N} for NGC 5846). The high-resolution observations by \textit{Chandra} revealed  an increasing number of structural features in the hot ISM , e.g., jets, cavities, cold fronts, putting the validity of the HE assumption in question \citep{2012ASSL..378..207S, 2012ASSL..378..235B}. {In addition, gas} bulk motions {on large scales} could cause a significant deviation from the HE, resulting in an incorrect mass profile if azimuthally averaged quantities are applied \citep{2007PhR...443....1M}. On small scales, comparisons between X-ray and optical measurements of total mass profiles have shown strong departures from HE, with additional non-thermal pressure components \citep[e.g.,][]{2008MNRAS.388.1062C, 2009ApJ...706..980J, 2013MNRAS.430.1516H, 2014ApJ...787..134P} that can be as large as \(\sim 40\%\) of the total pressure \citep{2004MNRAS.350..609C, 2006MNRAS.370.1797P, 2012A&A...539A..34D}.

The present investigation places itself within this line of research. We study differences between mass profiles derived from X-ray data in the HE assumption and those from more robust optical indicators, to investigate the physical state of the hot ISM and gather hints on galaxy interactions and nuclear feedback. {In the following analysis we assume the mass profile from the optical data as the baseline against which we compare the X-ray results.} In this first paper, we focus on two early-type galaxies (ETGs), NGC 4649 and NGC 5846. This paper presents in details the method that we have developed for this analysis and that will be applied to a larger sample of galaxies in the near future. NGC 4649 represents a good example of a relatively relaxed system with symmetric and smooth X-ray mass profiles in fair agreement with optical measurements {in the \(3-12\) kpc range}, with only some disturbances in the {inner regions, and evidence for ram pressure stripping in the outer part of the galaxy}. NGC 5846, {on the other hand, is a system} where bulk motion of gas introduces some measurable deviations in its mass profile from optical data both on small and large scales. For this analysis we use the entire available set of \textit{Chandra} and \textit{XMM-Newton} data, thus combining large radial extent with high angular resolution. When the hot gas distribution is not azimuthally symmetric, we take into account this asymmetry by analyzing the X-ray data in various pie sectors, instead of relying on azimuthally averaged quantities, to derive the most accurate estimate of the density and temperature profiles of the hot gas. We then compare with recent high quality total mass profiles from optical data, available thanks to integral field spectroscopy and the use of several mass tracers - stellar dynamics (SD), globular clusters (GCs) and planetary nebulae (PNe) - extending to many effective radii (e.g., \citealt{2012ApJ...748....2D, 2013MNRAS.436.1322C, 2015ApJ...804L..21C, 2016MNRAS.460.3838A, 2016MNRAS.457..147F}). We explore the {systematic uncertainties} on the X-ray mass profiles {from} different analysis {techniques, including} background subtraction, gas profile fit smoothing, and {free} element {abundances}.

The paper is organized as follows: in Sect. \ref{sec:sample} we briefly present the main properties of the objects selected for our analysis, in Sect. \ref{sec:analysis} we present the procedure we adopted for the reduction and analysis for \textit{Chandra} and \textit{XMM-Newton} data, in Sect. \ref{sec:results} we present and discuss the results of our analysis. Then in Sect. \ref{sec:summary} we draw our conclusions\footnote{In the following all angles are measured from the N direction {going counterclockwise, that is, 0 is N, 90 is E, 180 is S and 270 is W}.}. For brevity, in what follows we refer to X-ray mass profiles derived under the HE assumption as HE mass profiles.

\section{Sample Selection}\label{sec:sample}

We selected two well studied ETGs - NGC 4649 and NGC 5846 - in order to test our analysis method and compare the results in {the presence of} different ISM situations. The main properties of these two galaxies are listed in Table \ref{tab:sources}.

\begin{figure}
\centering
\includegraphics[scale=0.85]{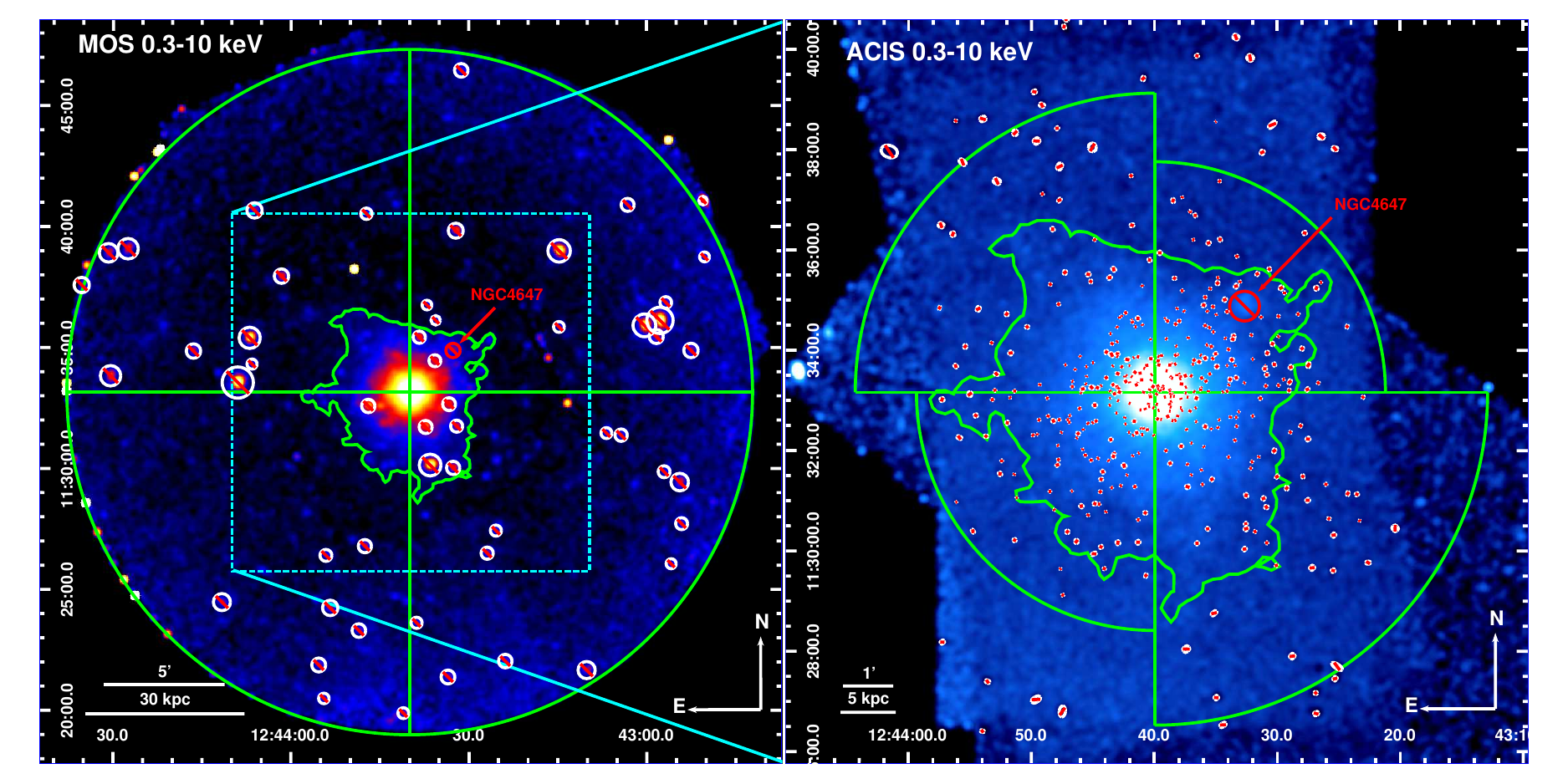}
\caption{(Left) \textit{XMM}-MOS merged, exposure corrected 0.3-10 keV image of NGC 4649. MOS1 and MOS2 data from OBSIDs 0021540201 (\(\sim 45\mbox{ ks}\)) and 0502160101 (\(\sim 70 \mbox{ ks}\)) are reduced excluding intervals of high background flares and scaling the blank-sky files to match the high-energy count rate of the event files \citep{2005ApJ...629..172N}. Point sources (marked with white circles) are detected on the merged event files with the sliding box method (\textsc{eboxdetect}), visually checked for spurious detections {and indicated with white circles}. The spectral extraction sectors used throughout the paper are shown in green, and the cyan dashed box represents the region expanded in the right panel. {The green contours highlight the ``stingray'' shaped large scale emission from NGC 4649, with the faint wing-like structures reported by \citet{2014AAS...22335808W} in the NE and SW directions. The location of the companion spiral galaxy NGC 4647 is indicated with a red circle.} (Right) {Same as left panel, but for the} \textit{Chandra} ACIS merged, exposure corrected 0.3-10 keV image. C6 and C7 data from OBSIDs 00785, 08182, 08507, 12975, 12976 and 14328 are reduced using the CGA pipeline, and the detected point sources are indicated with white ellipses.}\label{fig:N4649_mos}
\end{figure}

\begin{figure}
\centering
\includegraphics[scale=0.55]{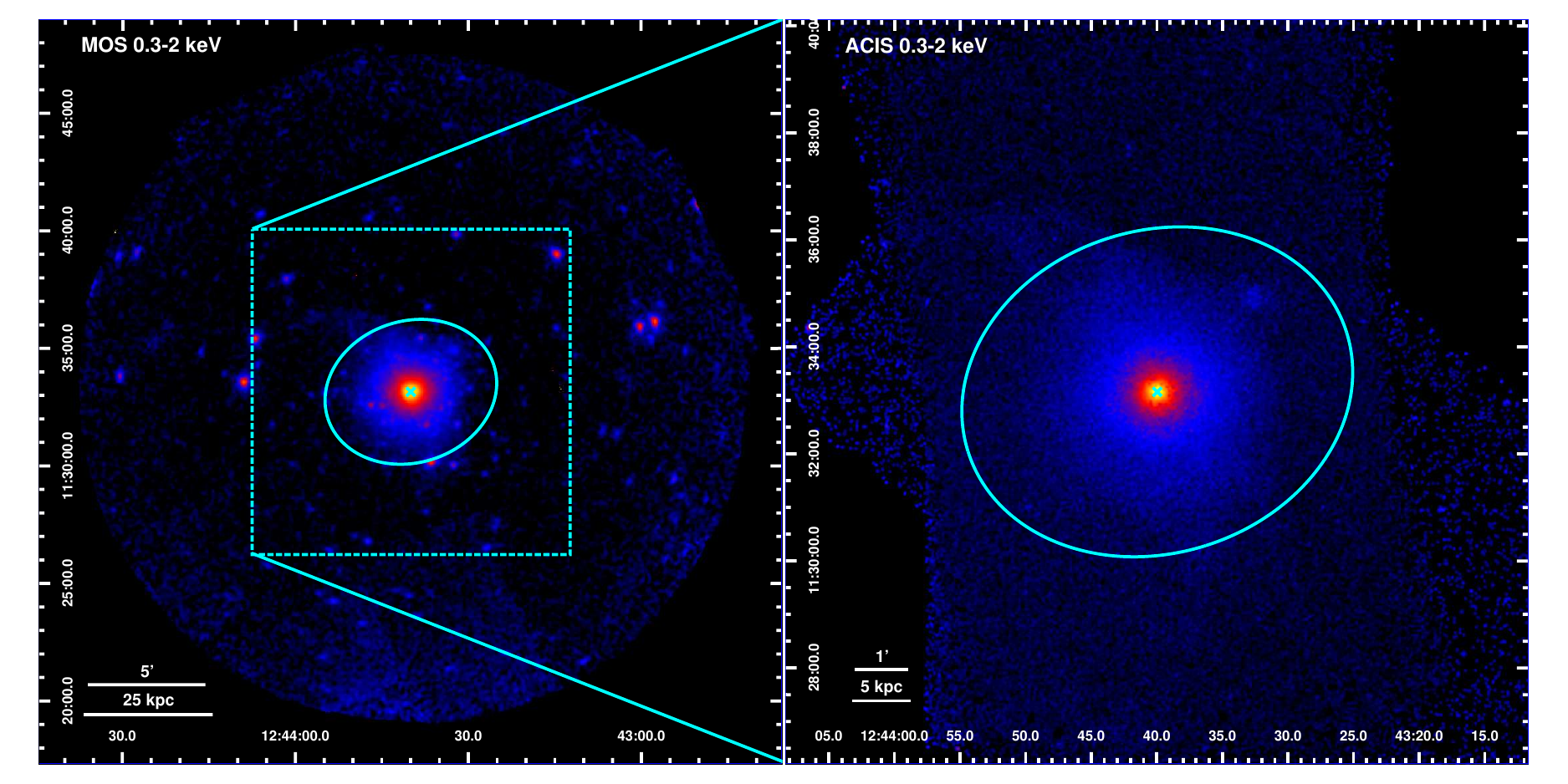}
\includegraphics[scale=0.55]{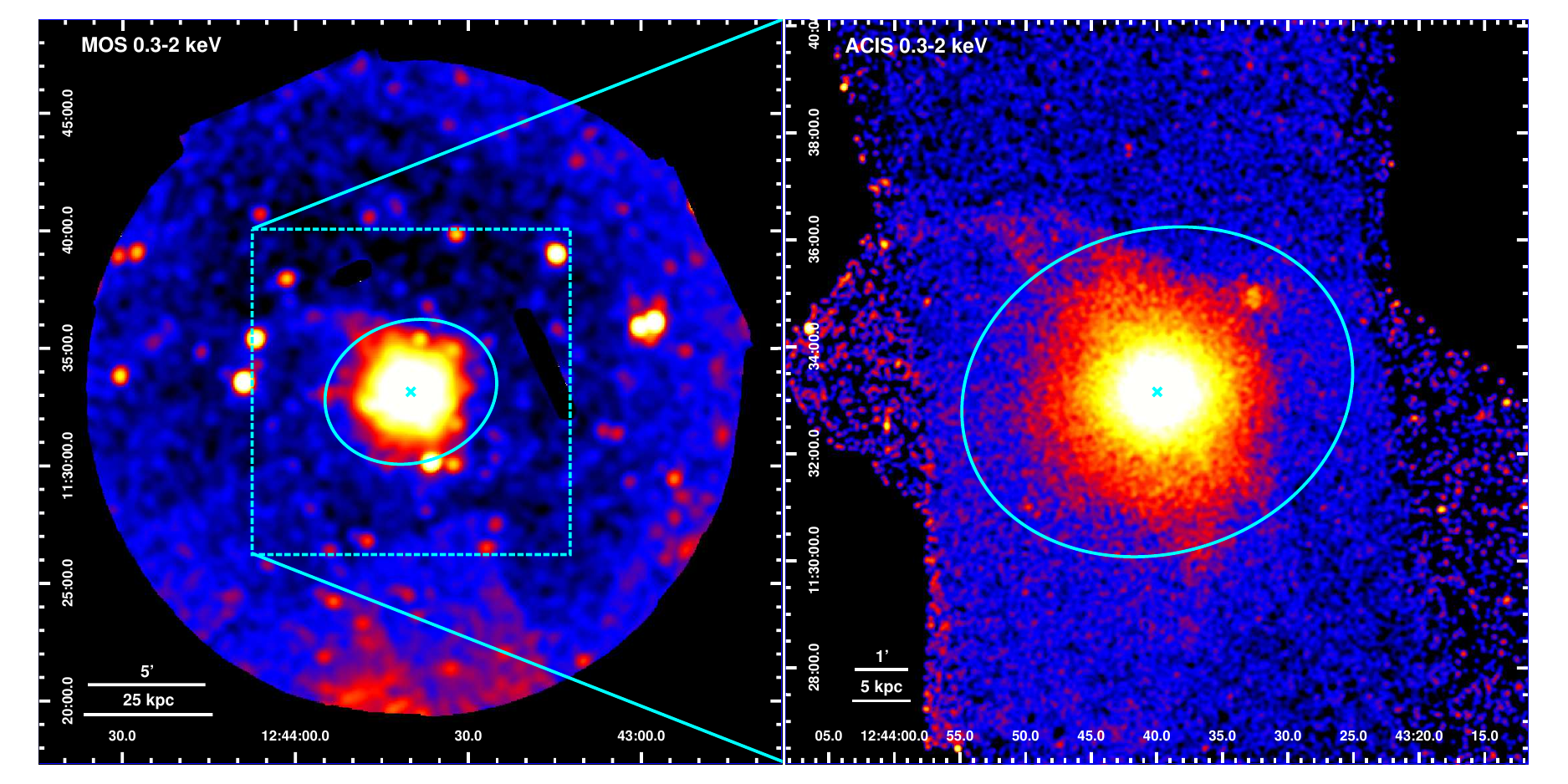}
\caption{Same as Fig. \ref{fig:N4649_mos} in the 0.3-2 keV band with the D25 ellipse shown in cyan. Small scale and large scale structures are shown in upper and lower panel, respectively.}\label{fig:N4649_mos2}
\end{figure}

\begin{figure}
\centering
\includegraphics[scale=0.33]{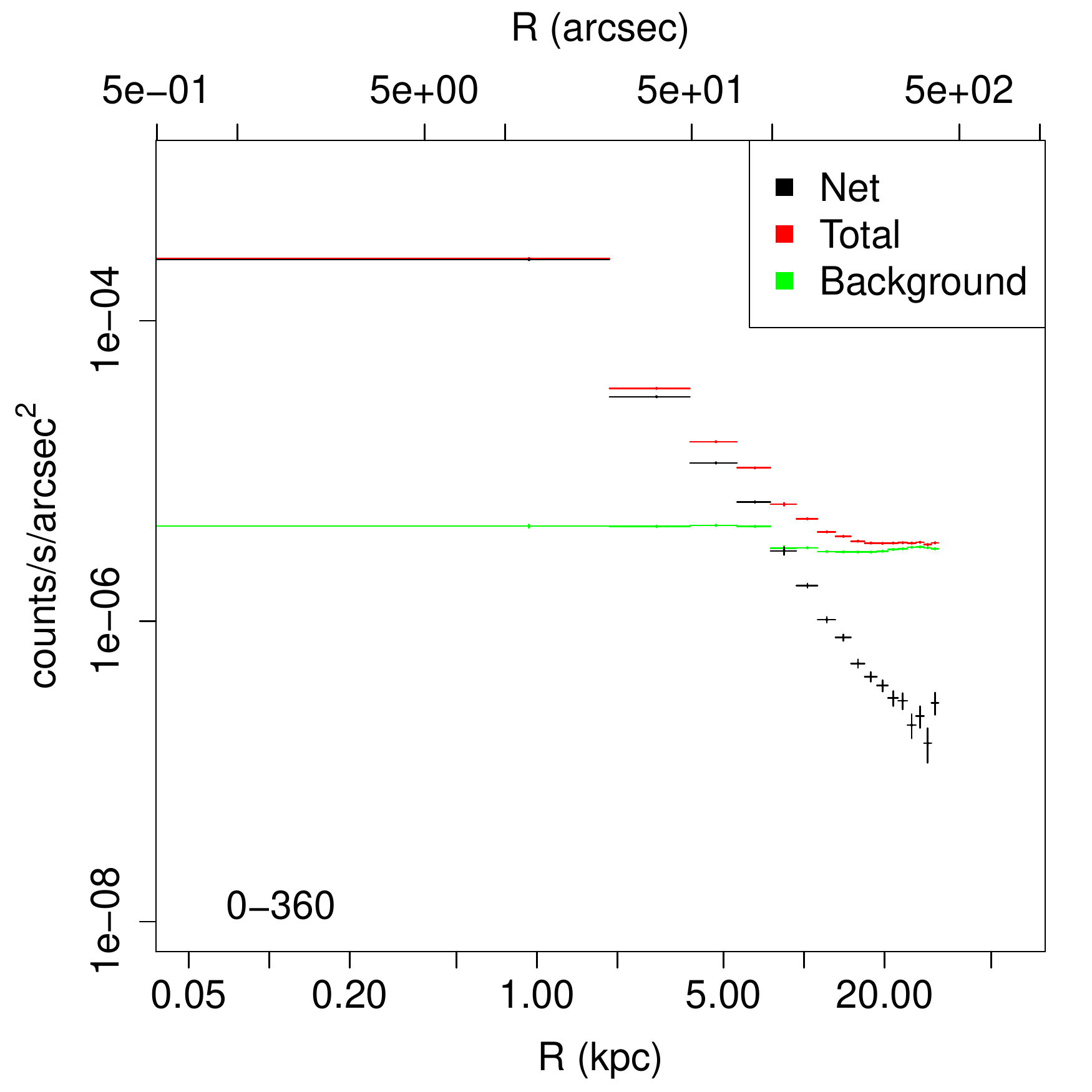}
\includegraphics[scale=0.33]{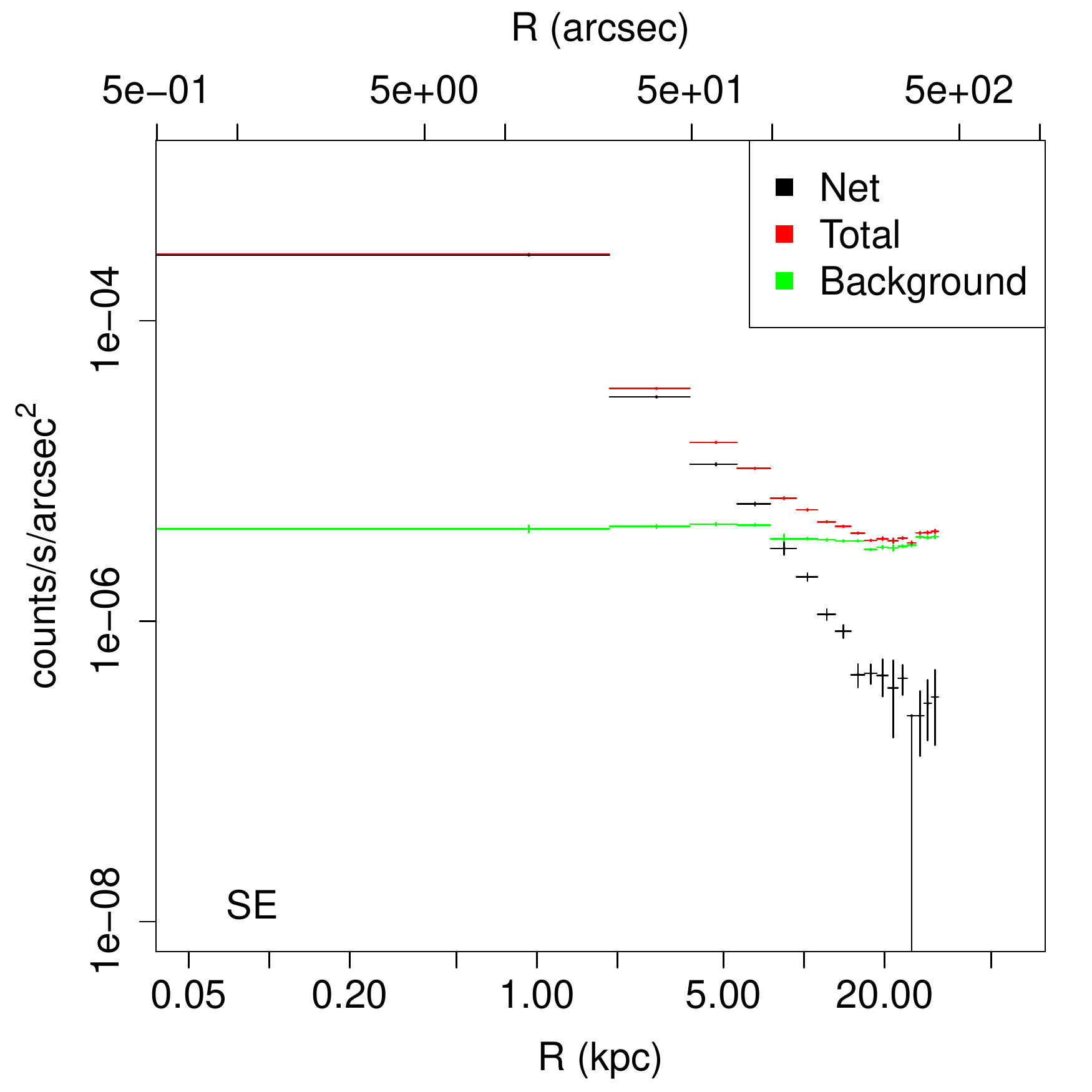}
\includegraphics[scale=0.33]{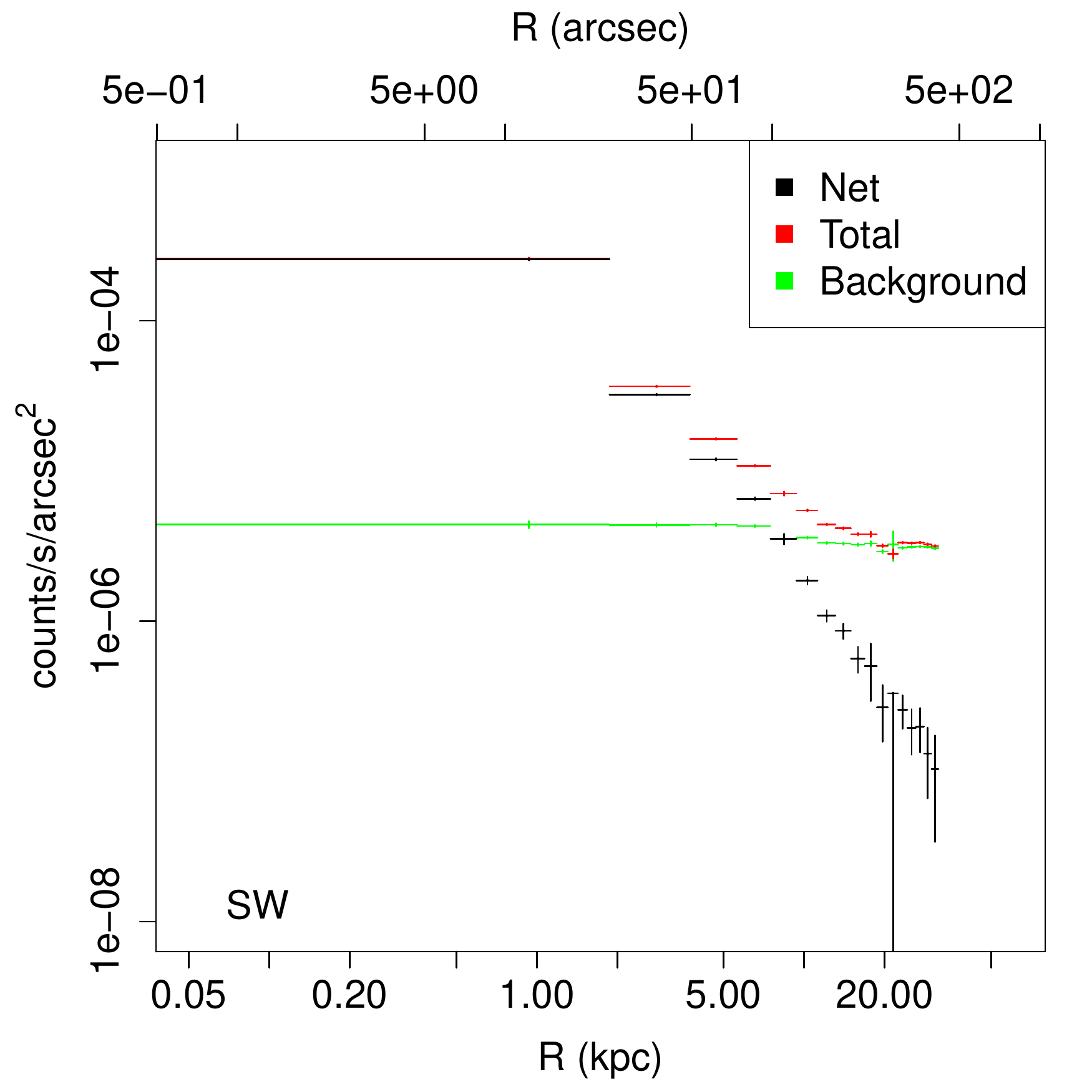}\\
\includegraphics[scale=0.33]{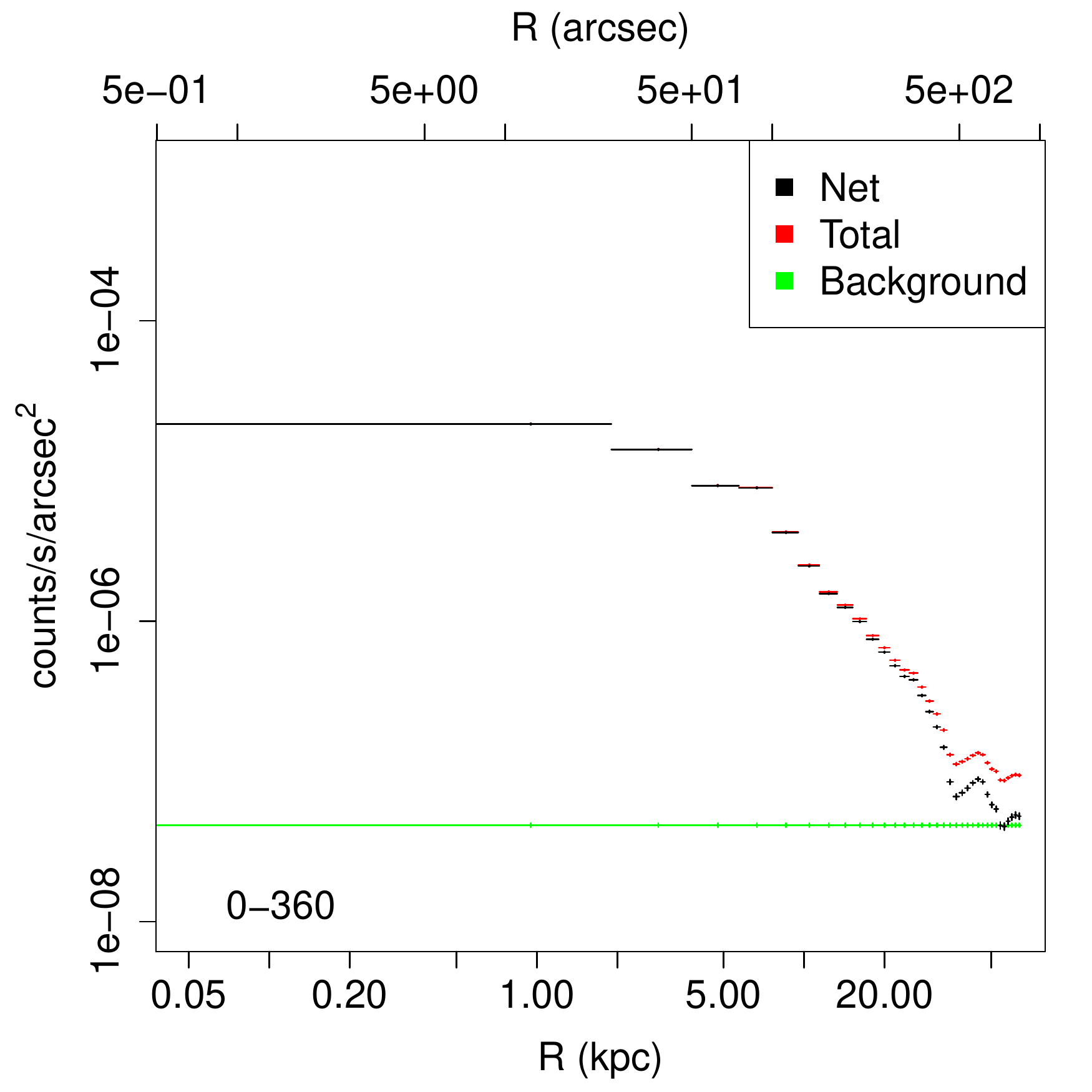}
\includegraphics[scale=0.33]{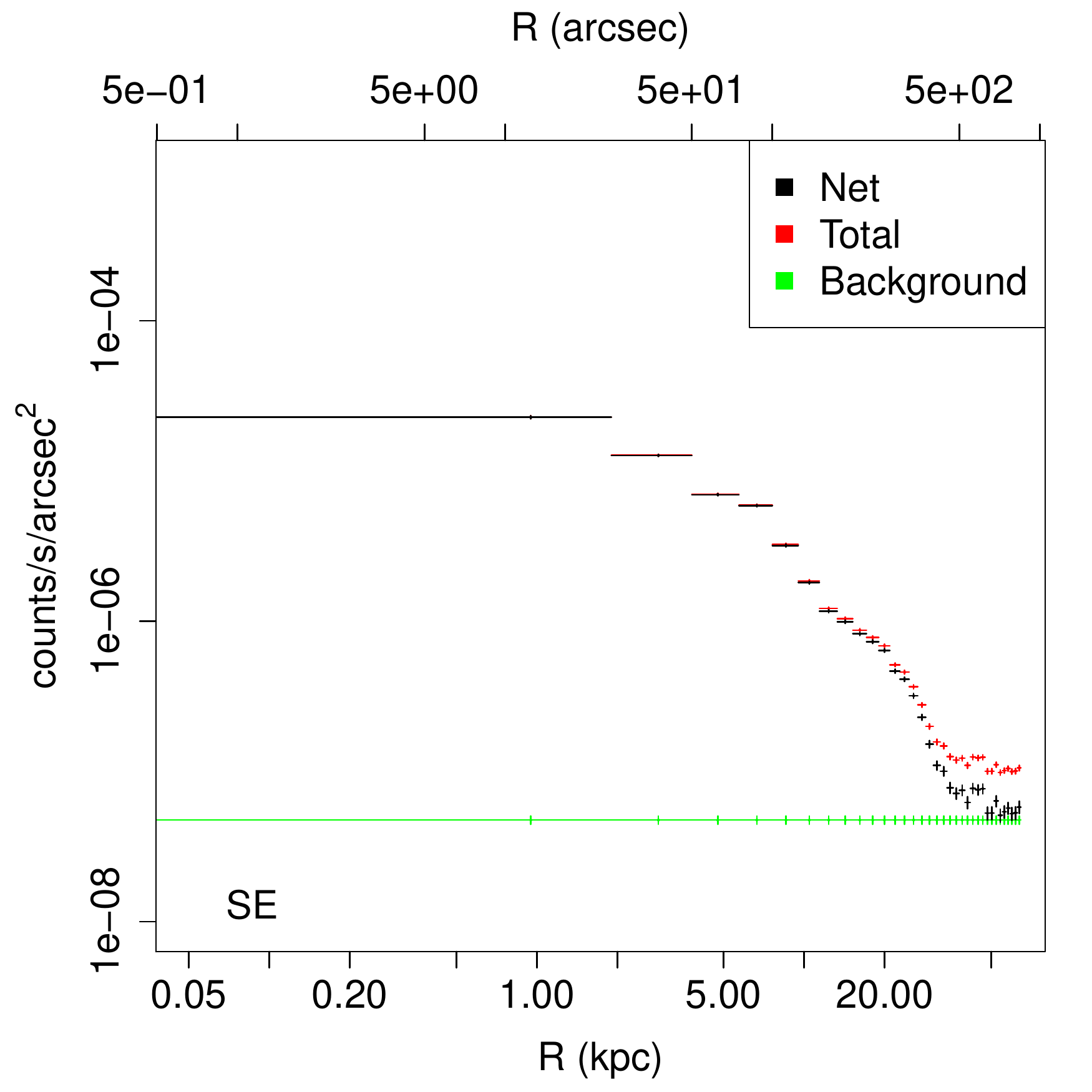}
\includegraphics[scale=0.33]{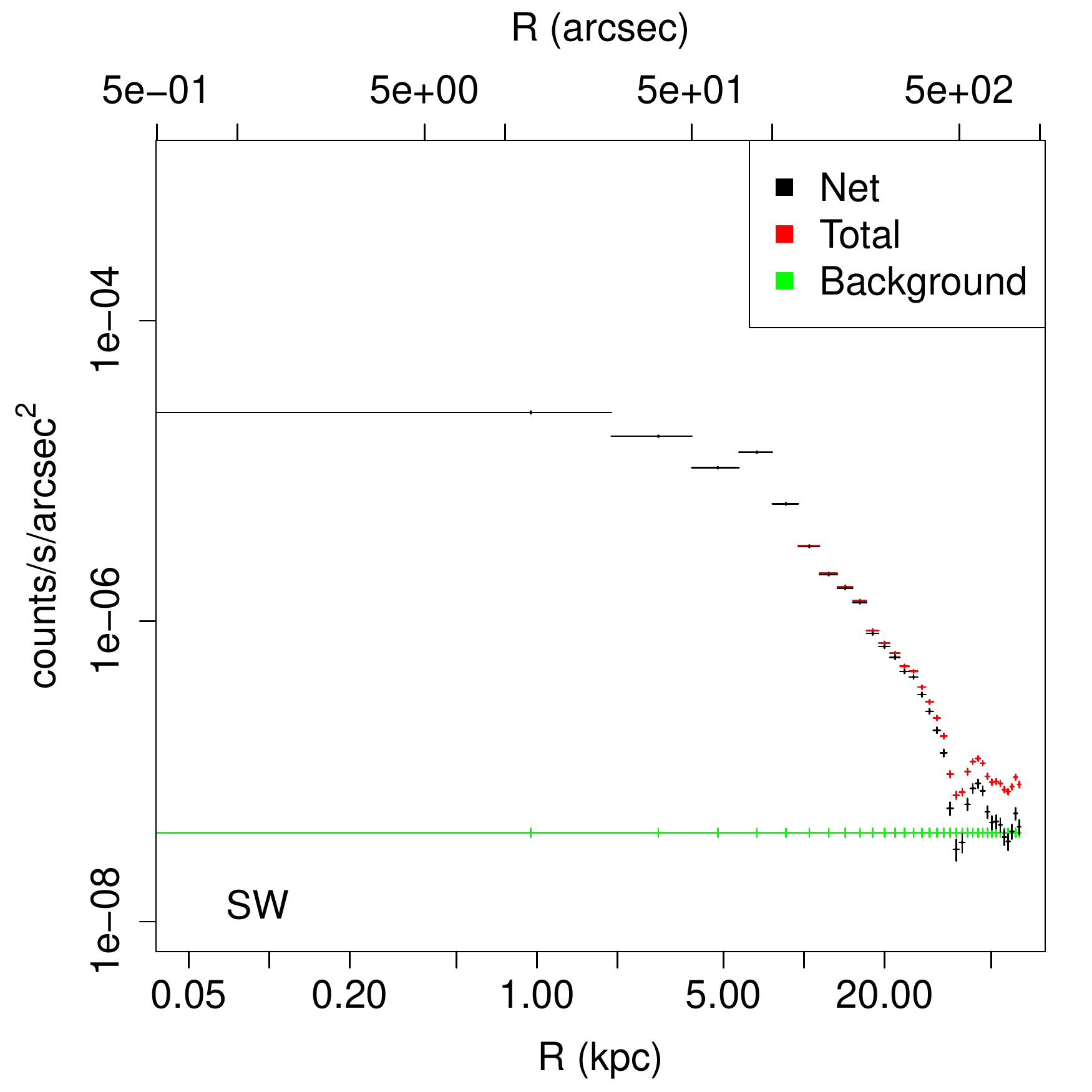}
\caption{Brightness profiles for NGC 4649 in the \(0.3-10\) keV band in the different sectors shown in Fig. \ref{fig:N4649_mos}, from left to right full (0-360), SE (90-180) and SW (180-270), respectively. In particular we show brightness profiles for \textit{Chandra} ACIS data {in the top row} and for the \textit{XMM}-MOS data obtained from the reduction procedure proposed by \citet{2005ApJ...629..172N} in the bottom row. The annuli width is \(\sim 25''\), 50 pixels for \textit{Chandra} ACIS data and 500 pixels for \textit{XMM}-MOS data. Red, black and green points represent total, net, and background brightness profiles, respectively.}\label{fig:N4649_bp_chandra}
\end{figure}

\begin{enumerate}
\item NGC 4649 {(M60)}, is a nearby {(\(\sim 16\) Mpc)} X-ray-bright giant elliptical galaxy located in a group at the eastern edge of the Virgo cluster {which} harbors a faint nuclear radio source \citep{2002AJ....124..675C}. {On} \(< 3\) kpc scales {(\(\sim 40''\))}, small disturbances in the X-ray emitting gas \citep{2008MNRAS.383..923S, 2010MNRAS.404..180D, 2013MNRAS.430.1516H} are related to the {nuclear} radio emission. These disturbances are also related to discrepancies between the mass profile presented by \citet{2010ApJ...711..484S} - evaluated from optical (SD and GCs) data - and the mass profile obtained from X-ray data. \citet{2014ApJ...787..134P} {making} use of the deepest \textit{Chandra} data available showed that the radio source contributes a non-thermal pressure component accounting {for} \(\sim 30\%\) of the observed gas pressure.

{A recent analysis of \textit{Chandra} data by \citet{2017arXiv170305883W} showed evidence for ram pressure stripping of the gas in NGC 4649 on scales \(\gtrsim 12\) kpc (\(\gtrsim 160''\)), as a consequence of its motion through the Virgo ICM. In particular, these authors show an edge in the surface brightness profile in the NW direction, consistent with a cold front, and two bent, wing-like structures in the NE and SW directions interpreted as Kelvin-Helmholtz instabilities caused by the motion of the galaxy through the Virgo ICM.} {In addition,} significant anisotropies in the projected two-dimensional spatial distributions of GCs and low-mass X-ray binaries in the outskirts of NGC 4649 have been reported by \citet{2014ApJ...783...18D}, \citet{2014ApJ...780..132M} and \citet{2015MNRAS.450.1962P}, suggesting that this galaxy has experienced mergers and/or multiple accretions of less massive satellite galaxies during its evolution. Recently, lower limits on the turbulence velocities in NGC 4649 ISM have been set by \citet{2017arXiv170204364O} making use of deep \textit{XMM-Newton} RGS observations.

Figure \ref{fig:N4649_mos} shows merged, exposure corrected, broad {band} (0.3-10 keV) images of NGC 4649 data from \textit{XMM-Newton} MOS (left panel) and \textit{Chandra} ACIS (right panel). The left panel shows the {XMM-Newton} merged data with detected point sources indicated with white circles. The right panel of the same figure {shows} the central region of the source as imaged by \textit{Chandra}. {In both panels, the regions used for spectral extraction are indicated in green.} Figure \ref{fig:N4649_mos2} shows the same data {in the 0.3-2 keV band} with different color scales to highlight small (upper panel) and large (lower panel) scale structures. 

{The surface brightness profiles within \(12\) kpc, that is, away from the disturbances reported by \citet{2017arXiv170305883W}, are smooth in all sectors, indicating that on such scales NGC 4649 features a {fairly} relaxed ISM distribution, with a smooth and symmetric morphology, and only some disturbances in the inner regions of the galaxy. As an example, Fig. \ref{fig:N4649_bp_chandra} shows a comparison between the surface brightness profiles in the full (0-360), SW (90-180) and SE (180-270) sectors. In the following analysis we will therefore restrict our analysis to the inner \(< 12\) kpc region in order to avoid the regions disturbed by NGC 4649 motion in the Virgo ICM. The complete set of figures is presented in Appendix \ref{app:profiles}.}

\item NGC 5846 is the central and brightest galaxy in a group \citep{2005AJ....130.1502M} located at a distance of \(\sim 23\) Mpc \citep{2001ApJ...546..681T}. It is nearly spherical (E0-1) in shape and is kinematically classified as a slow rotator \citep{2011MNRAS.414..888E}. NGC 5846 exhibits pronounced disturbances in its ISM morphology, hence posing a serious question as to whether it can be in HE. \textit{Chandra} data show a disturbed hot gas morphology on arcsecond scales, with similarity between X-ray emission and the H\(\alpha\)+[N \textsc{ii}] features extending into the inner regions of the galaxy \citep{2002A&A...386..472T}. \citet{2006ApJ...646..143F} in their 2D spectral analysis of \textit{XMM-Newton} data found a number of ringlike enhancements in the iron abundance map, pressure disturbances in the central region, and large-scale entropy elongations associated with the regions of enhanced metallicity. An analysis of deeper \textit{Chandra} data performed by \citet{2011ApJ...743...15M} {suggests} gas sloshing on \(\gtrsim 20\) kpc {scales} (\(\sim 180''\)), showing spiral-like tails and multiple cold fronts, possibly caused by interaction with the companion spiral galaxy NGC 5850. 

{``Sloshing" may arise when a gas distribution initially close to HE is slightly offset by the interaction with a similar, separate system, {(e.g., the gravitational influence of a large galaxy falling through a galaxy group)}, causing oscillations of the ISM and the formation of arc-like cold fronts \citep[e.g.][]{2006ApJ...650..102A, 2011MNRAS.413.2057R, 2010ApJ...717..908Z, 2013ApJ...764...83L}.} {X-ray ``bubbles" observed on sub-kiloparsec scale and} coincident with radio emission {suggest radio-mode feedback from the nuclear supermassive black hole. \citeauthor{2011ApJ...743...15M} also report} ram pressure stripping of the companion elliptical galaxy NGC 5846A during its infall toward NGC 5846. X-ray mass {profiles} for this source have been produced by \citet{2008AN....329..940D} {combining} the \textit{XMM-Newton} gas profiles presented by \citet{2006ApJ...646..143F} and long-slit SD data of \citet{2000A&AS..144...53K} together with PNe velocity dispersions presented by \citet{2008AN....329..912C}, and by \citet{2010MNRAS.409.1362D} making use of the de-projected gas profiles obtained by \citet{2010MNRAS.404.1165C} from \textit{Chandra} and \textit{XMM-Newton} data. The comparison between the azimuthally averaged \citet{2010MNRAS.409.1362D} X-ray mass profile and the {SLUGGS Survey} SD-GC mass profile \citep{2014MNRAS.439..659N} {shows that} the X-ray mass profile {falls} below and above the optical profiles at distances smaller and larger than \(\sim 10\) kpc {(\(\sim 90''\))}, respectively, {suggesting} that the hot ISM is {far from being in} HE. We note that \citet{2017arXiv170204364O} set lower limits on the turbulence velocities in ISM of NGC 5846 too.

In Figure \ref{fig:N5846_mos} we show the merged, exposure corrected, {broad band} (0.3-10 keV) images of NGC {5846 from} \textit{XMM-Newton} MOS and \textit{Chandra} ACIS in the left and right panel, respectively. The {sectors} used for the spectral extraction are indicated in green. Figure \ref{fig:N5846_mos2} shows the same data {in the 0.3-2 keV band} with different color scales to highlight small (upper panel) and large (lower panel) scale structures. The X-ray emission of NGC 5846 is clearly disturbed, showing the bubbles reported by \citet{2011ApJ...743...15M} (see Figure \ref{fig:N5846_mos2}, top-right panel) related with the central AGN activity. In addition, edges in the NE and SW directions likely due to interaction with the group companion spiral galaxy NGC 5850 are evident, while in the NW direction the emission appears extended (see Figure \ref{fig:N5846_mos2}, bottom panels). The surface brightness profiles presented in Figure \ref{fig:N5846_bp_chandra} show in the NE (30-90) direction {a clear perturbation corresponding to the} cold front features reported by \citet{2011ApJ...743...15M}, while this effect is less evident in the other sectors. In Figure \ref{fig:N5846_bp_chandra}, we only show the NW (250-30) sector, where the gas is most relaxed. {The} complete {set of} figures {is} presented in Appendix \ref{app:profiles}.
\end{enumerate}

\section{Data Analysis}\label{sec:analysis}

{For our analysis we use both} \textit{Chandra} ACIS and \textit{XMM-Newton} MOS data. The spatial resolution of ACIS {reveals} fine ISM features and disturbances in the inner regions of the galaxies. The large field of view of EPIC-MOS allows us to extend the analysis to larger radii {to constrain} the ISM features in lower surface brightness regions.

\subsection{\textit{Chandra} Data}\label{sec:data_chandra}

We make use of the \textit{Chandra} ACIS data products of the \textit{Chandra} Galaxy Atlas (CGA) project (Kim et al. in prep.). The CGA project aims at analyzing in an uniform way \textit{Chandra} ACIS archival data from a sample of 100 ETGs to systematically study the 2D spectral properties of the diffuse emission. While we generally follow the \textsc{CIAO} science threads, we have developed our own analysis pipelines. {Below} we briefly describe {the} key steps in data reduction.

Once ACIS data are retrieved from the \textit{Chandra} Data Archive\footnote{\href{http://cda.harvard.edu/chaser}{http://cda.harvard.edu/chaser}}, we run the ACIS level 2 processing {with} \textsc{chandra\_repro} to apply up-to-date {calibrations} (CTI correction, ACIS gain, bad pixels). After excluding time intervals of background flares exceeding 3\(\sigma\) with the \textsc{deflare} task, we obtain {the} low-background total exposures listed in Table \ref{tab:chandra_data}. For each observation, we generate a full resolution image in multiple energy bands. We then use \textsc{wavdetect} to {detect} point sources in each observation. Individual observations are corrected for a small positional inaccuracy by registering common sources ({with} \textsc{reproject\_aspect}), re-projected onto a single tangent plane ({with} \textsc{reproject\_obs}) and merged into a single event file ({with} \textsc{flux\_obs}). We then run \textsc{wavdetect} for {a second} time to detect point sources from the merged images in \(0.5-5\) keV and remove them before extracting spectra\footnote{In addition, data from the companion galaxy NGC 4647 ({located} at \(\sim 2.5'\) form NGC 4649 in the NW direction) were excluded for the present analysis.}.

{To estimate the} background emission, we download the blank sky data from the \textit{Chandra} archive, re-project them to the same tangent plane as each observation (see section 3.1) and rescale them to match the rate at higher energies (\(9-12\) keV) where the photons are primarily from particle background \citep{2006ApJ...645...95H}.

\subsection{\textit{XMM-Newton} Data}\label{sec:data_xmm}

EPIC data were retrieved from the \textit{XMM-Newton} Science Archive\footnote{\href{http://nxsa.esac.esa.int/nxsa-web}{http://nxsa.esac.esa.int/nxsa-web}} and reduced with the \textsc{SAS}\footnote{\href{http://www.cosmos.esa.int/web/xmm-newton/sas}{http://www.cosmos.esa.int/web/xmm-newton/sas}} 14.0.0 software. To avoid cross calibration problems and perform our analysis on as homogeneous data as possible, we restricted our analysis to MOS {data}. The main properties of the \textit{XMM-Newton} data used in this work are presented in Table \ref{tab:XMM_data}.

\subsubsection{Background Subtraction}

Since one of the biggest source of uncertainties {in} \textit{XMM-Newton} {data analysis} is the background evaluation, we {confronted} three different ways of reducing MOS data. Here we briefly sum up the main properties of these method, while a complete discussion of them can be found in Appendix \ref{app:background} and \ref{app:effects_background}.

\begin{description}
\item[Simple Background Subtraction] Following the prescriptions of \citet{2005ApJ...629..172N}, the background is evaluated from appropriate blank-sky files, normalized to match the \(9.5-12\) keV count rate of the event files. However, since the particle background and the sky background are independent, we apply this normalization only in the \(2-7\) keV band. The main advantage of this background subtraction procedure is represented by its simplicity both in the data reduction and in the spectral fitting phases. In addition, it allows for spectral extraction over the whole MOS field of view. However, the use of blank-sky files represents an approximation over a complete background modeling, and may not be accurate enough at surface brightness close to the background level.
\item[Double Background Subtraction] The second reduction method is that proposed by \citet{2001A&A...365L..80A, 2002A&A...390...27A}. To disentangle the cosmic X-ray background from the instrumental background we {made} use of vignetting corrected event files, since the cosmic X-ray background can be considered uniform on scales \(\sim 30'\) but is affected by vignetting, while the instrumental background is non uniform but is not vignetted. We first {subtracted from} the source spectrum the spectrum extracted in the same region from the blank-sky file. Then we {performed} the same procedure, but from a region of the FOV that we \textit{assumed} to be source free. Finally we {subtracted} these two net spectra taking into account the ratio of the extraction region areas. The {resulting} spectrum is expected to contain only source emission. This procedure {yields} a precise evaluation of the \textit{XMM-Newton} background, being able to {at least partially disentangle} the astrophysical and the instrumental background. It, however, requires the selection of a source-free region.
\item[Background Modeling] The third reduction method we adopted is the one proposed in the ``Cookbook for Analysis Procedures for \textit{XMM-Newton} EPIC Observations of Extended Object and the Diffuse Background\footnote{\href{https://heasarc.gsfc.nasa.gov/docs/xmm/esas/cookbook/xmm-esas.html}{https://heasarc.gsfc.nasa.gov/docs/xmm/esas/cookbook/xmm-esas.html}}'' \citep{2011AAS...21734417S}. The background for this reduction method is partly subtracted and partly modeled. In addition, we modeled the instrumental and the cosmic background. This procedure allows us to analyze the whole MOS FOV, and to subtract the QPB and to model the instrumental and the cosmic background. This {however requires a} more complex spectral modeling, with several free parameters that make it difficult to find stable fitting convergence in un-{supervised} analysis procedures (especially in conjunction with 3D-deprojection, see Sect. \ref{sec:spectra}).
\end{description}

\subsubsection{Common Reduction Steps}\label{sec:common_steps}

Regardless of the data reduction procedure adopted, the final steps we performed on the data sets were the same. In particular, we merged data from MOS1 and MOS2 detectors from all observations using the \textsc{merge} task, in order to detect the fainter sources that wouldn't be detected otherwise. Sources were detected on these merged images following the standard SAS sliding box task \textsc{edetect\_chain} that mainly consist of three steps: 1) source detection with local background, with a minimum detection likelihood of 8; 2) remove sources in step 1 and create a smooth background maps by fitting a 2-D spline to the residual image; 3) source detection with the background map produced in step 2 with a minimum detection likelihood of 10.

{The} task \textsc{emldetect} {was then} used to determine the parameters for each input source by means of a maximum likelihood fit to the input images, selecting sources with a minimum detection likelihood of 15 and a flux in the \(0.3-10\) keV band larger than \({10}^{-14}\mbox { erg}\mbox{ cm}^{-2}\mbox{ s}^{-1}\) (assuming an energy conversion factor of \(1.2\times {10}^{-11}\mbox{ cts}\mbox{ cm}^{2}\mbox{ erg}^{-1}\)). An analytical model of the point spread function (PSF) {was} evaluated at the source position and {normalized to the source brightness. The} source extent {was then} evaluated as the radius at which the PSF level equals half of  local background. We finally visually inspected the detected sources and removed evident spurious detections (i. e., at chip borders, in regions of diffuse emission, etc.). We then produced ``swiss-cheese" images for each detector array and observation.

\subsection{Spectral Analysis}\label{sec:spectra}

We extracted data for spectral analysis from circular annuli spanning an entire 360 deg or partial annuli with specific sectors, when the gas distribution {was} not spherically symmetric. The minimum width {of} each annulus {was} chosen to {avoid} that the {significant PSF scattering} lead to strong mixing between the spectra in adjacent annuli. In particular for \textit{Chandra} ACIS we have chosen a minimum {annulus} width of 2 pixels (corresponding to about 1 arcsec), while for \textit{XMM}-MOS we have chosen a minimum {annular} width of 600 pixels (corresponding to about 30 arcsec). In this way we obtained a finer grid in the inner regions of the galaxy with \textit{Chandra} data, while we were able to extend to larger radii with \textit{XMM-Newton} data.

The inner and outer radii of each annulus {were} adaptively determined, based on a given signal-to-noise ratio. In particular we produced bins corresponding to a minimum signal to noise {ratios} of 30, 50 and 100, both for \textit{Chandra} and \textit{XMM-Newton} data {(the signal to noise ratio used for \textit{XMM-Newton} and \textit{Chandra} data is reported in the captions of Figures \ref{fig:N4649_gas_profiles_merged} and \ref{fig:N5846_gas_profiles_merged}). If the selected signal to noise ratio could not be reached, the last spectral extraction bin was extended to the edge of the detector.} For \textit{Chandra} ACIS data the background is evaluated from blank sky files (see Sect. \ref{sec:data_chandra}). For \textit{XMM}-MOS data the background has been evaluated using the prescriptions described in Sects. \ref{sec:data_chandra} and \ref{sec:data_xmm}\footnote{Note that for the reduction procedure presented in \citet{2011AAS...21734417S} the background contribution cannot be estimated before the spectral fitting. Therefore for this method we used the annuli obtained from the reduction method of proposed by \citet{2005ApJ...629..172N}.}.

The source spectra {were} extracted from each spatial bin, {in each} observation. The corresponding arf and rmf files {were} also extracted per observation to take into account time and position dependent ACIS and MOS responses. Once source spectrum, background spectrum, arf and rmf per obsid {were} generated, we {used} \textsc{combine\_spectra} \textsc{CIAO} and \textsc{epicspeccombine} \textsc{SAS} tools to {create} a single data set per {each extraction} bin. This way the spectral fitting is simpler and quicker. We compared our results with those of joint fitting of individual spectra and found no significant difference.

To make use of the \(\chi^2\) fit statistic we binned the spectra to obtain a minimum of \(20\) counts per bin using the \textsc{specextract} and \textsc{evselect} tasks; in the following, errors correspond to the \(1\)-\(\sigma\) confidence level for one interesting parameter (\(\Delta\chi^2 = 1\)). In all the spectral fits we included photo-electric absorption by the Galactic column density along the line of sight reported in Table \ref{tab:sources}. To account for projection effects, we used the \textsc{projct} model implemented in \textsc{XSPEC} \citep[ver. 12.9.0o,][]{1996ASPC..101...17A} excluding data from any detected point source.

Spectral fitting {was} performed separately for \textit{Chandra} and \textit{XMM-Newton} data in the \(0.5-8\mbox{ keV}\) energy range, adopting Gehrels weighting \citep{1986ApJ...303..336G}, and using a model comprising a \textsc{apec} thermal component (with solar abundances from \citealt{1998SSRv...85..161G} and ATOMDB code v2.0.2\footnote{\href{http://www.atomdb.org}{http://www.atomdb.org}}), plus a thermal bremsstrahlung component with temperature fixed at \(7.3\) keV to account for undetected point sources \citep{2003ApJ...587..356I}. In order to obtain stable fits we first performed 2D projected fits to get reliable starting values for the 3D de-projected profiles, and then applied the \textsc{projct} model, and after fitting these profiles we froze the bremsstrahlung component normalizations to their best-fit values to reduce the uncertainties on the more relevant parameters of our analysis.

Since the outermost (and usually {larger}) {spectral extraction} bin does not reach the minimal signal to noise requirement, its spectral parameters are usually not well constrained and show large fluctuations. Moreover, due to the ``onion peeling" procedure used in the 3D de-projected process, this outermost bin influences all the inner shells and may yield unreliable results. To avoid this problem we excluded the outermost bin from the following analysis, and only considered bins that reach the selected minimum signal to noise. Regarding the \textsc{apec} component metal abundances we performed {spectral} fits both linking the abundances between all sectors (\textit{fixed} abundances) and allowing the element abundances to freely vary in each annulus (\textit{free} abundances). In the latter case, however we usually had to use binning with higher signal to noise ratio in order to be able to constrain the element abundances. {We note that in general we conservatively choose the same signal to noise ratio both for the fixed and free abundance models to compare the resulting profiles on the same grounding, while as a general rule the fixed abundance models can be pushed to lower signal to noise ratios.} Uncertainties on best-fit parameters (\textsc{apec} component temperatures, normalizations and element abundances) were then evaluated with the \textsc{error} command, and finally we converted the \textsc{apec} normalizations \(EM\) to the gas density \(\rho\) using the relation \(\rho=\sqrt{EM\cdot{10}^{14}/\eta\cdot V\cdot 4\pi {\left[{(1+z)\cdot D_A}\right]}^2}\), where \(\eta\) is the electron to hydrogen ratio (that we assume to be 1.2 for a fully ionized gas), \(V\) is the volume of the elliptical shell, \(z\) is the source redshift and \(D_A\) is the angular diameter distance to the source. From the gas temperature (\(T\)) and density (\(\rho\)) profiles we then produced gas pressure {\(P=k\,T\,n\) and entropy \(S=kT/n^{2/3}\), where} \(k\) is the Boltzmann constant, \(m_p\) is the proton mass, \(\mu\approx 0.62\) is the average molecular weight factor, and \(n=\rho/(\mu\,m_p)\) is the gas number density.

\subsection{X-ray HE Mass Profiles and Gas Profile Fitting}\label{sec:mass_profiles}

From the gas temperatures \(T\) and density \(\rho\) profiles that we obtain in the various sectors, we can evaluate the total mass profiles under the assumption of {HE:}
\begin{equation}\label{eq:hee}
M(<R) = - R\, \frac{k T(R)}{G\mu m_p}\left({\frac{d \log{\rho}}{d \log{R}}+\frac{d \log{T}}{d \log{R}}+\frac{P_{NT}}{P}\frac{d \log{P_{NT}}}{d \log{R}}}\right)
\end{equation}
{where \(P_{NT}\) is the non-thermal pressure component. All the X-ray mass profiles presented in the following analysis are obtained assuming \(P_{NT}=0\). To} make use of the {HE equation} we need to model \(T(R)\) and \(\rho(R)\) to have a good estimate of the (logarithmic) slopes that go into Eq. \ref{eq:hee}. {While} broken or double-broken power laws {have been used} \citep[e.g.][]{2008ApJ...683..161H}, for this work we decided to fit the gas profiles with a cubic smoothing spline (\textsc{smooth.spline} R language package, see for example \citealt{2009ApJ...706..980J}), {to avoid sudden variations in the profile slopes and so in the resulting mass profiles}. This {spline function's} smoothing parameter (typically in the \(0-1\) range) can be varied, with 0 yielding the interpolating spline and 1 yielding a monotone fit to the data. Fit models for gas pressure and entropy profiles were performed by combining fit models for the gas temperature and pressure profiles as indicated in Sect. \ref{sec:spectra}.

For each sector we fitted the merged \textit{Chandra} ACIS and \textit{XMM-Newton} MOS profiles with a cubic spline with increasing smoothing parameter from 0.5 to 1.0. To evaluate the uncertainties on the best fit function we performed Monte Carlo simulations drawing from the observed temperature and density distributions assuming a gaussian distribution. {To} take into account the asymmetric error bars that we get from the spectral fittings, we considered two gaussian distributions with different widths - for the upper and lower errors - and {drew} for each of them 1000 random realizations of the profiles, {which were then} fitted with the cubic spline. {From these results} we then evaluated the distribution of the best fit functions and of their slopes. We add that, at variance with the standard practice of producing mass profile points at the central radius of each bin, we instead decided to produce the mass profile points at the outermost radii of each bin. This is due to the fact that our main goal in this work is to estimate the DM distribution, which is better constrained at the largest radii. Finally, we notice that for NGC 4649 we also tried including in the spectral fitting an additional hot gas component, with temperature fixed at 2.5 keV \citep[e.g.][]{2002ApJ...572..160G}, to account for possible interloper emission from {the} Virgo cluster, but we did not find any significant effect of this additional model component on the final mass profiles.

\subsection{Factors affecting the HE mass profiles}\label{sec:effect}

In producing the total HE mass profile, there are a number effects that must be taken into account, namely the smoothing parameter of the spline function used to fit the gas density and temperature profiles, the different angular sections when the hot gas distribution is not azimuthally symmetric, the radial gradient of metal abundances and the accuracy in background subtraction. These effects are discussed in detail in Appendix \ref{app:effects}. Here we briefly sum up the main aspects of our assumptions.

\begin{enumerate}
\item Low values of the smoothing parameter of the spline function yield best fit models catching the finer details of the gas profiles, while large values of the smoothing parameter favor monotonic profiles. The choice of the smoothing parameter value is a trade-off between the fit function ability to represent the details of the gas profiles and the need to get rid of the noise to get meaningful physical results. In general we choose a value of the smoothing parameter between 0.6 and 0.8 (see captions of Figs. \ref{fig:N4649_gas_profiles_merged} and \ref{fig:N5846_gas_profiles_merged}).

\item The total HE mass obtained from Eq. \ref{eq:hee} is the total mass enclosed within a radius \(R\) assuming spherical symmetry, that is, the total mass assuming that the gas is in HE and distributed in a spherical shell with constant temperature. Therefore, even if we extract spectra in angular sectors, it will make sense to compare the resulting mass profiles with each other and with the mass profiles from optical markers.

\item Although the abundance profiles are rather noisy they tend to decrease at larger radii, and this yields flatter gas density profiles and therefore smaller enclosed mass. Allowing variable abundances yields in addition larger uncertainties (especially on gas density) which in turn translates to larger uncertainties on the enclosed mass profile. 

\item The different background subtraction methods yield similar gas and mass profiles, with only minor effects on the outer bins at fainter surface brightnesses. For the double subtraction method, however, we have to assume a source free region that will be excluded by our spectral extraction, so the main effect of the background subtraction proposed by \citet{2001A&A...365L..80A, 2002A&A...390...27A} is to restrict our analysis to smaller radii in certain direction. In the following, therefore, we will focus on results from \citeauthor{2005ApJ...629..172N}'s procedure.
\end{enumerate}

\section{Results}\label{sec:results}

\subsection{NGC 4649}\label{sec:results_N4649}

We extracted spectra and HE mass profiles in the four quadrant directions NW (270-360), NE (0-90), SE (90-180) and SW (180-270). {As mentioned in Sect. \ref{sec:sample}, for this source we restrict our analysis to the inner \(12\) kpc region in order to avoid the disturbed regions reported by \cite{2017arXiv170305883W}}. {On} sub-kiloparsec scales {the NE and SW} quadrants contain the weak radio jet \citep{2002AJ....124..675C} emerging from the central faint AGN as well as the associated cavities in the X-ray emitting gas, while in NW and SE quadrants lie the overdense regions where the gas is displaced on the sides of the jet \citep{2014ApJ...787..134P}.

In Figure \ref{fig:N4649_gas_profiles_merged} we show the gas profiles (temperature and density) in the full 360 sector (top row), {in the SE (middle row) and SW (bottom row)} obtained with \textit{Chandra} (black) and \textit{XMM-Newton} data (green). {The} complete set of gas and mass profiles is presented in the Appendix \ref{app:profiles}. These {gas} profiles are obtained with the fixed abundance model. In the same figures, we overplot the best fit models to these profiles in red, with the width of the curve indicating the corresponding \(1\sigma\) error. In the third column of the same figure we show the total HE mass profiles (in black) obtained in the various sectors by mean of Eq. \ref{eq:hee} from the best fits to gas temperature and density profiles. In the same panels these HE mass profiles are compared with the mass profile (in yellow) obtained from SD and GC kinematic \citep{2010ApJ...711..484S}. We also compare our HE mass profiles with the HE mass profiles (in light blue) obtained by \citet{2008ApJ...683..161H} making use of \textit{Chandra} data. In Figure \ref{fig:N4649_gas_profiles_merged_abund} we show the gas profiles obtained with the free abundance model together with the total HE mass profiles obtained from the best fits to gas temperature and density profiles. In addition, in the rightmost panels of the same figure we show the element abundances profiles in comparison with the fixed abundance model.

Our HE mass profiles in different sectors (at \(R = 3-12\) kpc) {are all relatively similar}, consistent with the optical measurement, and confirm the relaxed morphology of the ISM in NGC 4649 {at these scales,} without evidence for strong azimuthal asymmetry. The NE sector has the largest masses {in the} radius range of \(0.5-3\) kpc, almost consistent with the optical measurement at the same radius while the SW sector has the lowest masses, most significantly discrepant from the optical measurement. The uncertainties in the small scale HE mass profiles, as reflected by the range of mass profiles among different sectors, could have a significant impact when X-ray data are used to estimate the mass of the nuclear supermassive black hole (see Sect. \ref{sec:results_total_mass}).

The discrepancies between the optical and the X-ray HE mass profiles indicate deviations from the hydrostatic equilibrium between 0.5 and 3 kpc, and confirm the non-thermal pressure component accounting \(\sim 30\%\) of the observed gas pressure as reported by \citet{2008ApJ...683..161H} and \citet{2014ApJ...787..134P}.

The X-ray HE mass profiles of \citet{2008ApJ...683..161H} were obtained with a \textsc{VAPEC} component to model the ISM emission with free abundances, so they are closer to our HE mass profiles obtained with the free abundances model. However \citeauthor{2008ApJ...683..161H} used a broken power-law for modeling the gas profiles, which might explain the minor differences from our results (see 3rd column in Figure \ref{fig:N4649_gas_profiles_merged}). In addition \citeauthor{2008ApJ...683..161H} made use of ATOMDB v1.3.2 which yields systematically lower temperatures with respect to v2.0.2 in the temperature range of interest here \citep[see][]{2012ApJ...757..121L}.

These differences are highlighted in the velocity profiles (as evaluated from \(\varv = \sqrt{(G M_X / R)}\)) presented in Figure \ref{fig:N4649_velocity_profiles_all}, where in the left and right panel we show HE profiles from fixed and free abundance models, respectively, and compare them with those of \citet{2008ApJ...683..161H}, the velocity profile of \citet{2010ApJ...711..484S} and the measurement of \citet{2016MNRAS.460.3838A} at \(5 R_e\) (25 kpc). 
In the same figure we show the velocity profile from \citet{2011MNRAS.415.1244D} that best fits their SD, GC and PNe data. This profile follows \citeauthor{2010ApJ...711..484S}'s up to 7.6 kpc and is flat from that point onwards (their VC4 profile); it is compatible with \citeauthor{2016MNRAS.460.3838A}'s point, as well as with our X-ray HE profiles. In this figure the uncertainties on the velocity profiles obtained in this work are not shown, and the width of these profiles is fixed for clarity of representation (proper errors are presented in Figures \ref{fig:N4649_gas_profiles_merged} and \ref{fig:N4649_gas_profiles_merged_abund}).

\subsection{NGC 5846}\label{sec:results_N5846}

\begin{figure}
\centering
\includegraphics[scale=0.85]{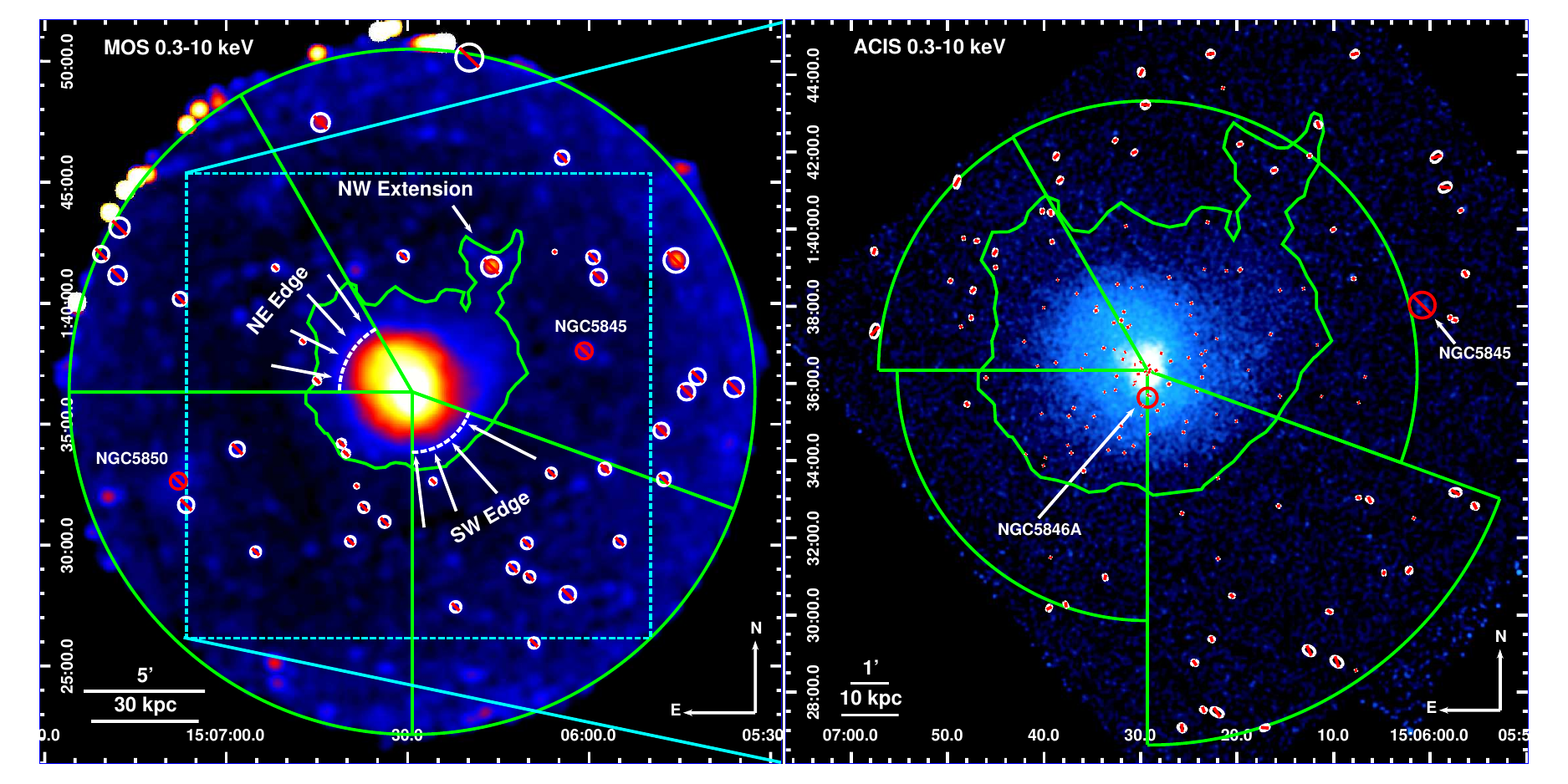}
\caption{(Left) \textit{XMM}-MOS merged, exposure corrected 0.3-10 keV image of NGC 5846. MOS1 and MOS2 data from OBSIDs 0021540501, 0723800101 and 0723800201 (\(\sim 165\mbox{ ks}\)) are reduced excluding intervals of high background flares and scaling the blank-sky files to match the high-energy count rate of the event files \citep{2005ApJ...629..172N}. Point sources (marked with white circles) are detected on the merged event files with the sliding box method (\textsc{eboxdetect}) and visually checked for spurious detections. The spectral extraction sectors used throughout the paper are shown in white. The cyan dashed box represents the region expanded in the right panel. {The green contours highlight the large scale emission from NGC 5846, with the edges associated with the cold fronts reported by \citet{2011ApJ...743...15M} in the NE and SW direction, as well as the extended emission in the NW sector. The location of the companion galaxies NGC 5845, NGC 5850 and NGC 5846A is indicated with red circles.} (Right) {Same as left panel, but for the} \textit{Chandra} ACIS merged, exposure corrected 0.3-10 keV image. C0, C1, C2, C3, C6 and C7 data from OBSIDs 00788 and 07923 (\(\sim 110\mbox{ ks}\)) are reduced using the CGA pipeline, and the detected point sources are indicated with white ellipses..}\label{fig:N5846_mos}
\end{figure}

To investigate the effects of {azimuthally asymmetric gas morphology} on HE mass profiles, we extracted spectra in four sectors in the NE (30-90), SE (90-180), SW (180-250) and NW (250-30), directions, as well as in the full (0-360) sector (see Figures \ref{fig:N5846_mos} and \ref{fig:N5846_mos2}). The NE and SW sectors are the most disturbed since they contain a surface brightness edge \citep{2011ApJ...743...15M}, while the SE sector shows a smaller amount of disturbance. \citet{2011ApJ...743...15M} suggest that these features may result from interactions with group galaxies. In the NW sector instead the gas emission appears extended, smooth and mostly relaxed.

In Figure \ref{fig:N5846_gas_profiles_merged} we show the gas profiles obtained with fixed element abundances from the \citet{2005ApJ...629..172N} reduction process for the \textit{XMM-Newton} data. Profiles are {in the full (0-360) sector and in the NW and NE directions, representative of the most disturbed and most relaxed ISM conditions (the complete set of gas and mass profiles is presented in the Appendix \ref{app:profiles}.)} In Figure \ref{fig:N5846_gas_profiles_merged_abund} we show the gas profiles obtained in the different sectors but with the free abundance model. We also compare our mass profiles with the optical mass profile (in yellow, \citealt{2014MNRAS.439..659N}) which makes use of SD and GC kinematics from the SLUGGS Survey, and with the X-ray mass profile  {(in light blue) previously determined by} \citet{2010MNRAS.409.1362D} from \textit{Chandra} and \textit{XMM-Newton} data. {Again, we highlight these} differences in the velocity profiles presented in Figure \ref{fig:N5846_velocity_profiles_all}, where in the left and right panel we show HE X-ray profiles from fixed and free abundance models, respectively, and compare them with the HE X-ray profile of \citet{2010MNRAS.409.1362D}, the mass profile of \citet{2014MNRAS.439..659N} and the mass measurement at \(5 R_e\) (33 kpc) by \citet{2016MNRAS.460.3838A}. The uncertainties on the velocity profiles obtained in this work are not shown and the width of the latter is fixed for clarity of representation  (proper errors are presented in Figures \ref{fig:N5846_gas_profiles_merged} and \ref{fig:N5846_gas_profiles_merged_abund}). A comprehensive mass analysis of NGC 5846 that includes SD, GC and also PNe data has been recently presented by \citet{2016MNRAS.462.4001Z}, yielding profiles close to those of \citeauthor{2014MNRAS.439..659N}.

The X-ray HE mass profiles of \citet{2010MNRAS.409.1362D} were obtained with element abundances fixed to 0.5 solar, {and as expected are} closer to our fixed abundance profiles than our free abundance profiles. Also, the HE mass profile of \citeauthor{2010MNRAS.409.1362D} extend down to \(\sim 0.8\) kpc while ours stop at \(\sim 1.6\) kpc because, as explained in Sect. \ref{sec:mass_profiles}, our HE mass profiles are evaluated at the outer radius of each bin rather than at its center. At the central region (\(R < 10\) kpc), the X-ray HE profiles lie below the optical profile. This discrepancy is seen in all sectors. As in the central region of NGC 4649, we interpret these small scale deviations from optical profiles as ISM disturbances (e.g,. cavities) in the inner regions of the galaxy connected with recent AGN activity \citep{2011ApJ...743...15M}.

As noted by \citet{2014MNRAS.439..659N}, the \citet{2010MNRAS.409.1362D} HE mass and velocity profiles fall below and above the optical profiles at distances smaller and larger than \(\sim 15\) kpc, respectively, thus indicating that the ISM in this source deviates from HE. When looking at the fixed abundance models we could reproduce the similar discrepancies in the NE sector (3rd column in the second row of Figure \ref{fig:N5846_gas_profiles_merged}). We interpret the deviations of the X-ray mass profiles from the optical one as a consequence of environmental effects, possibly due to sloshing of the hot ISM in NGC 5846 caused by the interaction with the group companion NGC 5850, compressing and decompressing the gas for \(R\geq 15\) kpc and \(R\leq 15\) kpc, respectively; in fact, a discrepancy with a mass jump similar to that shown here has been found in A1795, considered a typical example of sloshing in galaxy clusters \citep{2001ApJ...562L.153M}. In the NW sector (the last column in the top row) where the hot gas is most relaxed, our HE mass profile is smooth (the circular velocity increases slowly but monotonically) and close to the optical mass profile. However, the X-ray mass profile is still higher than the optical profile even in the NW sector. When allowing for variable element abundances, the large scale deviations between X-ray and optical derived mass profiles considerably decrease. This is because the negative abundance gradient (decreasing with increasing radii, see the last column in Figure \ref{fig:N5846_gas_profiles_merged_abund}) effectively reduces the pressure gradient, resulting in a flatter mass profiles (see also Figs. \ref{fig:N5846_gas_profiles_merged_app} and \ref{fig:N5846_gas_profiles_merged_abund_app}). In particular, in the NW sector (the 3rd column in the bottom row), the HE X-ray mass profile from the free abundance model is consistent with the optical profile.

\subsection{Total Mass}\label{sec:results_total_mass}

{To} assess the effect of the deviation from HE on the X-ray measurement of total mass, {we compare our results with those of SLUGGS \citep{2016MNRAS.460.3838A}. {We adopt their values for \(R_e\) and measure the mass withing 5 effective radii (roughly corresponding to the limit of optical mass tracers, \citealt{2012ApJ...748....2D,2016MNRAS.460.3838A}).} At \(5 R_e\) DM should also dominate the total mass.}

The HE masses from different sectors are summarized in Table \ref{tab:masses}, with the maximum radius \(R_{max}\) reached in the present analysis (column 3), the \(R_e\) values from \citet{2016MNRAS.460.3838A} (column 4), the total mass \(M_X\) enclosed in \(5 R_e\) (column 5) both for fixed and free element abundance models. If \(R_{max}\) does not reach \(5 R_e\) because of signal to noise ratio limitations (see Sect. \ref{sec:spectra}) {or because it exceeds the \(12\) kpc boundary imposed to exclude the Virgo ICM and regions affected by ram-pressure effects \citep{2017arXiv170305883W}, we extrapolate to this radius the mass profile fitting presented in Sect. \ref{sec:results_mass_decomposition} and indicate the value of \(M_X (< 5 R_e)\) with an asterisk. The value of} \(M_X (< 5 R_e)\) in boldface {is the one} that we consider reliable, {and it is derived} from the NW sector of NGC 5846 where we do not find strong deviations from HE from comparison with the optical mass profiles. {This} value {is} based only on X-ray data and {does not} depend on any particular {assumed model for the DM halo}.

We compare our results with the total mass enclosed in \(5 R_e\) as extrapolated from previous X-ray studies (column 6; \citealt{2008ApJ...683..161H} for NGC 4649 and \citealt{2010MNRAS.409.1362D} for NGC 5846), and with the total mass \(M_O\) enclosed in \(5 R_e\) from kinematic estimates (columns 7 and 8; compare \citealt{2016MNRAS.460.3838A} for both galaxies, \citealt{2010ApJ...711..484S} for NGC 4649, and \citealt{2014MNRAS.439..659N} for NGC 5846). All mass values are corrected for the same distance as in Table \ref{tab:sources}.

For NGC 4649, {the values of \(M_X\) extrapolated to \(5 R_e\) in the different sectors are consistent {within the uncertainties}, indicating that on \(3-12\) kpc scales the ISM distribution is relaxed and without strong anisotropies (see Fig. \ref{fig:N4649_gas_profiles_merged}).} The effect of variable abundance is to yield flatter gas density and pressure profiles, and then lower mass estimates. As expected, \(M_X\) obtained from the free abundance model is slightly lower than that from the fixed abundance model. However, within the measurement error, all models are generally consistent with the previous X-ray  \citep{2008ApJ...683..161H} and kinematic \citep{2016MNRAS.460.3838A} mass measurements. 
{The SE sector contains the diffuse tail reported by \citet{2017arXiv170305883W}, that appears faint when compared with simulations of gas stripping and Kelvin-Helmholtz instabilities in M89 \citep{2015ApJ...806..103R, 2015ApJ...806..104R}. According to these authors, the faintness of this structure could be due either to NGC 4649 larger distance from the cluster core (\(971\) kpc) with respect to M89 (\(390\) kpc), or to the fact that we may be observing this source shortly upon completing a turn around in its orbit in Virgo. In any case the state of the gas in the tail is unclear, so we extended our analysis in the SE direction beyond \(12\) kpc. Table \ref{tab:masses} includes the total mass value that we measure in this sector at \(5 R_e\) from X-ray data (without extrapolation). This value is indeed compatible with the extrapolated values obtained in this and other sectors, thus confirming that any deviation from HE in this region is small.} The free abundance models are barely consistent with the \citet{2010ApJ...711..484S} optical model as noted by \citet{2015MNRAS.450.1962P} (see their Sect. 8.3). More {up-to-date} data sets \citep{2016MNRAS.460.3838A} give a lower mass compatible with our X-ray HE mass profiles. {Averaging the free abundance model extrapolated to \(5 R_e\) we can estimate \(M_X (< 5R_e) = (1.0 \pm 0.1) \times {10}^{12} M_{\astrosun}\) for NGC 4649.}

For NGC 5846, the highest value of \(M_X\) (in the NE sector for the fix abundance model) is compatible with the previous X-ray measurement by \citet{2010MNRAS.409.1362D}, but considerably higher (by 3 sigma) than the optical measurements by both \citet{2014MNRAS.439..659N} and \citet{2016MNRAS.460.3838A}. As in NGC 4649, \(M_X\) obtained from the free abundance model is lower than that from the fixed abundance model. At the NW sector where the gas is most relaxed and extended, \(M_X\) is consistent with optical values. From this sector (using the free abundance model), we determine the total HE mass within 5 effective radius to be \(M_X (< 5 R_e) = (1.2 \pm 0.2) \times {10}^{12} M_{\astrosun}\) for NGC 5846. Since the X-ray data reach \(R_{max} \sim 6 R_e\) we also evaluate \(M_X (< 6 R_e) = (1.5 \pm 0.2) \times {10}^{12} M_{\astrosun}\) for free abundance models.

\subsection{Mass Decomposition}\label{sec:results_mass_decomposition}

We tried decomposing the HE mass into various components and extrapolate the results to virial radius, to evaluate the contribution of the DM halo to the total mass. To do so we fitted the observed mass profiles obtained in the different sectors with a model comprising the gas and stellar mass contribution, the central black hole and DM. The gas mass profile is determined from the hot gas density profile. For the stellar mass profile, we use the ATLAS\(^{3\text{D}}\) \textit{r}-band profiles presented by \citet{2013MNRAS.432.1894S}, and assume a constant \(M_*/L_*\) ratio in a given galaxy. We note that, although some stellar light profiles from ATLAS\(^{3\text{D}}\) show some problems at radii larger than \(\sim 250''\), these radii are in the region where DM {is the dominant mass component}, so they do not affect our results. To describe the DM distribution we assume the Navarro, Frenk \& White \citep[NFW,][]{1997ApJ...490..493N} profile. We show the best-fit results of NGC 4649 and NGC 5846 in Fig. \ref{fig:N4649_mass_fits} and Fig. \ref{fig:N5846_mass_fits}, respectively. The gas, stars, BH and DM are marked by a green dashed line, a blue dot-dashed line, a black dotted line and a black dot-dashed line. We also show in both figures the optical mass profiles (in yellow) by \citet{2010ApJ...711..484S} and \citet{2014MNRAS.439..659N} for NGC 4649 and NGC 5846, respectively and \(M_O(< 5R_e)\) (blue star) from \citet{2016MNRAS.460.3838A}.

Since X-ray mass profiles assume hydrostatic equilibrium, we expect this fitting procedure to yield results incompatible with similar estimates from optical derived profiles. In particular, for the mildly disturbed case of NGC 4649, we obtain \(\left<{M_{BH}}\right>=(5.5\pm 1.1)\times{10}^9 M_{\astrosun}\) and \(\left<{M_*/L_{*, r}}\right>=(4.4\pm 0.4) M_{\astrosun}/L_{\astrosun}\) for fixed abundance models, while for free abundance models we get \(\left<{M_{BH}}\right>=(5.1\pm 0.9)\times{10}^9 M_{\astrosun}\) and \(\left<{M_*/L_{*, r}}\right>=(4.7\pm 0.2) M_{\astrosun}/L_{\astrosun}\). The black hole masse so evaluated is compatible with the value of \(M_{BH} = (4.5 \pm 1.0) \times {10}^9 M_{\astrosun}\) reported by \citet{2010ApJ...711..484S}, while our mass to light ratios \(M_*/L_{*, r}\) are smaller than the value of \(6.5\pm 0.4\) (corrected for the distance used in this work) presented by \citet{2013MNRAS.432.1862C}.

The discrepancies become more severe for the strongly disturbed case of NGC 5846, where we obtain \(\left<{M_{BH}}\right>=(1.7\pm 0.8)\times{10}^{10} M_{\astrosun}\) and \(\left<{M_*/L_{*, r}}\right>=(2.8\pm 0.6) M_{\astrosun}/L_{\astrosun}\) for fixed abundance models, while for free abundances we get \(\left<{M_{BH}}\right>=(1.9\pm 0.6)\times{10}^{10} M_{\astrosun}\) and \(\left<{M_*/L_{*, r}}\right>=(2.3\pm 0.5) M_{\astrosun}/L_{\astrosun}\). The black hole mass so evaluated is one order of magnitude larger than that reported by \citet{2008PASA...25..167G} \(M_{BH}={1.1}\pm 0.2\times{10}^9 M_{\astrosun}\) and estimated through velocity dispersion measurement, while our mass to light ratios are about one third the value of \(7.3\pm 0.5\) presented by \citet{2013MNRAS.432.1862C} (again corrected for the distance used in this work), although consistent with the range of mass to light ratios found for ellipticals in these authors' study.

We then tried fixing the \(M_*/L_{*, r}\) and \(M_{BH}\) to their literature values and evaluating the DM component as the difference between the total X-ray mass profile and the stellar, black hole and gas mass components. As mentioned before, {the} values of \(M_*/L_{*, r}\) obtained from optical data are higher than those we obtain from X-ray data, due to the {previously} discussed deviations from HE that may affect this technique, and this difference is more severe in the case of NGC 5846 where the ISM is more disturbed. {As a consequence, the DM profile can appear to be negative at small radii where the ISM is away from HE}, and it can be considered reliable only at large radii (where the total X-ray and optical mass profile are compatible).

This analysis shows that mass measurement and characterization only based on X-ray data can lead to misleading results due to the very nature of the ISM physics resulting in the X-ray emission, and it requires a selection of azimuthal directions in which the ISM can be considered fairly relaxed as a consequence of comparisons with optical profiles. Other assumptions - like abundance or DM profiles - make the extrapolation of these results to radii {much larger than those attained by X-ray analysis and up to the virial radius} even more problematic.

\section{Summary and Conclusions}\label{sec:summary}

{We present} the results of a joint \textit{Chandra} \textit{XMM-Newton} analysis {of two ETGs. NGC 4649 shows a relaxed and symmetric X-ray morphology in the range \(3-12\) kpc, while on larger scales there are evidences of gas stripping due to the galaxy motion in the cluster ICM. NGC 5846, on the other hand, shows signatures of mild sloshing at all radii}. We make use of X-ray based mass profiles in various azimuthal directions to investigate these effects on the ISM by comparison with optically based mass profiles, and studied the effects of different background subtraction/modeling procedures and element abundance gradients on the total mass evaluation. The main results of this analysis are:

\begin{enumerate}
\item The X-ray mass profiles of NGC 4649 appear to be fairly {smooth in the range \(3-12\) kpc} and do not show azimuthal asymmetries. {The values of \(M_X\) extrapolated to \(5 R_e\) in different directions are consistent with each other and with \(M_X\) measured at \(5 R_e\) in the SE sector that contains the faint tail structure.}
\item In NGC 5846 the X-ray derived mass profiles show significant asymmetries in the ISM distribution, especially in the NE direction where the halo shows evidences of interaction with the group companion spiral galaxy NGC 5850. The comparison with optical mass profiles shows in this direction evidence of gas compression and decompression due to gas sloshing on scales larger and smaller than \(\sim 15\) kpc, respectively, and these effects disappear in the NW direction where the emission is smooth and extended.
\item Deviations between the optical and X-ray profiles for \(r < 10\) kpc are observed in all sectors of NGC 5846, and they are interpreted as the consequence of ISM disturbances connected with recent AGN activity. On larger scales, instead, the match between X-ray (from the NW sector) and optical data is recovered after allowing for element abundance gradients, which shows decreasing abundances in the outer regions of the galaxy.
\item Mass measurement solely based on X-ray data can lead to misleading results if not coupled with the selection of azimuthal sectors in which the ISM can be considered fairly relaxed. Using the sectors with relaxed gas distribution, we measured the total mass (based on X-ray data) of \((1.2 \pm 0.2) \times {10}^{12} M_{\astrosun}\) for NGC 5846 at 5 effective radii, {while an extrapolation to \(5 R_e\) of the profile of NGC 4649 yields \((1.0 \pm 0.1) \times {10}^{12} M_{\astrosun}\)}. These values are consistent with {those} from previous optical measurements \citep{2011MNRAS.415.1244D, 2014MNRAS.439..659N, 2016MNRAS.460.3838A}.
\item Using the NW sector of NGC 5846 for which there is good agreement with optical mass profiles, we measure a masse at \(6 R_e\) of \(M_X (< 6 R_e) = (1.5 \pm 0.2) \times {10}^{12} M_{\astrosun}\).
\item We also performed fits of the mass profiles to evaluate the contributions of BH, gas, stars and DM to the total mass. While the DM dominates at large radii, as expected, the measurements of the BH mass is complicated by the difficulty in disentangling it from the stellar mass contribution. In addition, the X-ray derived mass profiles rely on hydrostatic equilibrium assumption that, as shown here, can not be the case due to either nuclear activity or galactic interactions. Therefore, virial mass evaluation is subject to significant uncertainties due to uncertain extrapolation of the observed gas profiles to the virial radius, which usually lies far beyond the maximum extent attained by the X-ray detectors.
\end{enumerate}

\acknowledgments

This work was supported by the Chandra grant AR5-16007X and by 2014 Smithsonian Competitive Grant Program for Science. GF thanks the Aspen Center for Physics, funded by NSF grant \#1066293, for their hospitality while this paper was completed. This work was also supported by NSF grants AST-1211995, AST-1616598, and AST-1616710.

\begin{table}
\begin{center}
\caption{General properties of the sources discussed in this work. (1) Source name; (2, 3) Source coordinates; (4) Distance in Mpc; (5) Spatial scale in pc/arcsec at the distance reported in column 4; (6) Galactic hydrogen column density in the source direction; (7) 5 effective radii \(R_e\) as reported by \citet{2016MNRAS.460.3838A}.}\label{tab:sources}
\resizebox{\textwidth}{!}{
\begin{tabular}{|c|c|c|c|c|c|c|}
\hline
\hline
(1) SOURCE NAME & (2) RA & (3) DEC & (4) DISTANCE & (5) SPATIAL SCALE & (6) \(N_H\) & (7) \(5 R_e\)\\
& hms & dms & Mpc & pc/arcsec & \({10}^{20}\mbox{cm}^{-2}\) & kpc \\
\hline
NGC 4649 & 12:43:40.0 & +11:33:10 & 15.7 & 76  & 2.04 & 25.1 \\
NGC 5846 & 15:06:29.3 & +01:36:20 & 23.1 & 112 & 4.29 & 33.0 \\
\hline
\hline
\end{tabular}
}
\end{center}
\end{table}

\begin{table}
\begin{center}
\caption{Summary of \textit{Chandra} data used in this work. (1) Source name; (2) OBSID; (3) Observation date; (4) ACIS array and chip used for the observation; (5) Clean exposure times after excluding time intervals of strong background flares.}\label{tab:chandra_data}
\resizebox{\textwidth}{!}{
\begin{tabular}{|c|c|c|c|c|}
\hline
\hline
(1) SOURCE NAME & (2) OBSID & (3) OBS-DATE & (4) ARRAY & (5) EXPOSURE \\
& & & & ksec \\
\hline
\multirow{12}{*}{NGC 4649} & \multirow{2}{*}{00785} & \multirow{2}{*}{2000-04-20} & ACIS-S 6 & 34.6 \\
\cline{4-5}
& & & ACIS-S 7 & 22.2 \\
\cline{2-5}
& \multirow{ 2}{*}{08182} & \multirow{2}{*}{2007-01-30} & ACIS-S 6 & 48.3 \\
\cline{4-5}
& & & ACIS-S 7 & 47.0 \\
\cline{2-5}
& \multirow{ 2}{*}{08507} & \multirow{2}{*}{2007-02-01} & ACIS-S 6 & 15.0 \\
\cline{4-5}
& & & ACIS-S 7 & 17.3 \\
\cline{2-5}
& \multirow{ 2}{*}{12975} & \multirow{2}{*}{2011-08-08} & ACIS-S 6 & 80.9 \\
\cline{4-5}
& & & ACIS-S 7 & 77.3 \\
\cline{2-5}
& \multirow{ 2}{*}{12976} & \multirow{2}{*}{2011-02-24} & ACIS-S 6 & 99.2 \\
\cline{4-5}
& & & ACIS-S 7 & 98.2 \\
\cline{2-5}
& \multirow{ 2}{*}{14328} & \multirow{2}{*}{2011-08-12} & ACIS-S 6 & 13.2 \\
\cline{4-5}
& & & ACIS-S 7 & 14.0 \\
\cline{2-5}
\hline
\multirow{6}{*}{NGC 5846}  & \multirow{2}{*}{00788} & \multirow{2}{*}{2000-05-24} & ACIS-S 6 & 22.2 \\
\cline{4-5}
& & & ACIS-S 7 & 23.2 \\
\cline{2-5}
& \multirow{4}{*}{07923} & \multirow{4}{*}{2007-06-12} & ACIS-I 0 & 87.5 \\
\cline{4-5}
& & & ACIS-I 1 & 88.2 \\
\cline{4-5}
& & & ACIS-I 2 & 88.0 \\
\cline{4-5}
& & & ACIS-I 3 & 88.2 \\
\hline
\hline
\end{tabular}
}
\end{center}
\end{table}

\begin{table}
\begin{center}
\caption{Summary of \textit{XMM-Newton} data used in this work. (1) Source name; (2) OBSID; (3) Observation date; (4) MOS array used for the observation; (5) Clean exposure times for the reduction procedures described in Sect. \ref{app:nevalainen}/\ref{app:arnaud}/\ref{app:snowden}.}\label{tab:XMM_data}
\resizebox{\textwidth}{!}{
\begin{tabular}{|c|c|c|c|c|}
\hline
\hline
(1) SOURCE NAME & (2) OBSID & (3) OBS-DATE & (4) ARRAY & (5) EXPOSURE \\
& & & & ksec \\
\hline
\multirow{4}{*}{NGC 4649} & \multirow{2}{*}{0021540201} & \multirow{2}{*}{2001-01-02} & MOS1 & 45.3/50.5/45.5 \\
\cline{4-5}
& & & MOS2 & 45.3/50.3/45.9 \\
\cline{2-5}
& \multirow{2}{*}{0502160101} & \multirow{2}{*}{2007-12-19} & MOS1 & 71.0/73.2/68.4 \\
\cline{4-5}
& & & MOS2 & 71.0/72.5/68.6 \\
\hline
\multirow{6}{*}{NGC 5846} & \multirow{2}{*}{0021540501} & \multirow{2}{*}{2001-08-26} & MOS1 & 13.4/15.4/13.3 \\
\cline{4-5}
& & & MOS2 & 13.4/15.7/13.2 \\
\cline{2-5}
& \multirow{2}{*}{0723800101} & \multirow{2}{*}{2014-01-21} & MOS1 & 69.6/77.8/65.9 \\
\cline{4-5}
& & & MOS2 & 70.6/79.3/67.7 \\
\cline{2-5}
& \multirow{2}{*}{0723800201} & \multirow{2}{*}{2014-01-17} & MOS1 & 84.7/87.6/83.0 \\
\cline{4-5}
& & & MOS2 & 86.0/87.6/84.0 \\
\hline
\hline
\end{tabular}
}
\end{center}
\end{table}

\begin{table}[ht!]
\begin{center}
\caption{Results of the mass profile fitting {(with uncertainties reported in parenthesis)}. (1) Name of the source; (2) angular sector; (3) maximum radius \(R_{max}\) reached in the present analysis. {Values indicated with an asterisk are truncated to avoid disturbances in ISM due to the motion of NGC 4649 in the Virgo ICM as reported by \citep{2017arXiv170305883W}. In addition, for the SE direction of NGC 4649 we also report the value measured directly at \(R_{max}\) without truncation}; (4) 5 effective radii \(R_e\) as reported by \citet{2016MNRAS.460.3838A}; (5) total mass enclosed in \(5 R_e\) {calculated} from the mass profile fits shown in Figures \ref{fig:N4649_mass_fits} and \ref{fig:N5846_mass_fits} (fixed/free abundance model). {The values indicated with an asterisk are extrapolated to \(5 R_e\) from the fitting presented in Sect. \ref{sec:results_mass_decomposition}. In addition, for the SE direction we also report the value of the total mass measured at \(5 R_e\). In boldface we indicate the value in the NW sector of NGC 5846 where where \(R_{max} > 5 R_e\) and we do not see strong deviations from HE.}; (6) total mass enclosed in \(5 R_e\) {extrapolated} from previous X-ray profiles (\citealt{2008ApJ...683..161H} for NGC 4649 and \citealt{2010MNRAS.409.1362D} for NGC 5846), (7) total mass enclosed in \(5 R_e\) as reported by \citet{2016MNRAS.460.3838A}, (8) total mass enclosed in \(5 R_e\) from previous optical profiles (\citealt{2010ApJ...711..484S} for NGC 4649 and \citealt{2014MNRAS.439..659N} for NGC 5846); (9) total mass enclosed in \(R_{6 R_e}\) from the mass profiles presented in this work. {Mass values are corrected for different distance adopted. In boldface we indicate the value in the NW sector of NGC 5846 where where \(R_{max} > 6 R_e\) and we do not see strong deviations from HE.}}\label{tab:masses}
\resizebox{\textwidth}{!}{
\begin{tabular}{|c|c|c|c|c|c|c|c|c|}
\hline
\hline
(1) SOURCE NAME & (2) SECTOR & (3) \(R_{max}\) & (4) \(5R_e\) & (5) \(M_X(<5R_e)\) & (6) \(M_{X,old}(<5R_e)\) & (7) \(M_O(<5R_e)\) & (8) \(M_O(<5R_e)\) & (9) \(M_X(<6 R_e)\) \\
& & kpc & kpc & \({10}^{11}M_{\astrosun}\) & \({10}^{11}M_{\astrosun}\) & \({10}^{11}M_{\astrosun}\) & \({10}^{11}M_{\astrosun}\)  & \({10}^{11}M_{\astrosun}\)\\
\hline
\multirow{5}{*}{NGC 4649} & 0-360 (FULL) & 12.0* & \multirow{5}{*}{25.1} & 12.4(0.3)* / 11.2(1.7)* &\multirow{5}{*}{11.2(0.9)} & \multirow{5}{*}{11.0(0.9)} & \multirow{5}{*}{16.2(5.7)} & \\
& 270-360 (NW)   & 11.4 & & 12.1(0.5)* / 8.5(1.7)* & & & & \\
& 0-90 (NE) & 12.0* & & 12.0(0.4)* / 10.9(2.6)* & & & & \\
& 90-180 (SE) & 12.0* \(\mid\) 25.5 & & 12.0(0.5)* / 12.3(3.6)* \(\mid\) 12.4(0.9) / 10.5(1.0) & & & & \\
& 180-270 (SW) & 12.0* & & 10.8(0.5)* / 9.8(1.5)*  & & & & \\
\hline
\multirow{5}{*}{NGC 5846} & 0-360 (FULL) & 45.9 & \multirow{5}{*}{33.0} & 16.0(0.6) / 15.6(2.7) & \multirow{5}{*}{22.2(2.2)} & \multirow{5}{*}{12.1(1.7)} & \multirow{5}{*}{10.1(1.9)} & \\
& 30-90 (NE) & 42.0 & & 18.1(2.0) / 10.5(3.0) & & & & \\
& 90-180 (SE) & 38.6 & & 15.8(0.9) / 12.9(1.9) & & & & \\
& 180-250 (SW) & 28.6 & & 13.9(0.7)* / 12.8(1.6)* & & & & \\
& 250-30 (NW) & 39.2 & & \textbf{13.1(0.3) / 12.1(1.5)} & & & & \textbf{16.4(0.5) / 15.0(1.8)} \\
\hline
\hline
\end{tabular}
}
\end{center}
\end{table}

\begin{figure}
\centering
\includegraphics[scale=0.54]{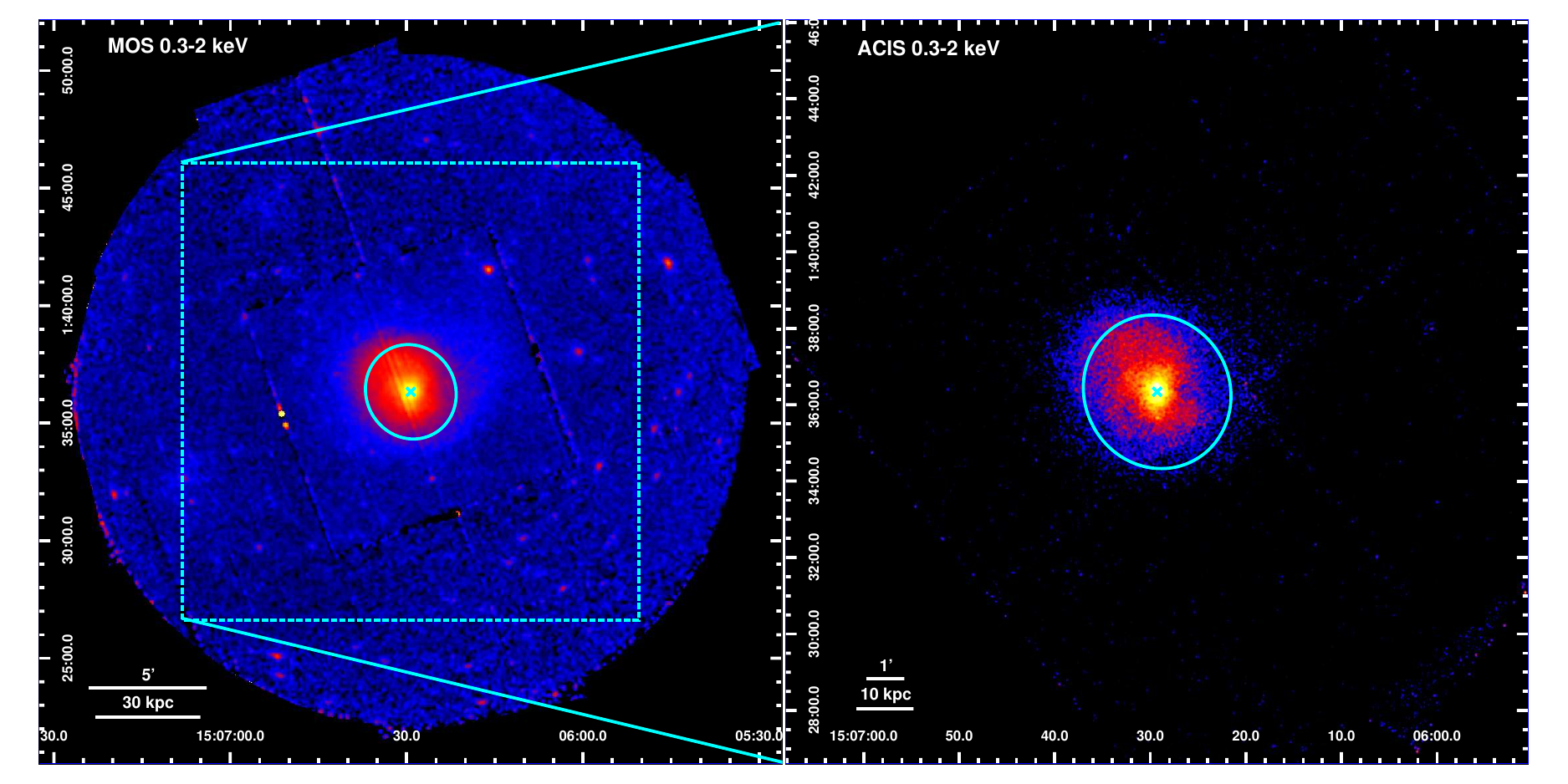}
\includegraphics[scale=0.54]{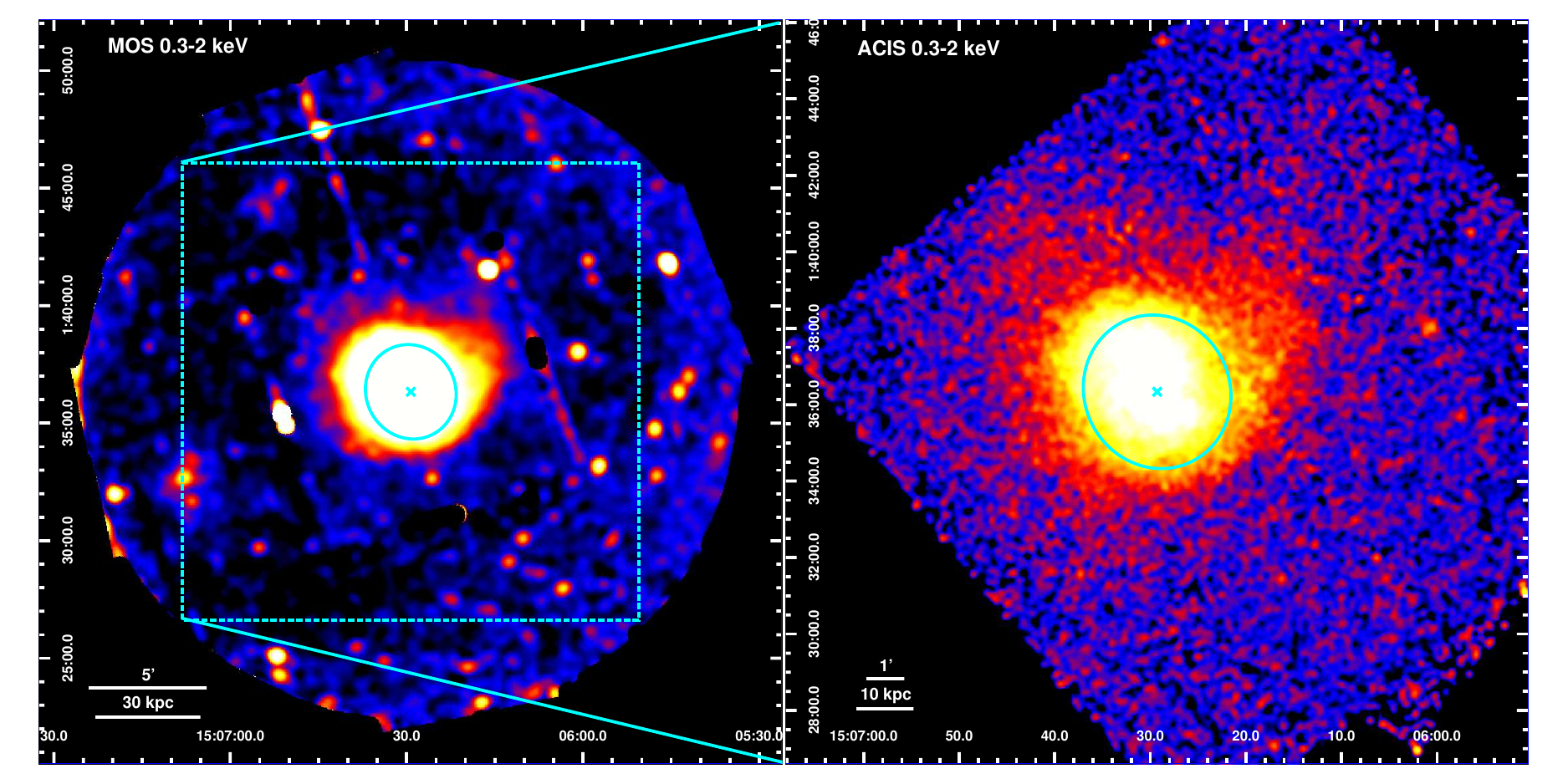}
\caption{Same as Fig. \ref{fig:N5846_mos} in the 0.3-2 keV band with the D25 ellipse shown in cyan. Small scale structures and large scale structures are shown in upper and lower panel, respectively.}\label{fig:N5846_mos2}
\end{figure}

\begin{figure}
\centering
\includegraphics[scale=0.33]{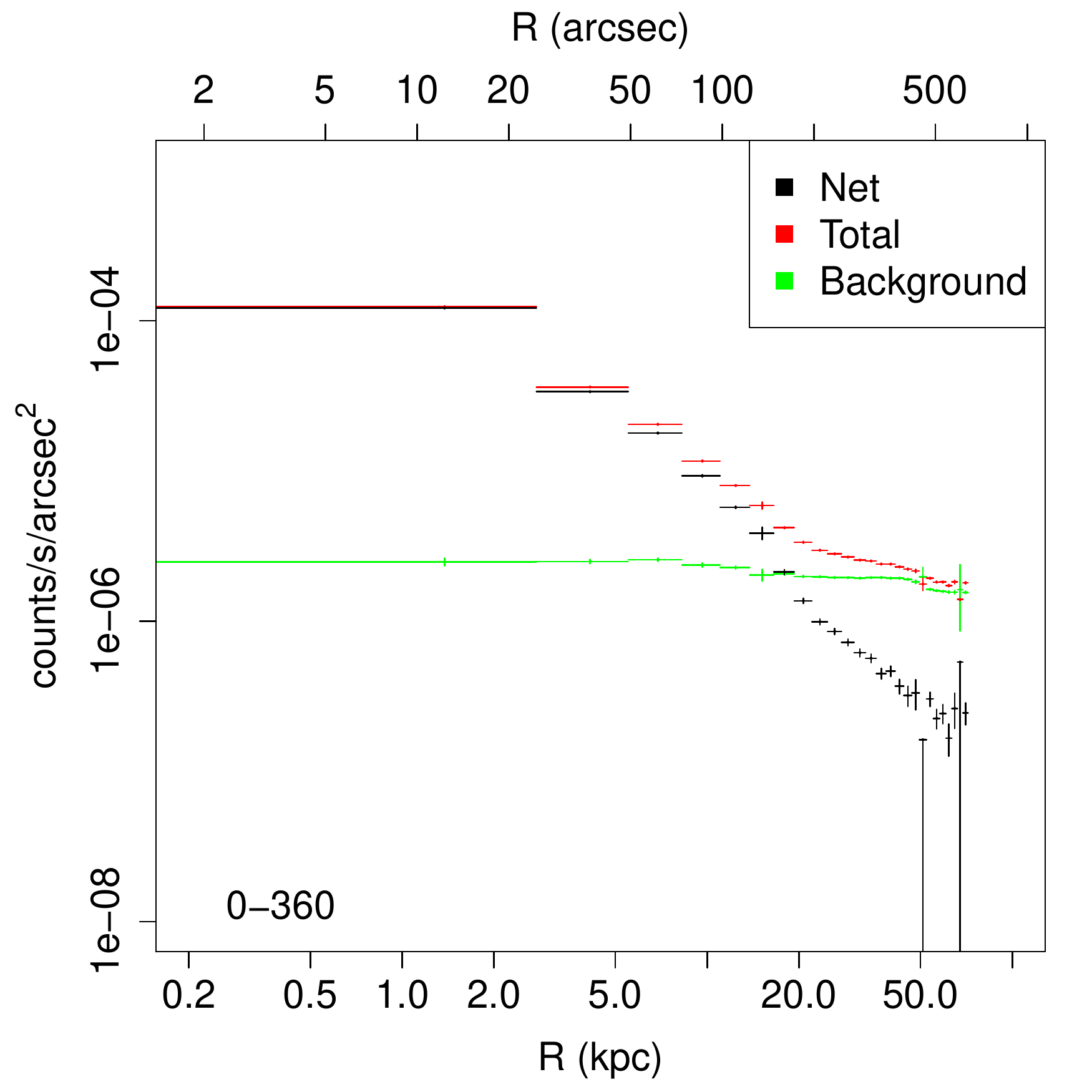}
\includegraphics[scale=0.33]{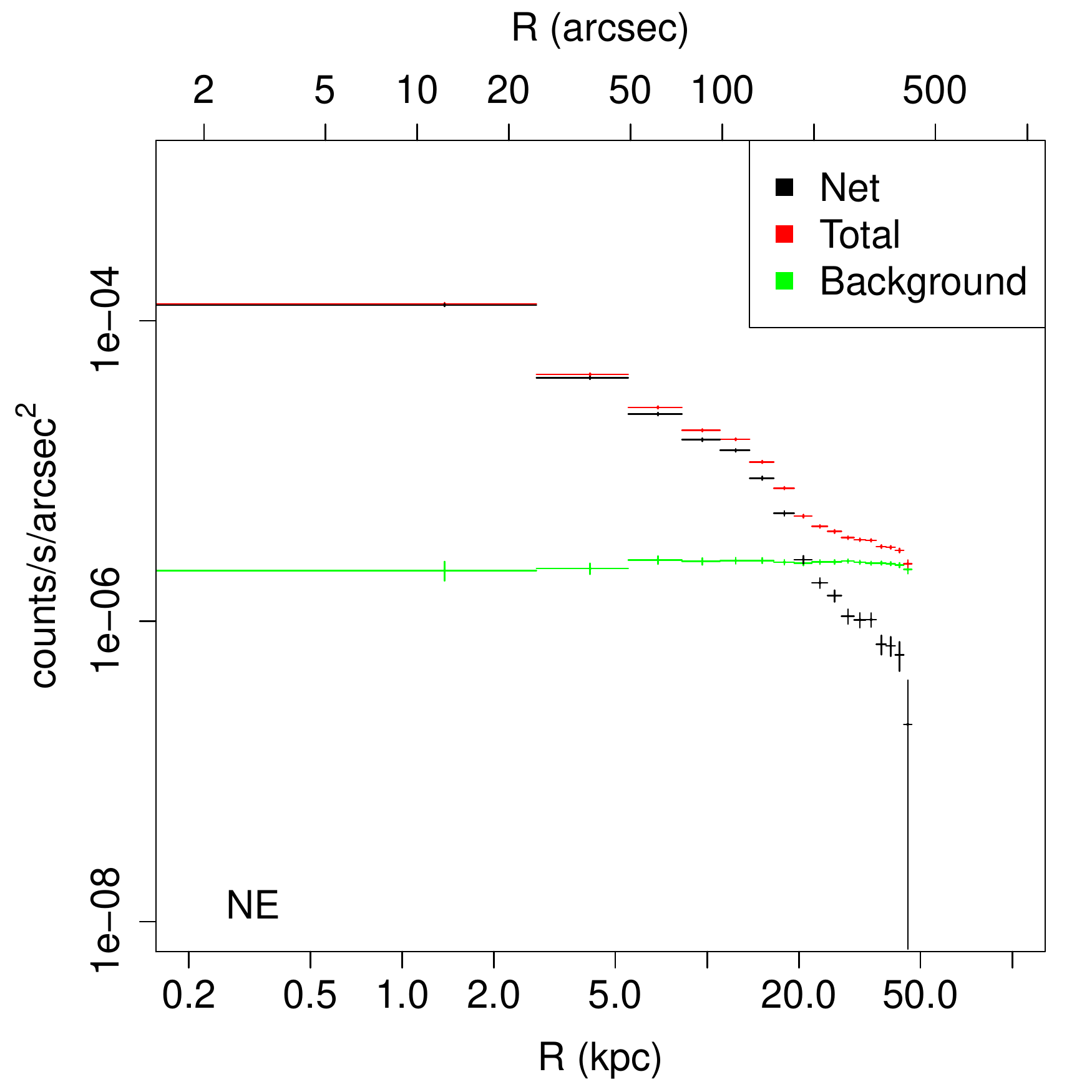}
\includegraphics[scale=0.33]{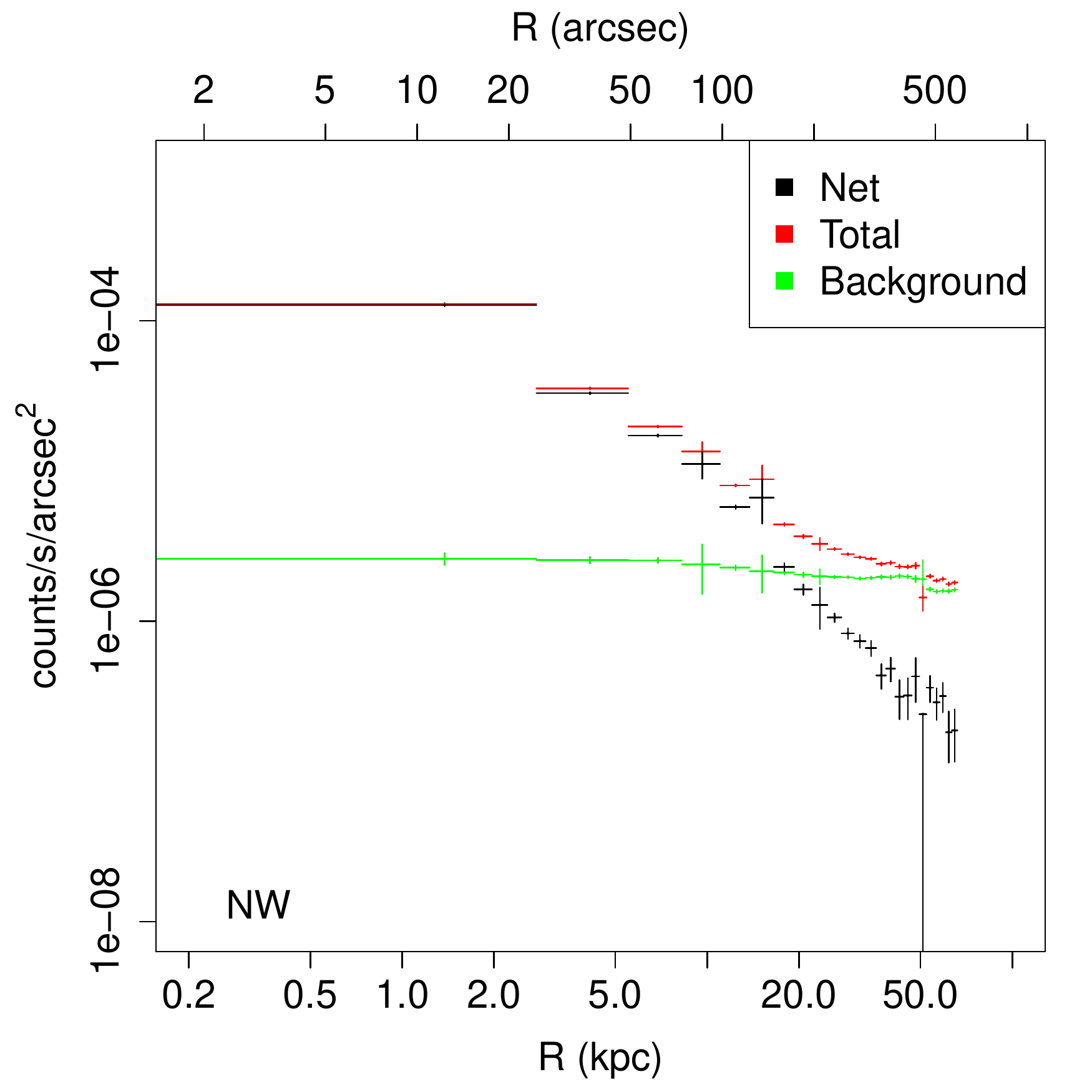}\\
\includegraphics[scale=0.33]{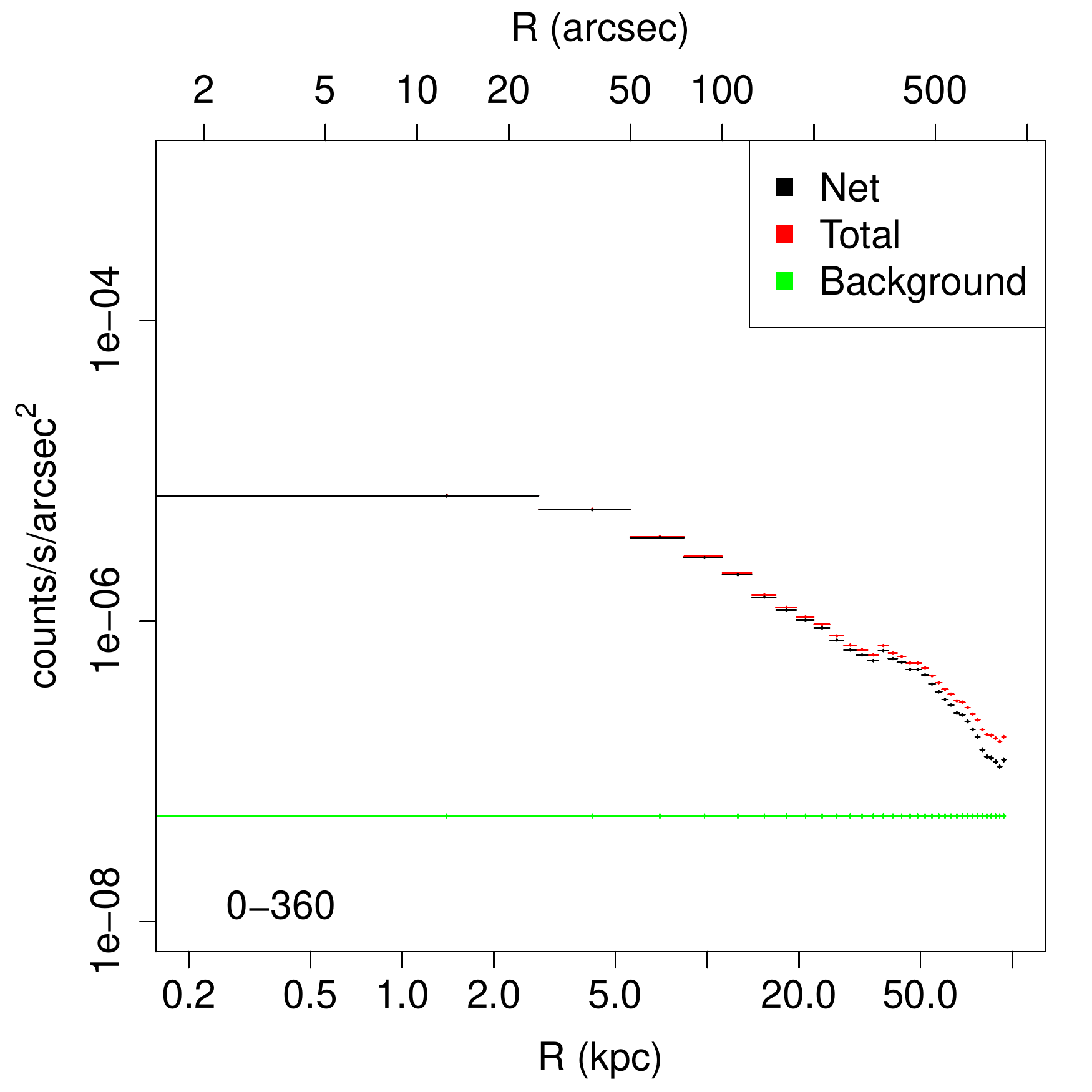}
\includegraphics[scale=0.33]{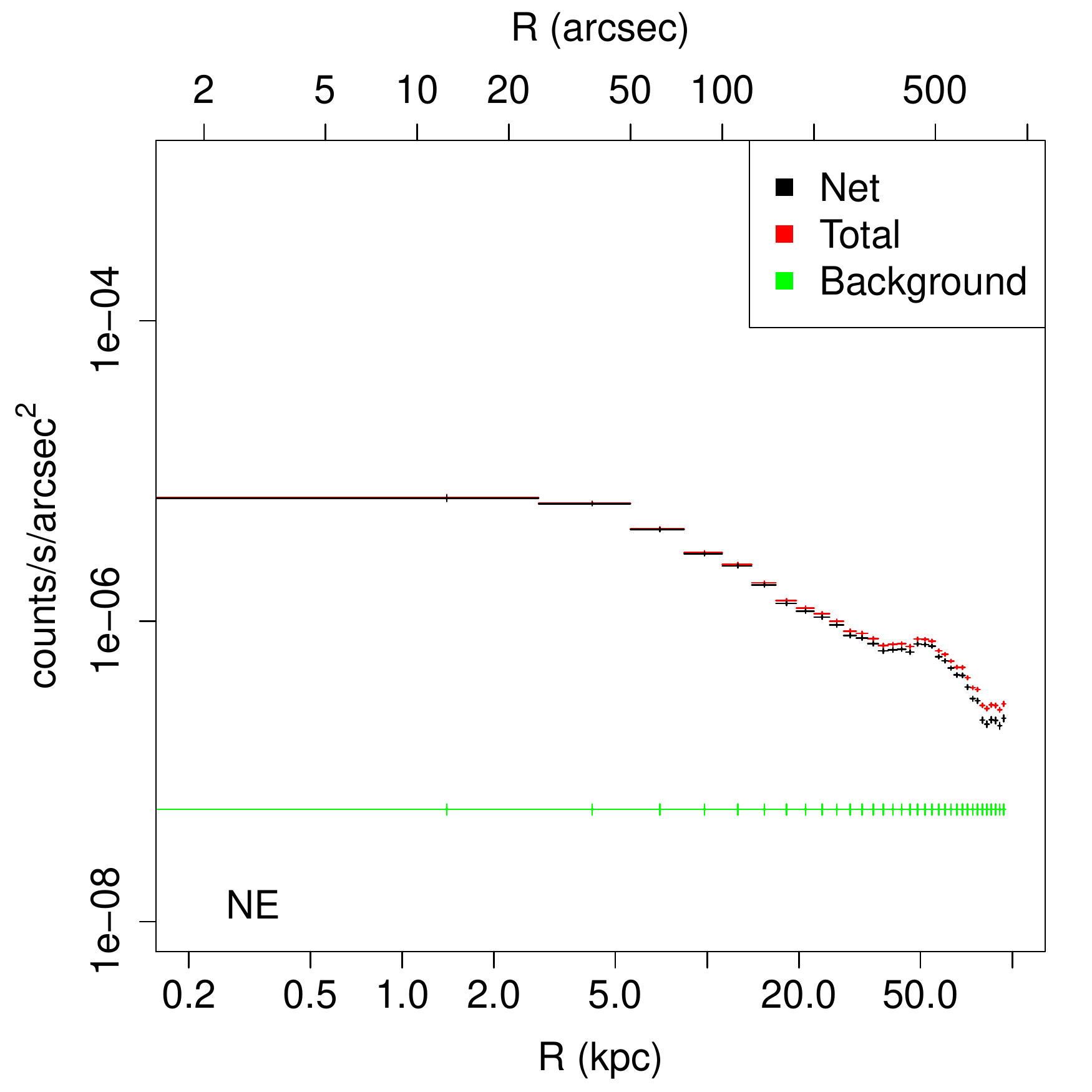}
\includegraphics[scale=0.33]{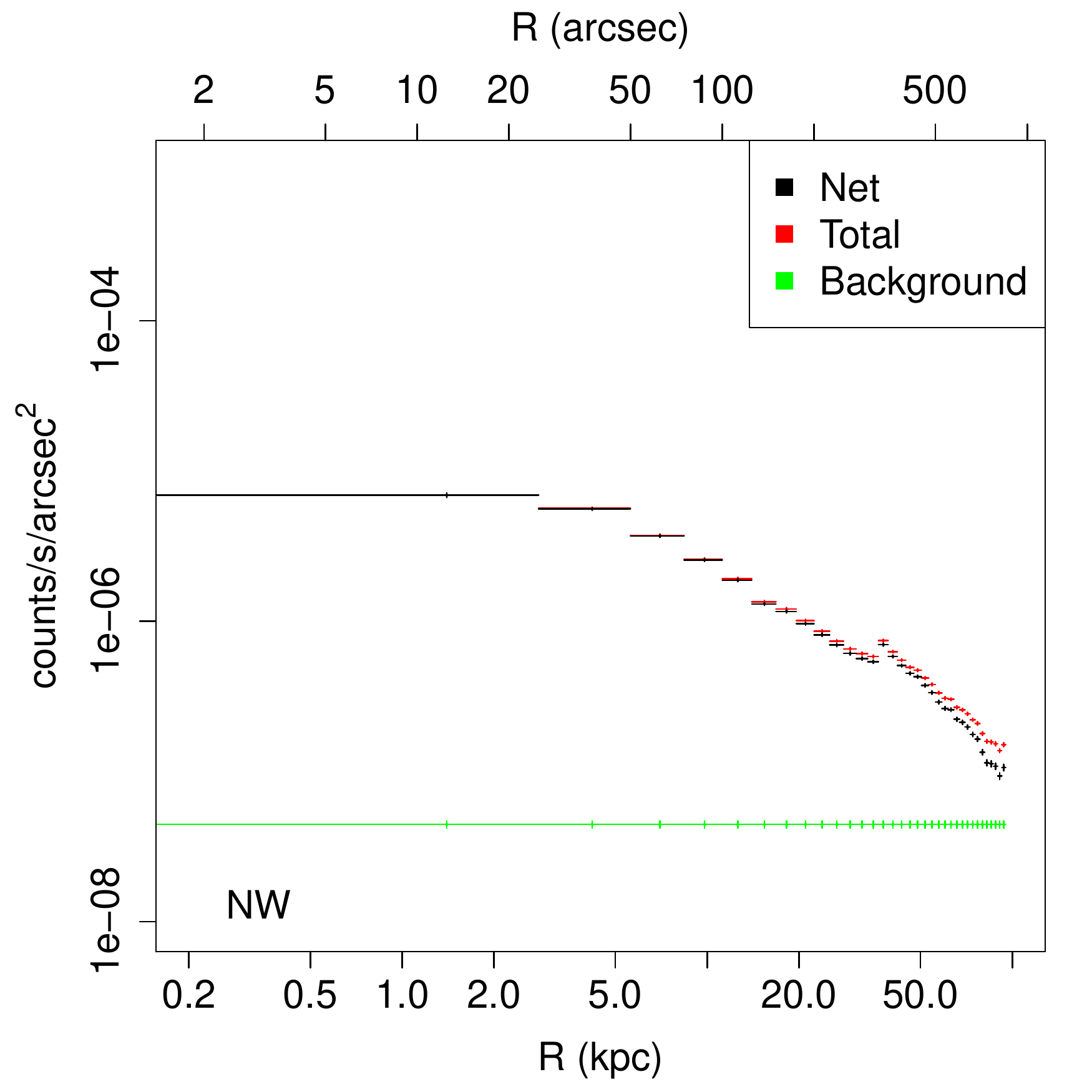}
\caption{Brightness profiles for NGC 5846 in the \(0.3-10\) keV band in the different sectors shown in Fig. \ref{fig:N5846_mos}, from left to right full (0-360), NE (30-90) and NW (250-30), respectively. In particular we show brightness profiles for \textit{Chandra} ACIS data {in the top row} and for the \textit{XMM}-MOS data obtained from the reduction procedure proposed by \citet{2005ApJ...629..172N} in the bottom row. The annuli width is \(\sim 25''\), 50 pixels for \textit{Chandra} ACIS data and 500 pixels for \textit{XMM}-MOS data. Red, black and green points represent total, net, and background brightness profiles, respectively.}\label{fig:N5846_bp_chandra}
\end{figure}

\begin{figure}
\centering
\includegraphics[scale=0.33]{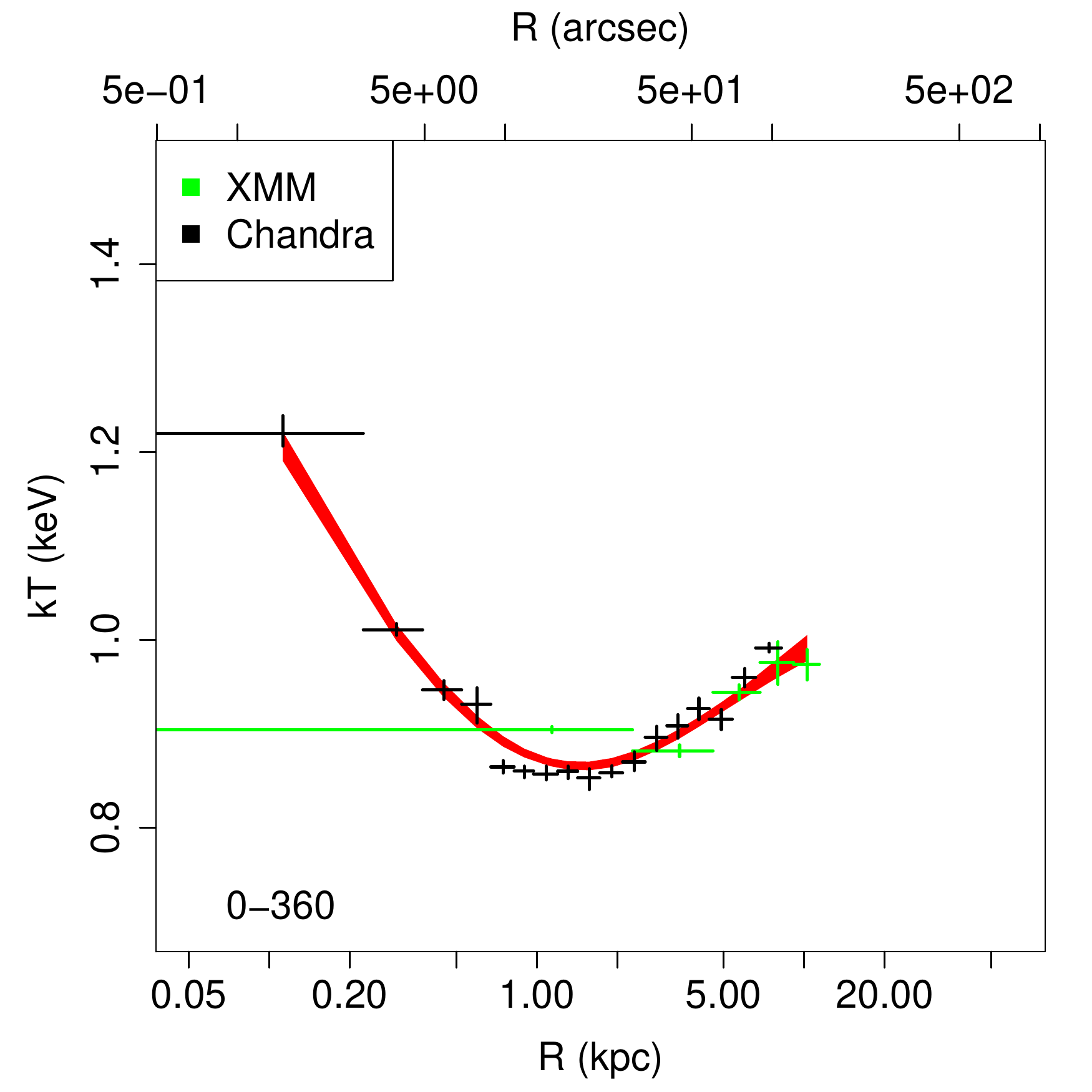}
\includegraphics[scale=0.33]{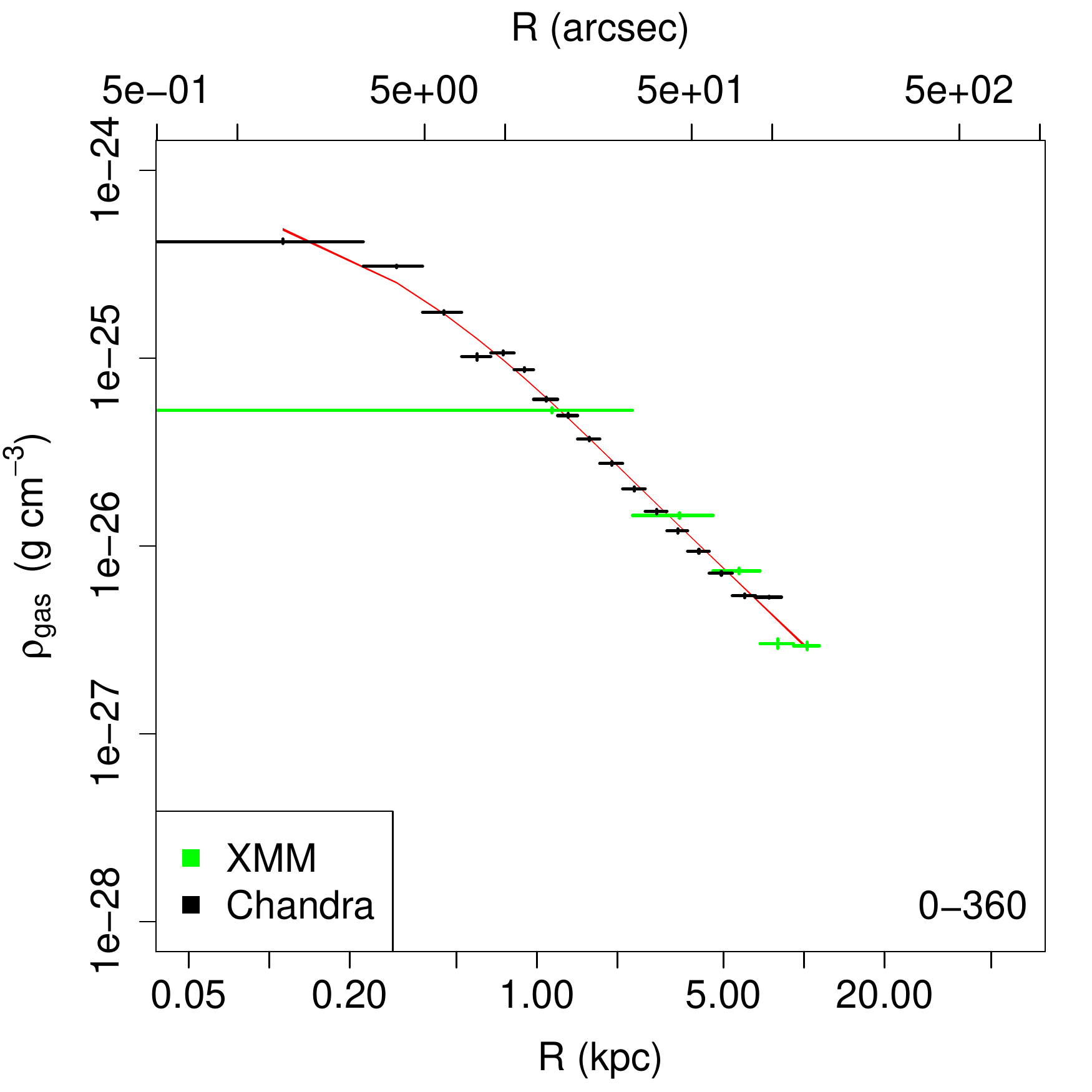}
\includegraphics[scale=0.33]{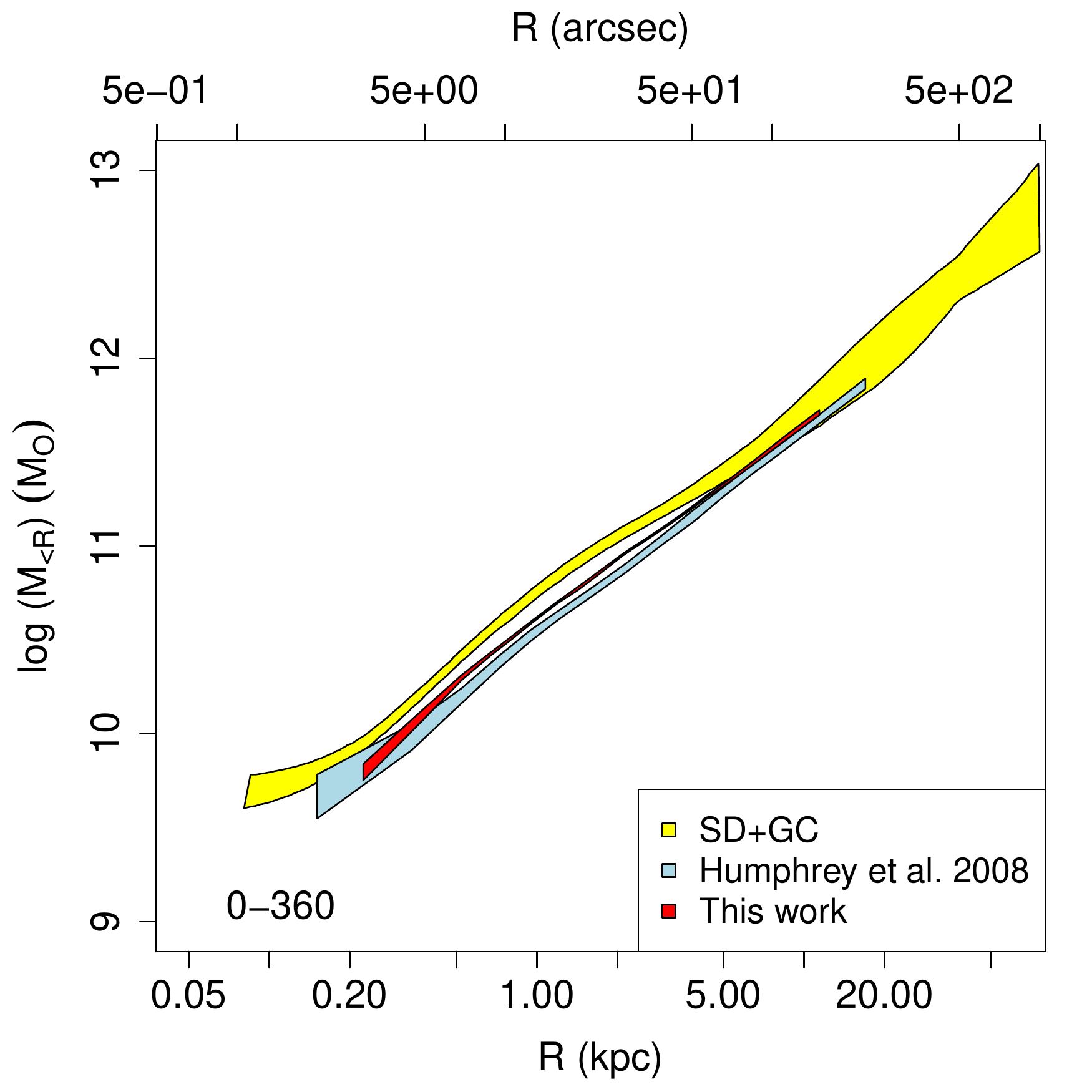}\\
\includegraphics[scale=0.33]{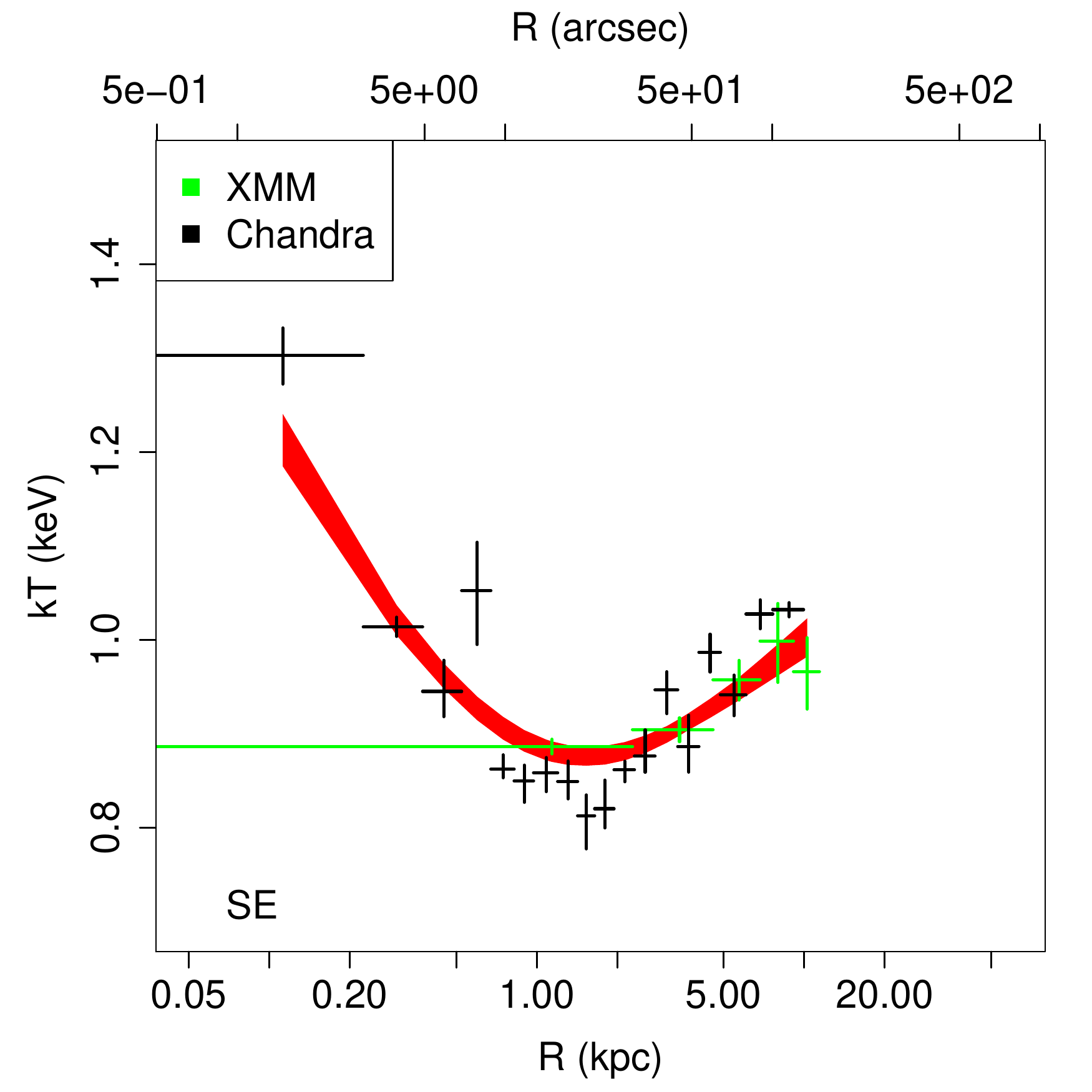}
\includegraphics[scale=0.33]{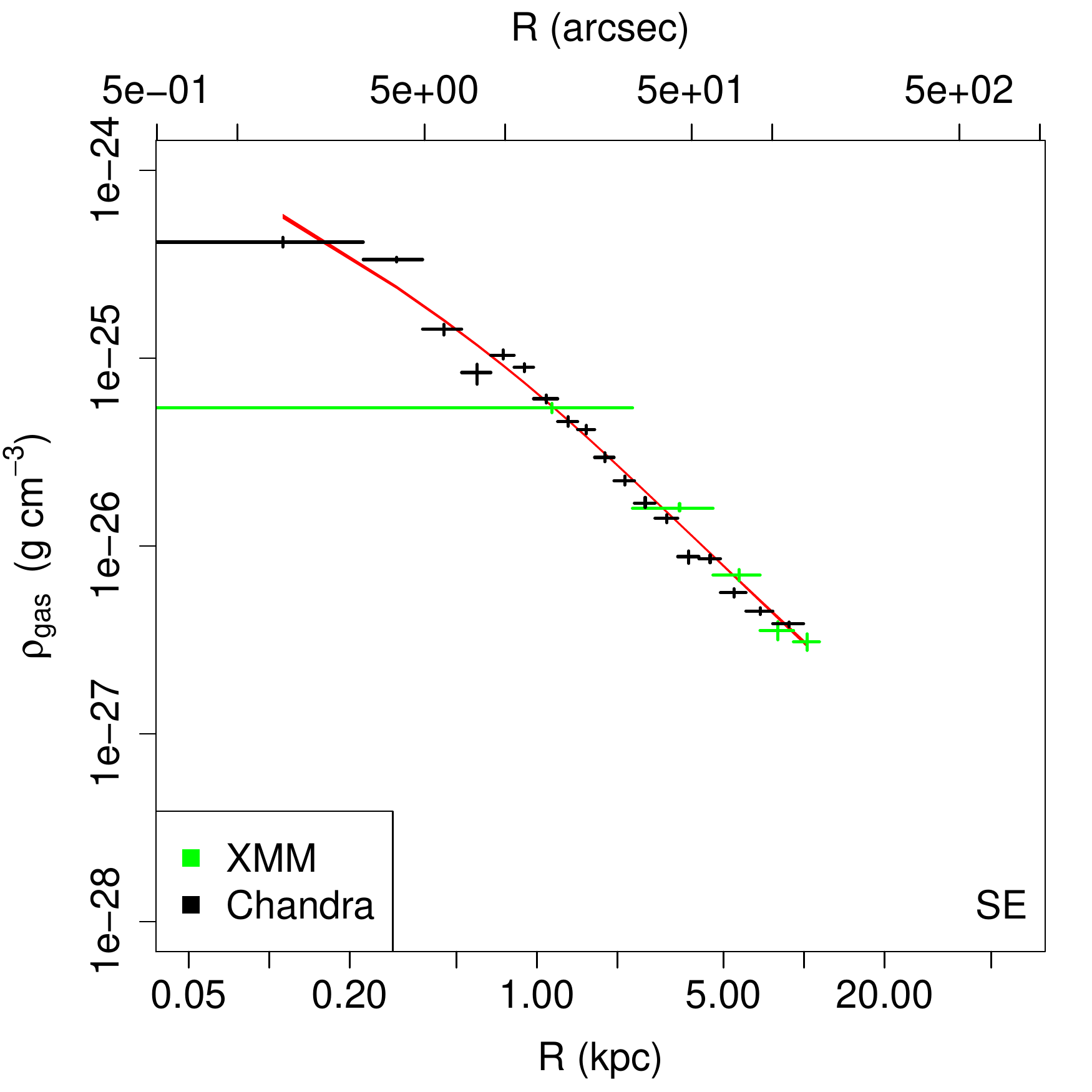}
\includegraphics[scale=0.33]{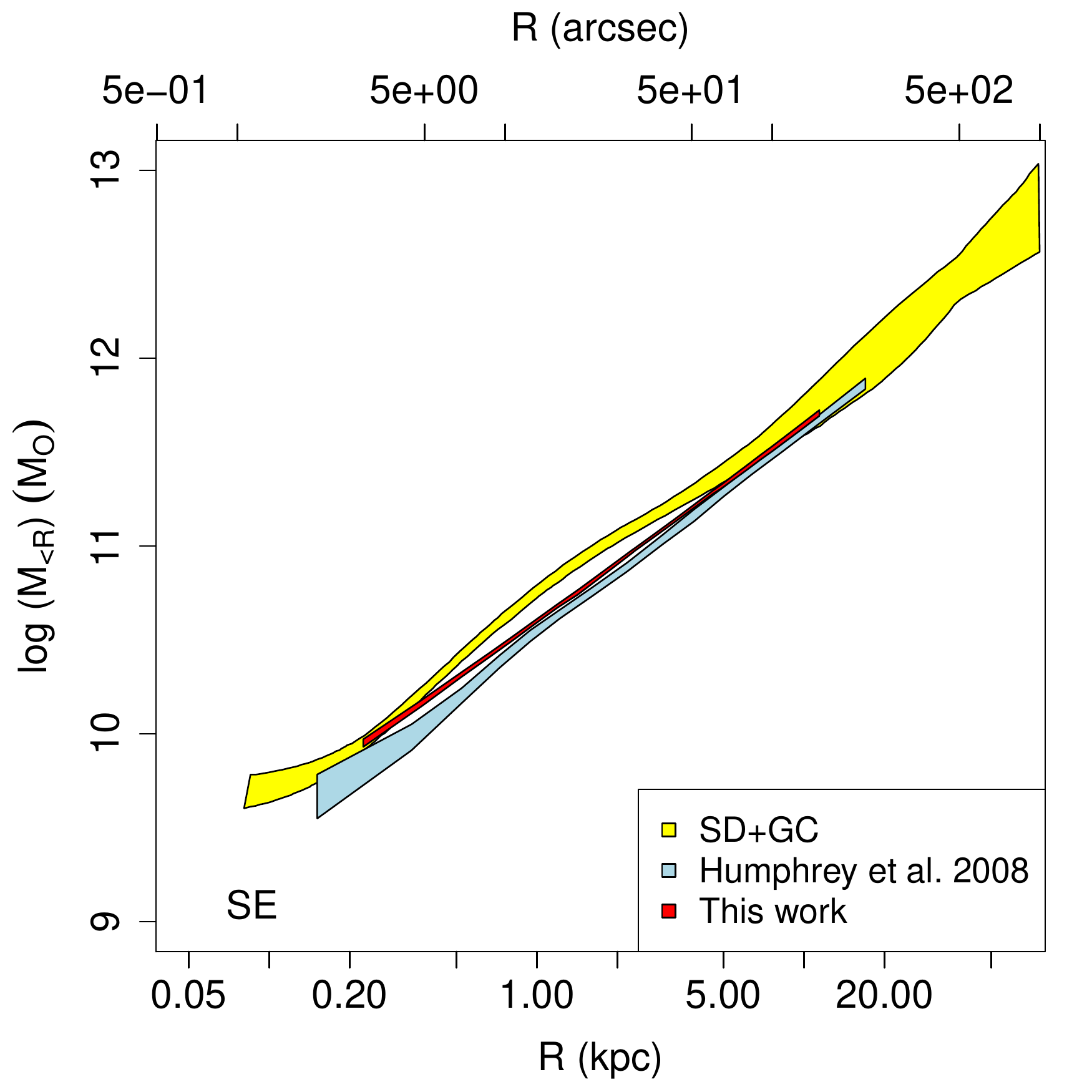}\\
\includegraphics[scale=0.33]{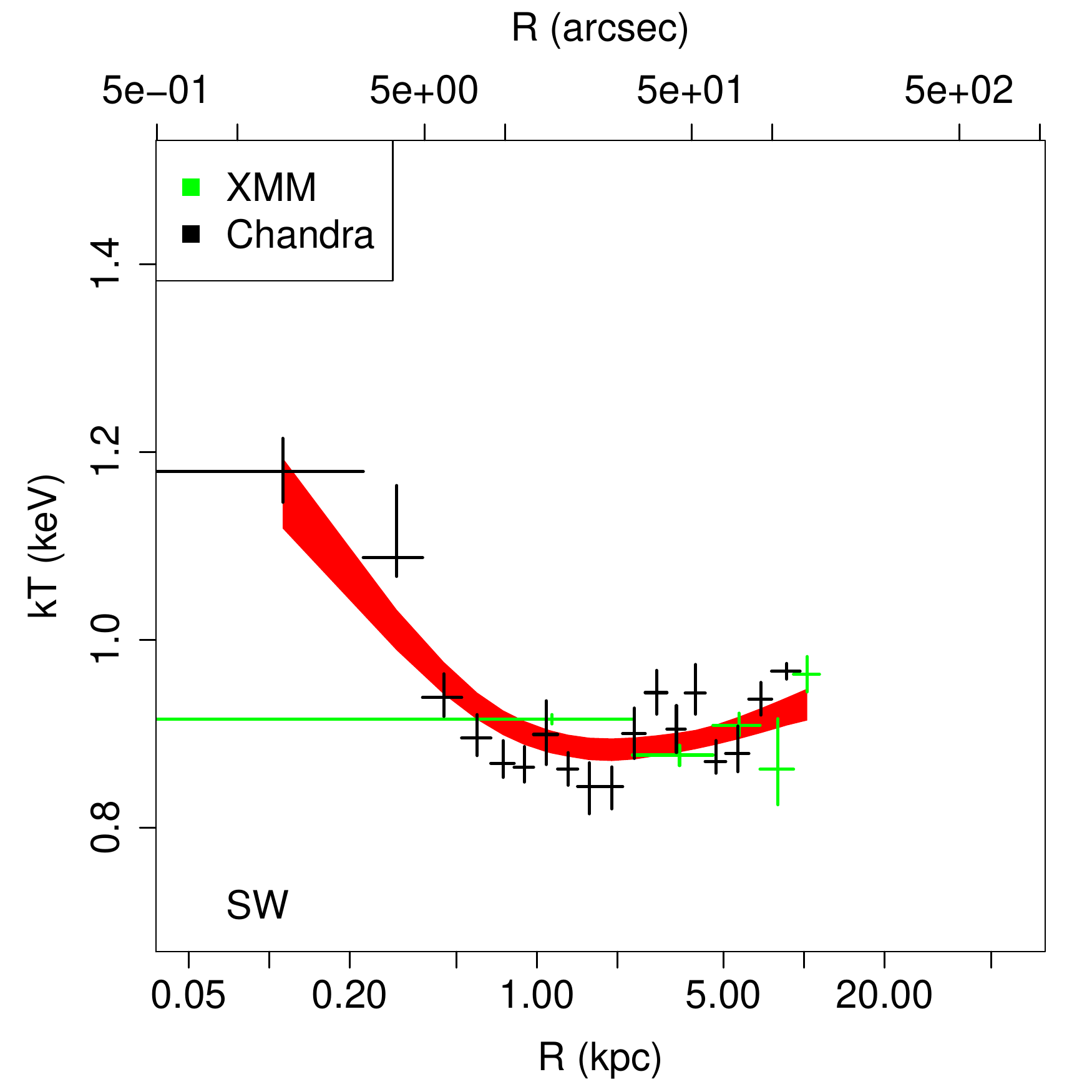}
\includegraphics[scale=0.33]{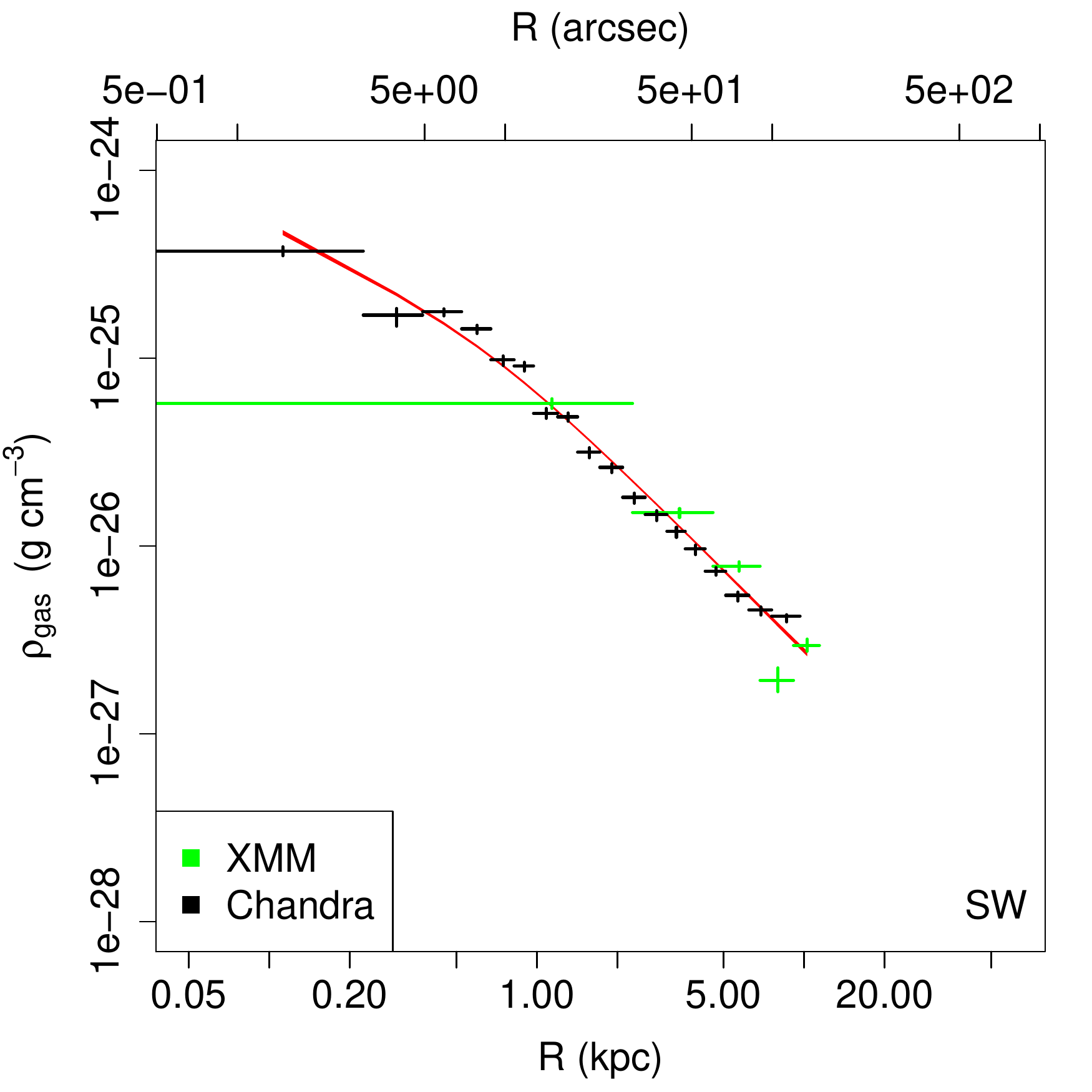}
\includegraphics[scale=0.33]{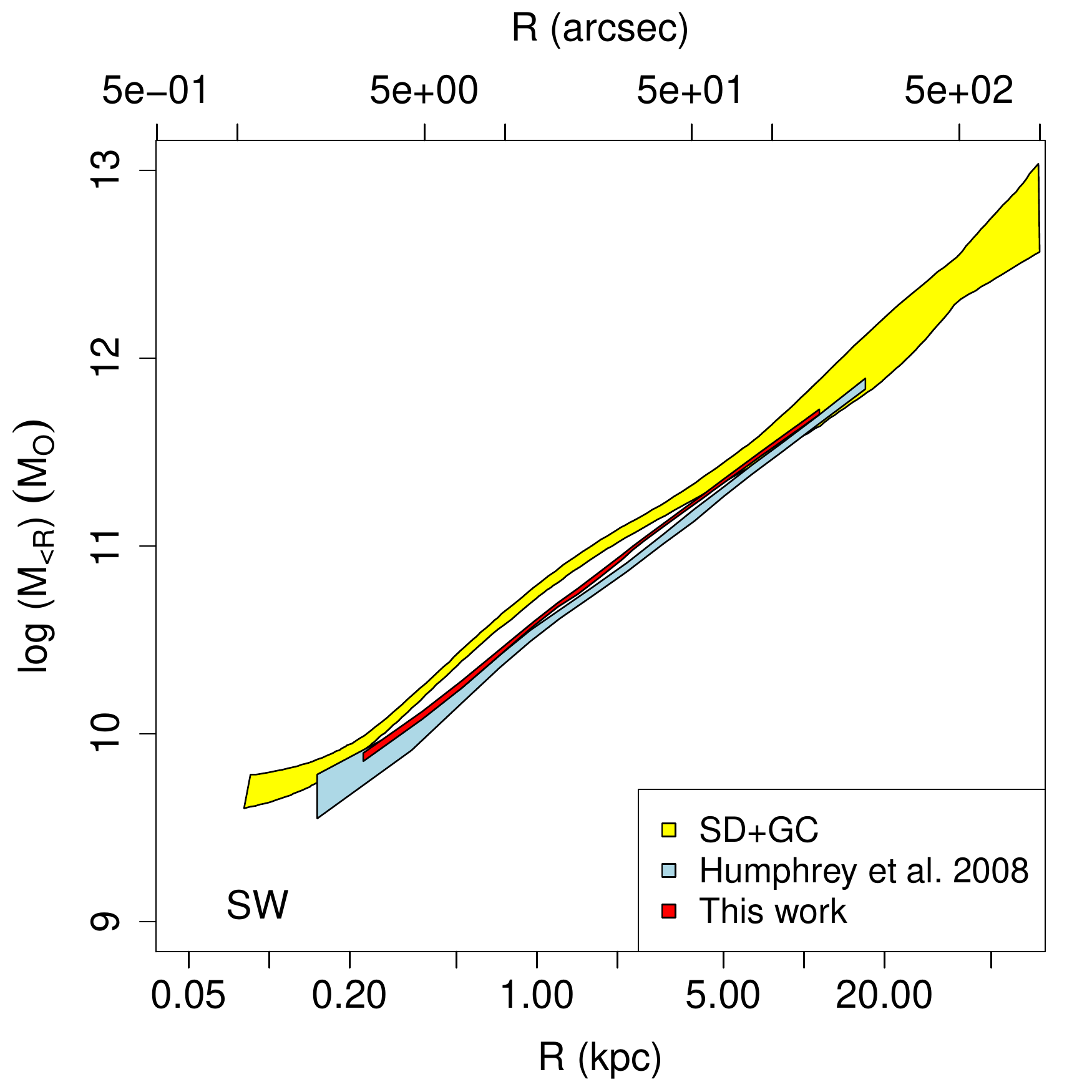}
\caption{Gas profiles obtained in NGC 4649 with the reduction procedure proposed by \citet{2005ApJ...629..172N}. From top to bottom we show the profiles obtained in the full (0-360), SE (90-180) and SW (180-270) sector, respectively. Spectra extracted in the annuli are then simultaneously fitted (separately for \textit{XMM} and \textit{Chandra} data) with the fixed abundance model, and de-projected using \textsc{projct} model. The annuli width is chosen to reach a signal to noise ratio of 30 for \textit{XMM}-MOS data (represented in red) and of 50 for \textit{Chandra} ACIS data (represented in black) with the exception of the full (0-360) sector for which we chose a signal to noise ratio of 50 for \textit{XMM} data and 100 for \textit{Chandra} data. Best fits of a smooth cubic spline are presented in red, with smoothing parameter from top to bottom of 0.7, 0.8 and 0.8. In each row we show the gas temperature (first column) and density (second column) profiles. In the third column we present in red the total mass profiles obtained by mean of Eq. \ref{eq:hee} from the best fits to gas temperature and density profiles. In the same panels we show in yellow the optical mass profile obtained from SD and GC reported by \citep{2010ApJ...711..484S}, and in light blue the X-ray mass profiles obtained by \citet{2008ApJ...683..161H}.}\label{fig:N4649_gas_profiles_merged}
\end{figure}

\begin{figure}
\centering
\includegraphics[scale=0.33]{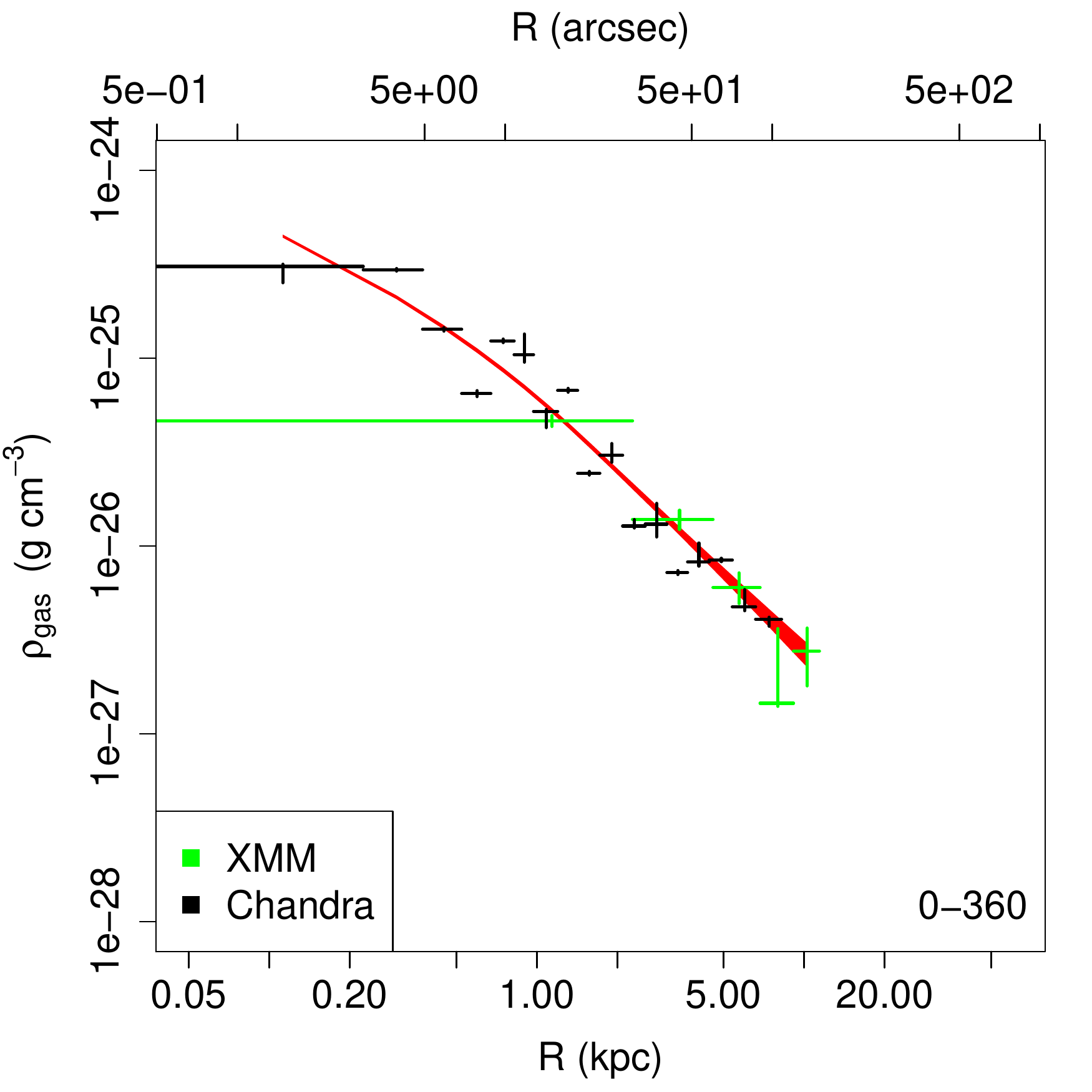}
\includegraphics[scale=0.33]{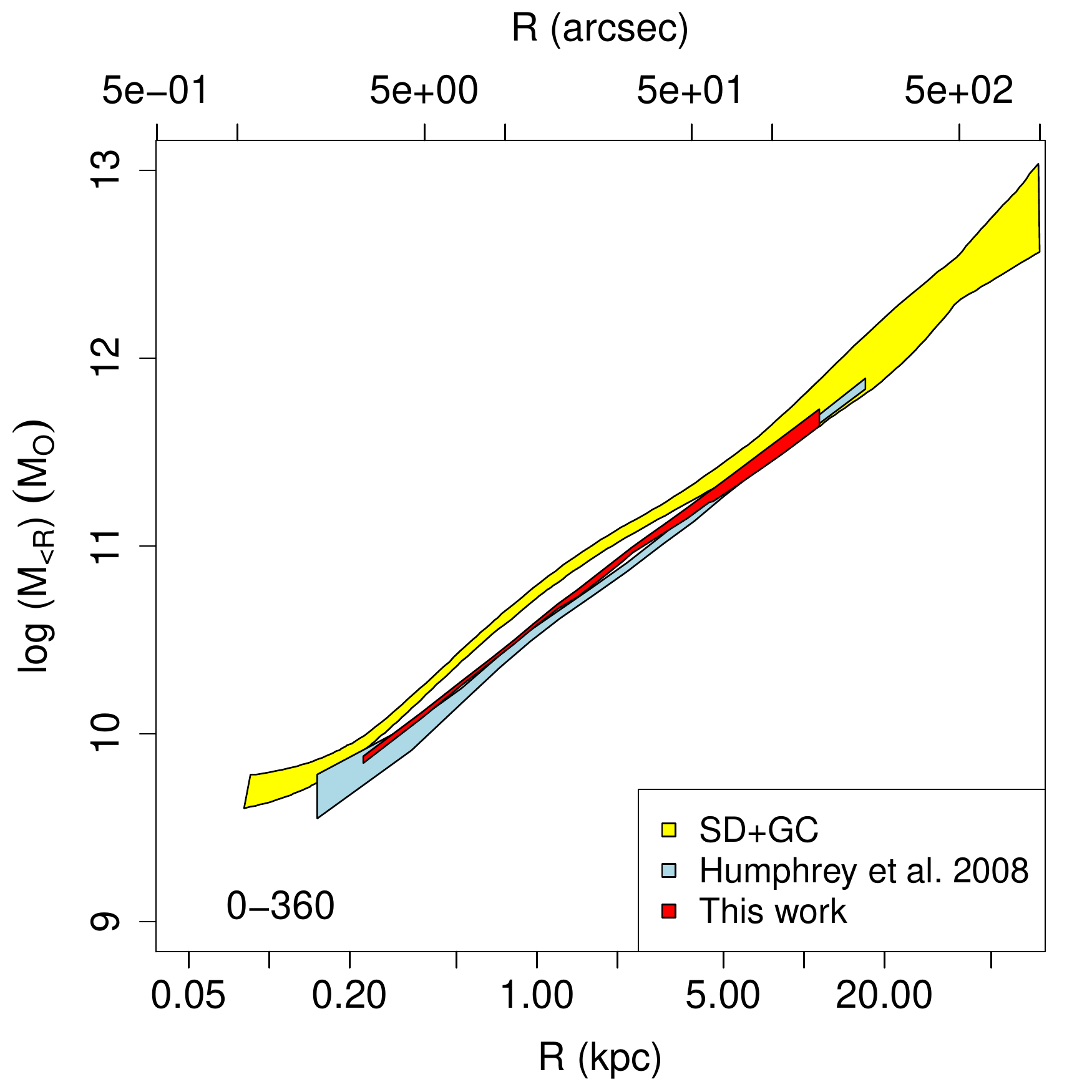}
\includegraphics[scale=0.33]{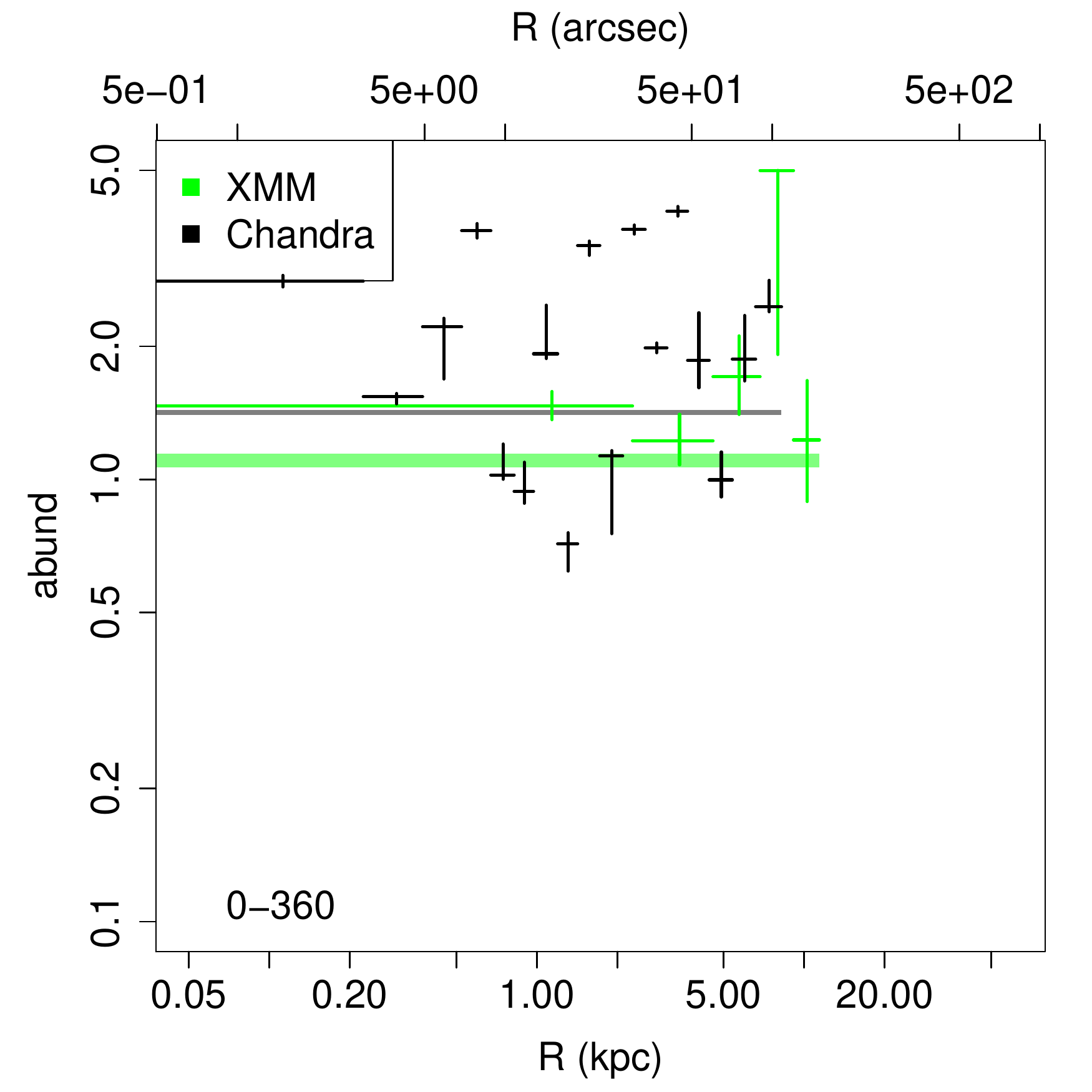}\\
\includegraphics[scale=0.33]{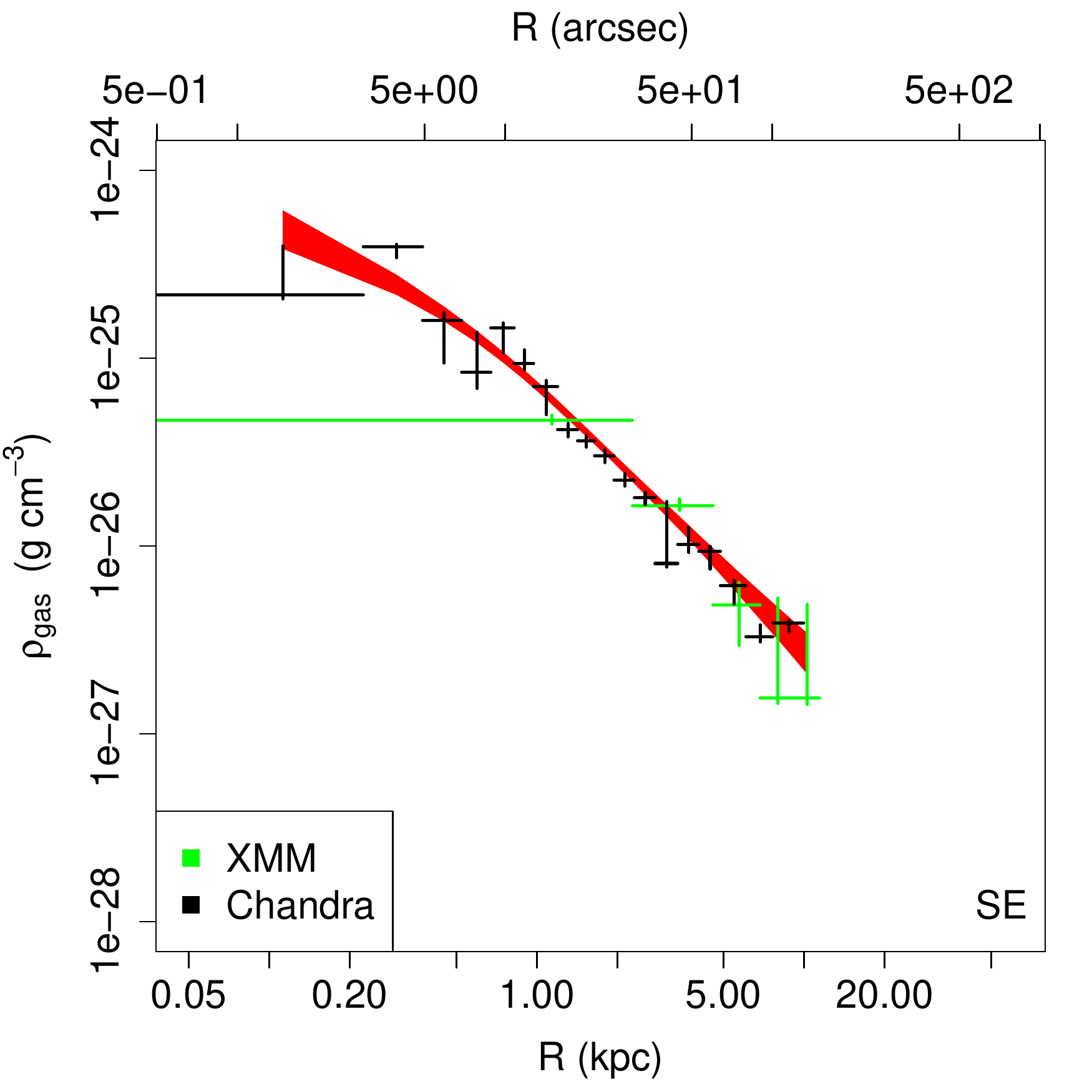}
\includegraphics[scale=0.33]{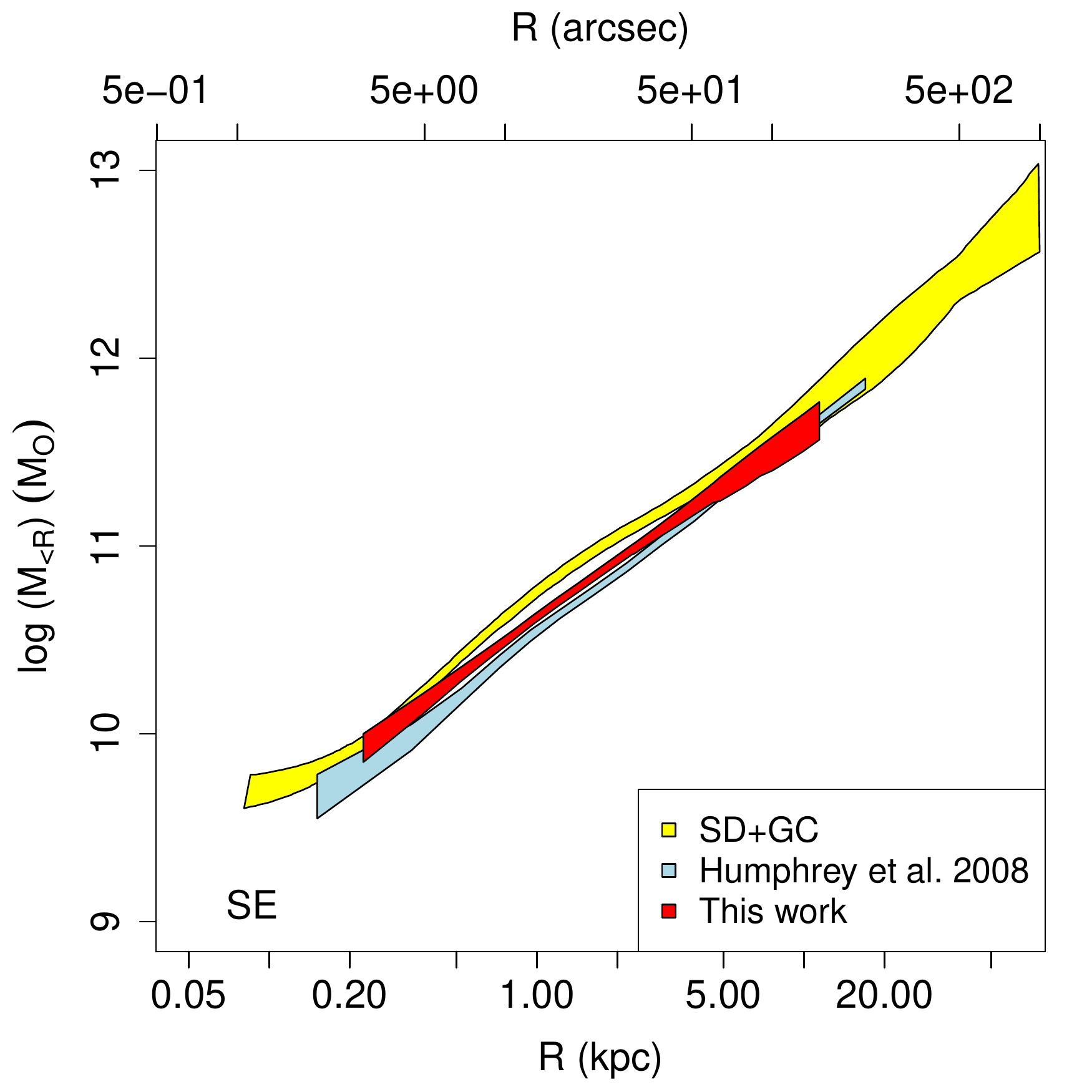}
\includegraphics[scale=0.33]{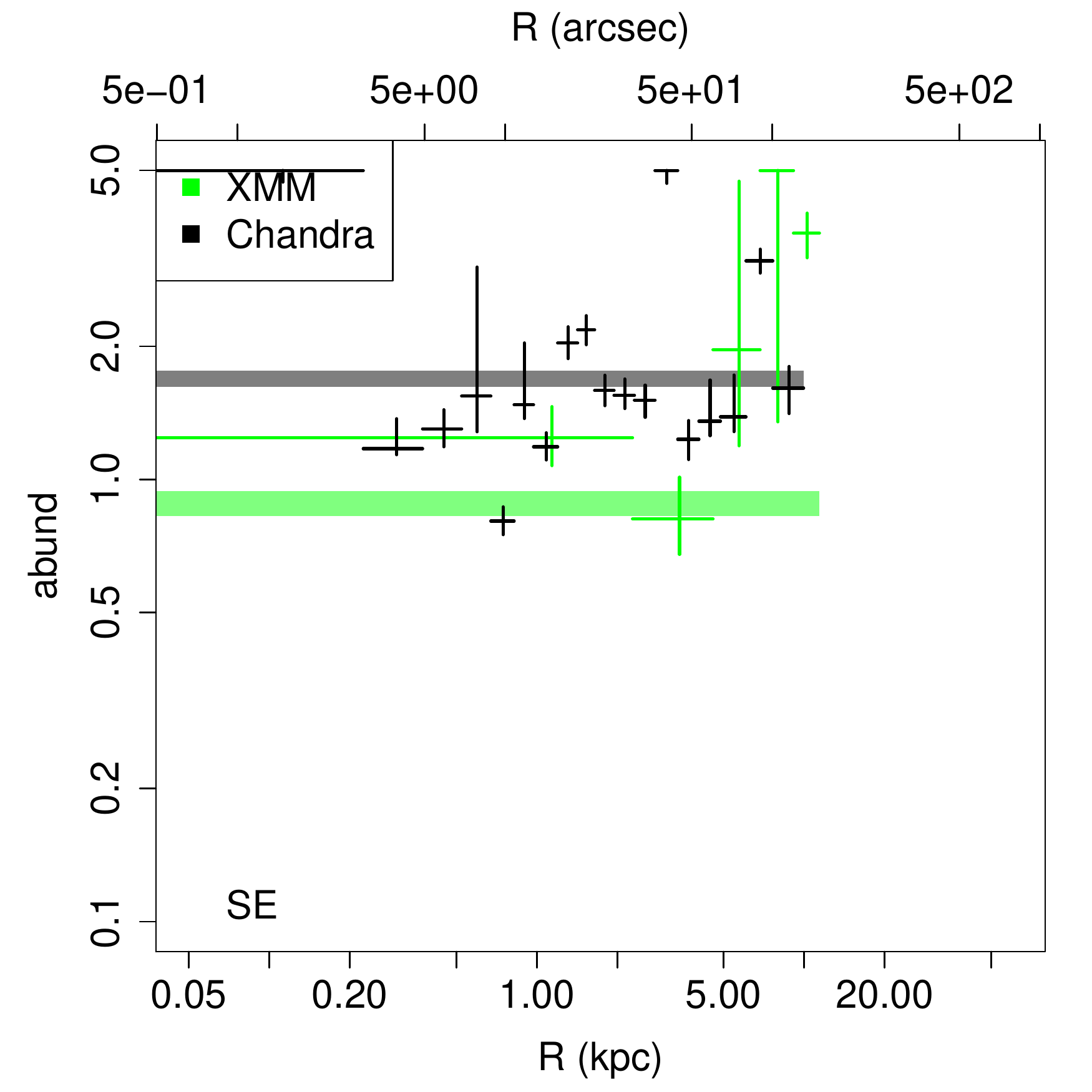}\\
\includegraphics[scale=0.33]{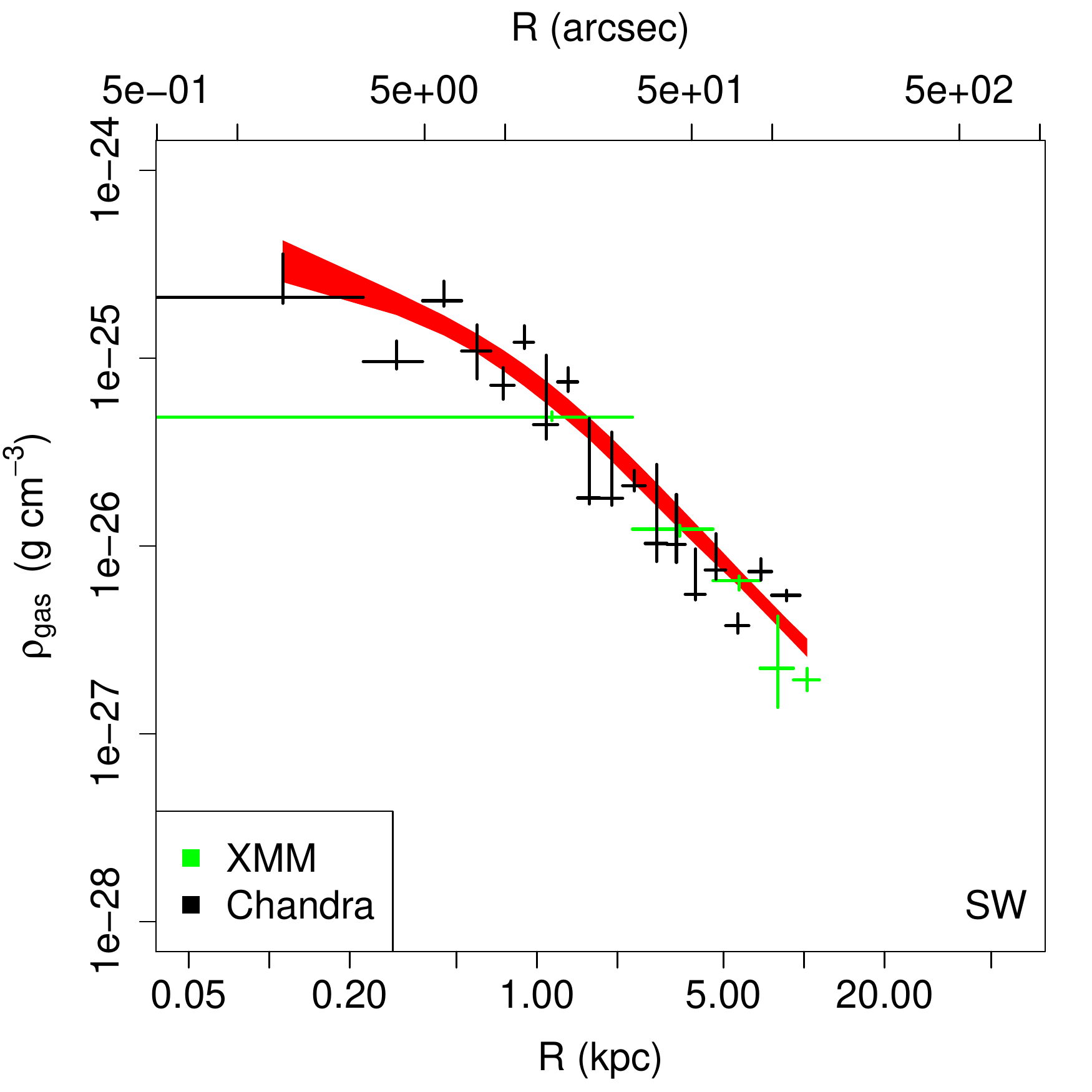}
\includegraphics[scale=0.33]{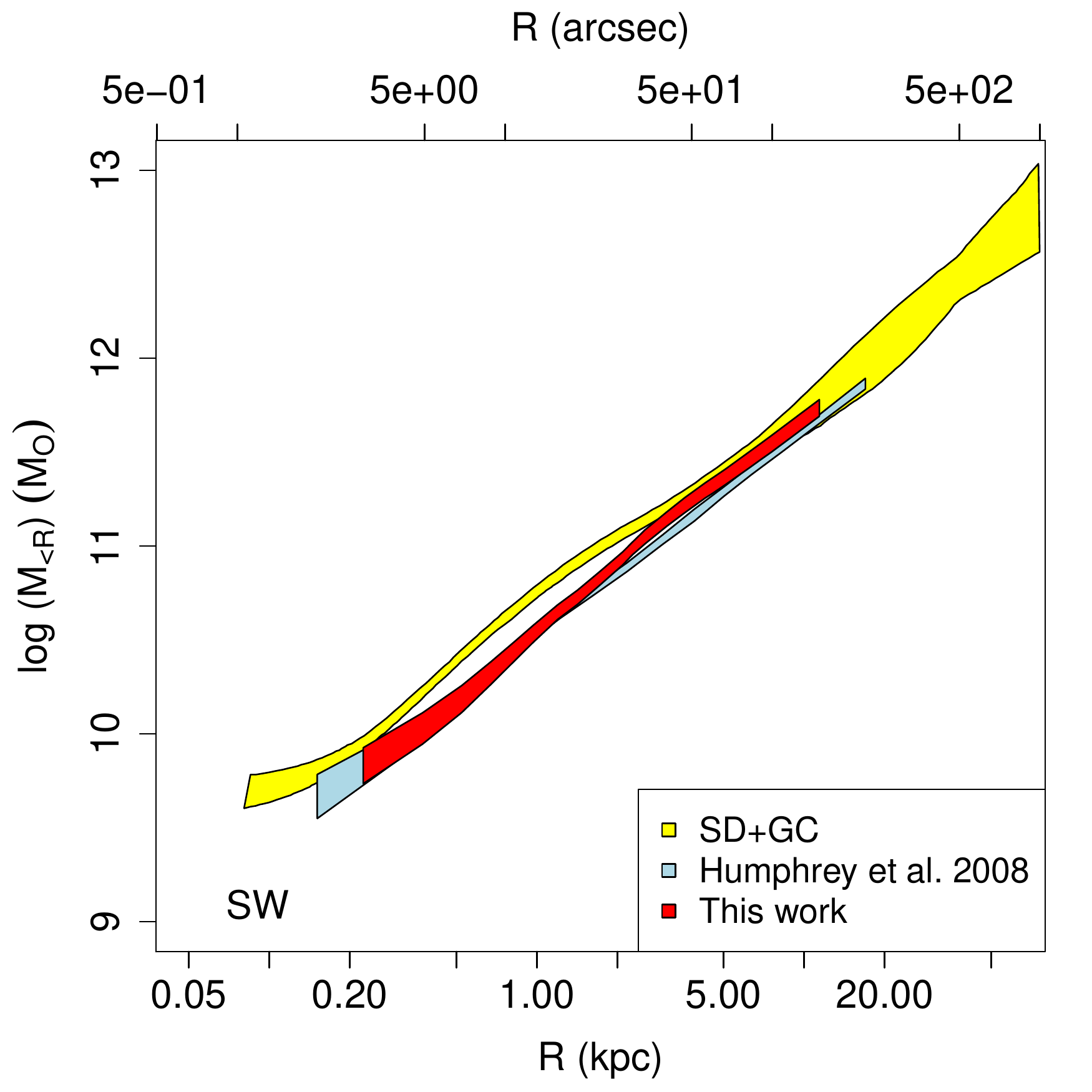}
\includegraphics[scale=0.33]{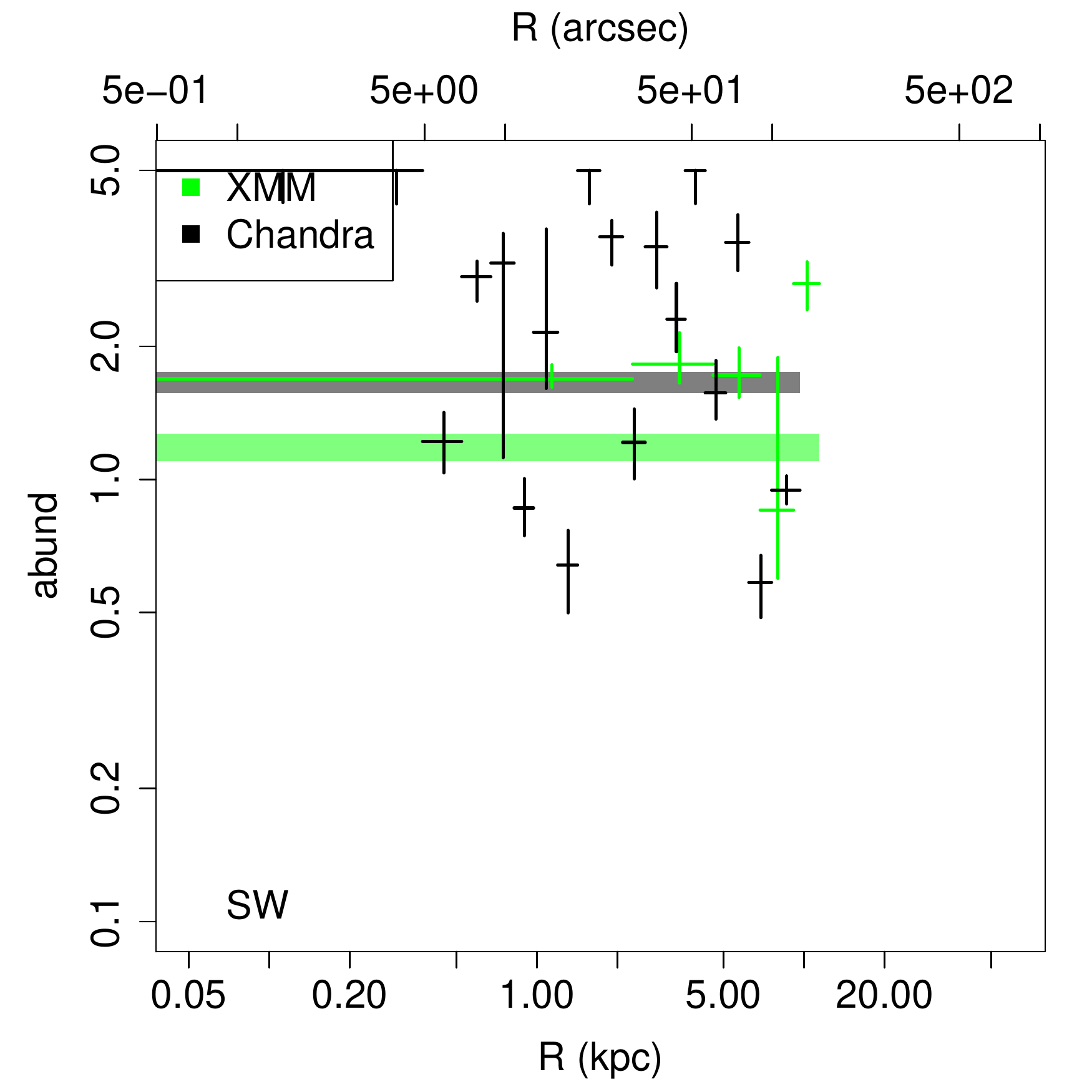}
\caption{Same as Figure \ref{fig:N4649_gas_profiles_merged} but with the free abundance model (temperature profiles are similar to the fixed abundance model case and therefore omitted). In addition, on the rightmost panel of each row we show the element abundances profiles for \textit{XMM}-MOS data (in green) and for \textit{Chandra} ACIS data (represented in black). In the same panels we overplot with green and black rectangles the values of the element abundances obtained with the fixed abundance model for \textit{XMM} and \textit{Chandra} data, respectively.}\label{fig:N4649_gas_profiles_merged_abund}
\end{figure}

\begin{figure}
\centering
\includegraphics[scale=0.50]{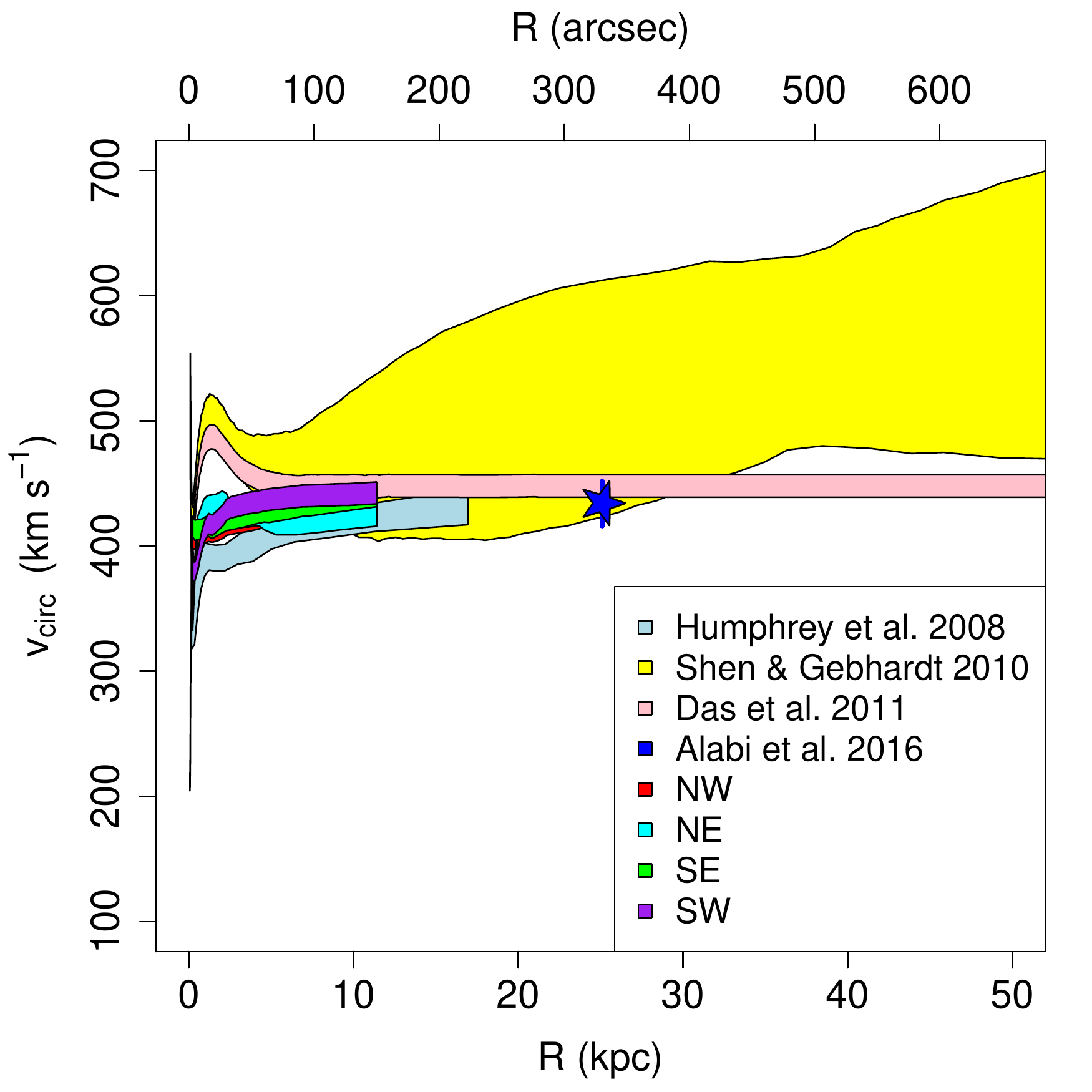}
\includegraphics[scale=0.50]{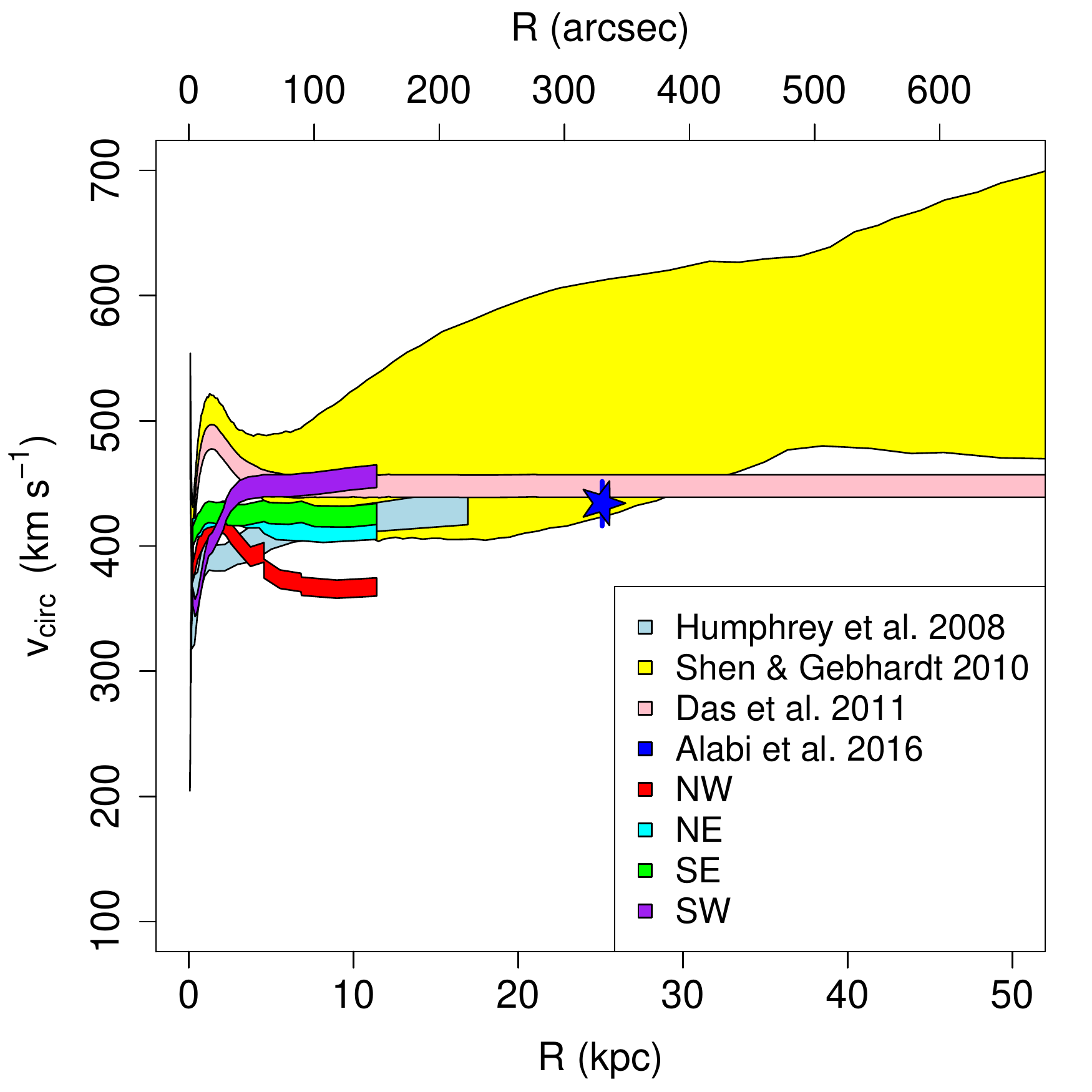}
\caption{{(left) Comparison of velocity profiles of NGC 4649 from fixed abundance models. The profiles for different angular sectors are shown in different colors indicated in the legend, while the X-ray profile presented by \citet{2008ApJ...683..161H} is shown in light blue, the SD+GC profile presented by \citet{2010ApJ...711..484S} is shown in yellow, the profile from \citet{2011MNRAS.415.1244D} obtained with X-ray and PNe data is presented in pink, and the blue star represents the measurement by \citet{2016MNRAS.460.3838A} with the corresponding uncertainty shown with a vertical blue line. (right) Same as left panel, but for free abundance models.}}\label{fig:N4649_velocity_profiles_all}
\end{figure}

\begin{figure}
\centering
\includegraphics[scale=0.33]{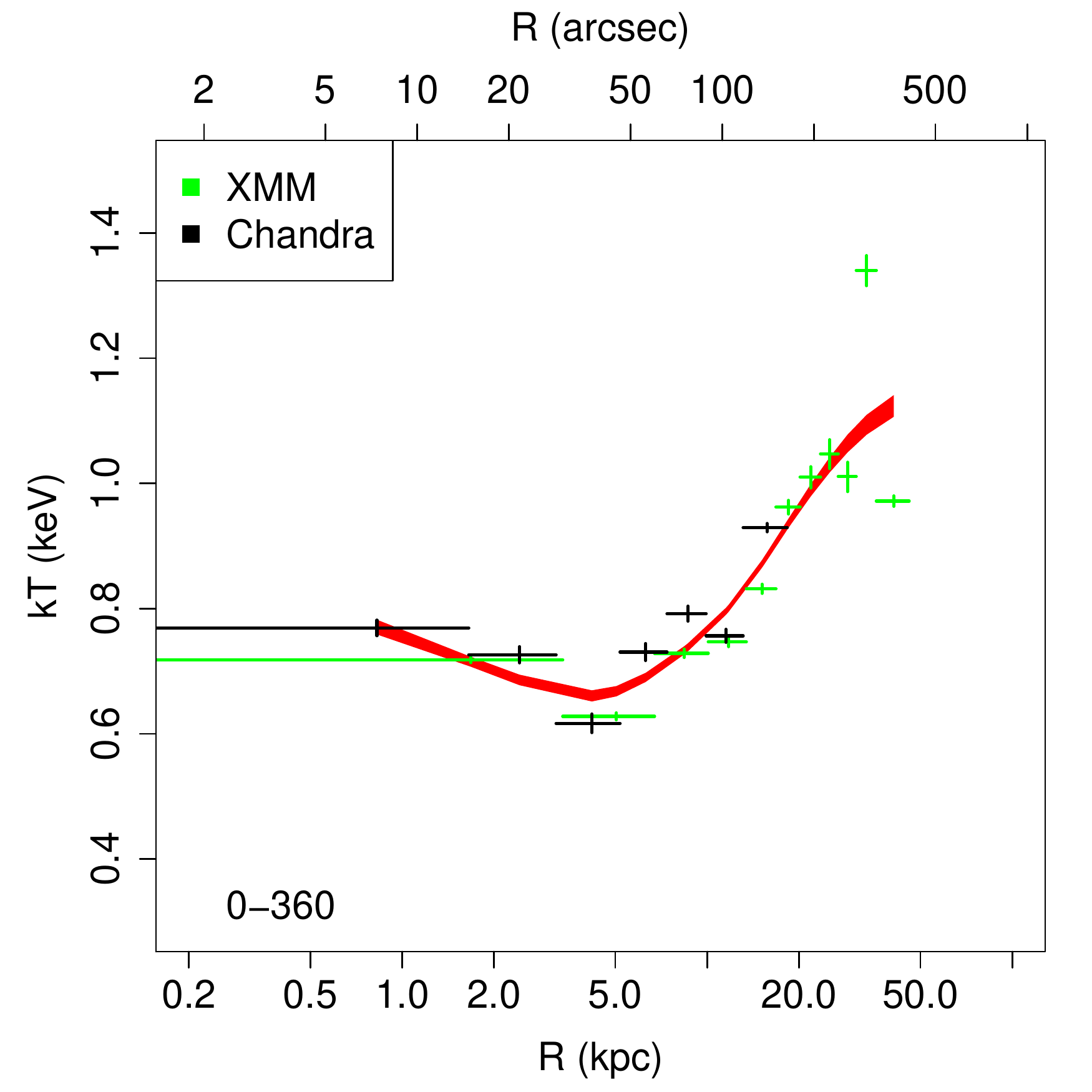}
\includegraphics[scale=0.33]{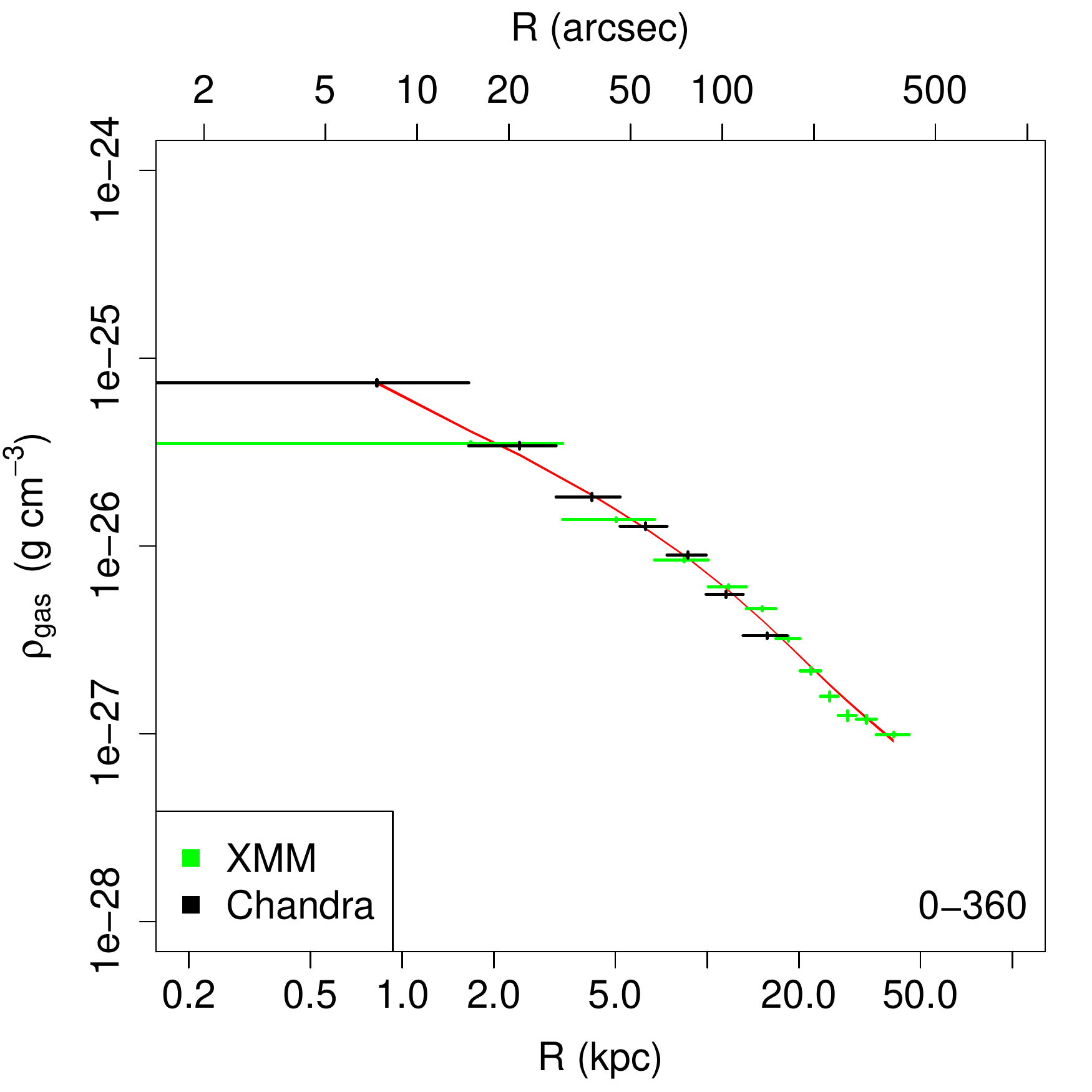}
\includegraphics[scale=0.33]{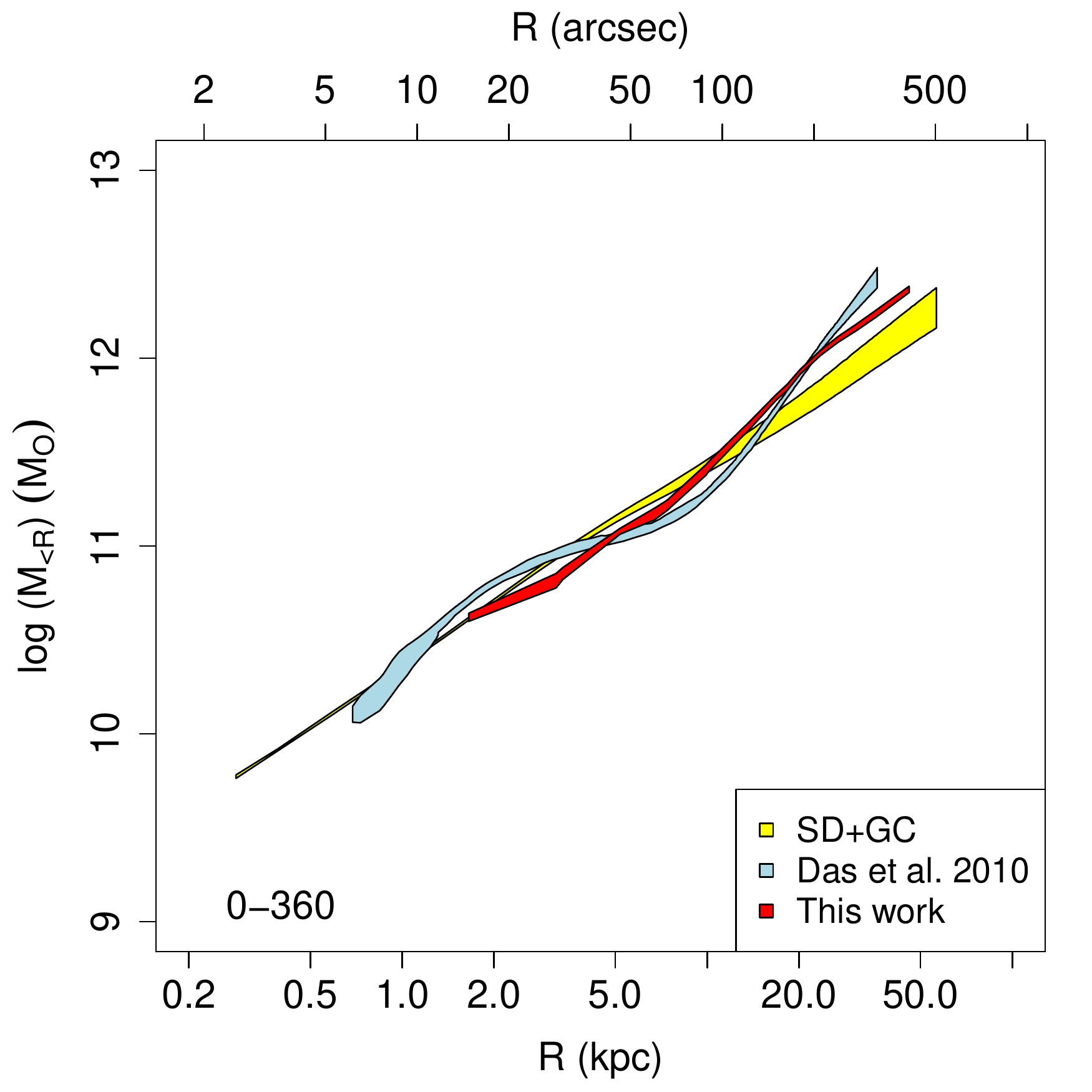}\\
\includegraphics[scale=0.33]{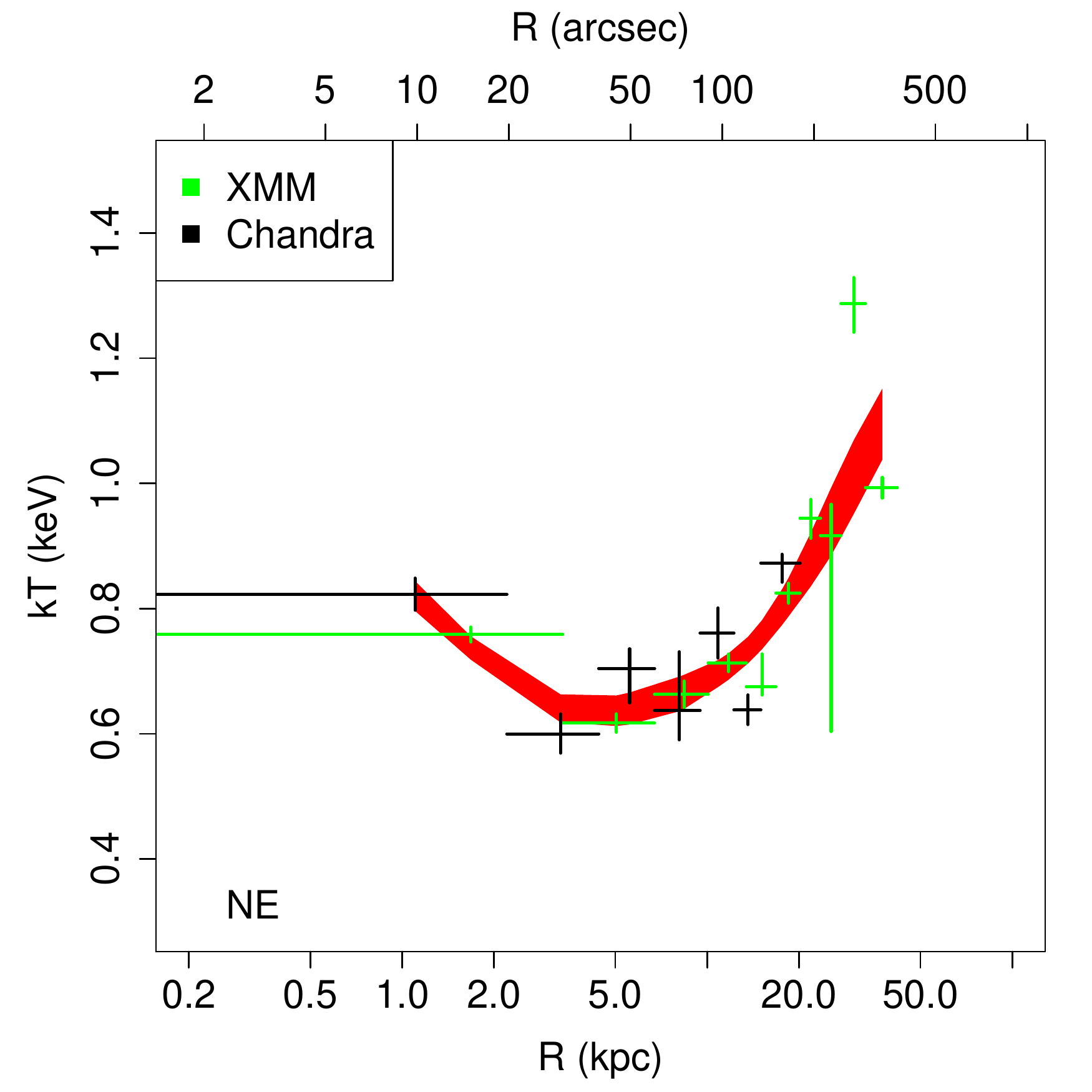}
\includegraphics[scale=0.33]{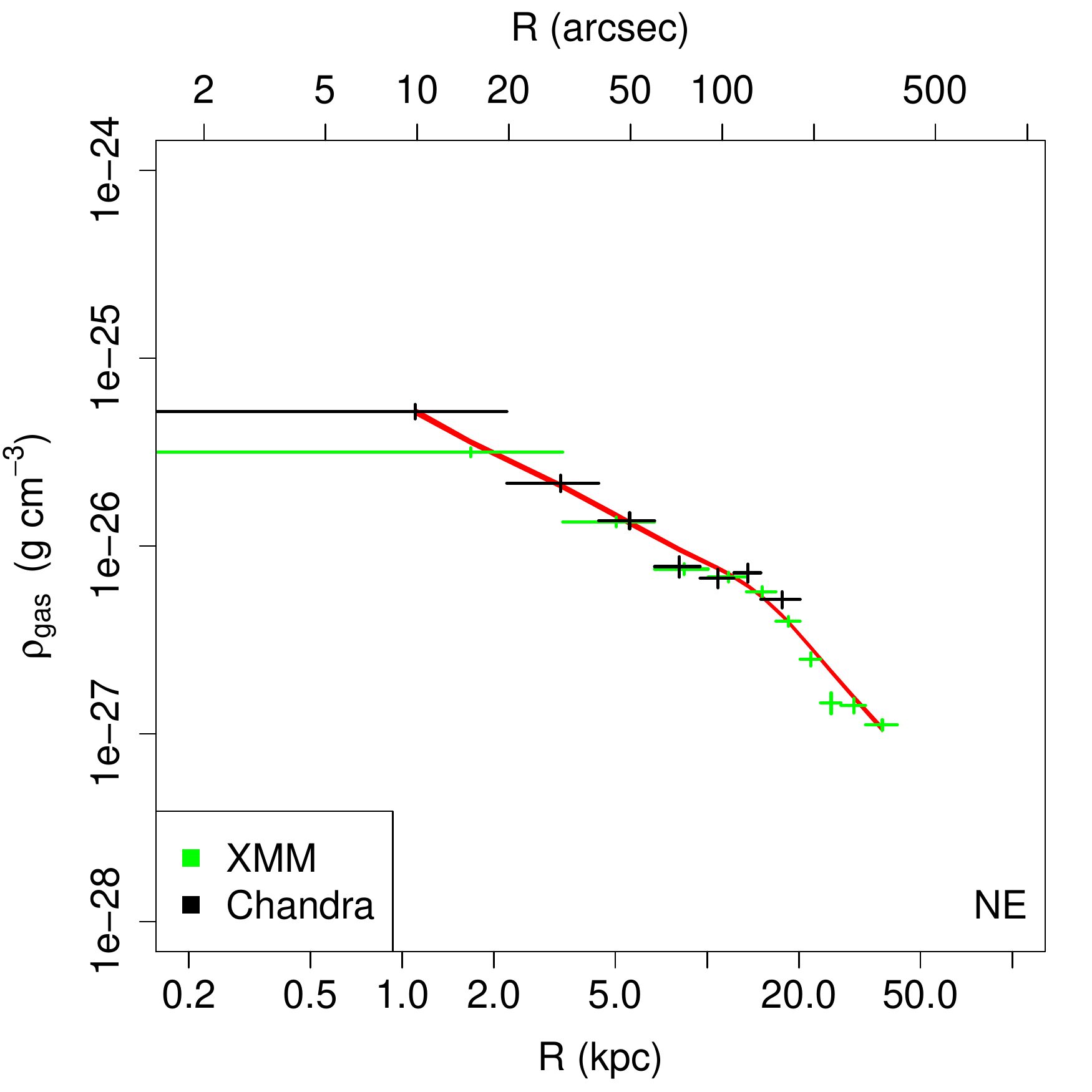}
\includegraphics[scale=0.33]{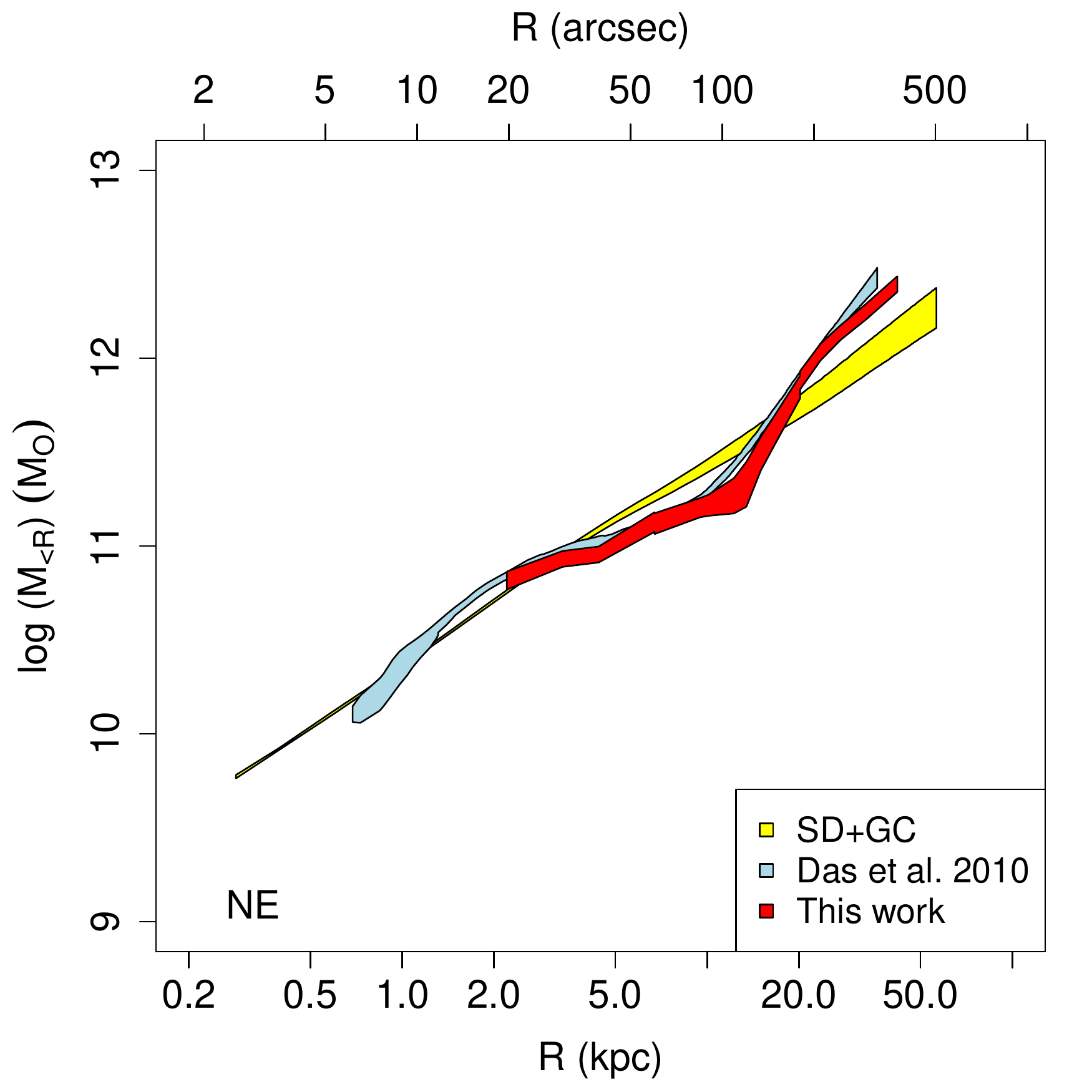}\\
\includegraphics[scale=0.33]{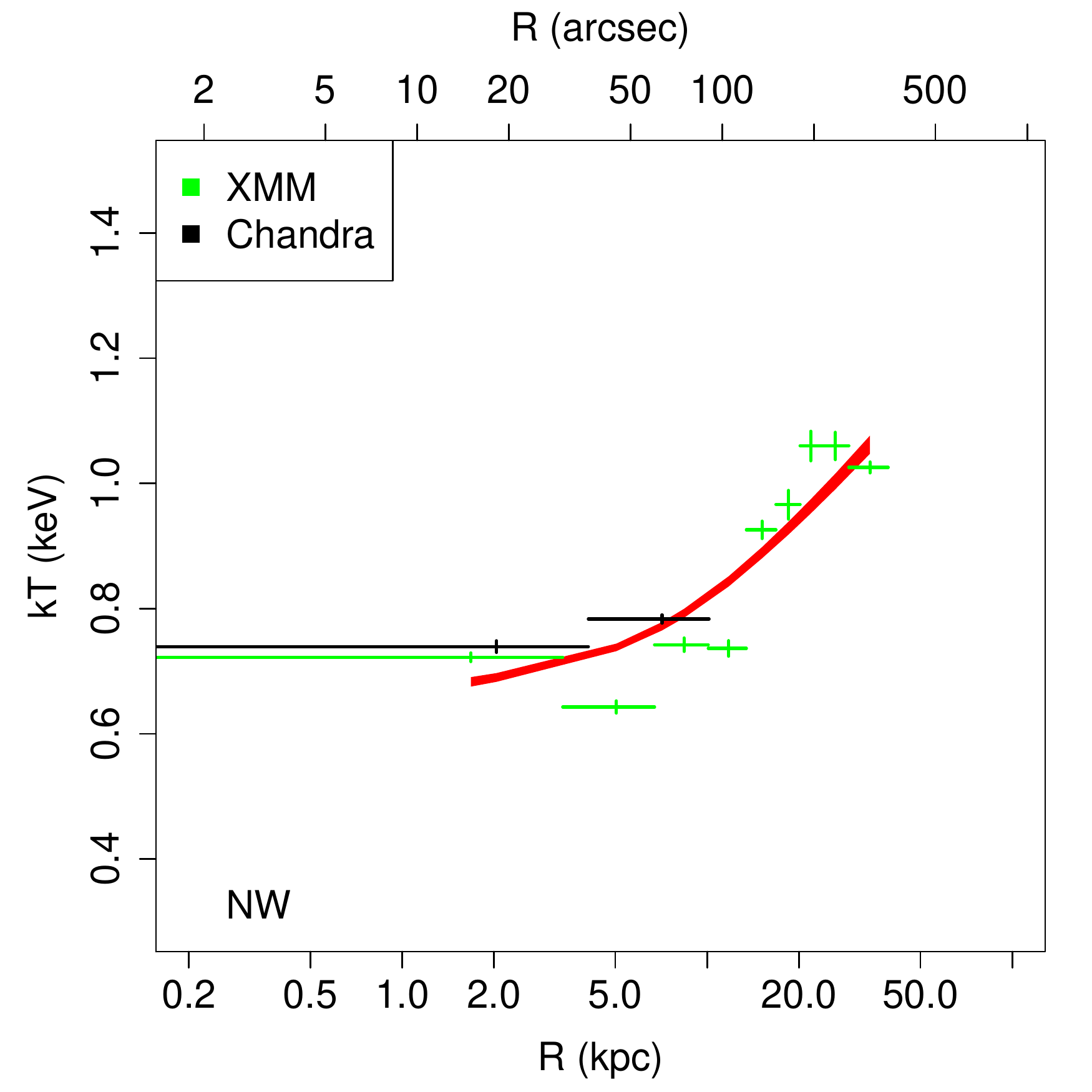}
\includegraphics[scale=0.33]{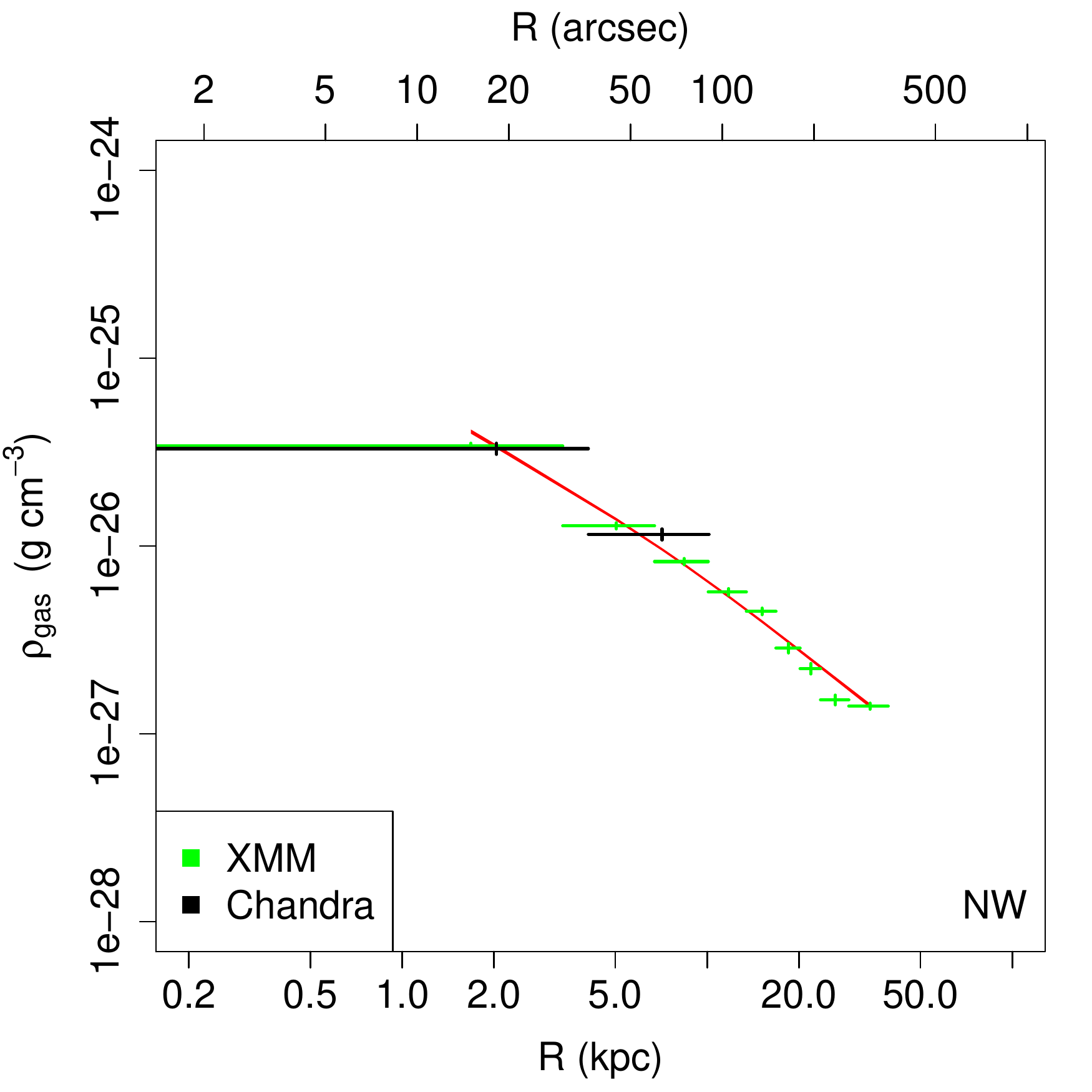}
\includegraphics[scale=0.33]{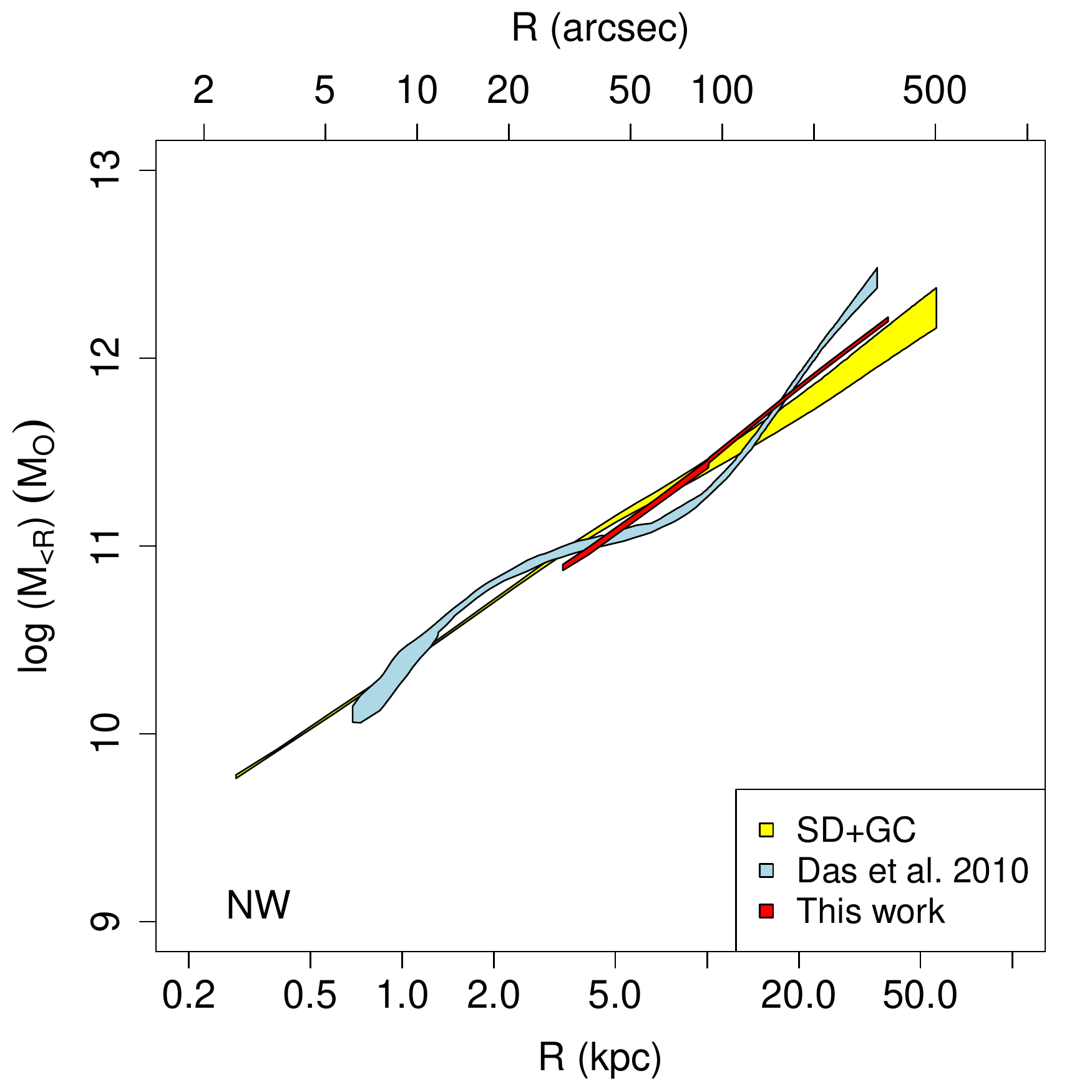}
\caption{Gas profiles obtained in NGC 5846 with the reduction procedure proposed by \citet{2005ApJ...629..172N}. From top to bottom we show the profiles obtained in the full (0-360), NE (30-90) and NW (250-30) sector, respectively. Spectra extracted in the annuli are then simultaneously fitted (separately for \textit{XMM} and \textit{Chandra} data) with the fixed abundance model, and de-projected using \textsc{projct} model. The annuli width is chosen to reach a signal to noise ratio of 50, 30 and 50 for \textit{XMM}-MOS data (represented in red) and of 100, 50 and 50 for \textit{Chandra} ACIS data (represented in black) for the full (0-360), NE (30-90) and NW (250-30) sectors, respectively. Best fits of a smooth cubic spline are presented in red, with smoothing parameter from top to bottom of 0.6, 0.6 and 0.7. In each row we show the gas temperature (first column) and density (second column) profiles. In the third column we present in red the total mass profiles obtained by mean of Eq. \ref{eq:hee} from the best fits to gas temperature and density profiles. In the same panels we show in yellow the optical mass profile obtained from SD and GC reported by \citep{2014MNRAS.439..659N}, and in light blue the X-ray mass profiles obtained by \citet{2010MNRAS.409.1362D}.}\label{fig:N5846_gas_profiles_merged}
\end{figure}

\begin{figure}
\centering
\includegraphics[scale=0.33]{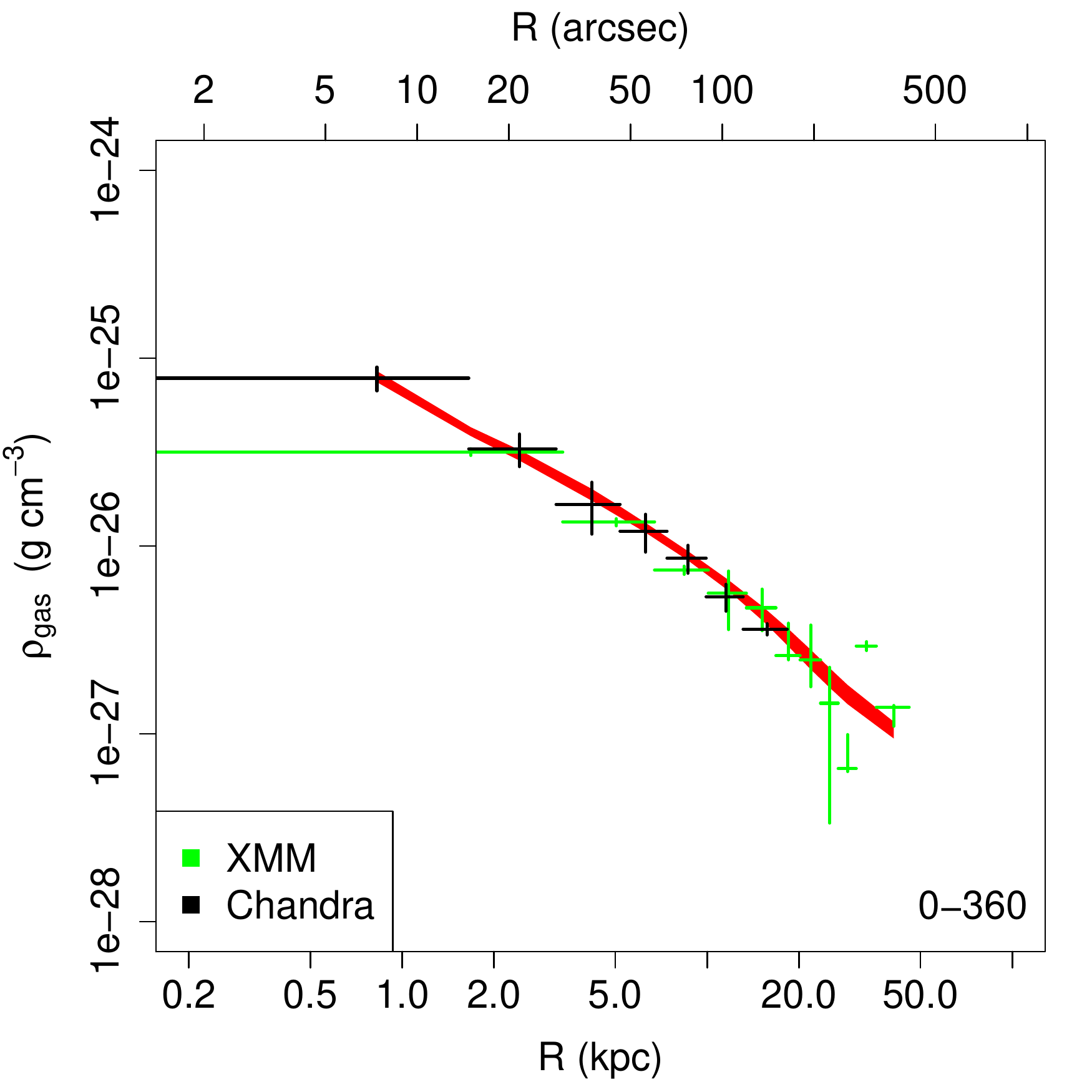}
\includegraphics[scale=0.33]{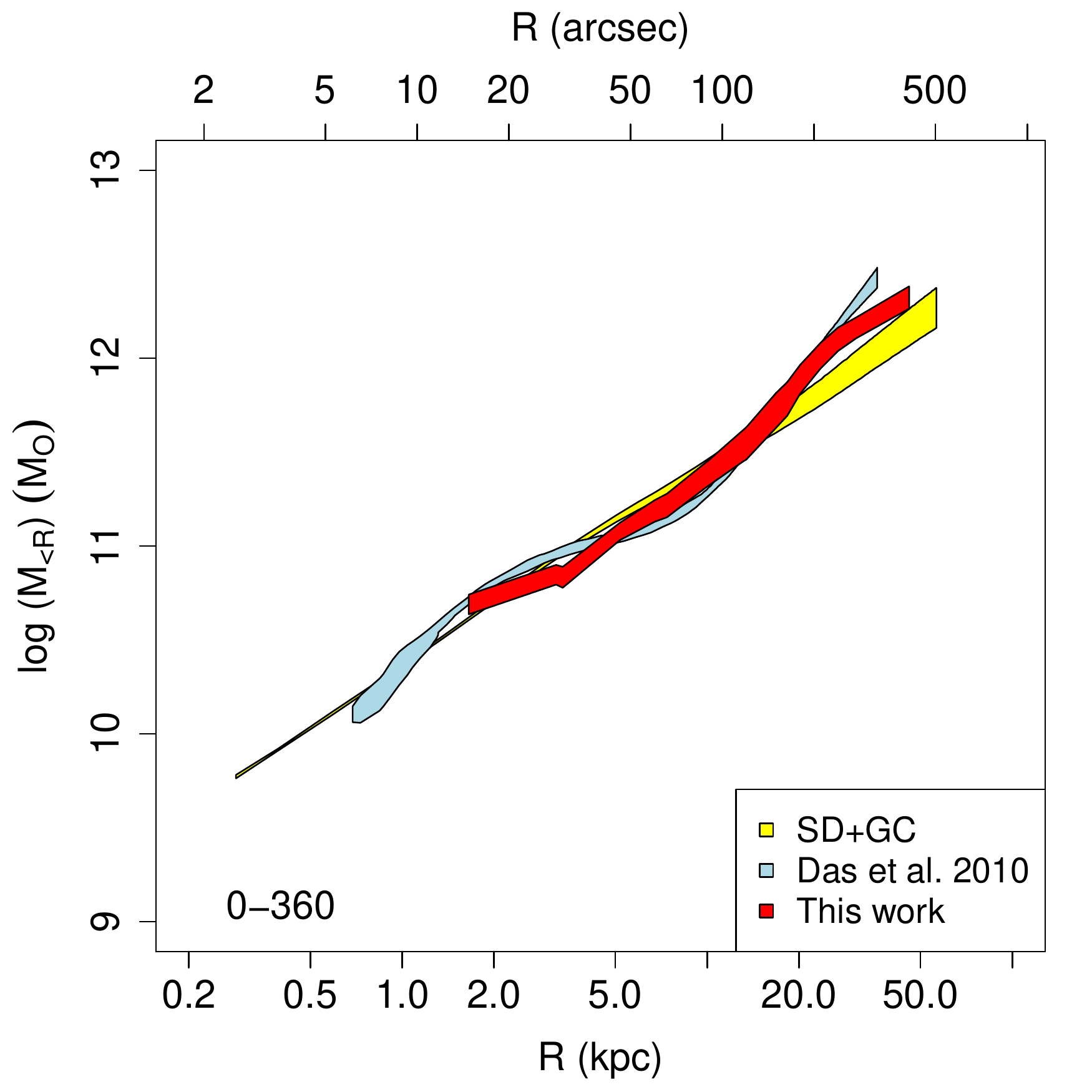}
\includegraphics[scale=0.33]{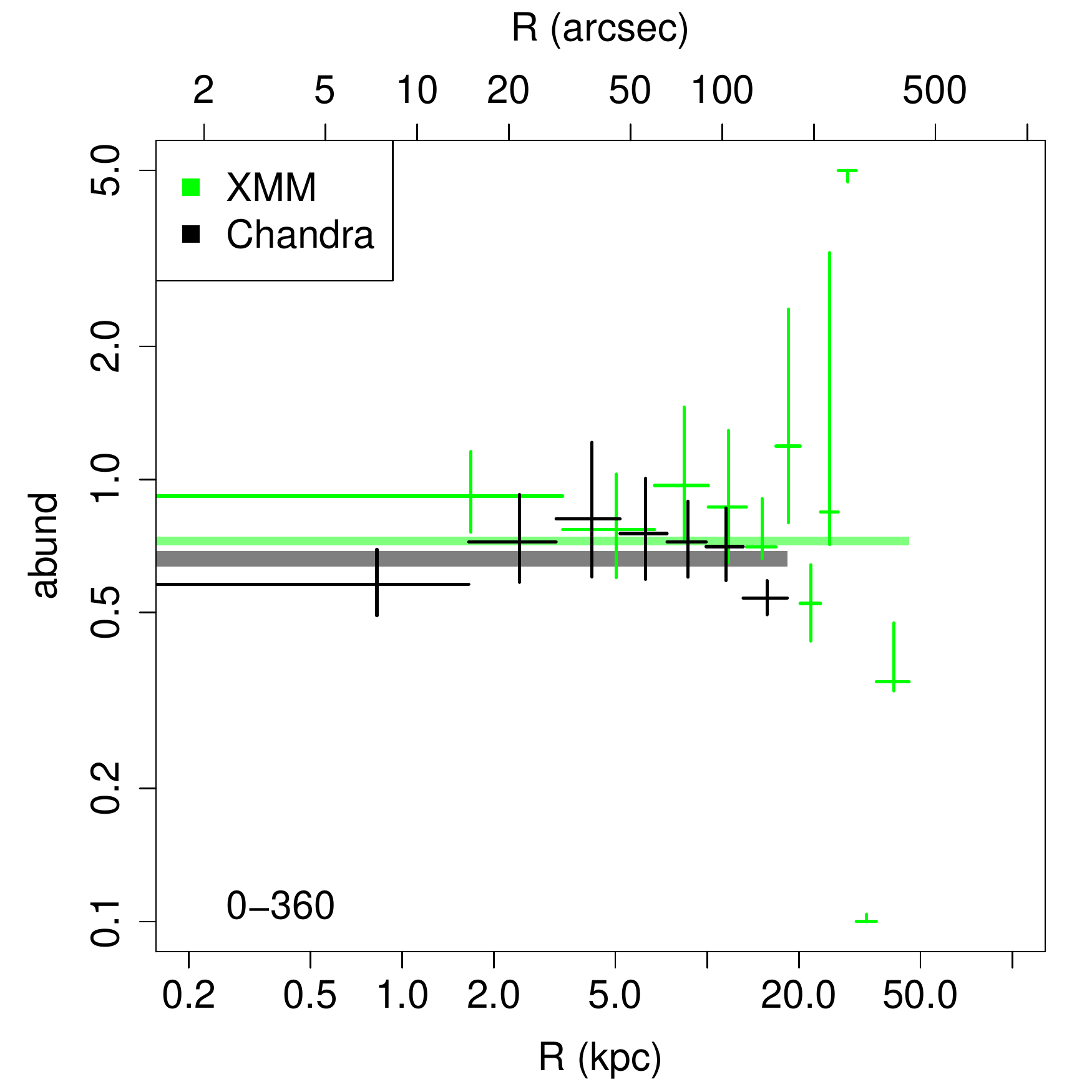}\\
\includegraphics[scale=0.33]{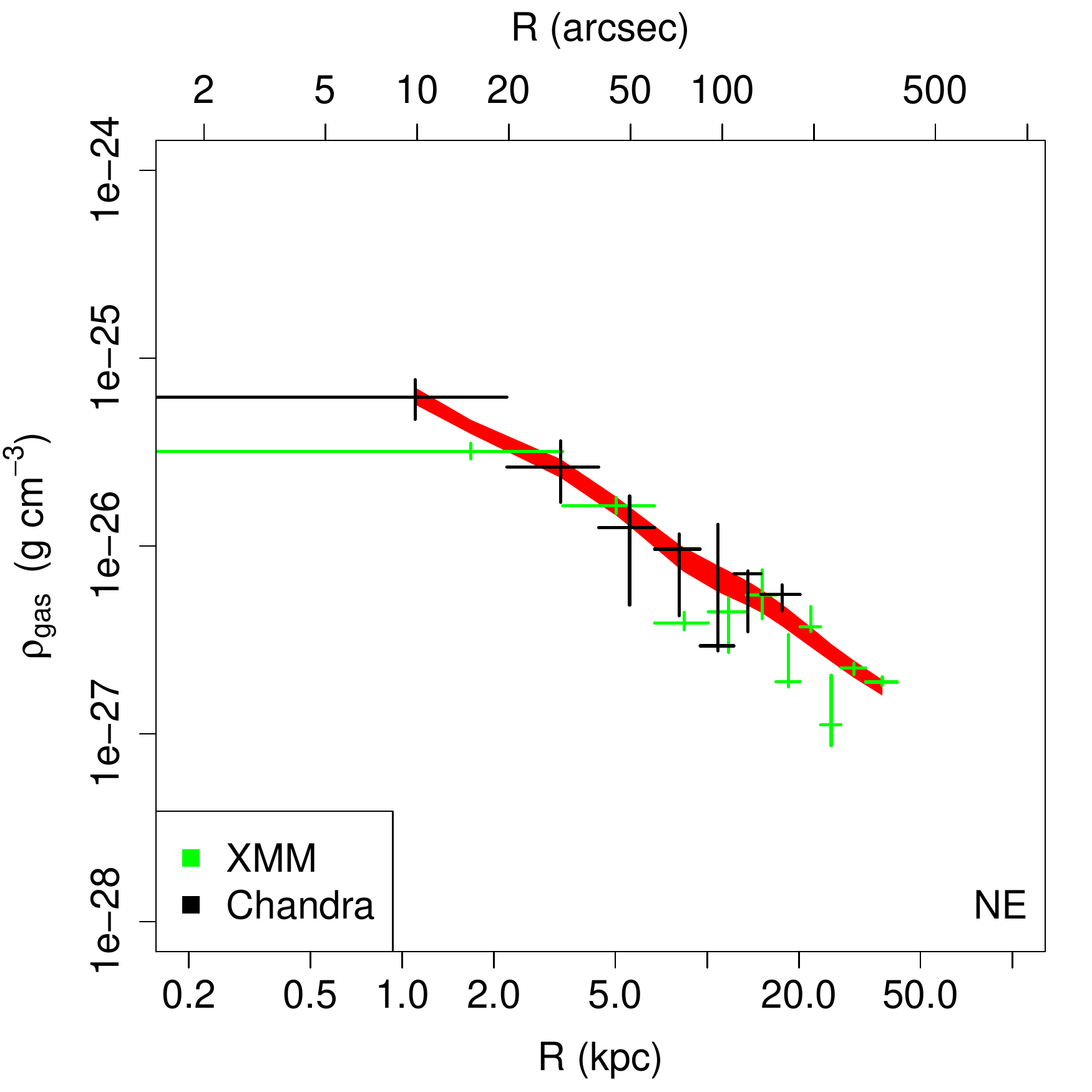}
\includegraphics[scale=0.33]{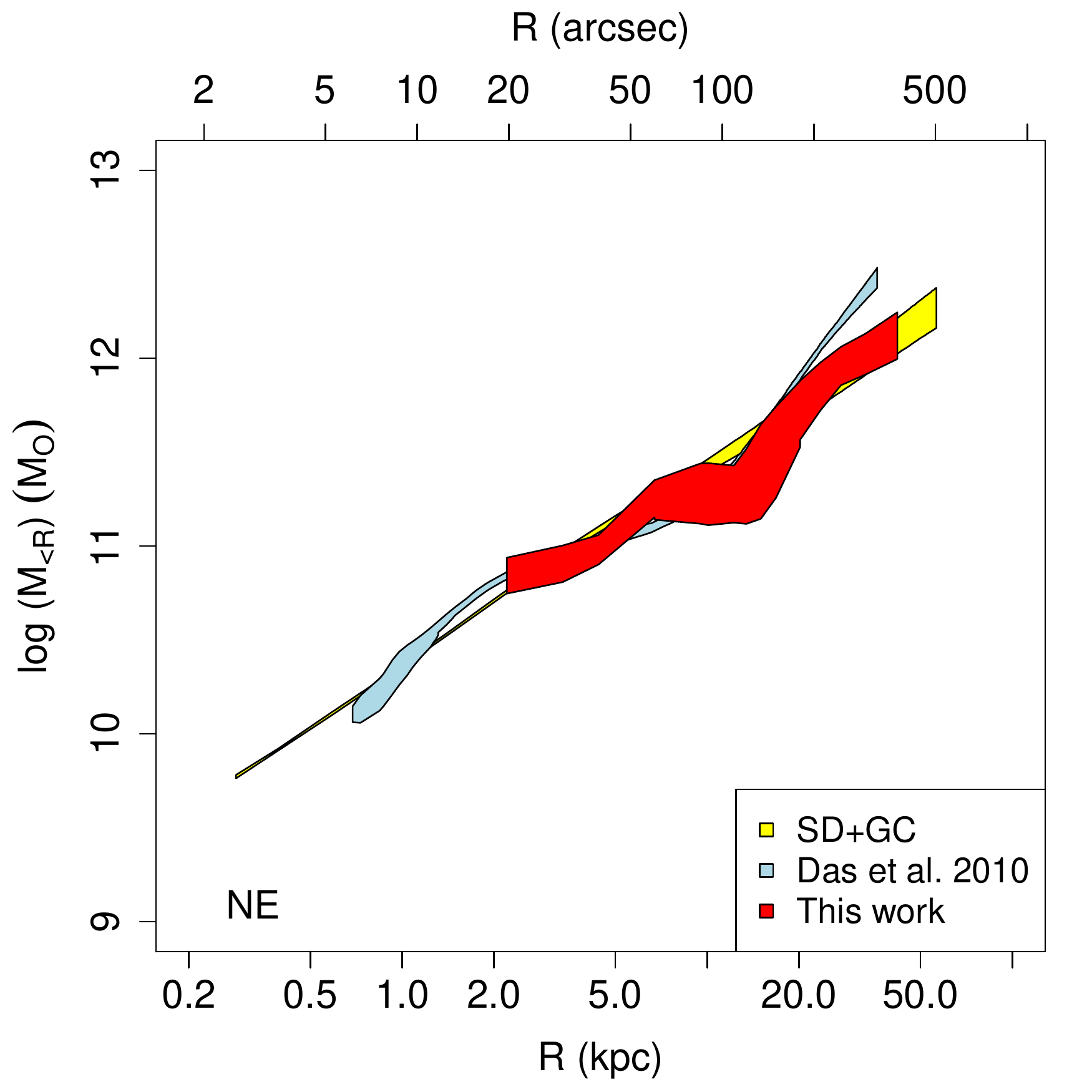}
\includegraphics[scale=0.33]{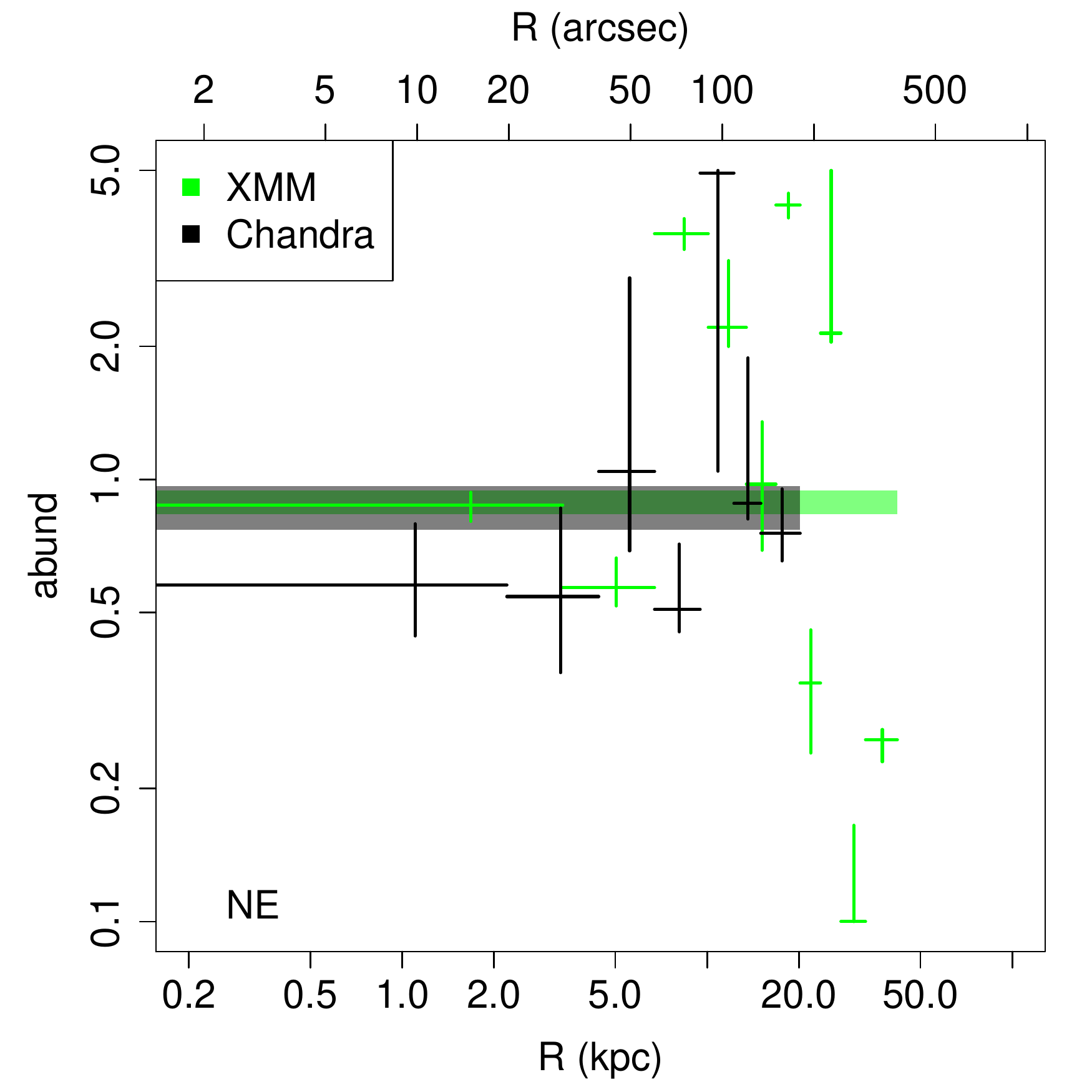}\\
\includegraphics[scale=0.33]{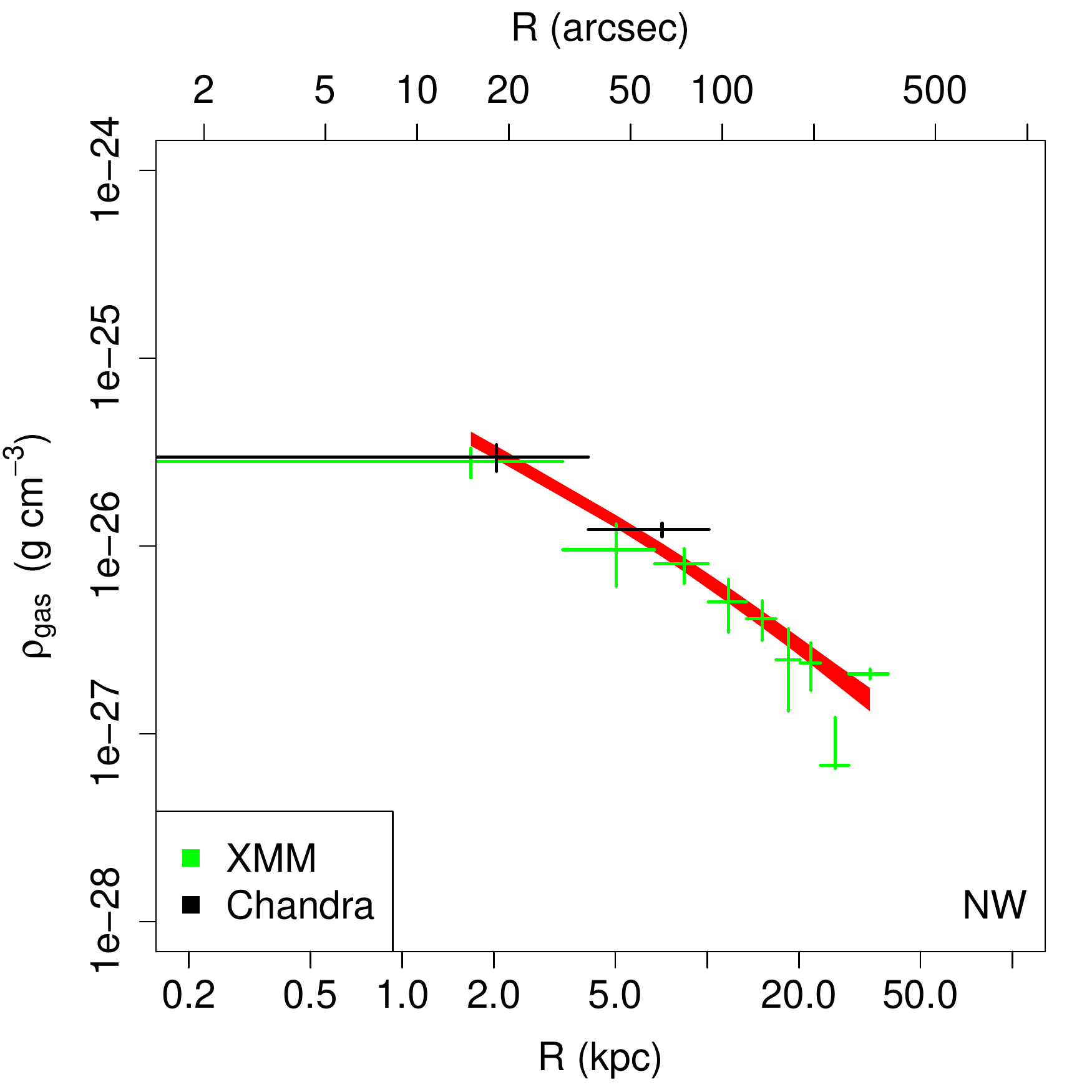}
\includegraphics[scale=0.33]{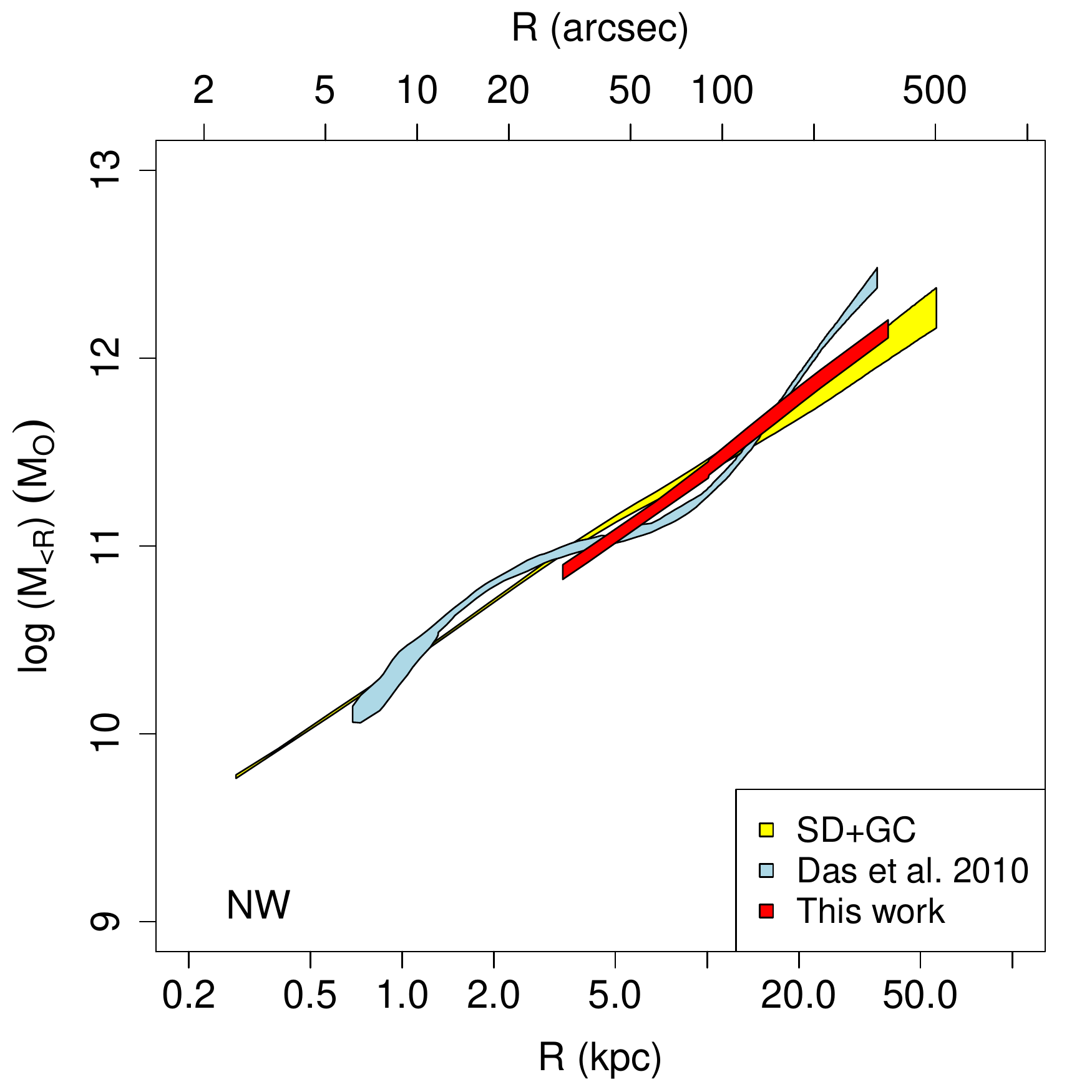}
\includegraphics[scale=0.33]{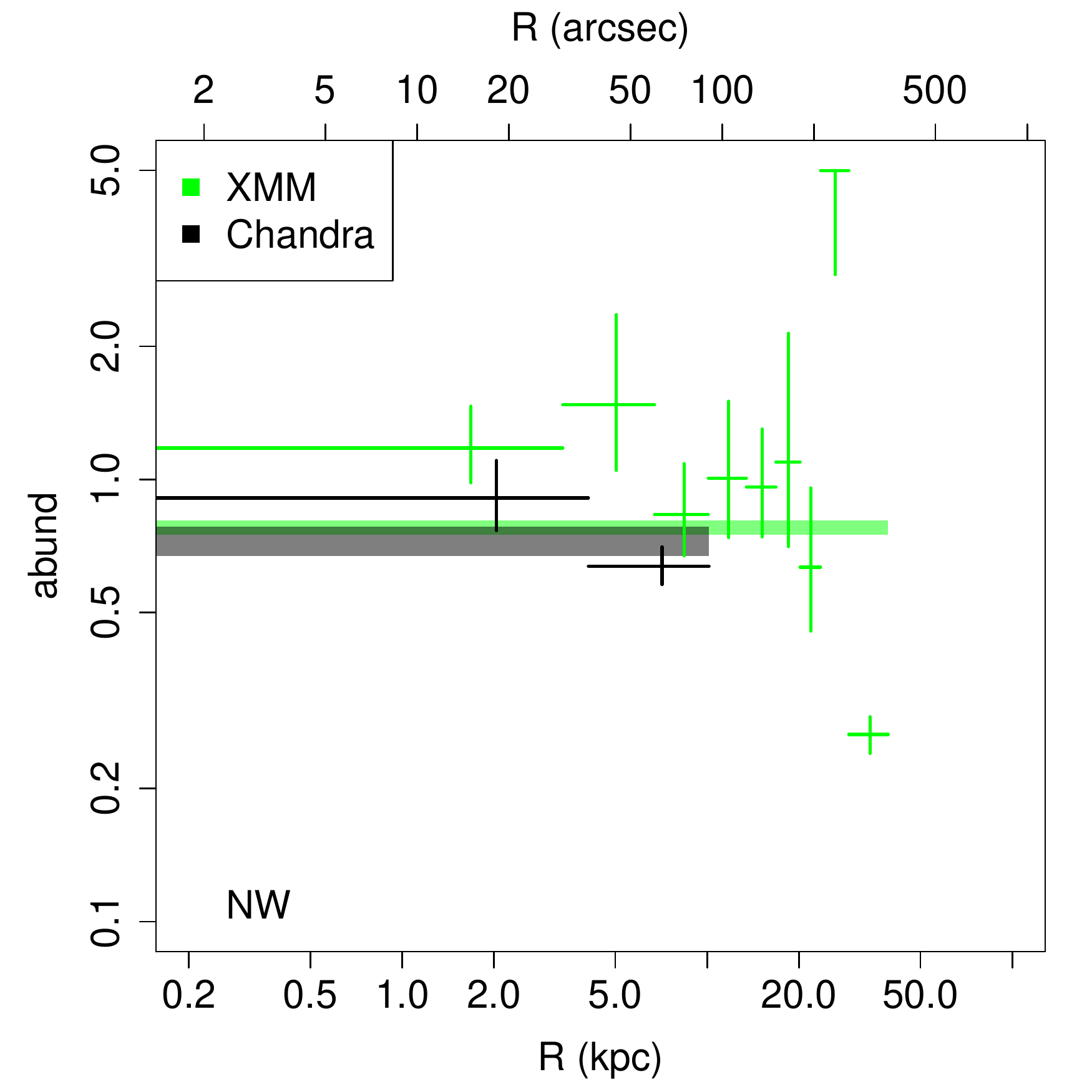}
\caption{Same as Figure \ref{fig:N5846_gas_profiles_merged} but with the free abundance model (temperature profiles are similar to the fixed abundance model case and therefore omitted). In addition, on the rightmost panel of each row we show the element abundances profiles for \textit{XMM}-MOS data (in green) and for \textit{Chandra} ACIS data (represented in black). In the same panels we overplot with green and black rectangles the values of the element abundances obtained with the fixed abundance model for \textit{XMM}  and \textit{Chandra} data, respectively.}\label{fig:N5846_gas_profiles_merged_abund}
\end{figure}

\begin{figure}
\centering
\includegraphics[scale=0.50]{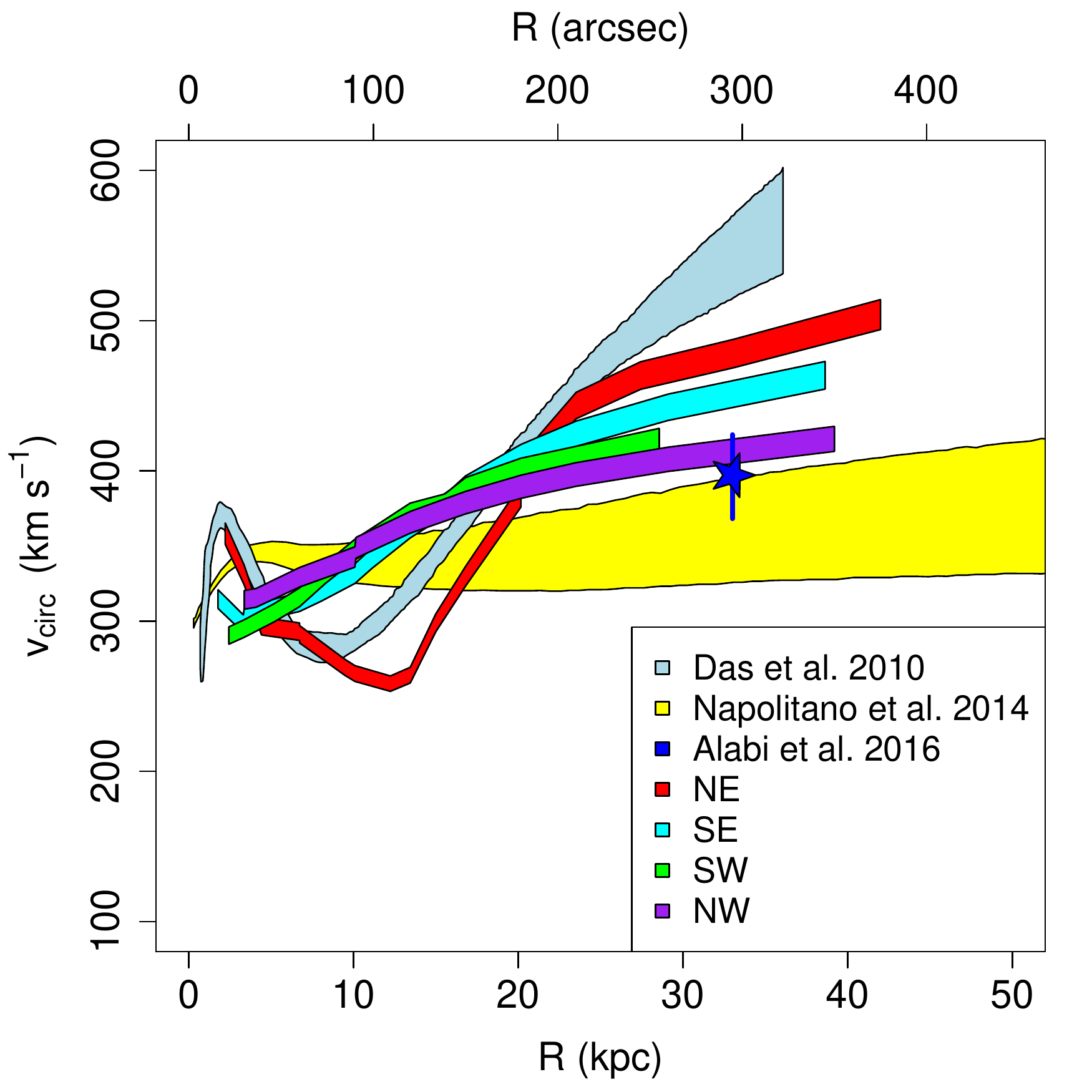}
\includegraphics[scale=0.50]{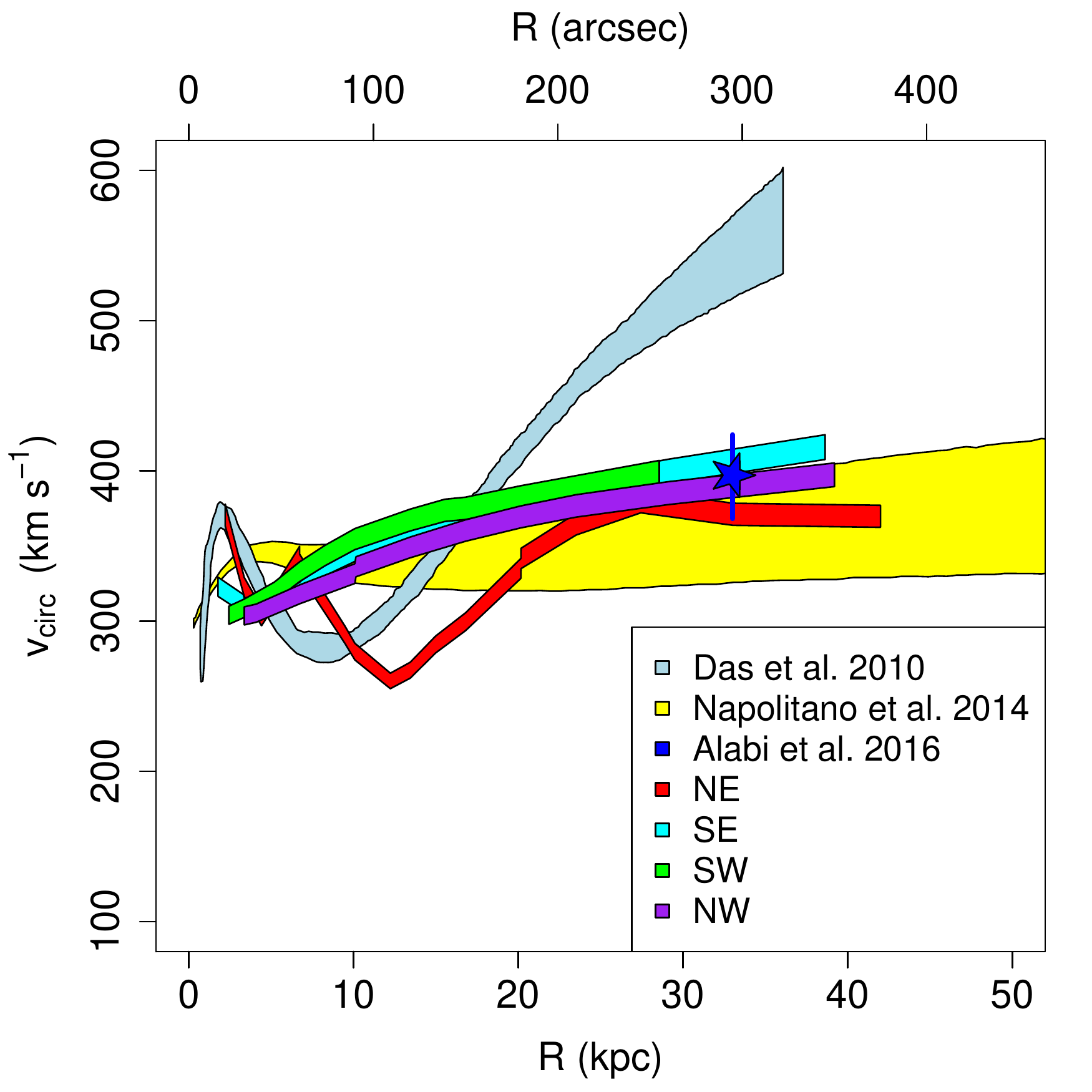}
\caption{{(left) Comparison of velocity profiles of NGC 5846 from fixed abundance models. The profiles for different angular sectors are shown in different colors indicated in the legend, while the X-ray profile presented by \citet{2010MNRAS.409.1362D} is shown in light blue, the SLUGGS SD+GC profile presented by \citet{2014MNRAS.439..659N} is shown in yellow, and the blue star represents the measurement by \citet{2016MNRAS.460.3838A} with the corresponding uncertainty shown with a vertical blue line. (right) Same as left panel, but for free abundance models.}}\label{fig:N5846_velocity_profiles_all}
\end{figure}

\begin{figure}
\centering
\includegraphics[scale=0.5]{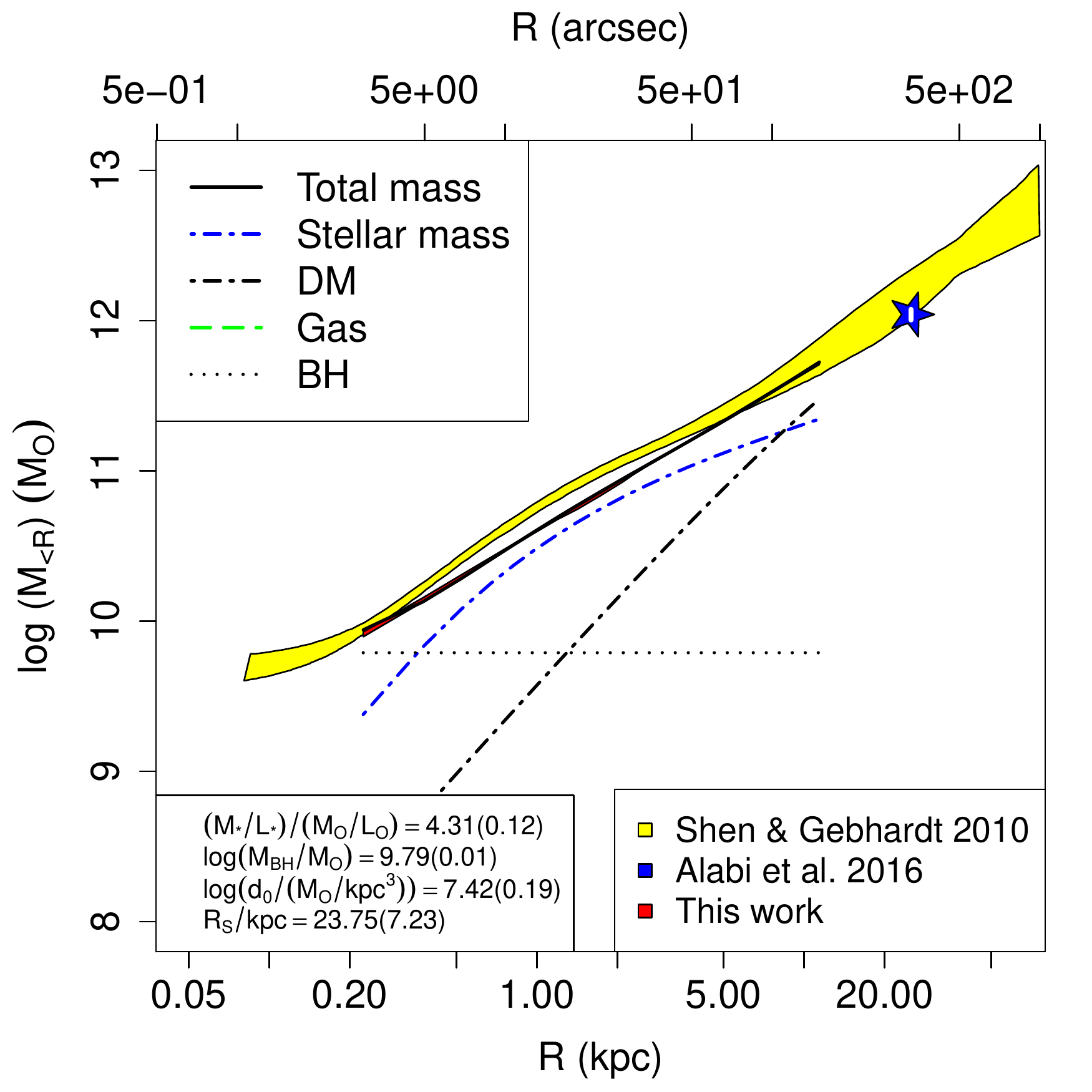}
\includegraphics[scale=0.5]{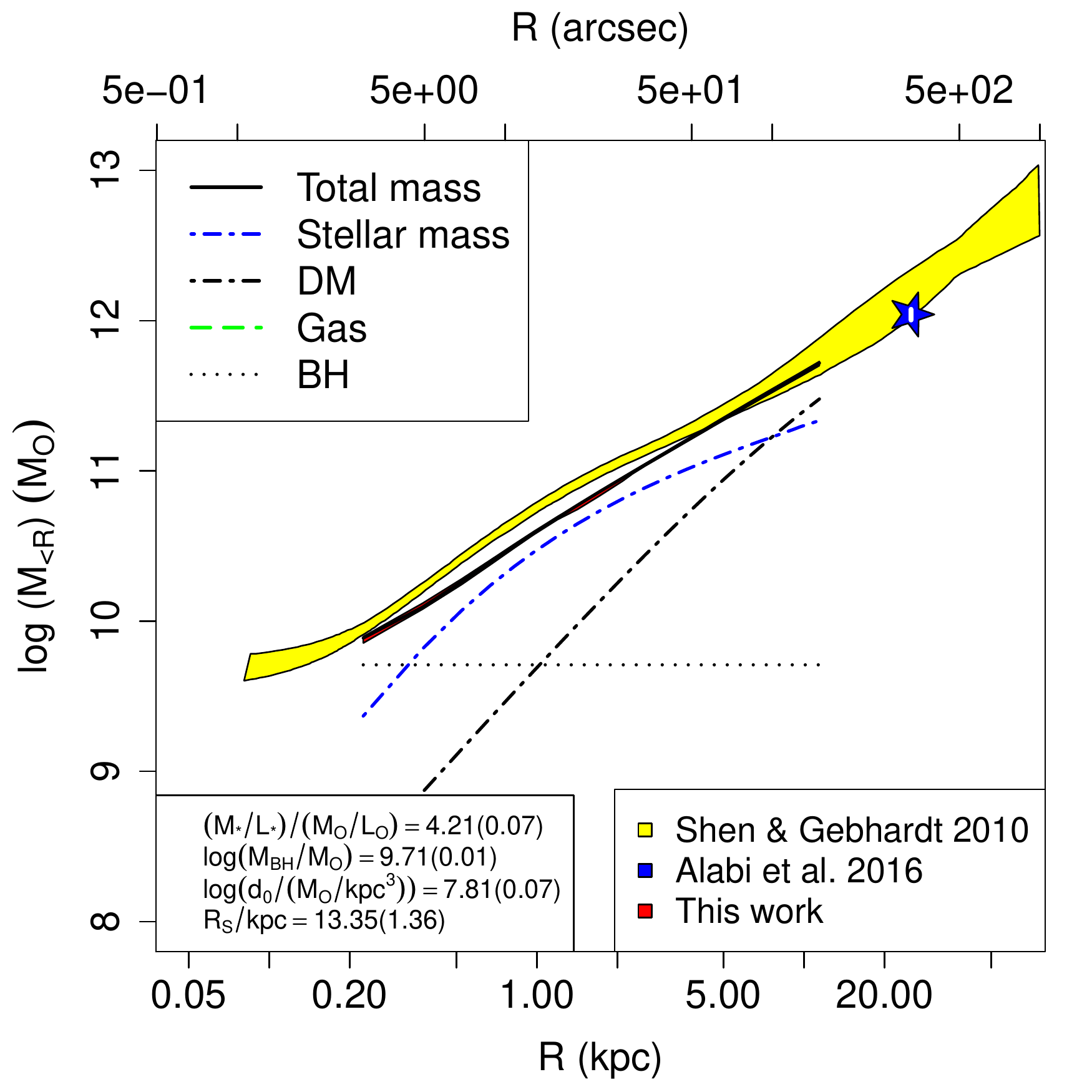}\\
\includegraphics[scale=0.5]{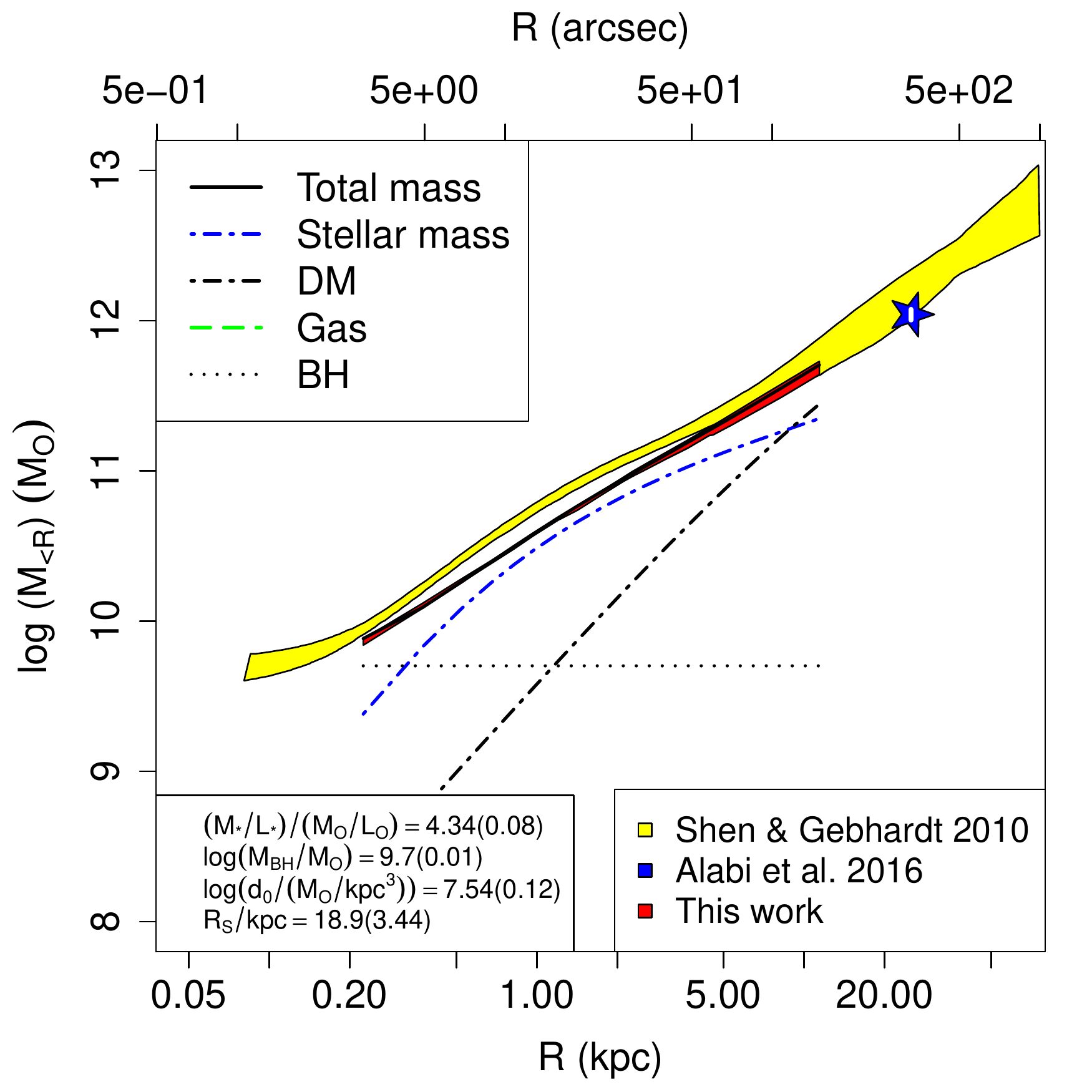}
\includegraphics[scale=0.5]{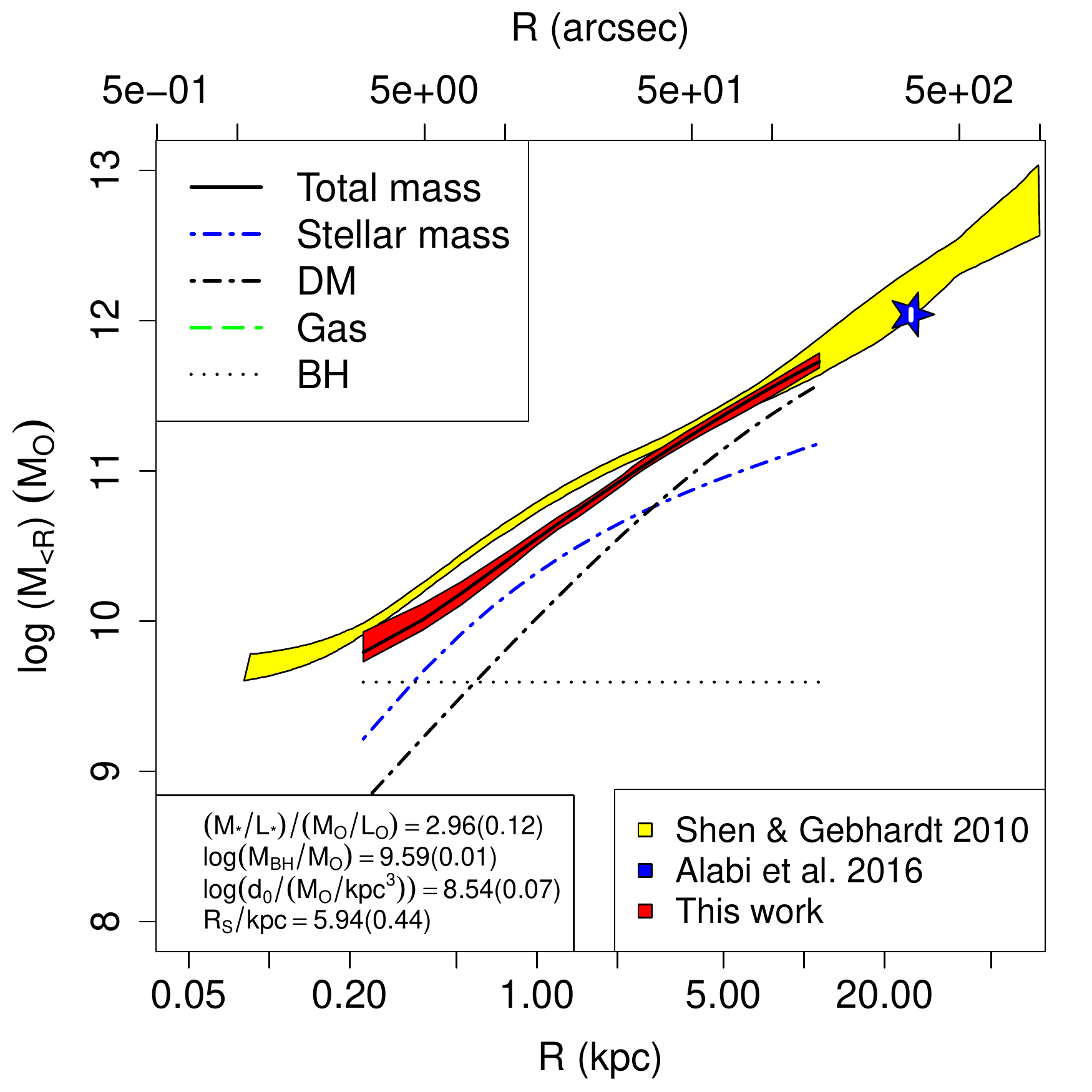}
\caption{Fit to mass profiles of NGC 4649 obtained with fixed (top row) and variable (bottom row) element abundances, in the full (0-360, left panels) and SW (180-270, right panels) sector. The mass profile form the {HE} equation is presented in red, and the best fit contributions of the various mass components (gas mass, stellar mass, black hole and NFW DM profile) are presented with different colors as reported in the legend. The best fit parameters are reported in lower left box. In yellow we show the optical mass profile obtained from SD and GC reported by \citep{2010ApJ...711..484S}, while the blue star represents the measurement by \citet{2016MNRAS.460.3838A} with the corresponding uncertainty shown as a white vertical line side the star itself.}\label{fig:N4649_mass_fits}
\end{figure}

\begin{figure}
\centering
\includegraphics[scale=0.5]{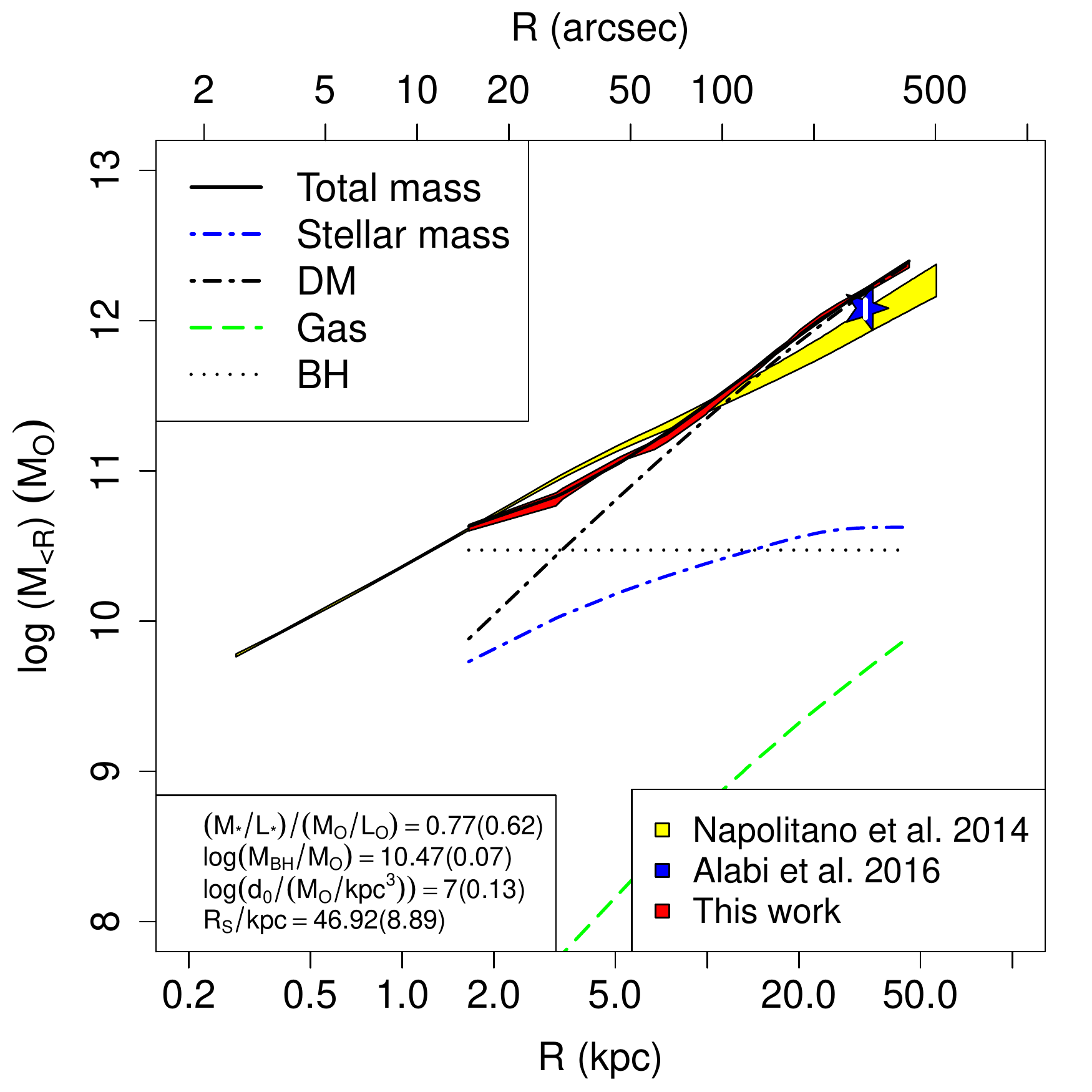}
\includegraphics[scale=0.5]{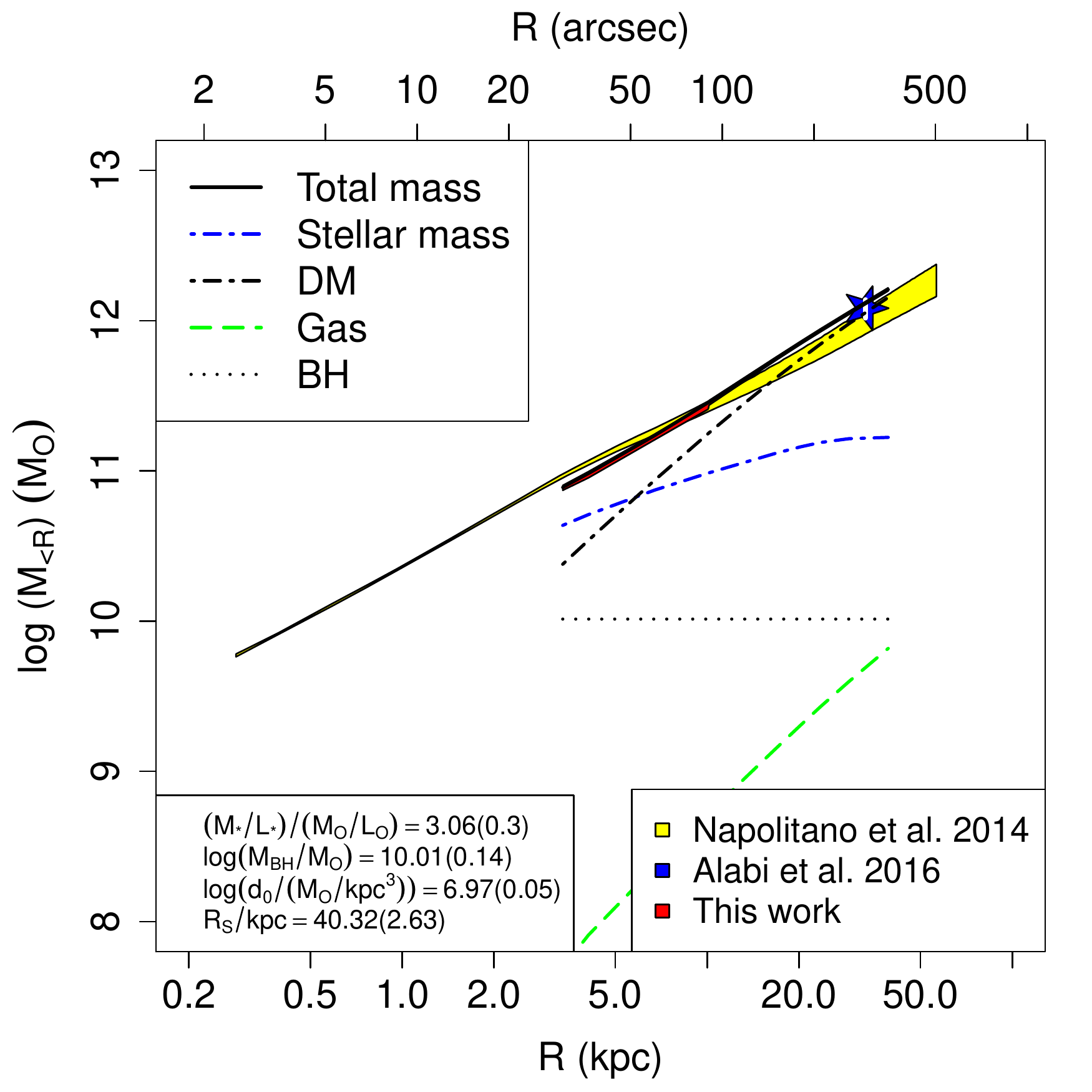}\\
\includegraphics[scale=0.5]{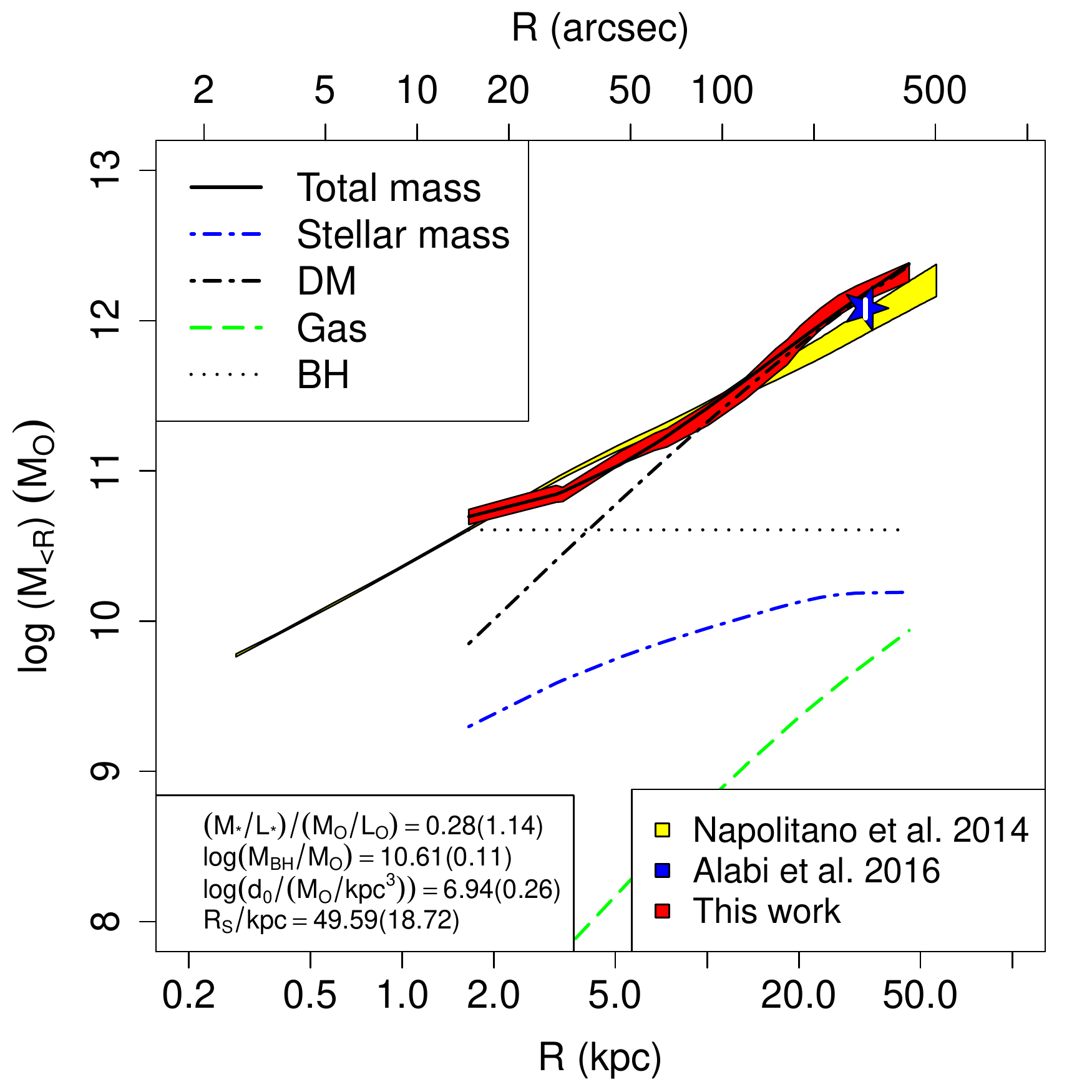}
\includegraphics[scale=0.5]{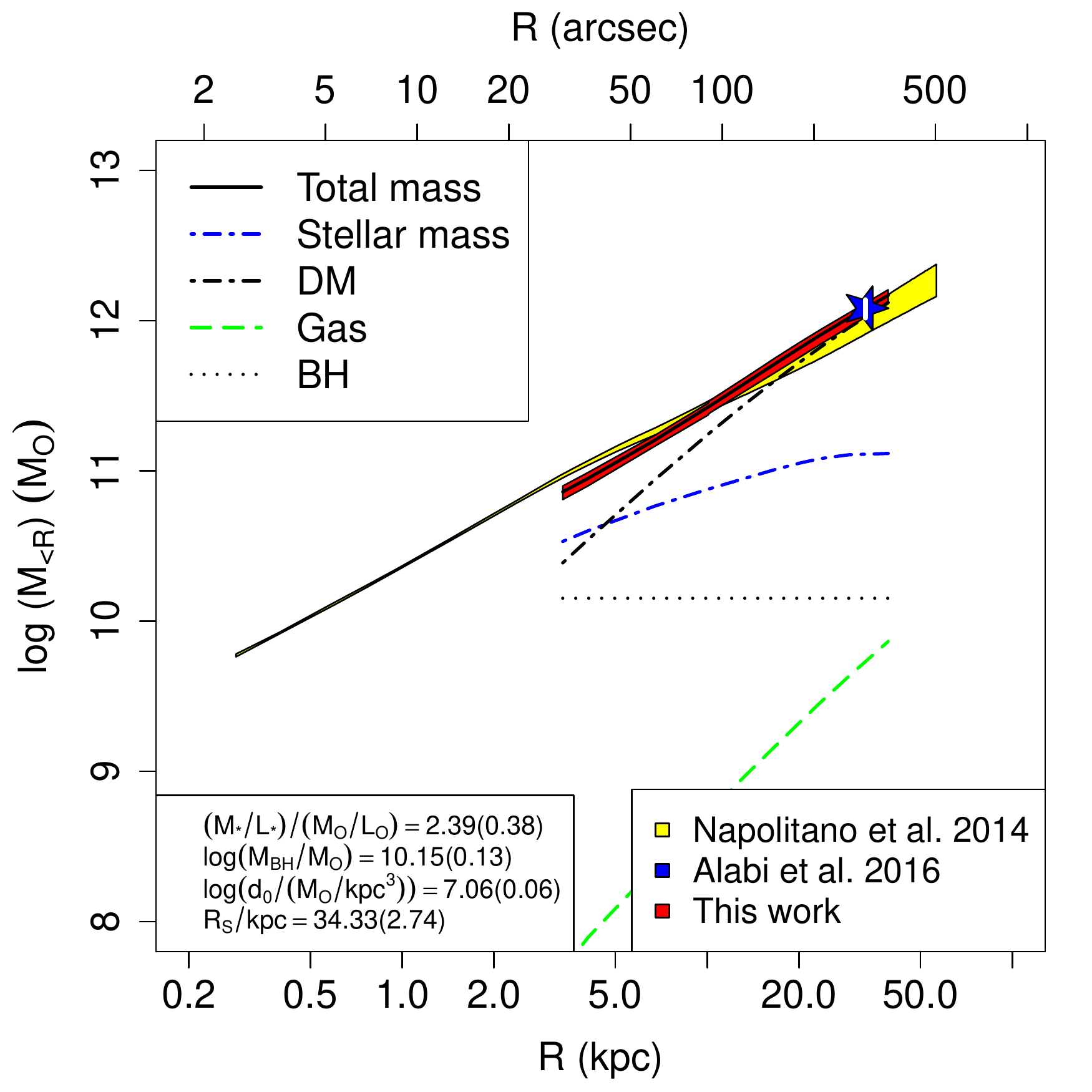}
\caption{Fit to mass profiles of NGC 5846 obtained with fixed (top row) variable (bottom row) element abundances, in the full (0-360, left panels) and NW (250-30, right panels) sector. The mass profile form the {HE} equation is presented in red, and the best fit contributions of the various mass components (gas mass, stellar mass, black hole and NFW dark matter profile) are presented with different colors as reported in the legend. The best fit parameters are reported in lower left box. In yellow we show the optical mass profile obtained from SD and GC reported by \citep{2014MNRAS.439..659N}, while the blue star represents the measurement by \citet{2016MNRAS.460.3838A} with the corresponding uncertainty shown as a white vertical line side the star itself.}\label{fig:N5846_mass_fits}
\end{figure}

\afterpage{
\begin{appendix}\label{app:appendix}

\section{Background subtraction}\label{app:background}

{In this section we present the details of the different reduction procedures tested in this work for the \textit{XMM-Newton} MOS data. In particular, our main aim here is to compare the impact of different ways of evaluating the background (both astronomical and instrumental) for these data.}

\subsection{Simple Background Subtraction}\label{app:nevalainen}

{Following} \citet{2005ApJ...629..172N} we filtered MOS1 and MOS2 data for hard-band flares by excluding {the} time intervals where the \(9.5-12\) keV count rate evaluated on the whole detector FOV was more than 3\(\sigma\) away from its average value. To achieve a tighter filtering of background flares, we iteratively repeated this process two more times, re-evaluating the average hard-band count-rate and excluding time intervals away more than 3\(\sigma\) from this value. The same procedure {was} applied to soft \(1-5\) keV band restricting the analysis to an annulus with inner and outer radii of 12' and 14', where the emission from the galaxy is expected to be small.

The background was evaluated from appropriate blank-sky files \citep{2007A&A...464.1155C} filtered in the same way as the event files. Following \citet{2005ApJ...629..172N} we normalized the blank-sky file exposures to match the \(9.5-12\) keV count rate of the event files, since at these energies we expect to see only the particle background. However, since the particle background and the sky background are independent, we apply this normalization only in the \(2-7\) keV band - where the sky background contribution is expected to be only \(\sim 20\%\) - and not to the \(0.5-2\) keV band - where the sky background is expected to dominate. We then excluded from the spectral fitting (see Sect. \ref{sec:spectra}) the \(1.45-1.55\) keV band due to variable Al K lines.

\subsection{Double Background Subtraction}\label{app:arnaud}

The second reduction method is that proposed by \citet{2001A&A...365L..80A, 2002A&A...390...27A}. The GTI filtering is performed on the whole detector FOV excluding time intervals where the \(10-12\) keV count rate was larger than \(0.15\) cts/s. {We} made use of the blank-sky files, again filtered in the same way as the event files, and with exposure normalized to match the \(10-12\) keV count rate of the event files.

To disentangle the cosmic X-ray background from the instrumental background we {made} use of vignetting corrected event files ({produced} with the \textsc{evigweight} \textsc{SAS} task), since the cosmic X-ray background can be considered uniform on scales \(\sim 30'\) but is affected by vignetting, while the instrumental background is non uniform but is not vignetted. Following \citeauthor{2002A&A...390...27A}'s ``double subtraction'' procedure, we first {subtracted from} the source spectrum (extracted for a region of area \(A\)) the spectrum extracted in the same region from the blank-sky file, obtaining the net spectrum \(I\). Then we {performed} the same procedure, but from a region (of area \(A'\)) of the FOV that we \textit{assumed} to be source free, obtaining the net spectrum \(I'\). Finally we {subtracted} these two net spectra taking into account the ratio of the extraction region areas. The {resulting} spectrum \(S=I-I'(A/A')\) is expected to contain only source emission.

\subsection{Background Modeling}\label{app:snowden}

The third reduction method we adopted is the one proposed in the ``Cookbook for Analysis Procedures for \textit{XMM-Newton} EPIC Observations of Extended Object and the Diffuse Background\footnote{\href{https://heasarc.gsfc.nasa.gov/docs/xmm/esas/cookbook/xmm-esas.html}{https://heasarc.gsfc.nasa.gov/docs/xmm/esas/cookbook/xmm-esas.html}}'' \citep{2011AAS...21734417S}. The GTI filtering for this procedure is preformed extracting \(8-12\) keV lightcurves both in annulus away from the central source in the exposed FOV and in the unexposed detector corners, and then excluding the time intervals for which the former countrate exceeds 1.2 times the latter countrate.

The background for this reduction method is partly subtracted and partly modeled. In particular, {quiescent} particle background (QPB) spectra are created through the \textsc{mos\_back} task and subtracted from the source spectra. In addition, we modeled the instrumental and the cosmic background as follows. We modeled the Al K\(\alpha\) and Si K\(\alpha\) lines with gaussian lines at \(\sim 1.49\) keV and \(\sim 1.75\) keV, and the soft proton component as a power law which is not folded through the instrumental effective area (adding in \textsc{XSPEC} a separate model with a diagonal unitary matrix). As for the cosmic background, we modeled the local hot bubble with a cool (\(\sim 0.1\) keV) unabsorbed thermal component, the cooler halo with a \(\sim 0.1\) keV absorbed thermal component, and the hotter halo and/or intergalactic medium with a higher temperature \(\sim 0.25-0.7\) keV absorbed thermal component. 

As mentioned in the main text, this method requires a complex spectral modeling, with several free parameters that make it difficult to find stable fitting convergence in un-{supervised} analysis procedures (especially in conjunction with 3D-deprojection, see Sect. \ref{sec:spectra}). To avoid this problem, in the spectral fitting procedure we used as starting values of the gas component (see Sect. \ref{sec:spectra}) the best fit value obtained with the procedure discussed in Sect. \ref{app:nevalainen}.

\section{Model Effects}\label{app:effects}

Here we show in detail the effects of the various parameters in our modelization can have on the final mass profile, namely the smoothing paramter of the spline function used to fit the gas profiles (\ref{app:spline}), the different angular section used for the spectral extraction (\ref{app:pies}), the element abundances (\ref{app:abundances}) and the different background subtraction (\ref{app:effects_background}).

\subsection{Effects of the spline function smoothing}\label{app:spline}

As {shown} in Figure \ref{fig:effects_smooth} lower values of the smoothing parameter yield best fit models that are able to represent all the finer details of the gas profiles, while higher values of the smoothing parameter favor monotonic profiles. In the same figure we show the effects of the increasing smoothing parameter on the mass profile resulting from Eq. \ref{eq:hee}. The choice of the smoothing parameter value is a trade-off between the best fit function ability to represent the finer details of the gas profiles and the need to get rid of the noise in these profiles in order to get meaningful physical results, that is, a combination of temperatures and slopes that yields decreasing or even negative values in the mass profiles. On the other hand, however, these apparently unphysical results may be the consequence of gas conditions away from {HE}. In the case presented in Figure \ref{fig:effects_smooth} for example, - that is gas profiles for NGC 4649 in the full (0-360) sector for the \citeauthor{2005ApJ...629..172N} reduction procedure - a value of the smoothing parameter of 0.5 yields noisy mass profiles, particularly steep in the inner bins. On the other hand, for values of the smoothing parameter larger or equal to 0.8 the mass profile is very smooth, but the best fit models deviate significantly from the data. In this case, therefore, the optimal choice for the smoothing parameter is 0.7. The smoothing parameters chosen in the other cases are reported in the captions of Figures \ref{fig:N4649_gas_profiles_merged} and \ref{fig:N5846_gas_profiles_merged}.
 
\subsection{Effects of the sector direction}\label{app:pies}

We note that the total mass obtained from Eq. \ref{eq:hee} is the total mass enclosed within a radius \(R\) assuming spherical symmetry, that is, the total mass assuming that the gas is in HE and distributed in a spherical shell with constant temperature. Therefore, even if we extract spectra in angular sectors, it will make sense to compare the resulting mass profiles with each other and with the mass profiles from optical markers.

In Figure \ref{fig:effects_pie} we show how the X-ray derived mass profiles depend on the sector chosen for the spectral extraction in the source NGC 5846 (with a smoothing parameter of 0.7). As recalled in Sect. \ref{sec:sample} this source shows evidence of gas sloshing as a consequence of the interaction with the group companion NGC 5850 in the NE (30-90) direction, reflecting in a disturbed X-ray morphology. As a matter of fact, the full (0-360) mass profile shows a break between 1-2 kpc followed by a sudden increase at \(\sim 10\) kpc. When we isolate the mass profiles extracted in different directions, we can see that these effects are mainly driven by the gas in the NE sector - that is, in the NE direction - {while this effects weaken in other sectors where the profiles are smoother and without indication of strong disturbances in the ISM.}

\subsection{Effects of the variable element abundances}\label{app:abundances}

In Figure \ref{fig:effects_abundances} we show the effects of metal abundances on best fit model to the gas profiles (and the resulting mass profiles) for the NW (250-30) sector of NGC 5846 (both with smoothing parameter of 0.7). In particular, in the left and central panels of the top and bottom row we show the gas temperature and density profiles resulting from the free and fixed abundance models, respectively. In addition, in the right panel of the top row we show the abundance profiles compared with the averaged values of the abundances obtained with the fixed abundance model. Although the abundance profile is rather noisy, we see that it tends to decrease at larger radii, and this yields flatter gas density profiles and therefore smaller enclosed mass as estimated from Eq. \ref{eq:hee}, as shown in the right panel of the bottom row of the same figure. In addition, allowing variable abundances yields in general larger uncertainties (especially on gas density) which in turn translates to larger uncertainties on the enclosed mass profile. 

\subsection{Effects of the background subtraction method}\label{app:effects_background}

In Figure \ref{fig:effects_background} we show the effects of the different background subtractions on the gas and mass profiles of NGC 4649 in the NE (0-90) direction (all the profiles have a smoothing parameters of 0.7). As evident, {the} different background subtraction methods yield similar gas and mass profiles, with only minor effects on the outer bins at fainter surface brightnesses. For the double subtraction method we have to assume a source free region that will be excluded by our spectral extraction. In particular, we selected a sector between 13' and 14' in the SE direction both for NGC 4649 and NGC 5846, so the main effect of the background subtraction proposed by \citet{2001A&A...365L..80A, 2002A&A...390...27A} is to restrict our analysis to smaller radii in this sector with respect to the others methods.

\begin{figure}
\centering
\includegraphics[scale=0.22]{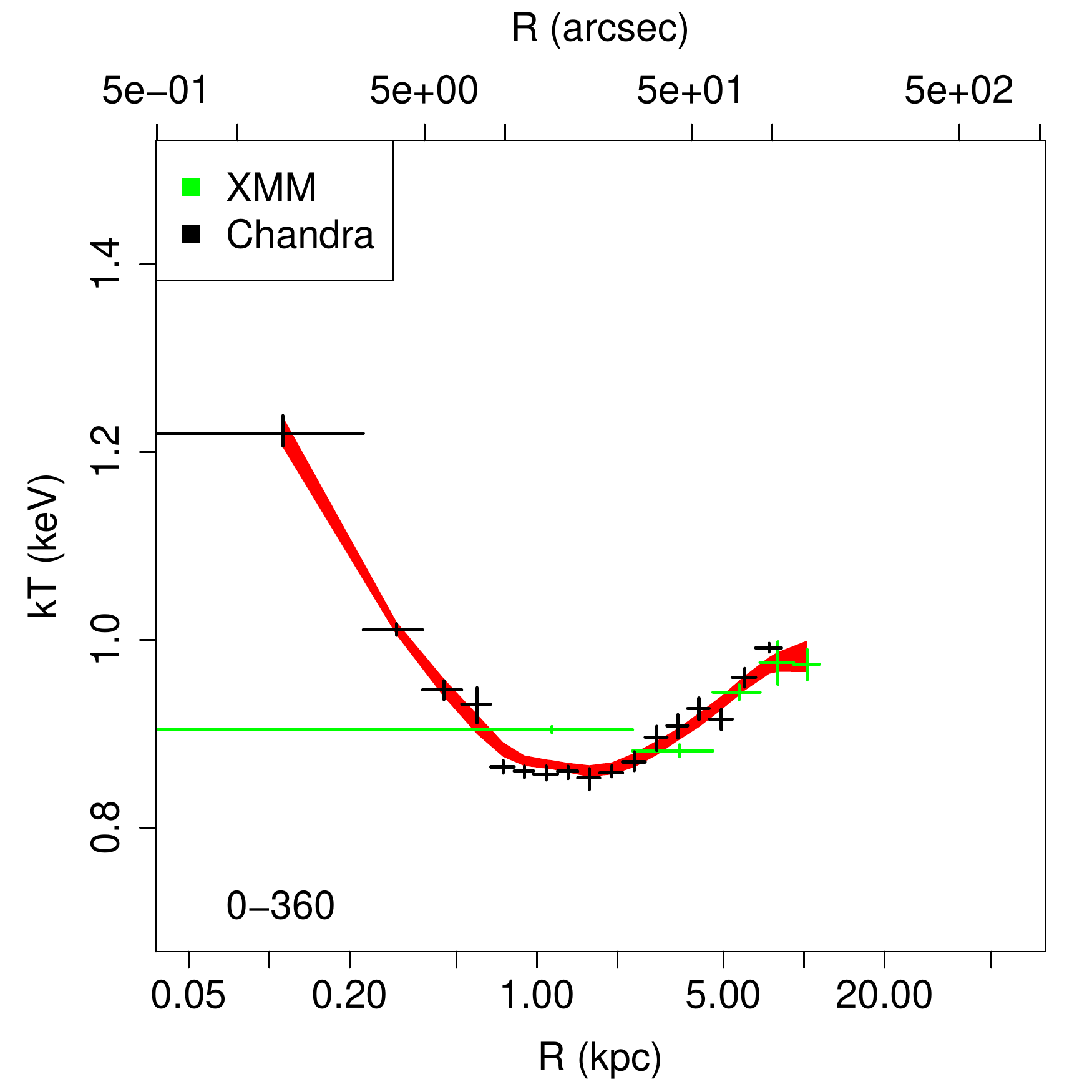}
\includegraphics[scale=0.22]{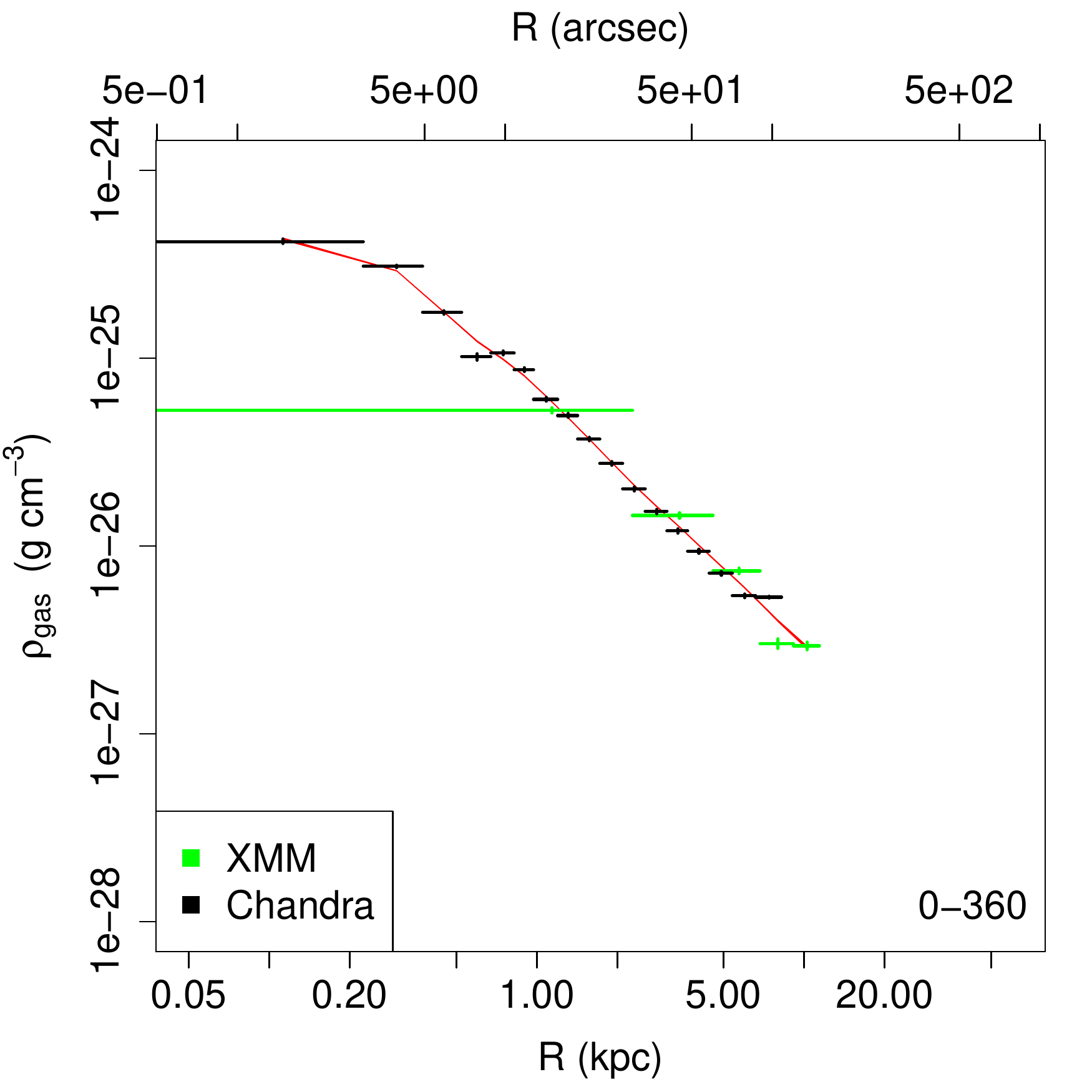}
\includegraphics[scale=0.22]{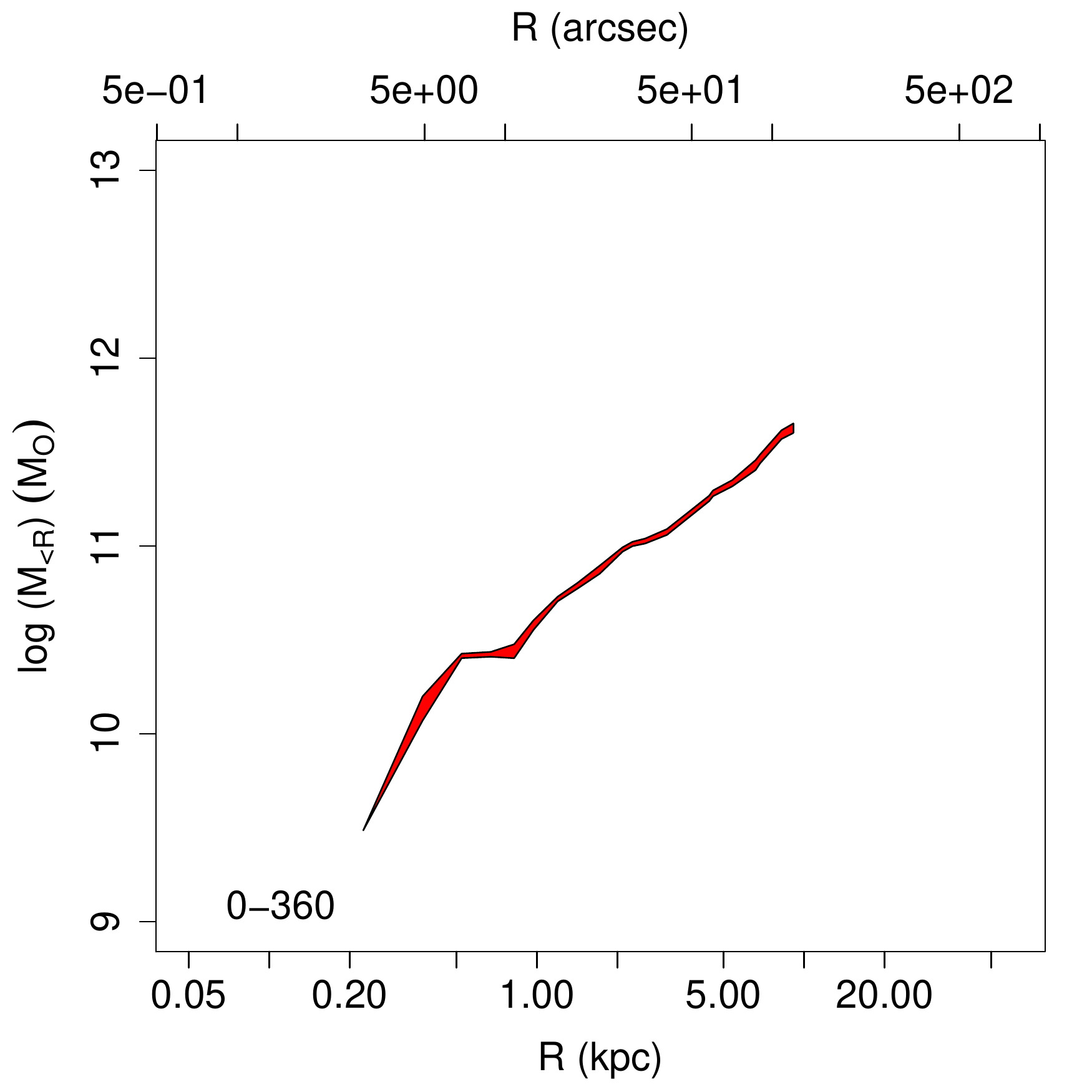}\\
\includegraphics[scale=0.22]{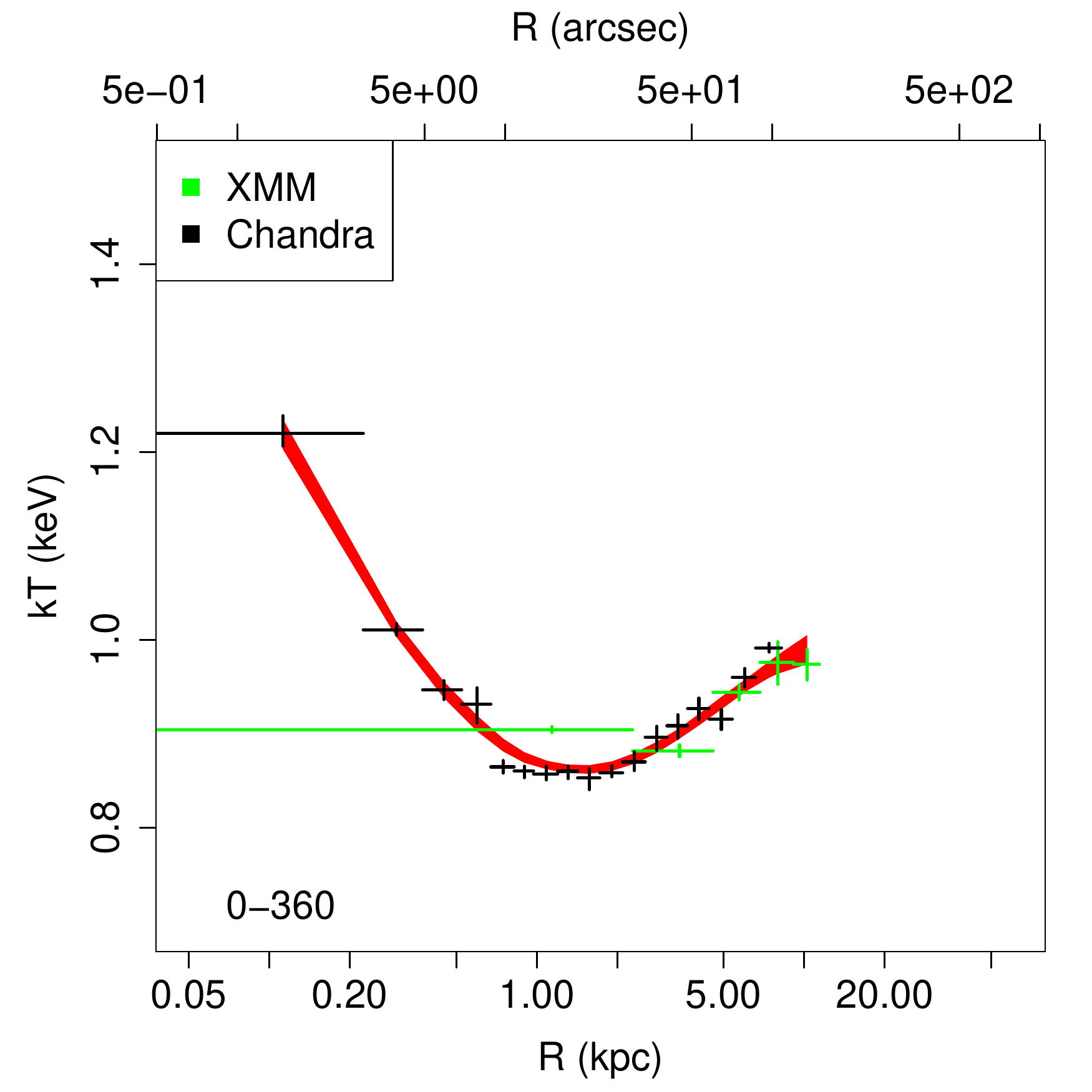}
\includegraphics[scale=0.22]{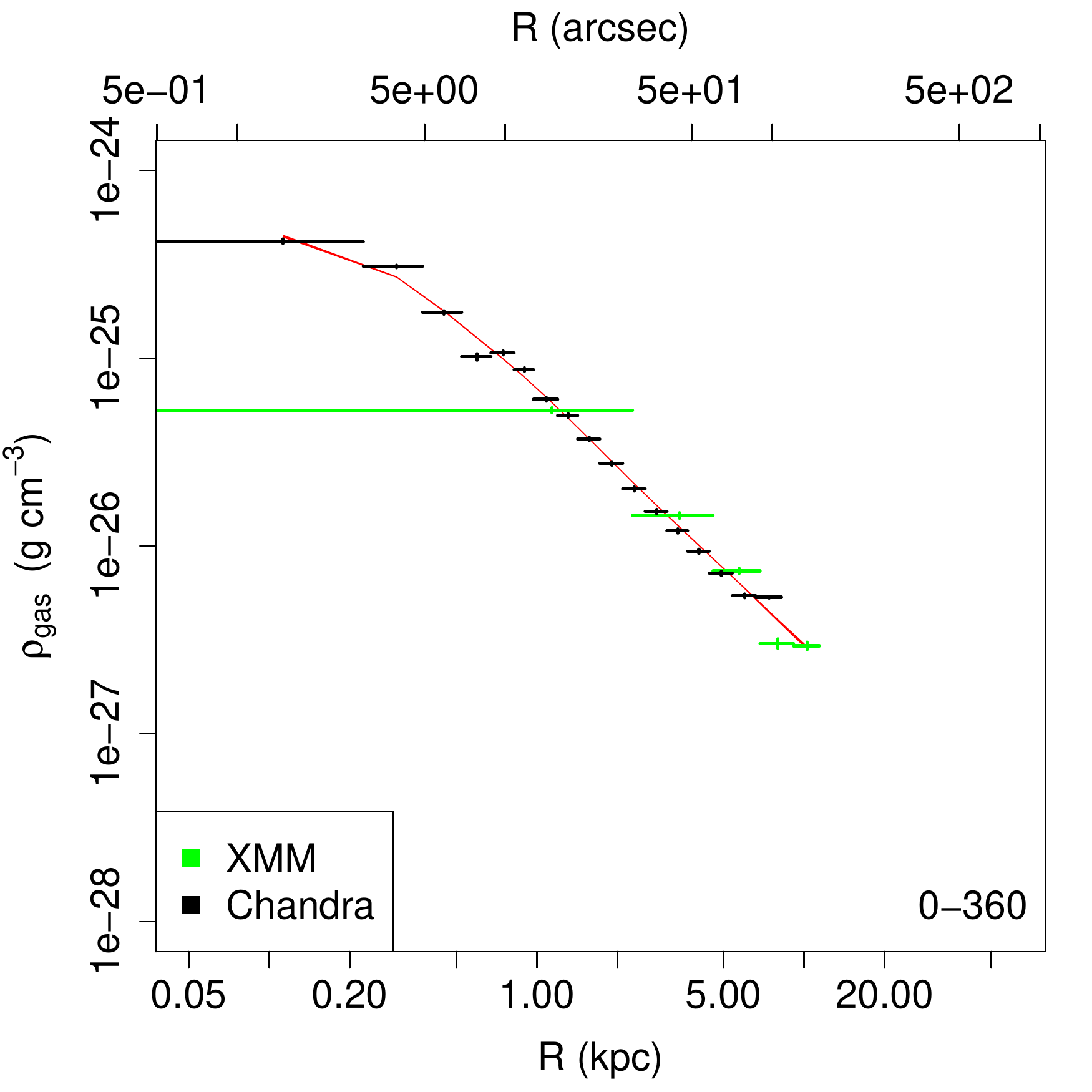}
\includegraphics[scale=0.22]{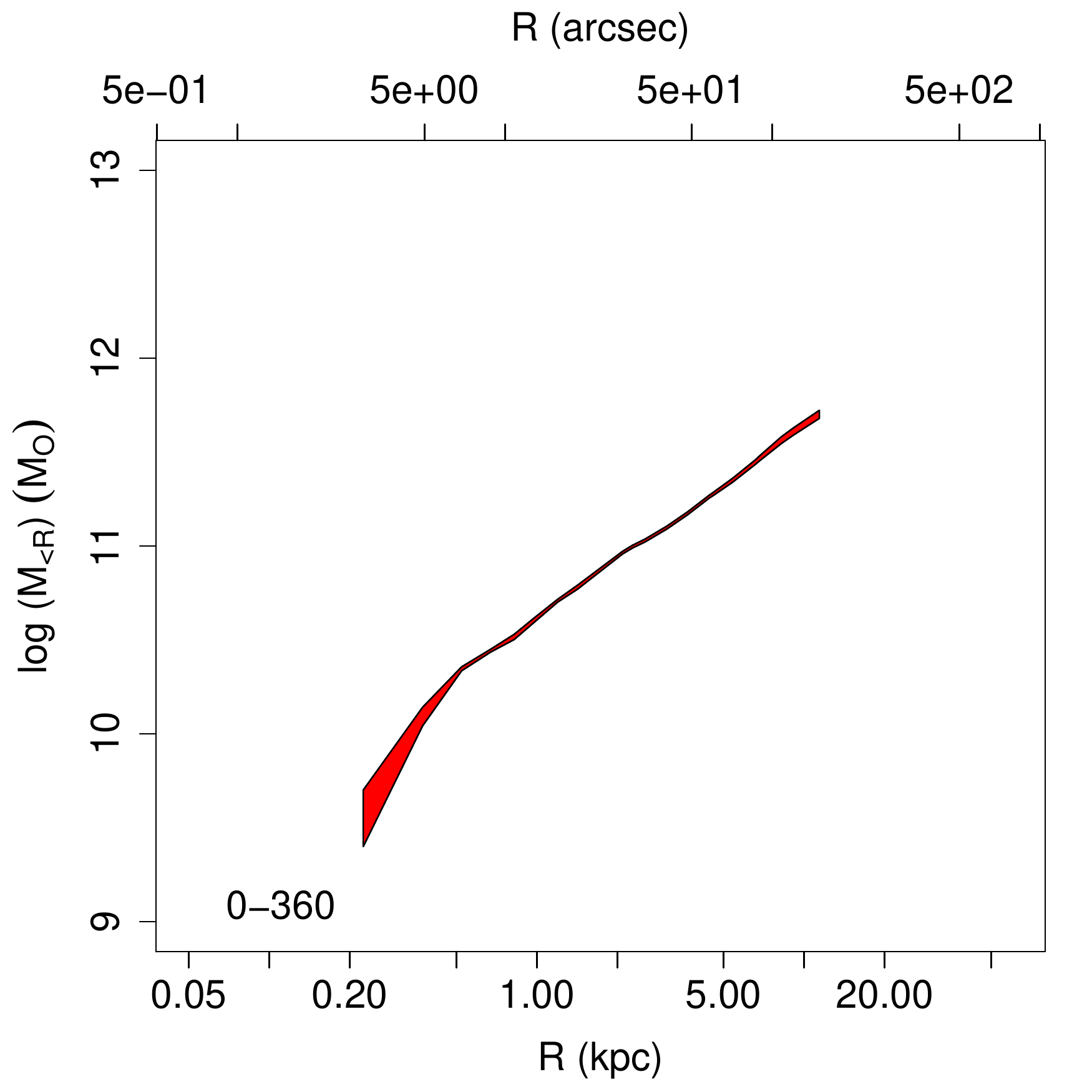}\\
\includegraphics[scale=0.22]{{N4649_temp_profile_merged_0_360_0_0_fit_0.7}.pdf}
\includegraphics[scale=0.22]{{N4649_nh_profile_merged_0_360_0_0_fit_0.7}.pdf}
\includegraphics[scale=0.22]{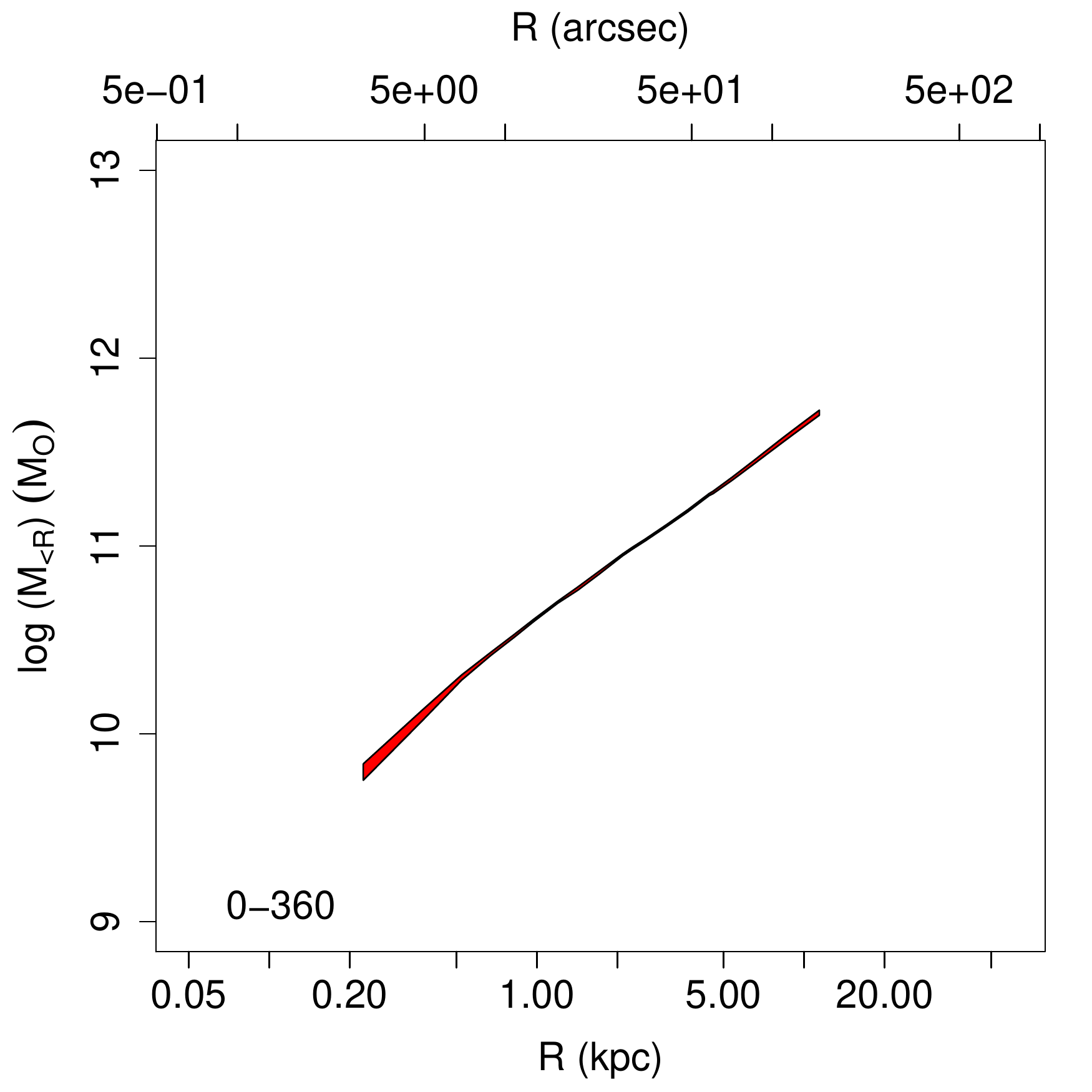}\\
\includegraphics[scale=0.22]{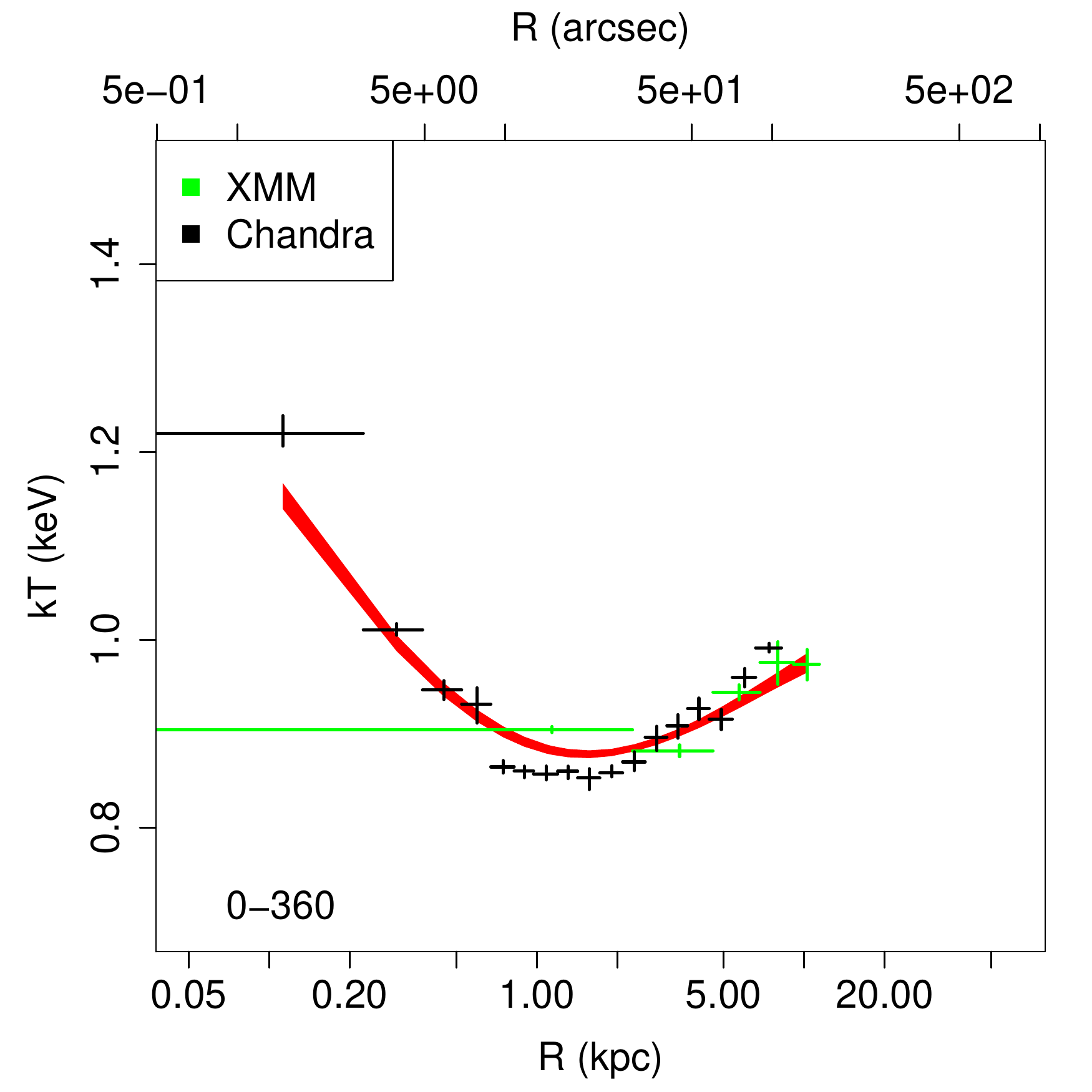}
\includegraphics[scale=0.22]{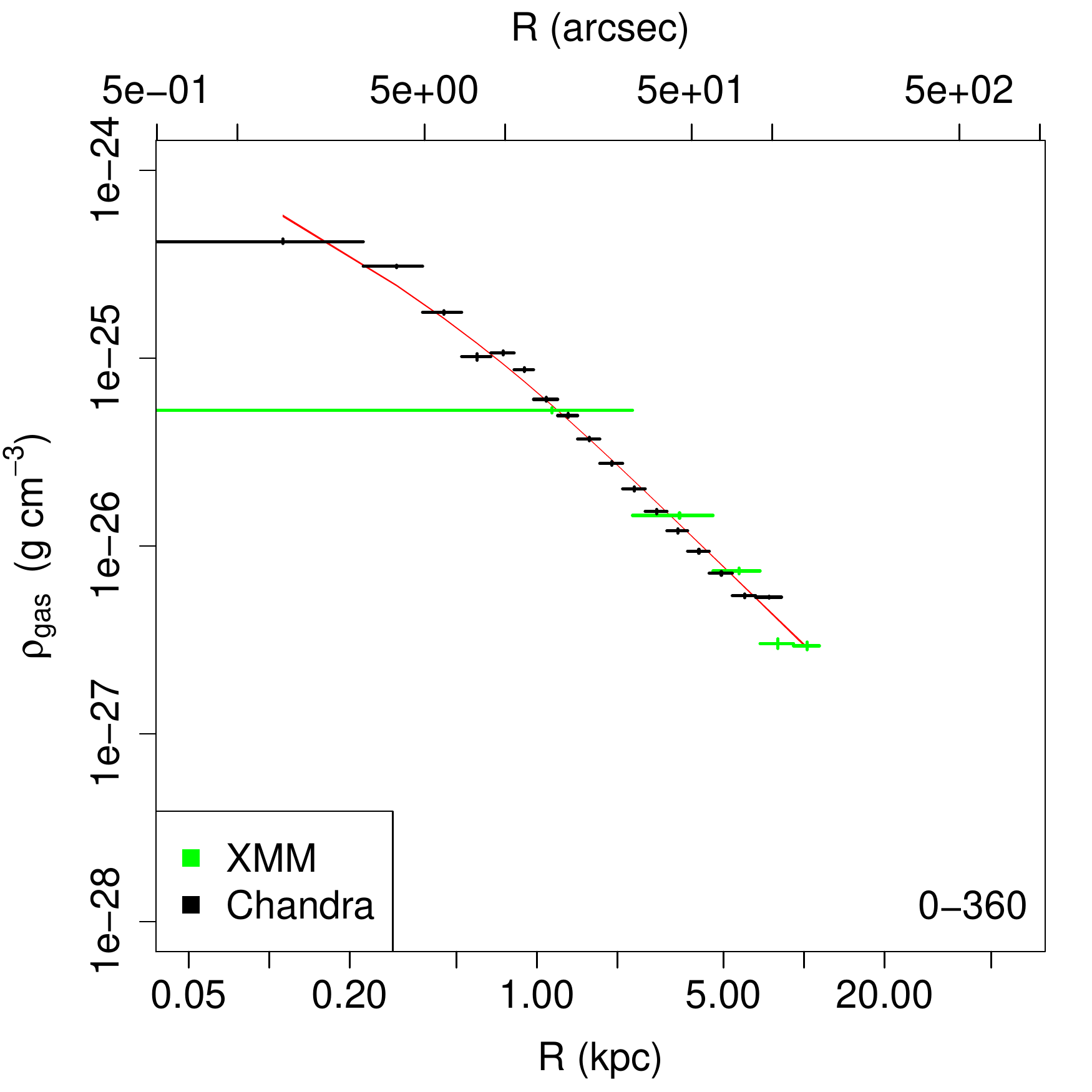}
\includegraphics[scale=0.22]{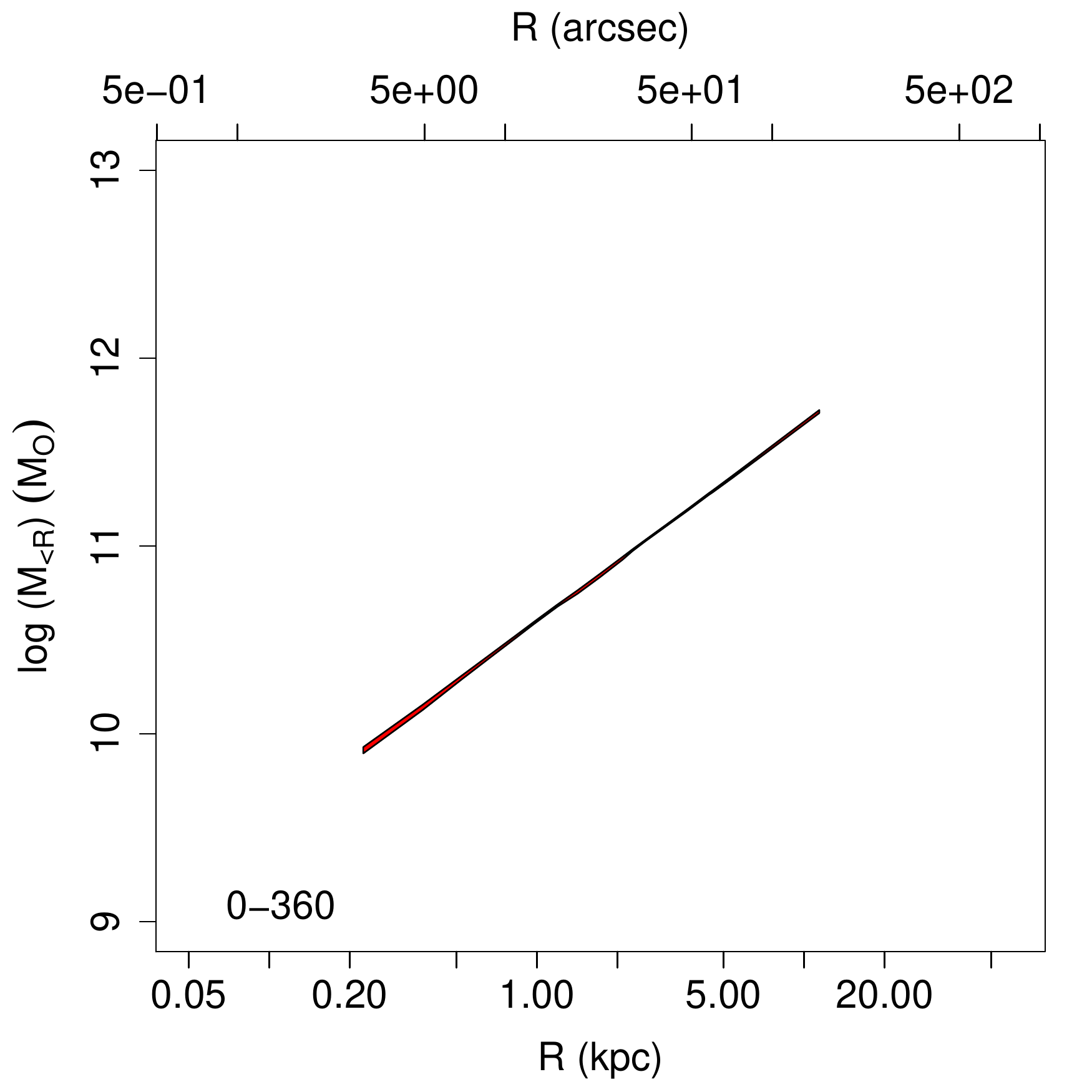}\\
\includegraphics[scale=0.22]{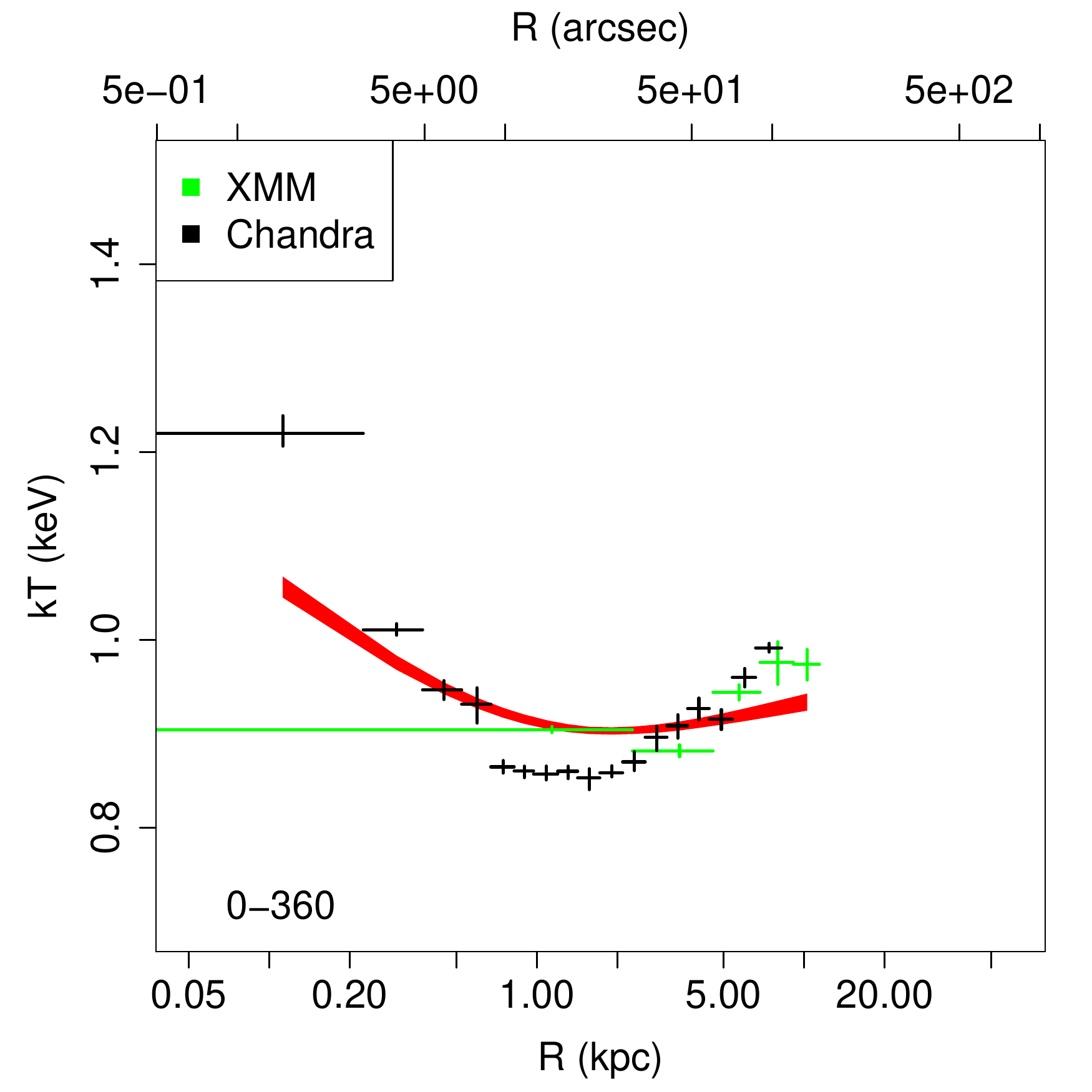}
\includegraphics[scale=0.22]{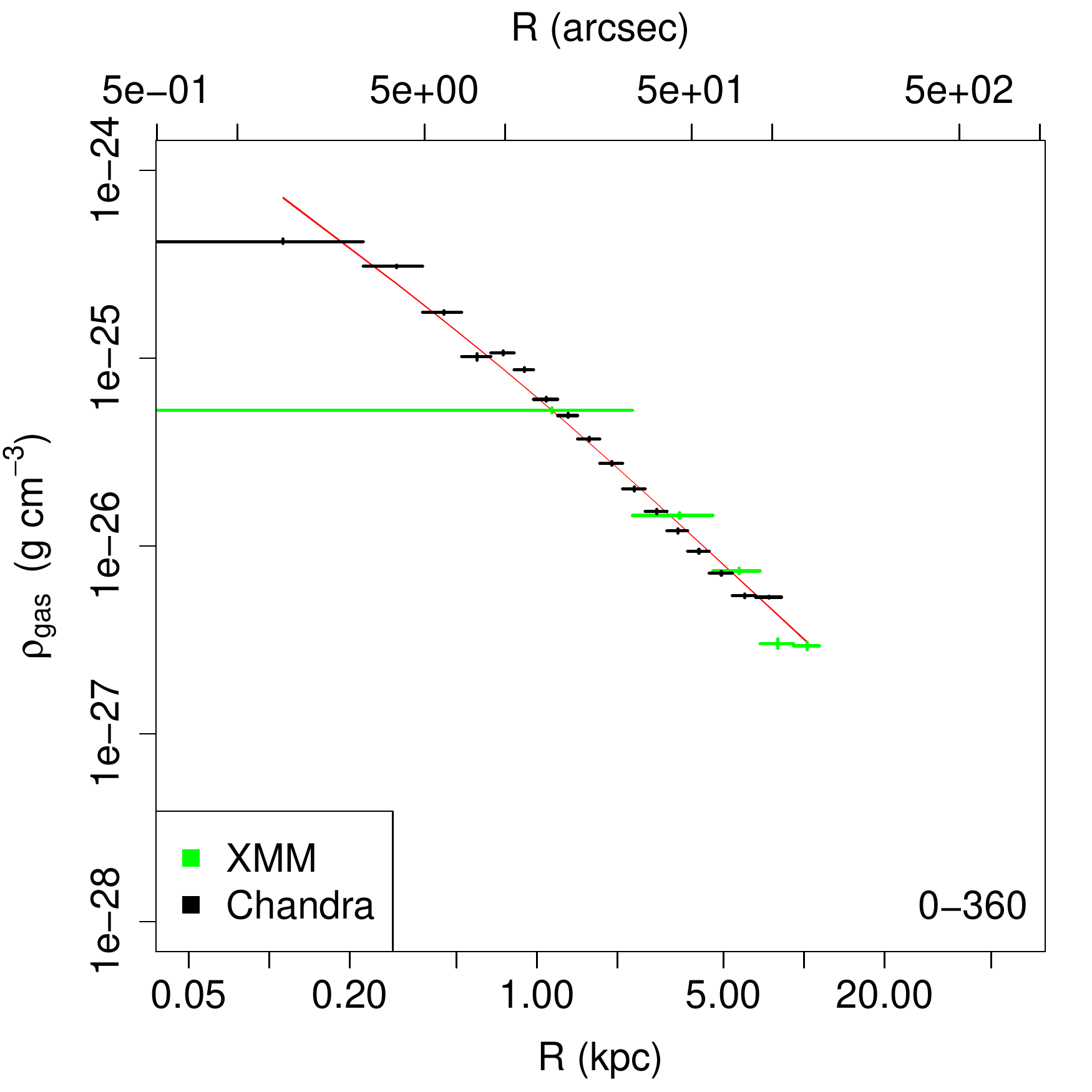}
\includegraphics[scale=0.22]{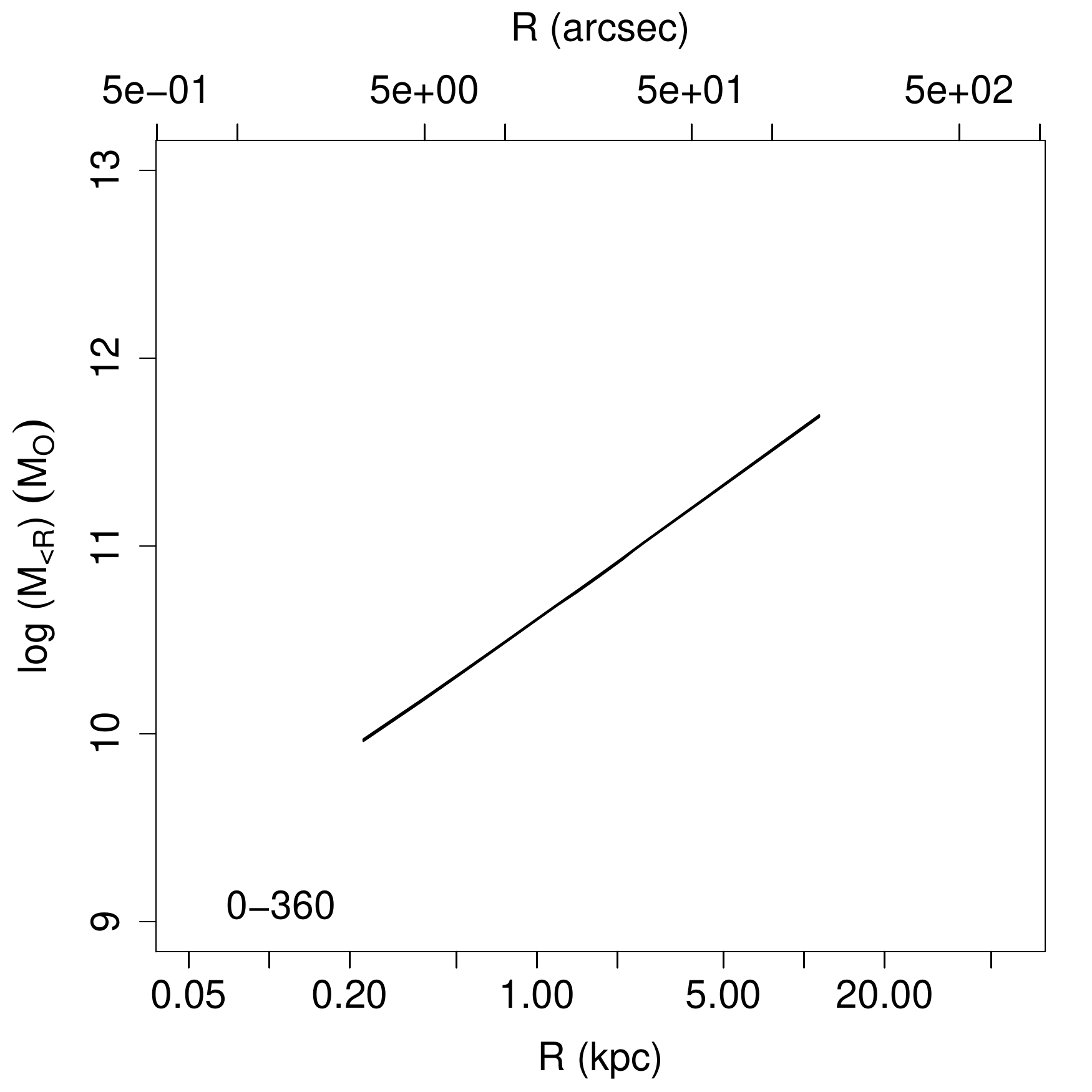}\\
\includegraphics[scale=0.22]{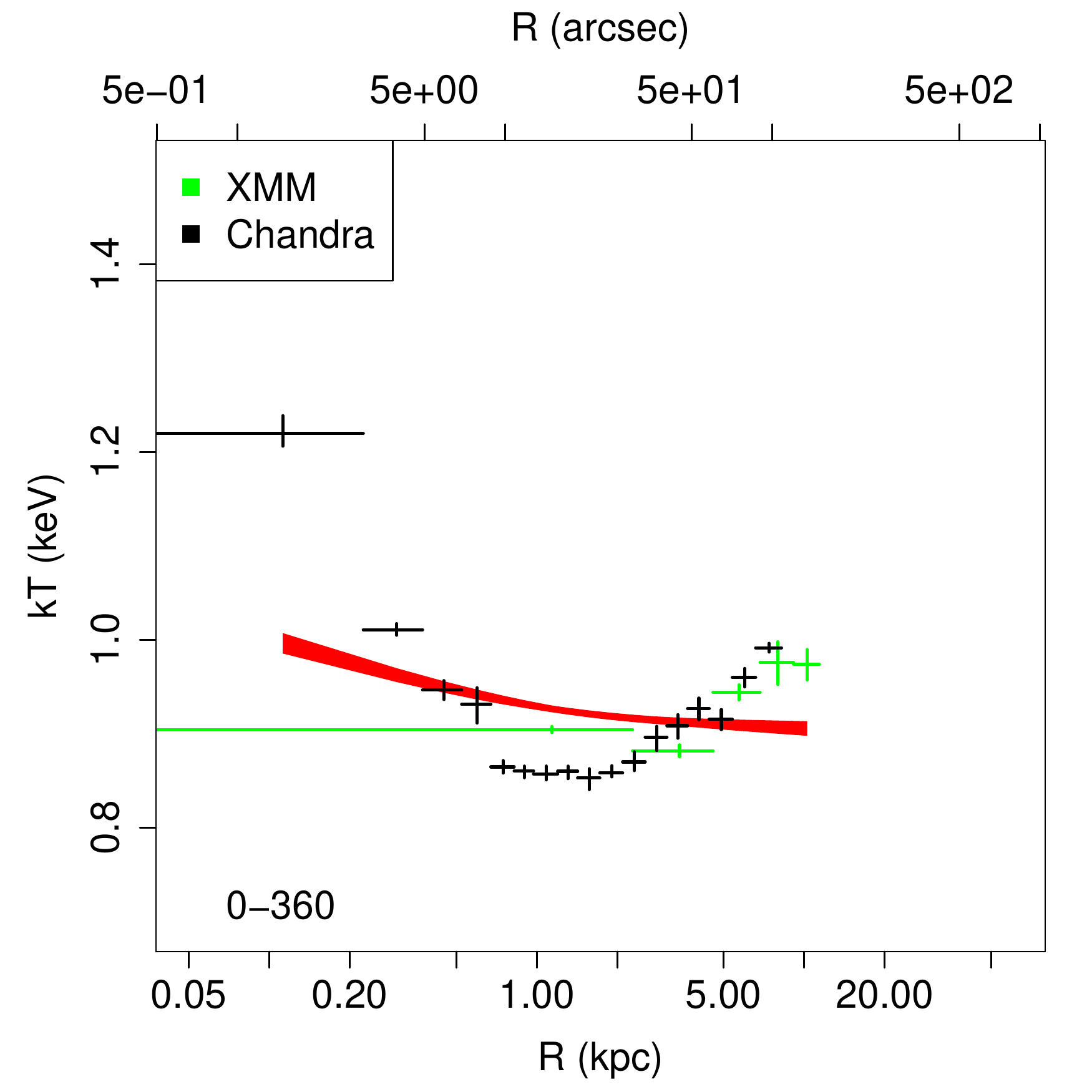}
\includegraphics[scale=0.22]{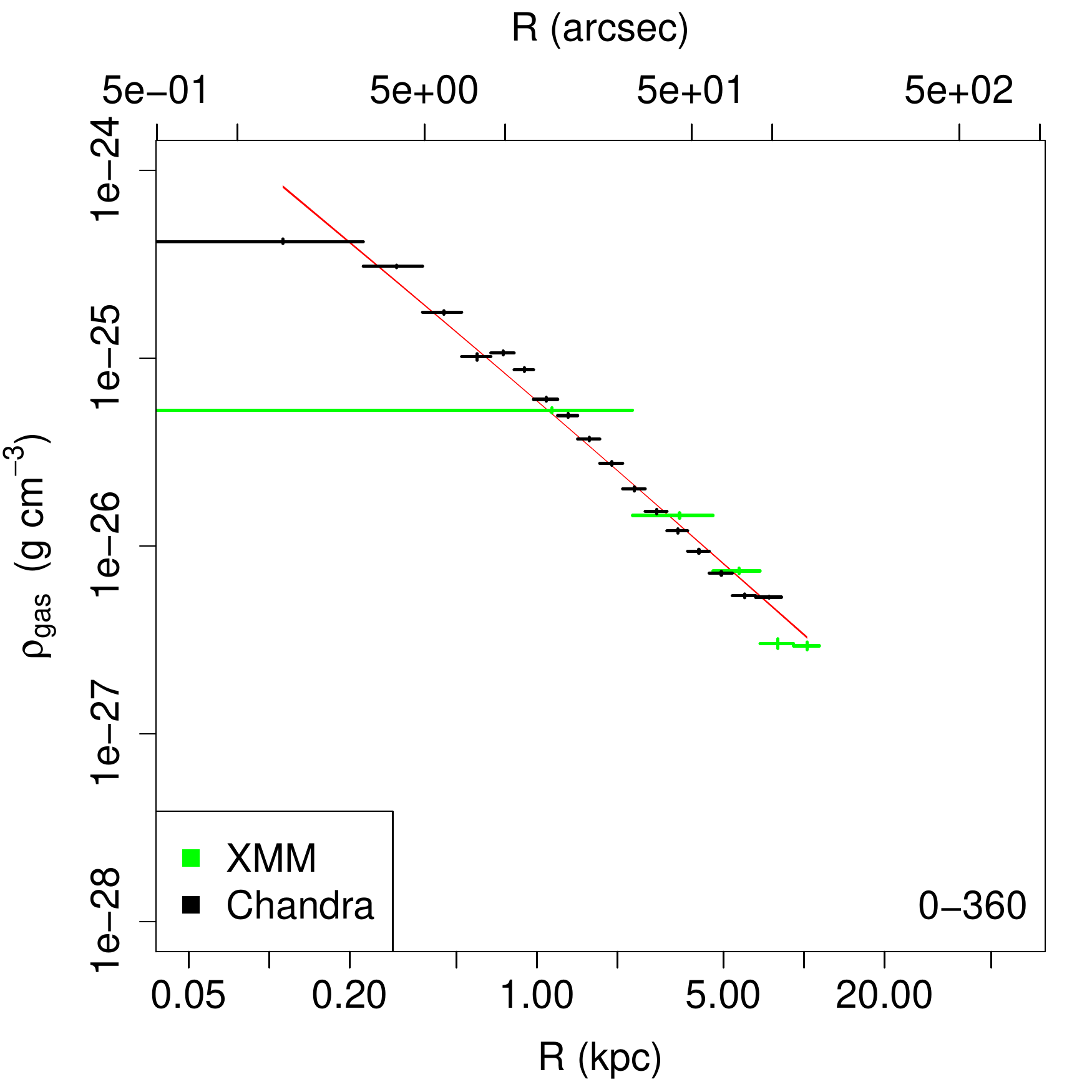}
\includegraphics[scale=0.22]{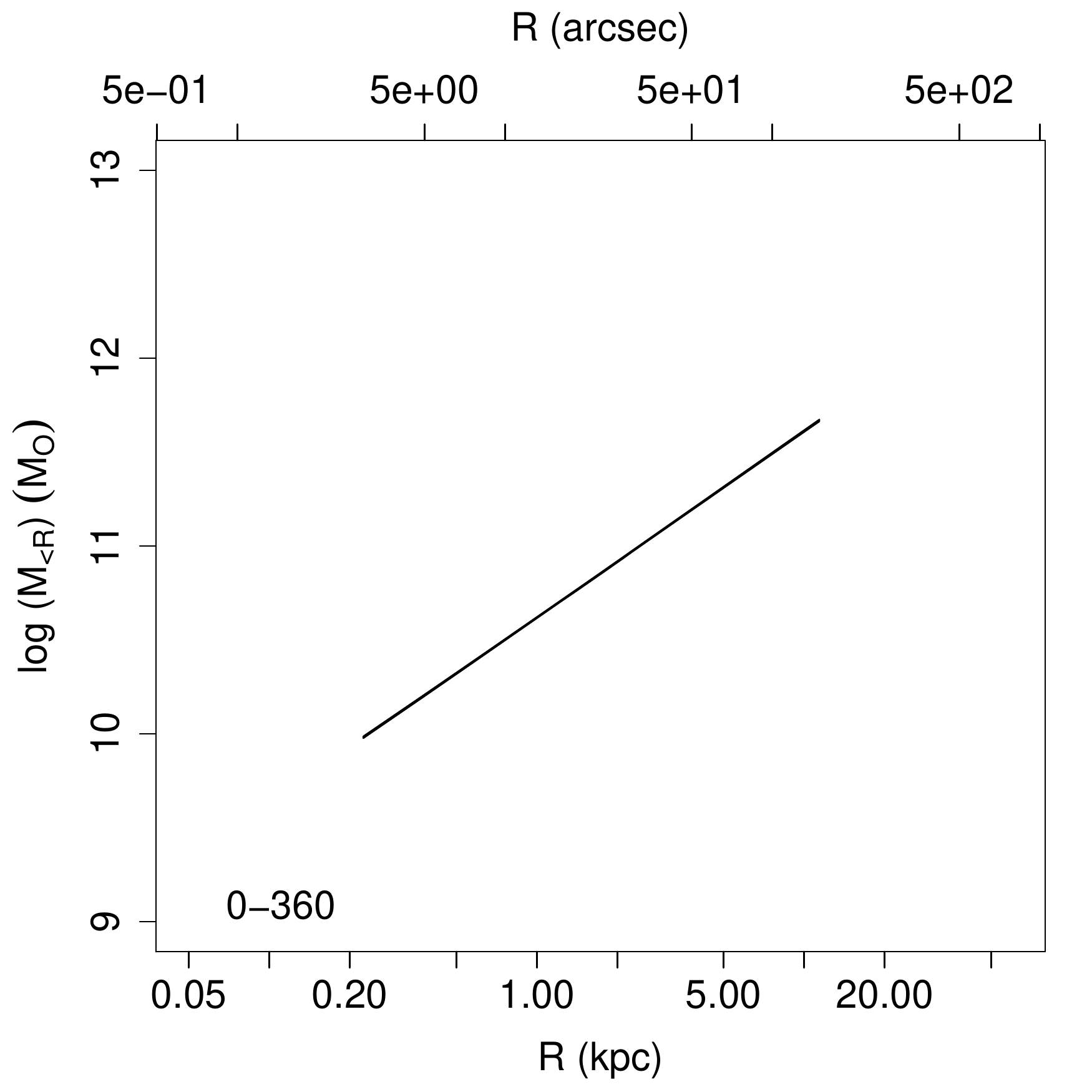}
\caption{Effect on the X-ray mass profiles of the increasing smoothing parameter of the spline model used for fitting the gas profiles. From left to right we present the gas temperature, gas density and resulting mass profile for the full (0-360) sector of NGC 4649 {for a minimum signal to noise ratio of 50 for \textit{XMM-Newton} and 100 for \textit{Chandra} data, respectively. From} top to bottom we show increasing smoothing parameters from 0.5 to 1 with 0.1 increase.}\label{fig:effects_smooth}
\end{figure}

\begin{figure}
\centering
\includegraphics[scale=0.25]{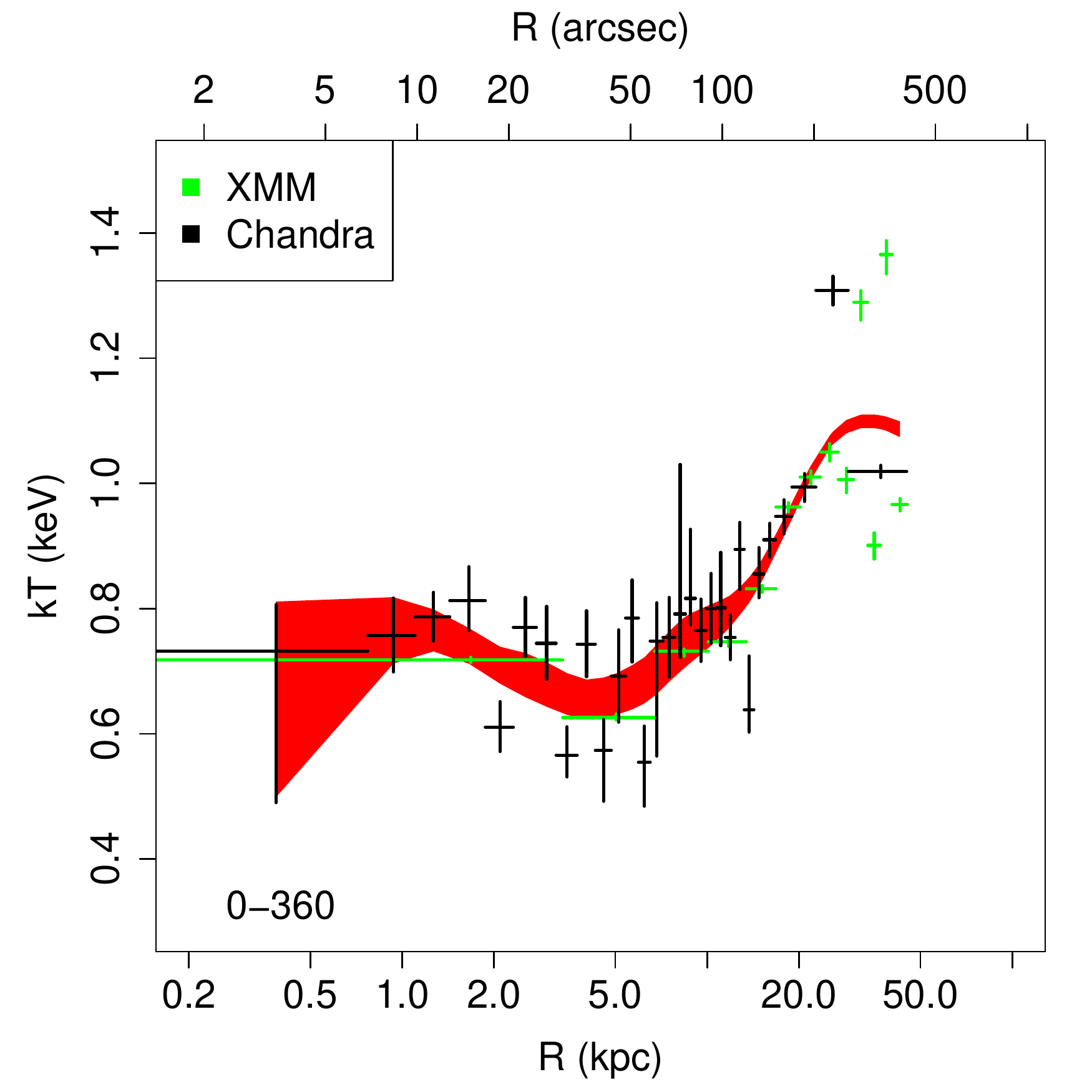}
\includegraphics[scale=0.25]{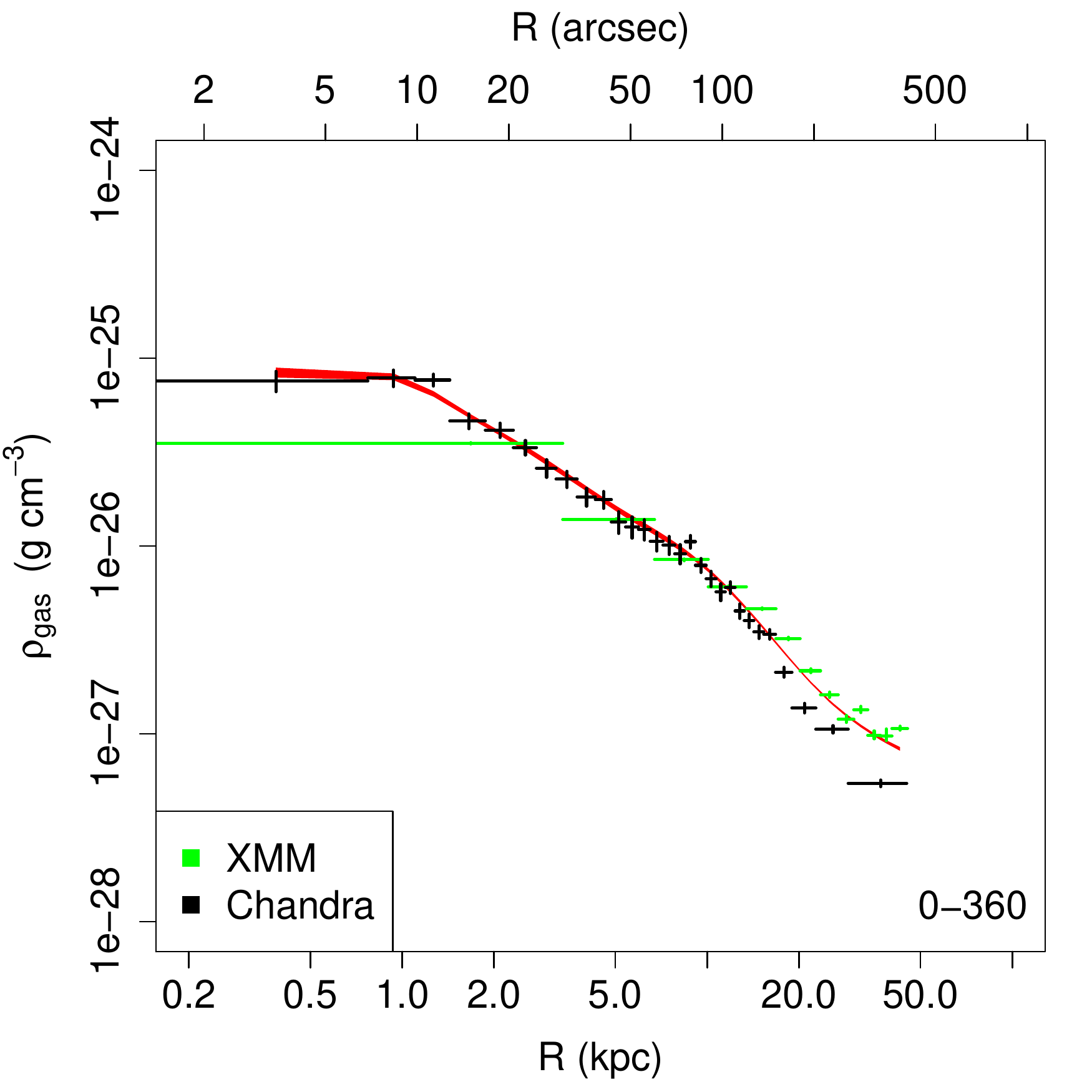}
\includegraphics[scale=0.25]{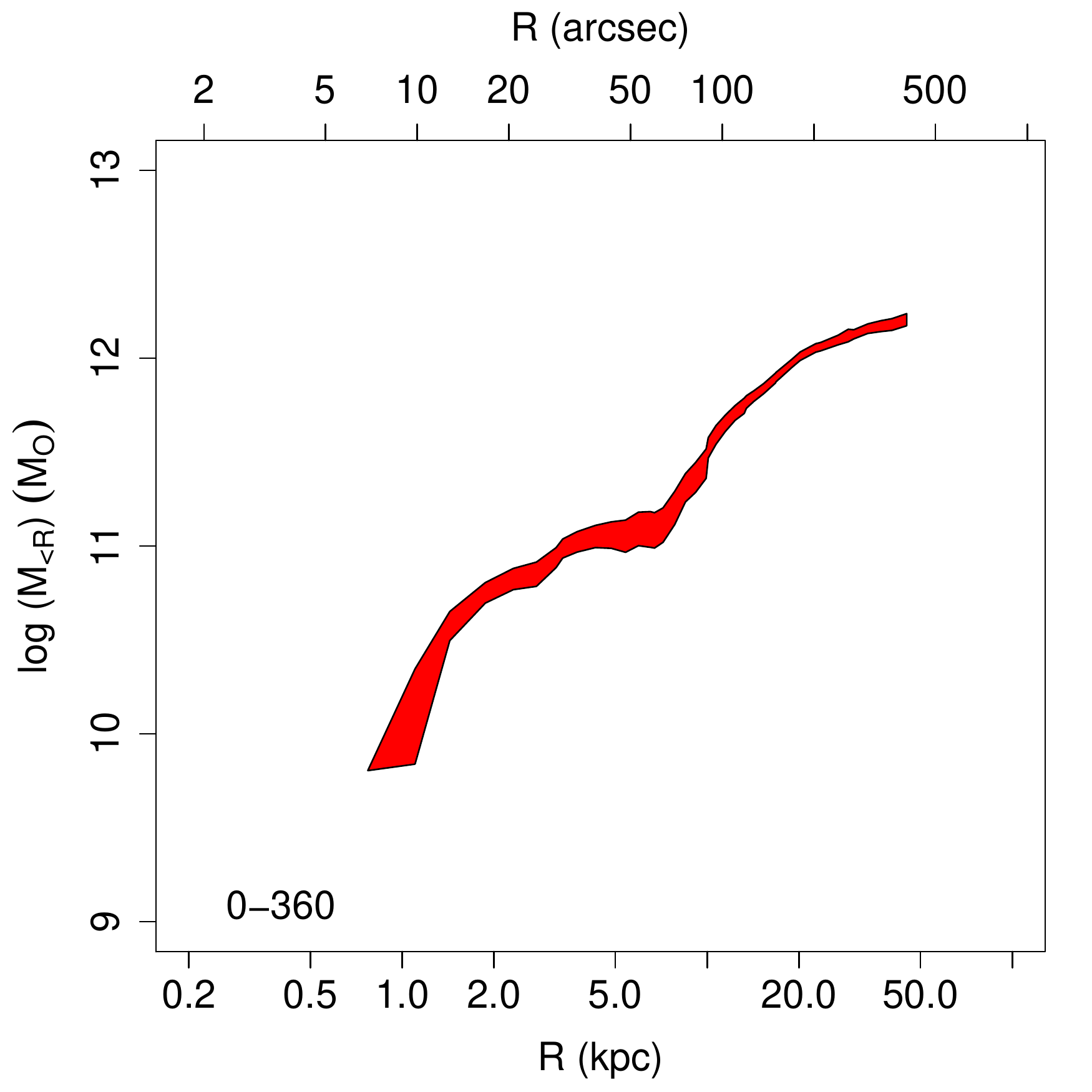}\\
\includegraphics[scale=0.25]{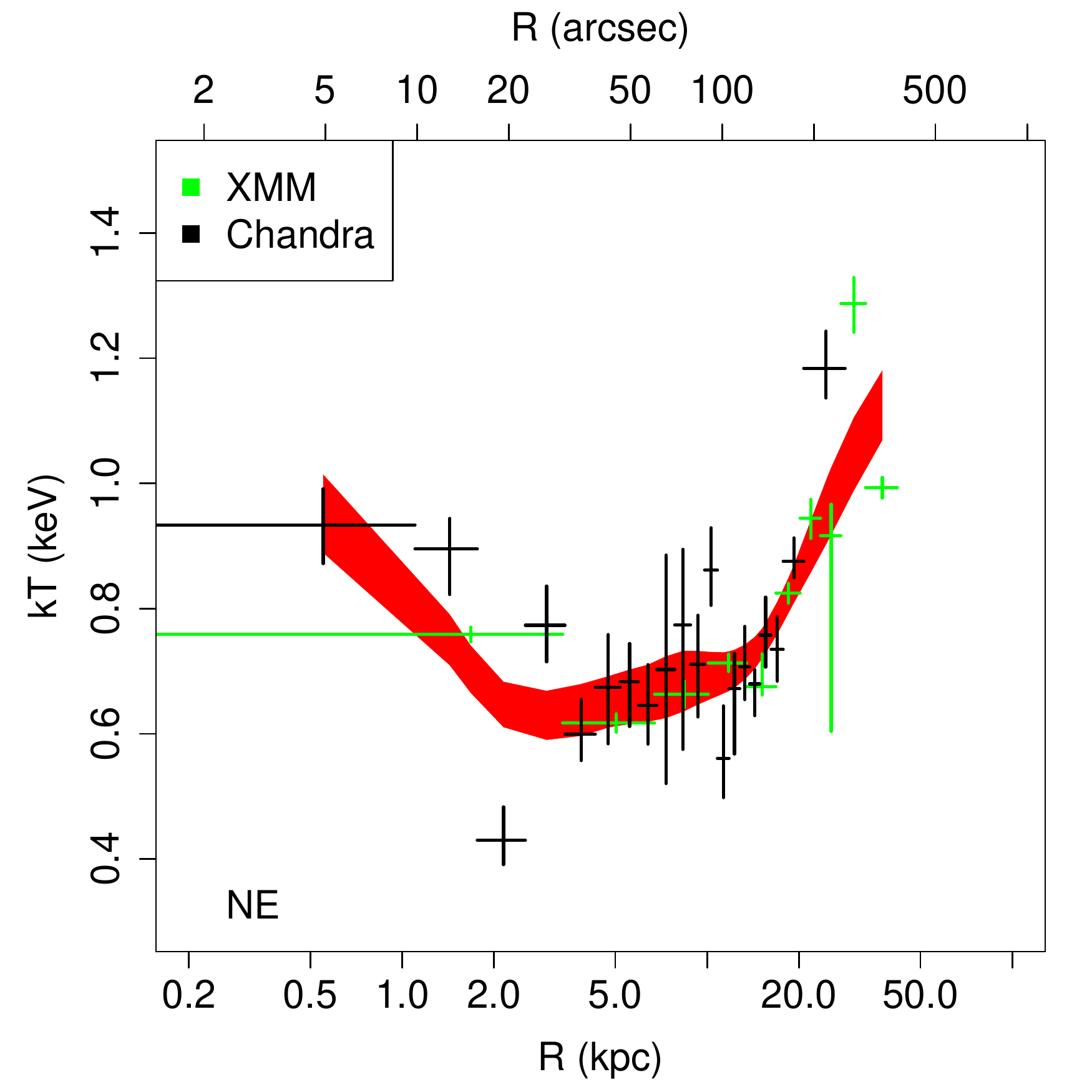}
\includegraphics[scale=0.25]{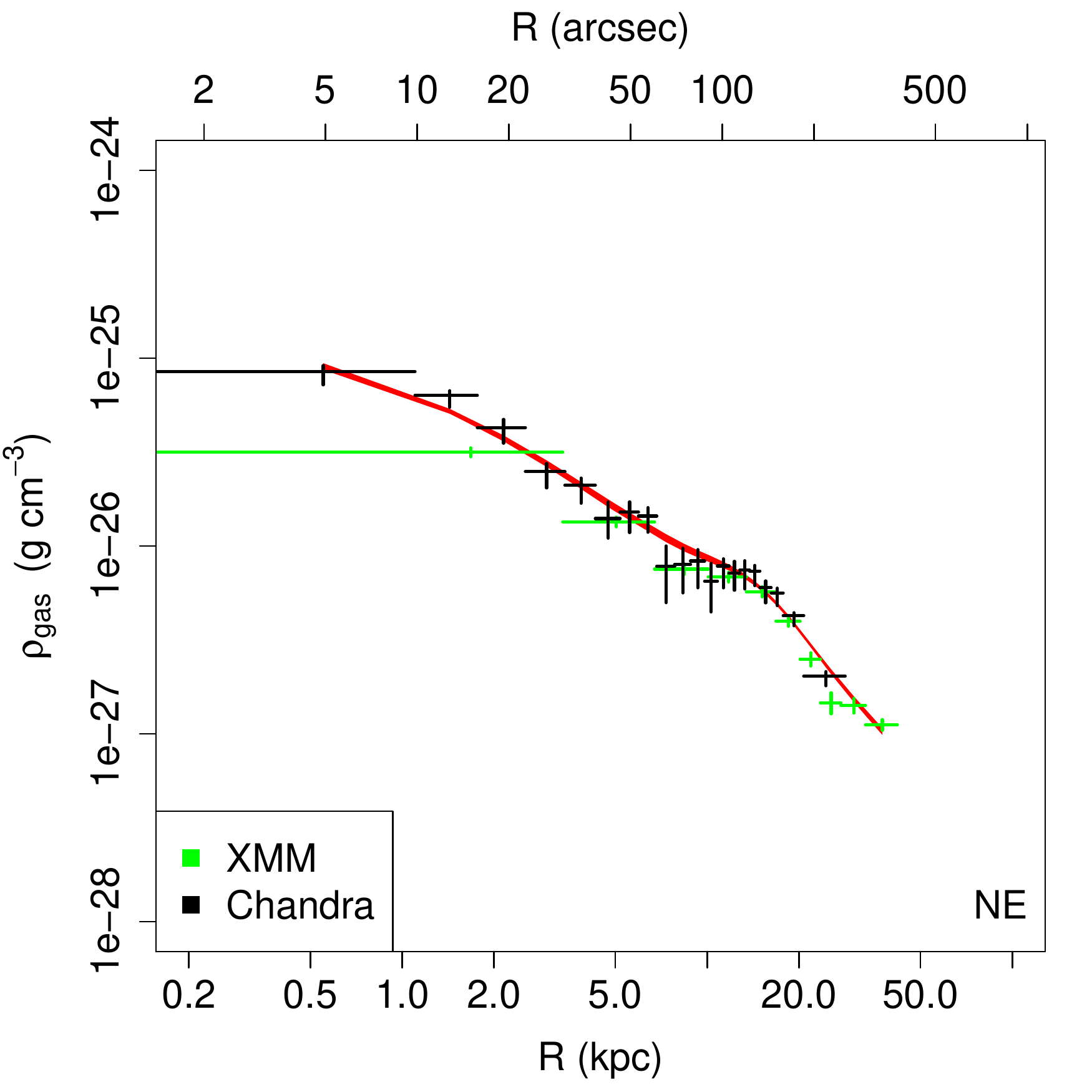}
\includegraphics[scale=0.25]{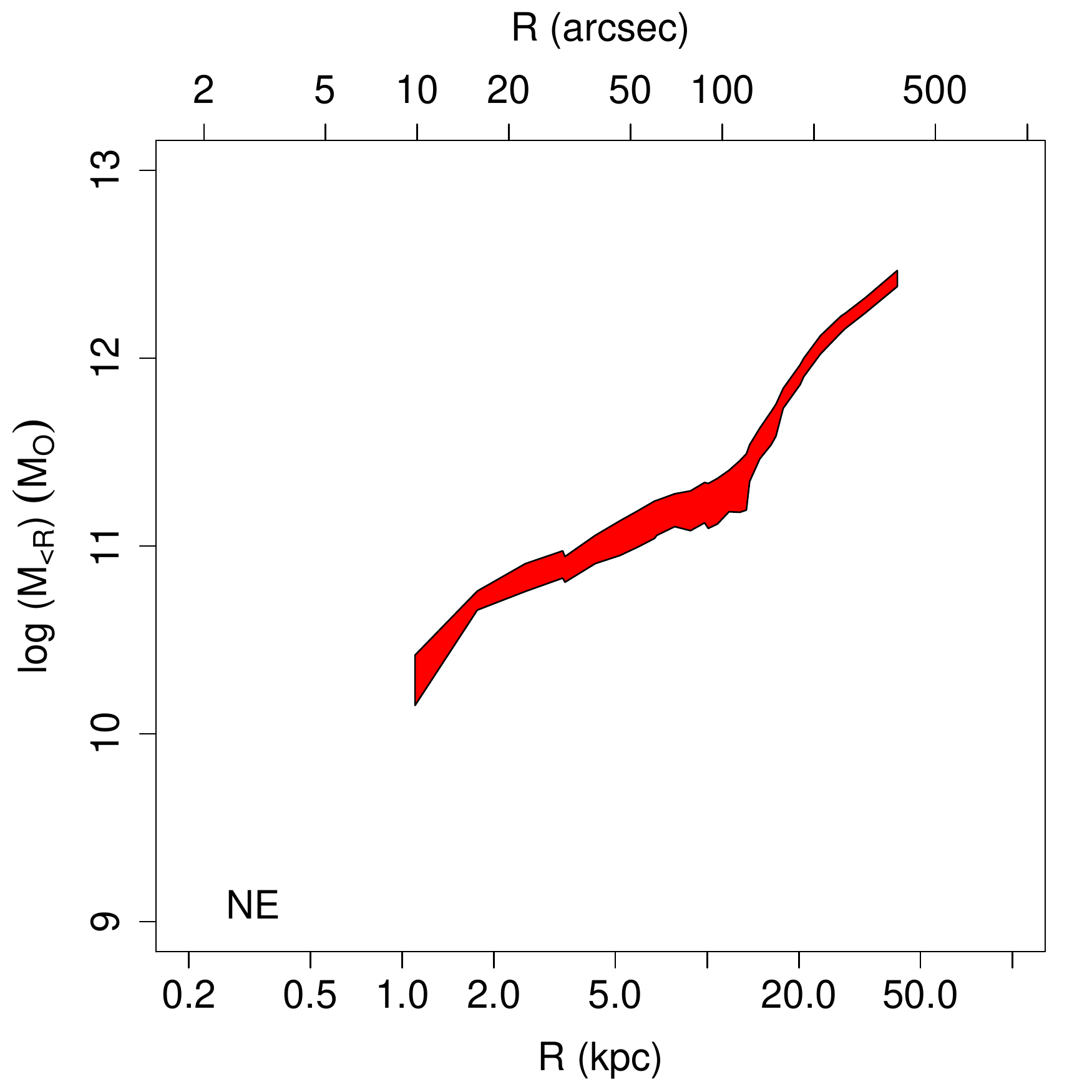}\\
\includegraphics[scale=0.25]{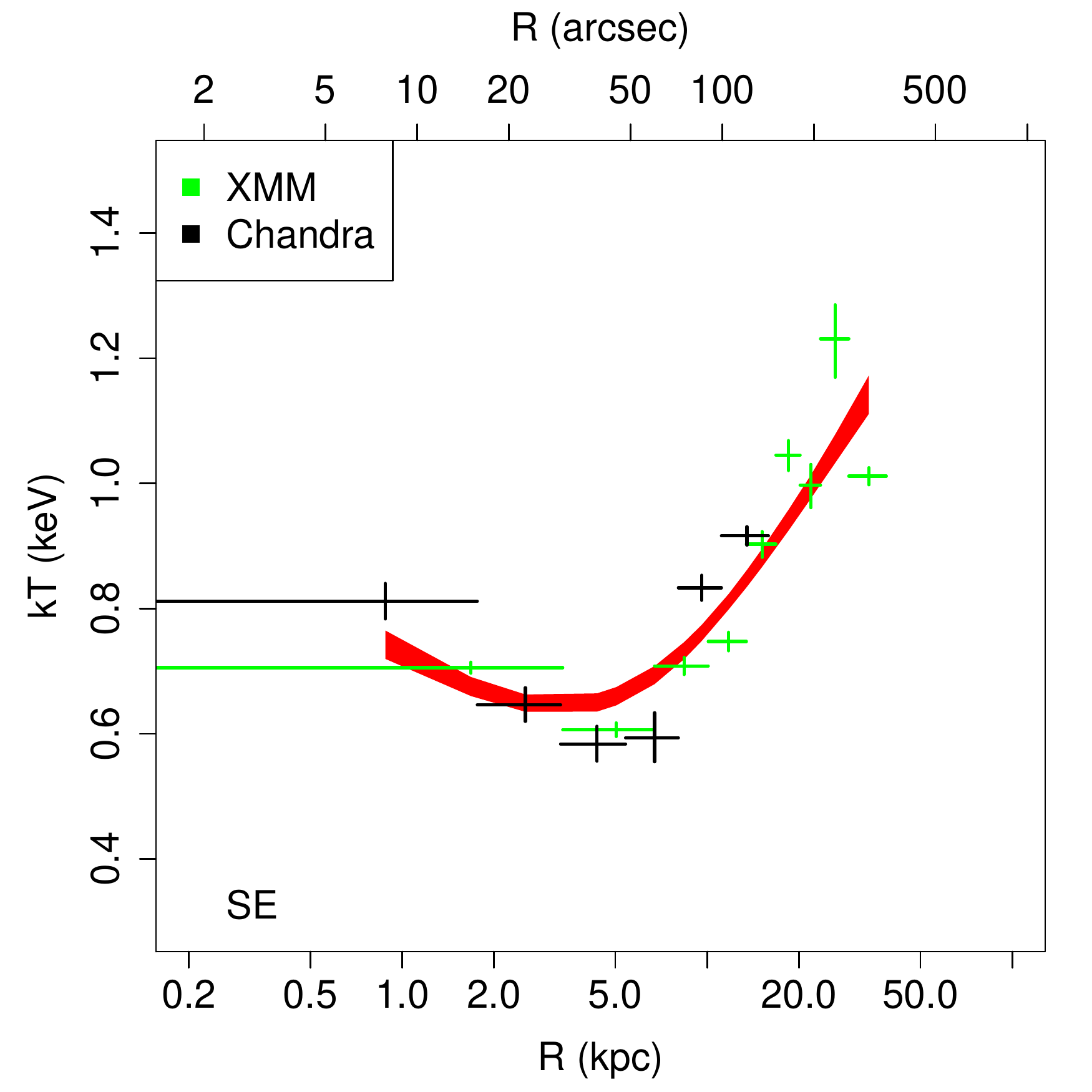}
\includegraphics[scale=0.25]{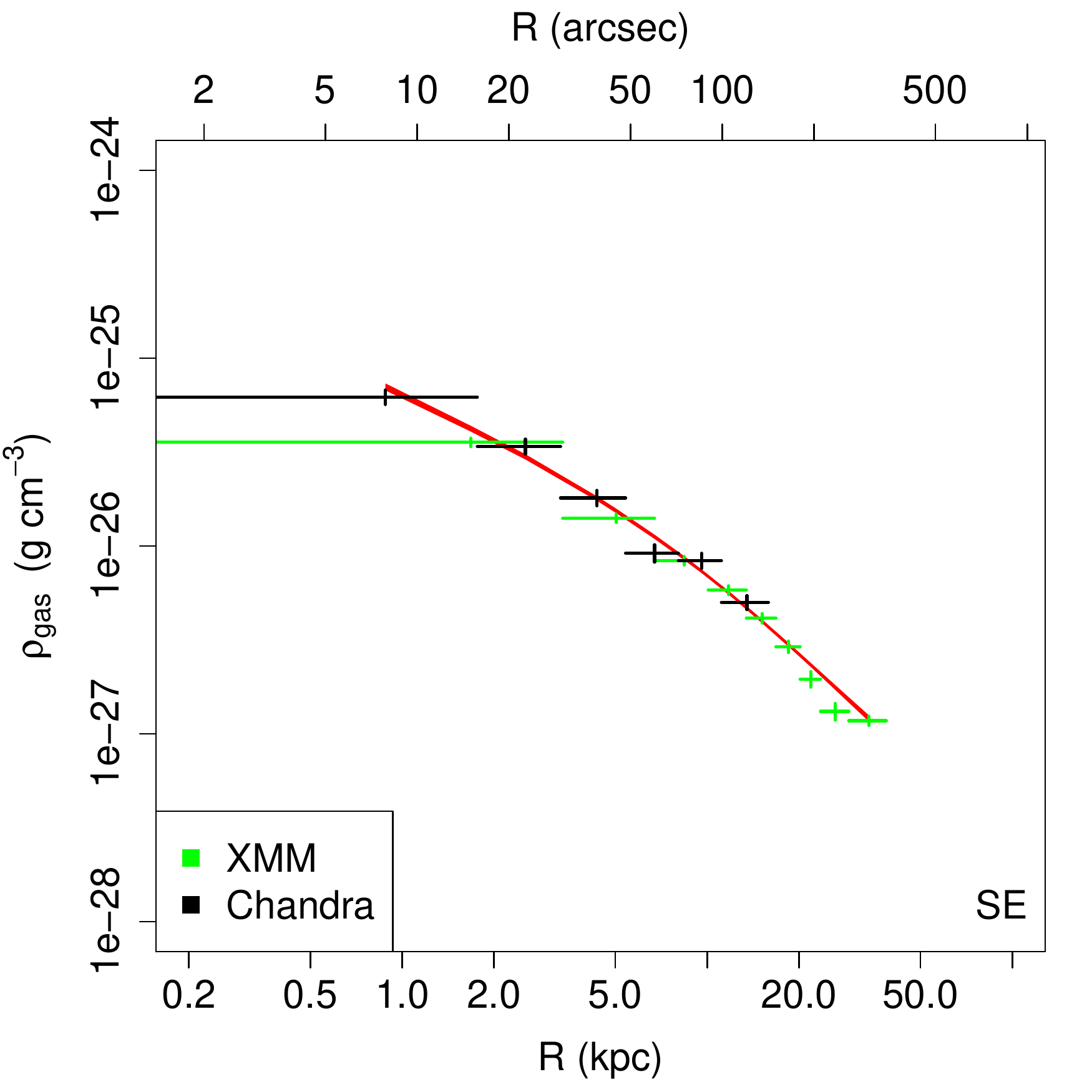}
\includegraphics[scale=0.25]{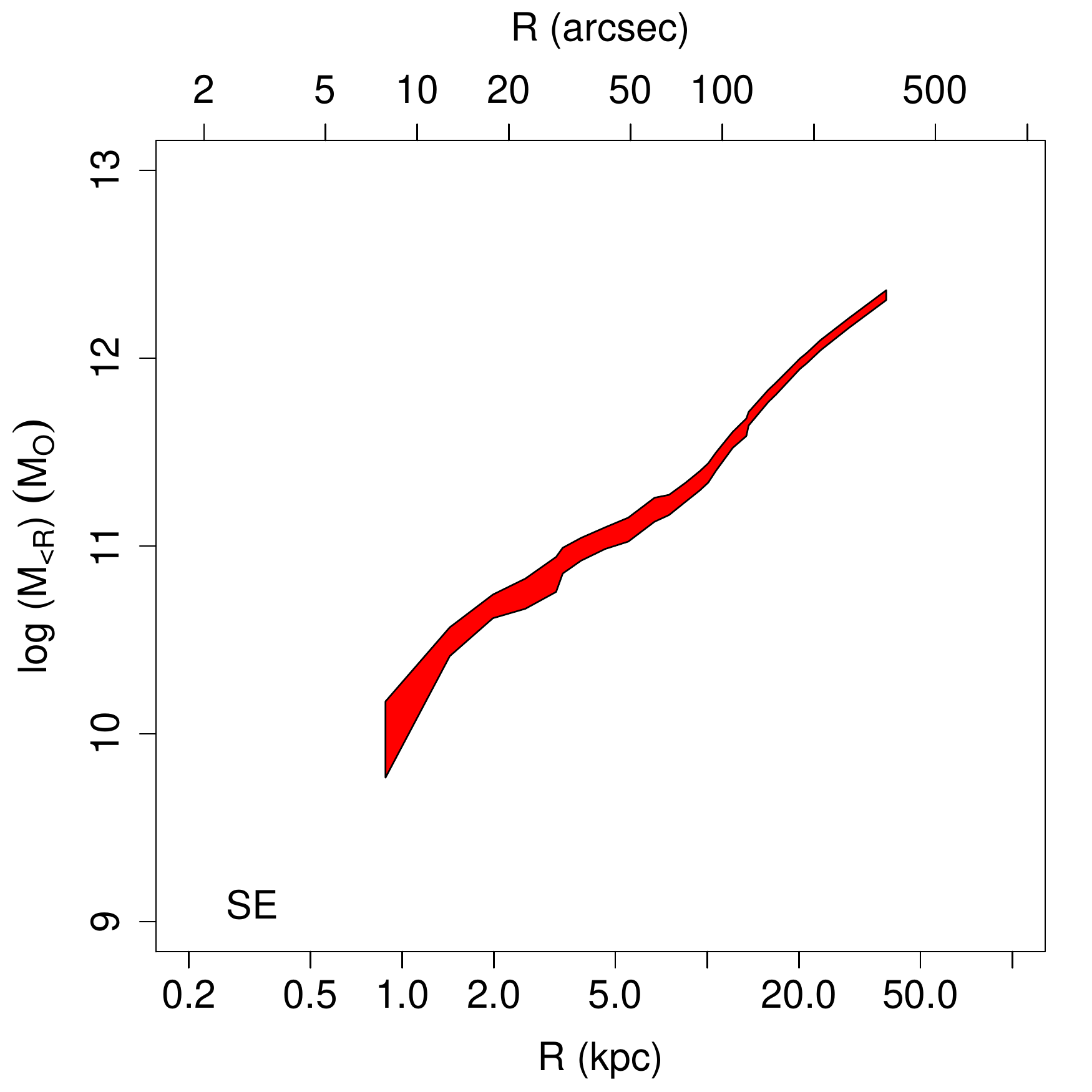}\\
\includegraphics[scale=0.25]{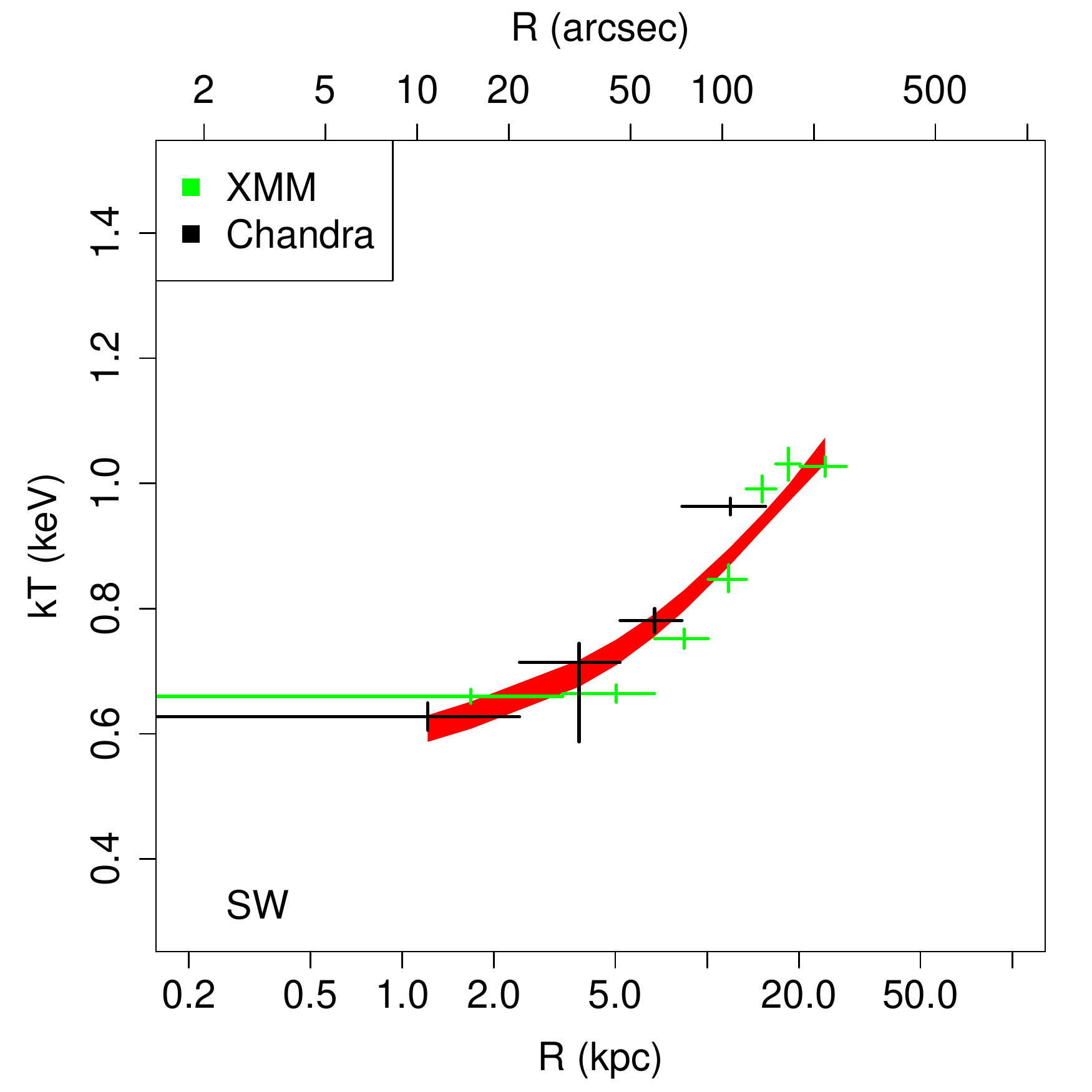}
\includegraphics[scale=0.25]{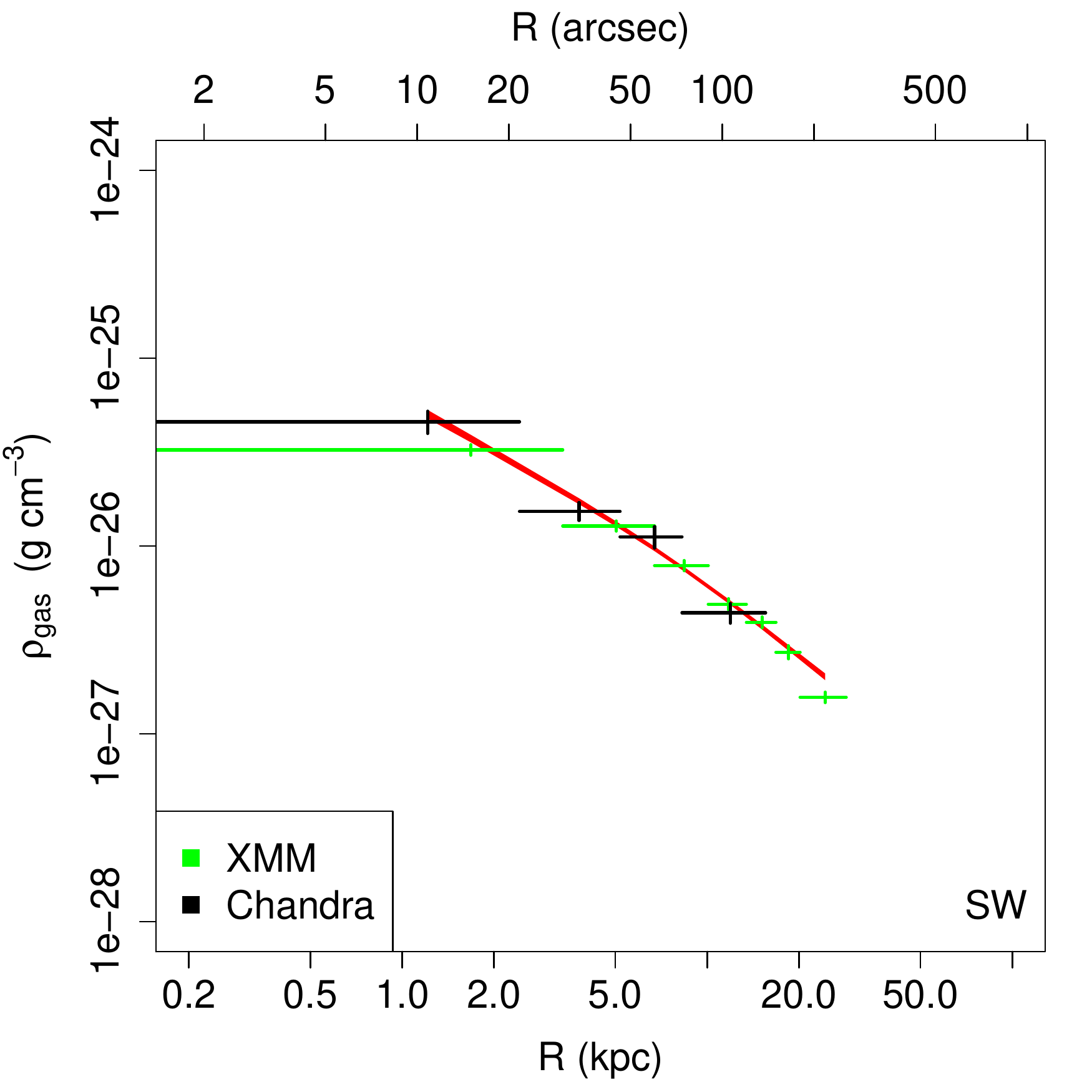}
\includegraphics[scale=0.25]{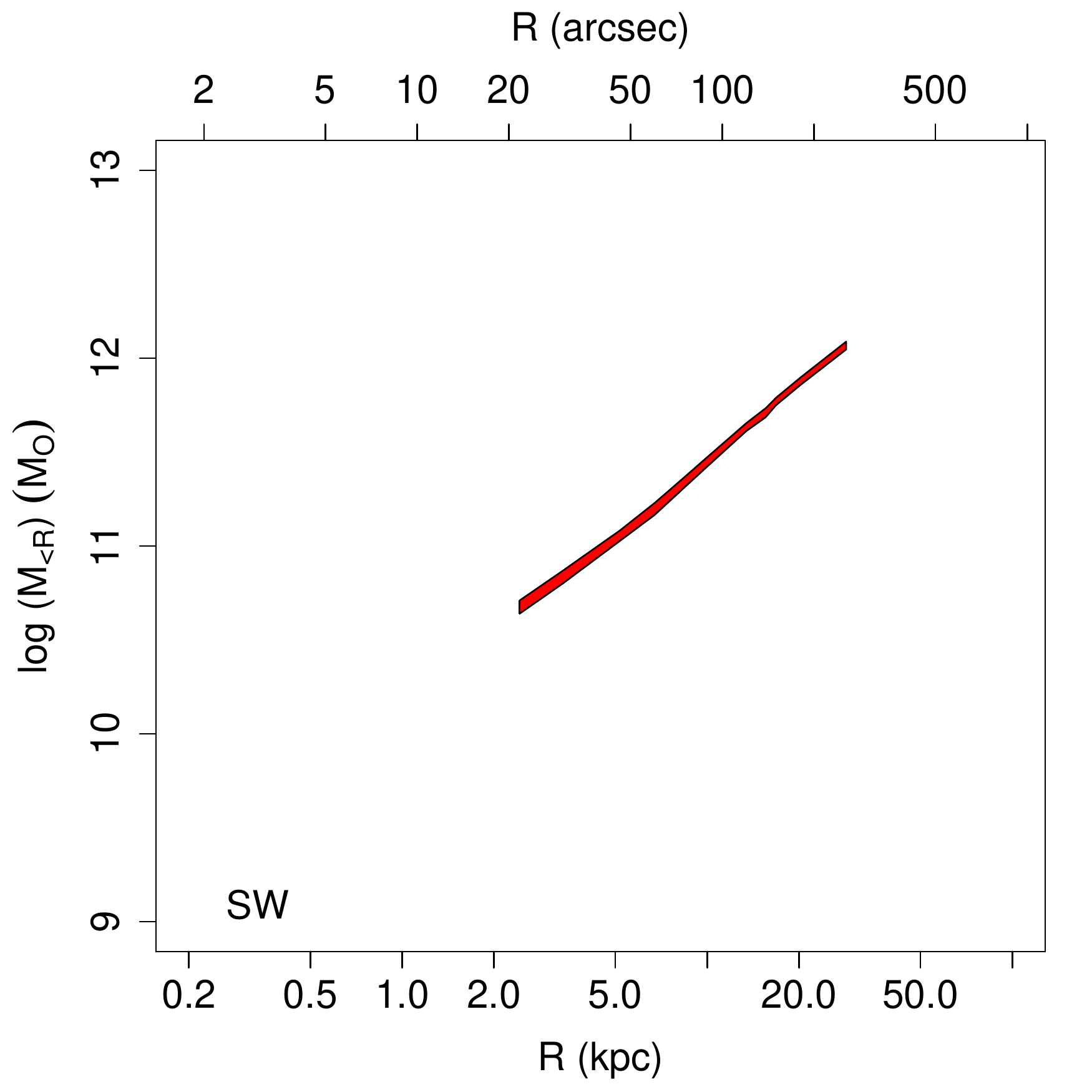}\\
\includegraphics[scale=0.25]{{N5846_temp_profile_merged_340_480_0_0_fit_0.7}.pdf}
\includegraphics[scale=0.25]{{N5846_nh_profile_merged_340_480_0_0_fit_0.7}.pdf}
\includegraphics[scale=0.25]{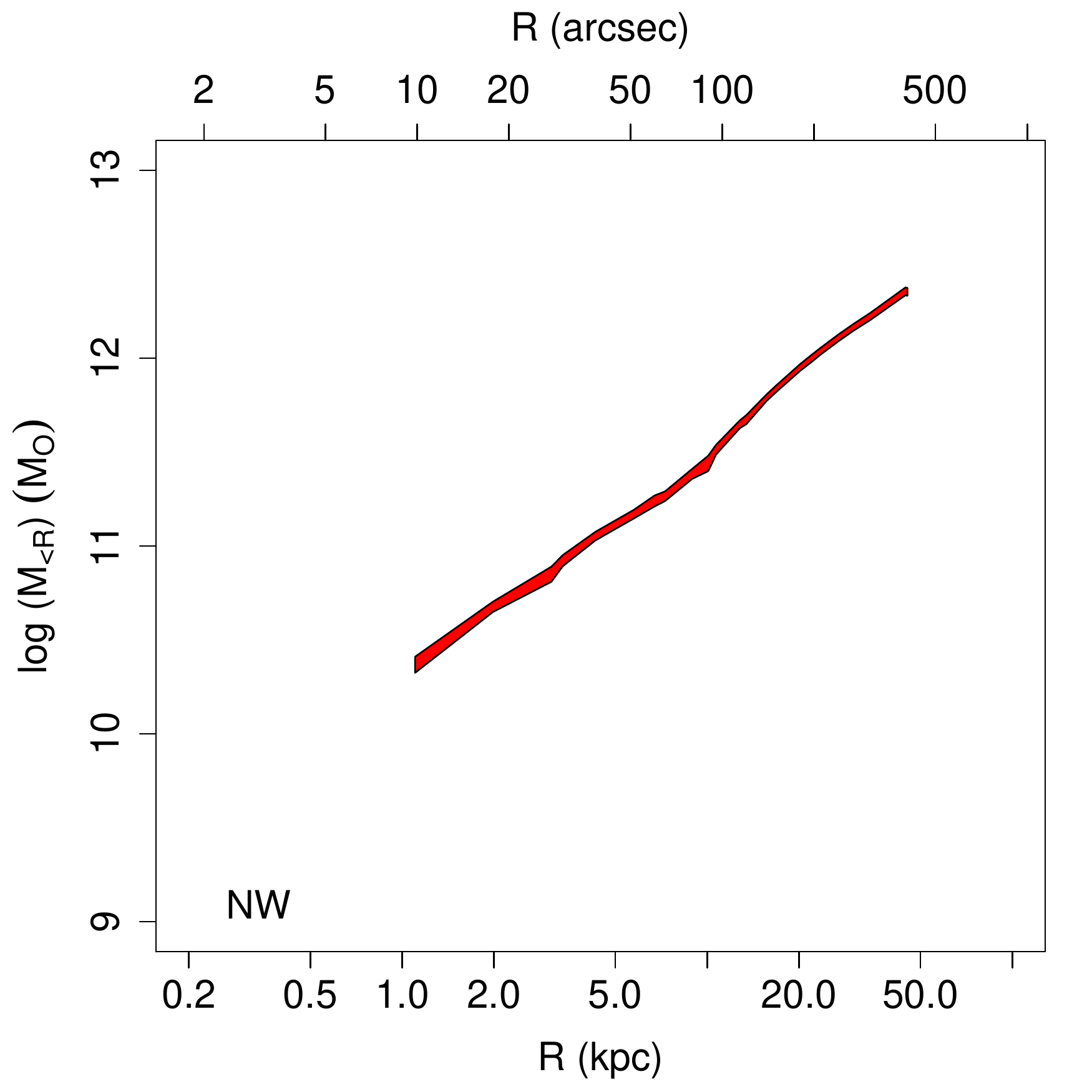}
\caption{Effect on the X-ray mass profiles of the different sectors selected for the spectral extraction. From left to right we present the gas temperature, gas density and resulting mass profile for NGC 5846 (with a smoothing parameter of 0.7), while from top to bottom we show the results in the full (0-360), NE (30-90), SE (90-180), SW (180-250) and NW (250-30) sectors, respectively. {The minimum signal to noise ratio is 30 for \textit{XMM-Newton} data in all sectors, while for \textit{Chandra} data the minimum signal to noise ratio is 50 in full (0-360), SW (180-250) and NW (250-30) sectors and 30 in NE (30-90) and SE (90-180) sectors.}}\label{fig:effects_pie}
\end{figure}

\begin{figure}
\centering
\includegraphics[scale=0.22]{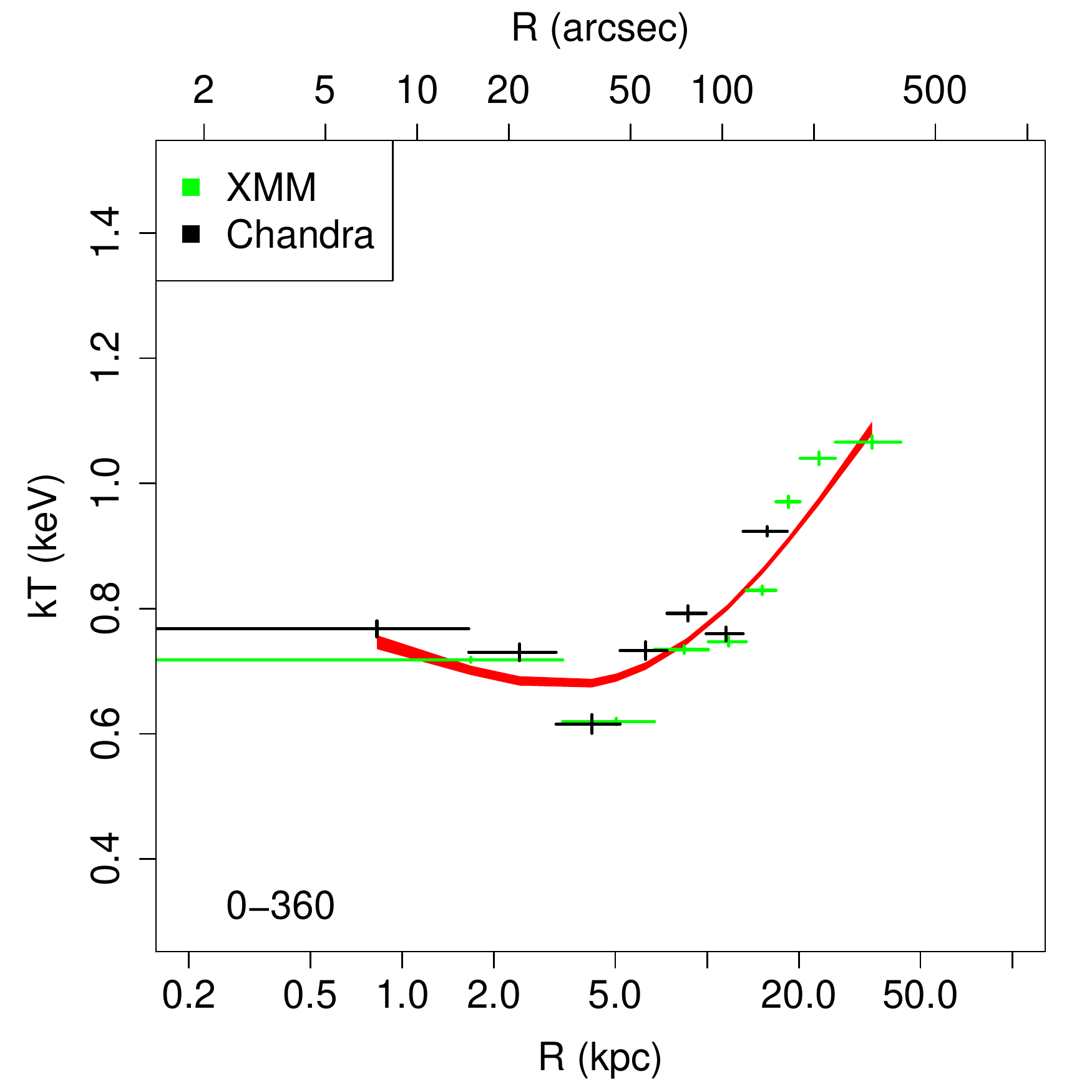}
\includegraphics[scale=0.22]{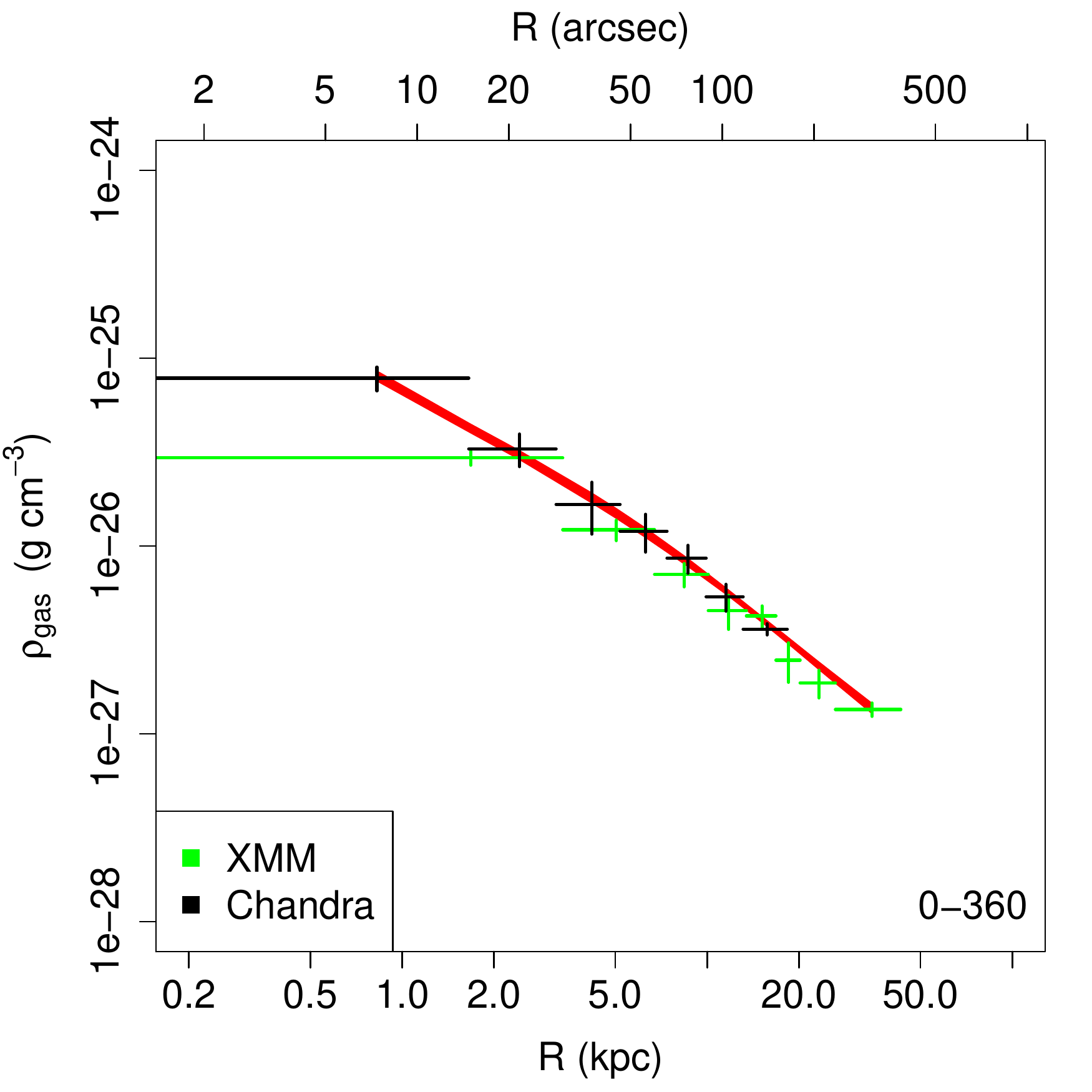}
\includegraphics[scale=0.22]{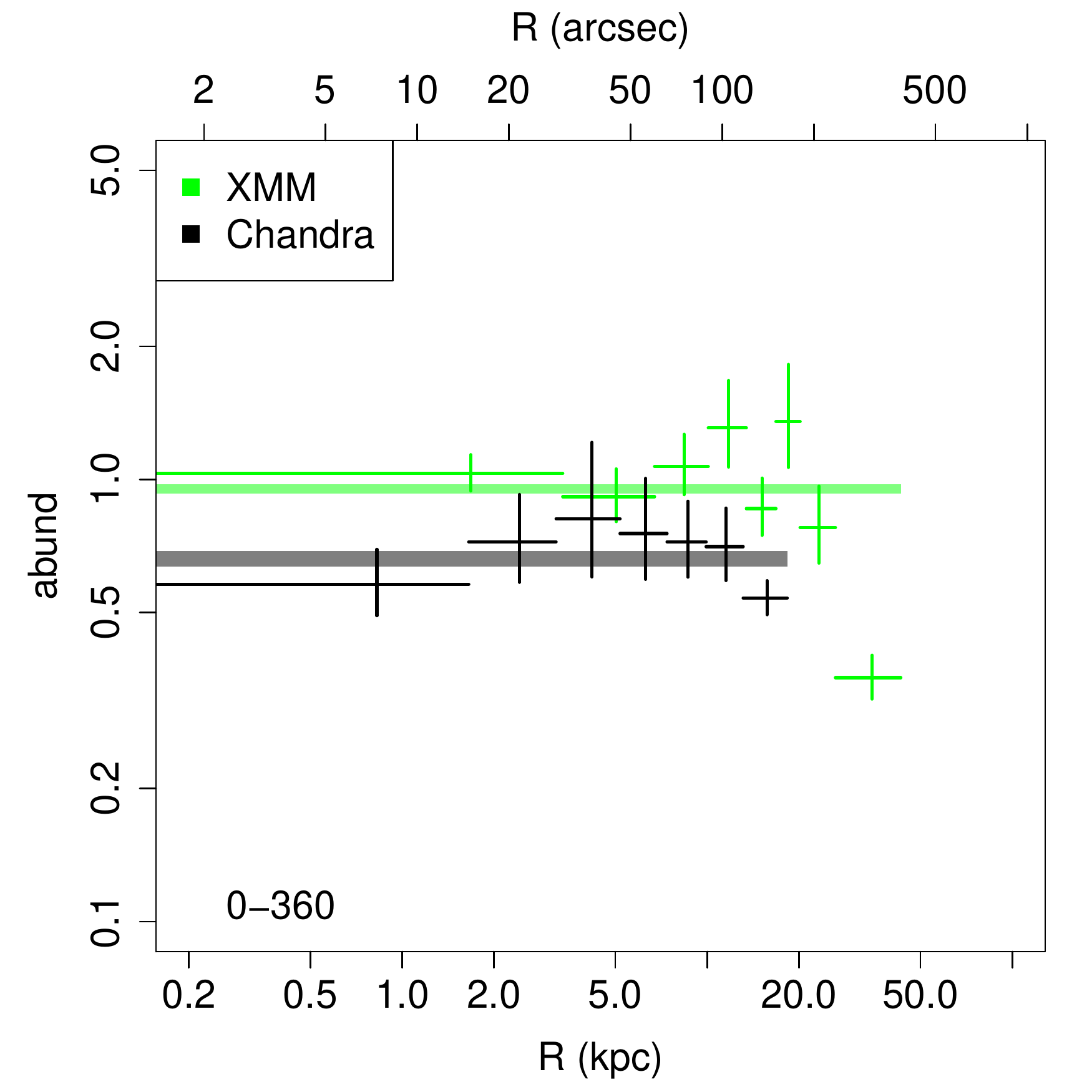}\\
\includegraphics[scale=0.22]{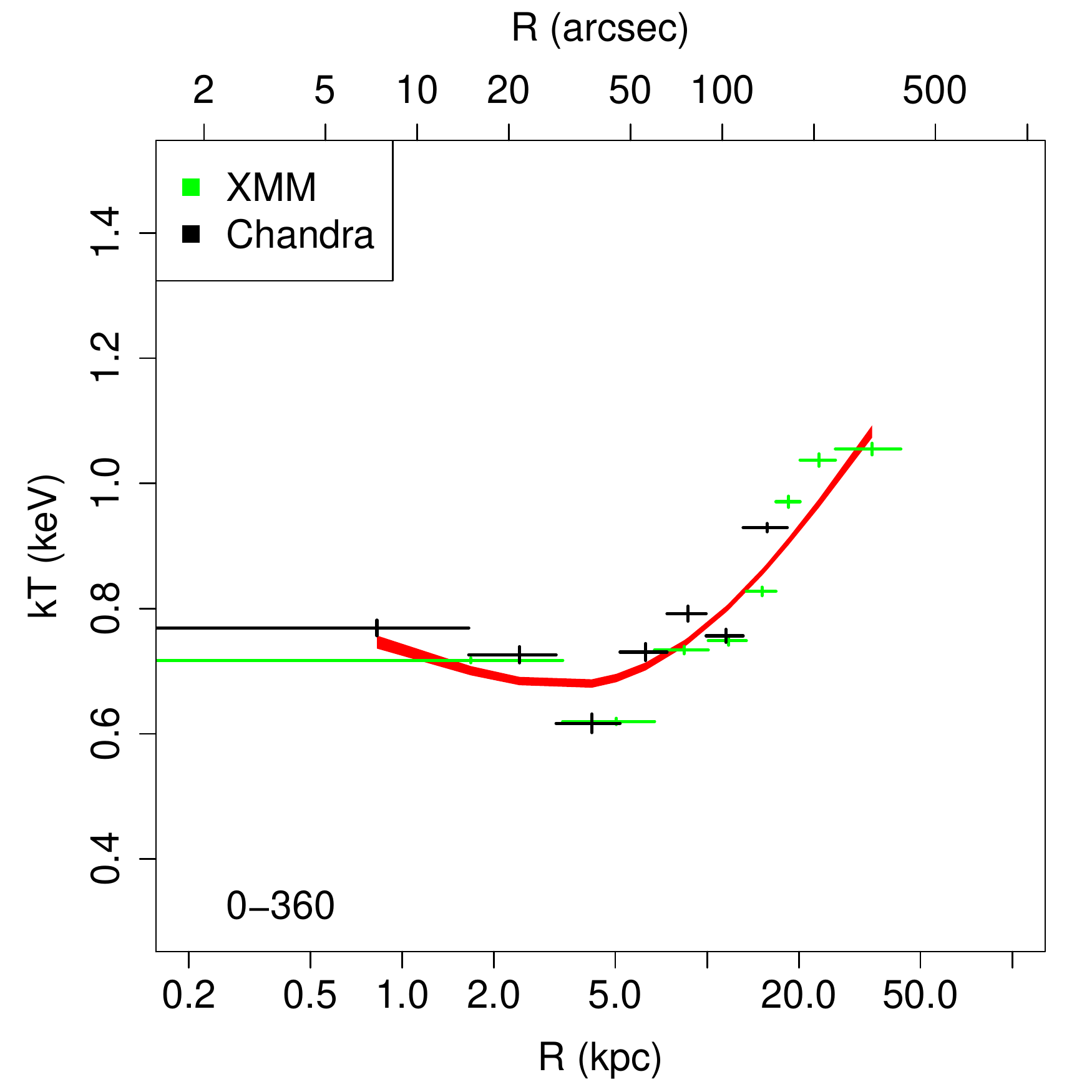}
\includegraphics[scale=0.22]{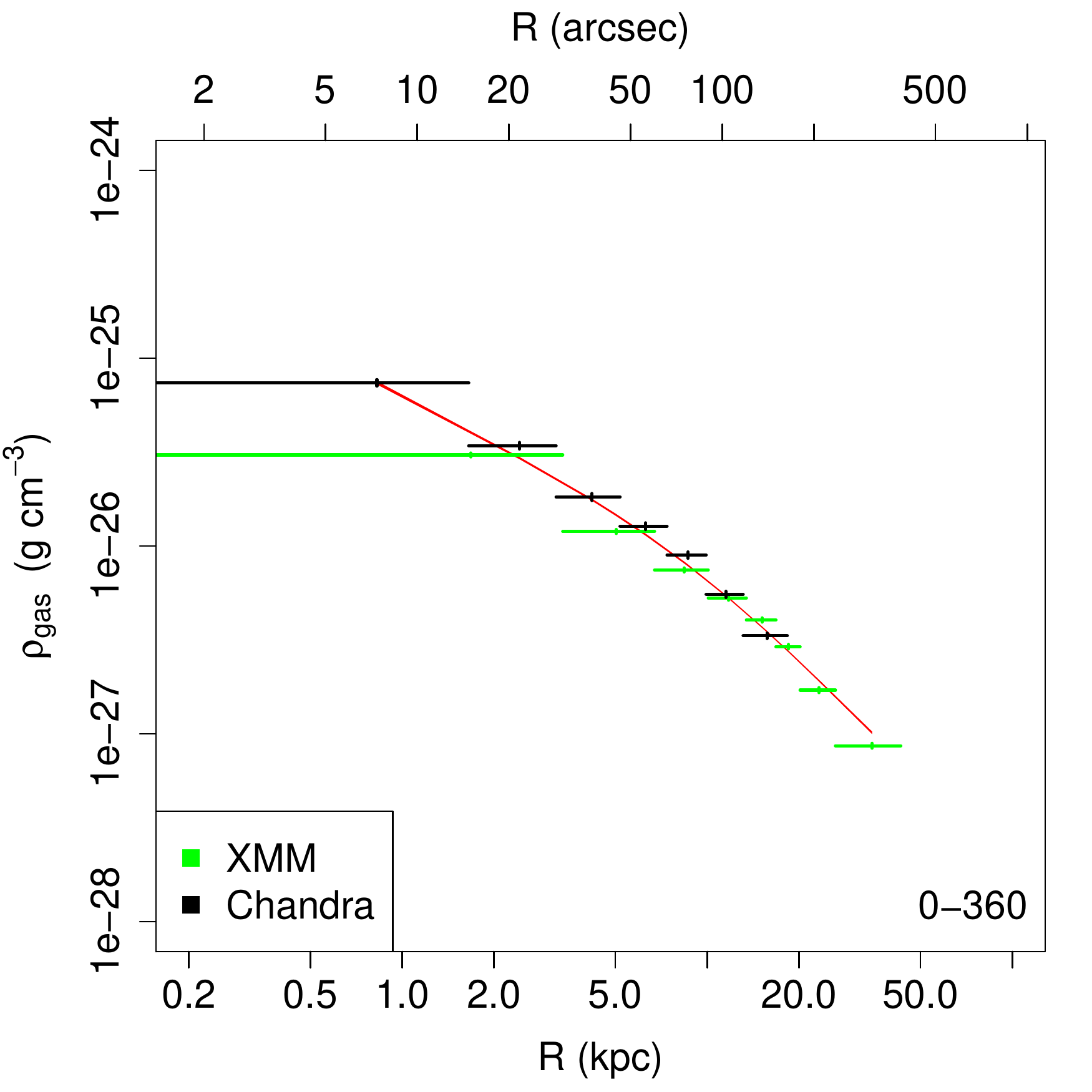}
\includegraphics[scale=0.22]{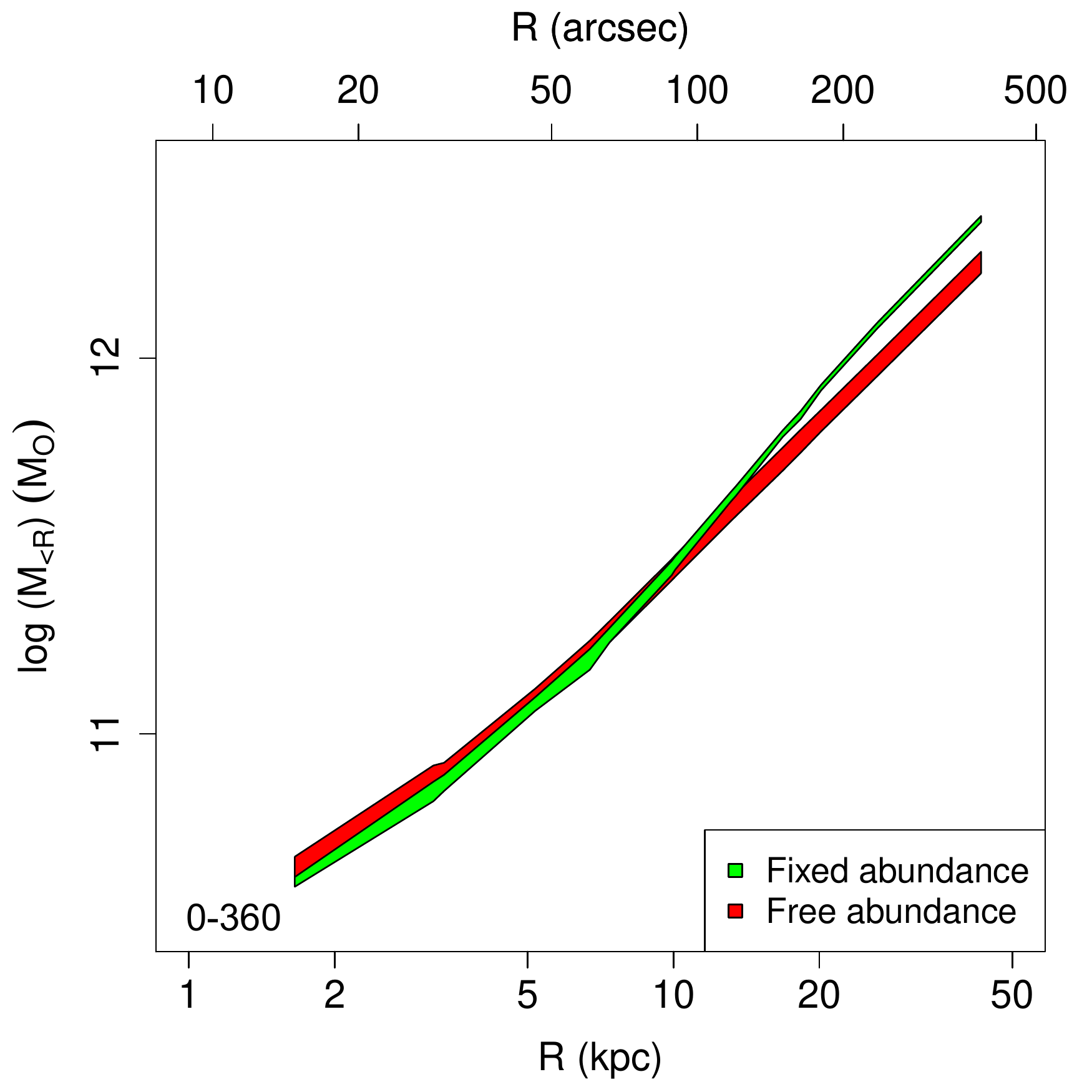}
\caption{Effect on the X-ray mass profiles of the free element abundance. In the top panels we present results from free abundance model for NGC 5846 in the full (0-360) sector, in particular from left to right we show the gas temperature, gas density (with overplotted the best fit spline model with a smoothing parameter of 0.7) and abundance profiles. In the bottom panels we present results from fixed abundance model, from left to right we show the gas temperature profile, gas density profile, and a comparison between the mass profiles obtained from the fixed (green) and free (red) abundance model. {The minimum signal to noise ratio is 100 both for \textit{XMM-Newton} and \textit{Chandra} data.}}\label{fig:effects_abundances}
\end{figure}

\begin{figure}
\centering
\includegraphics[scale=0.2]{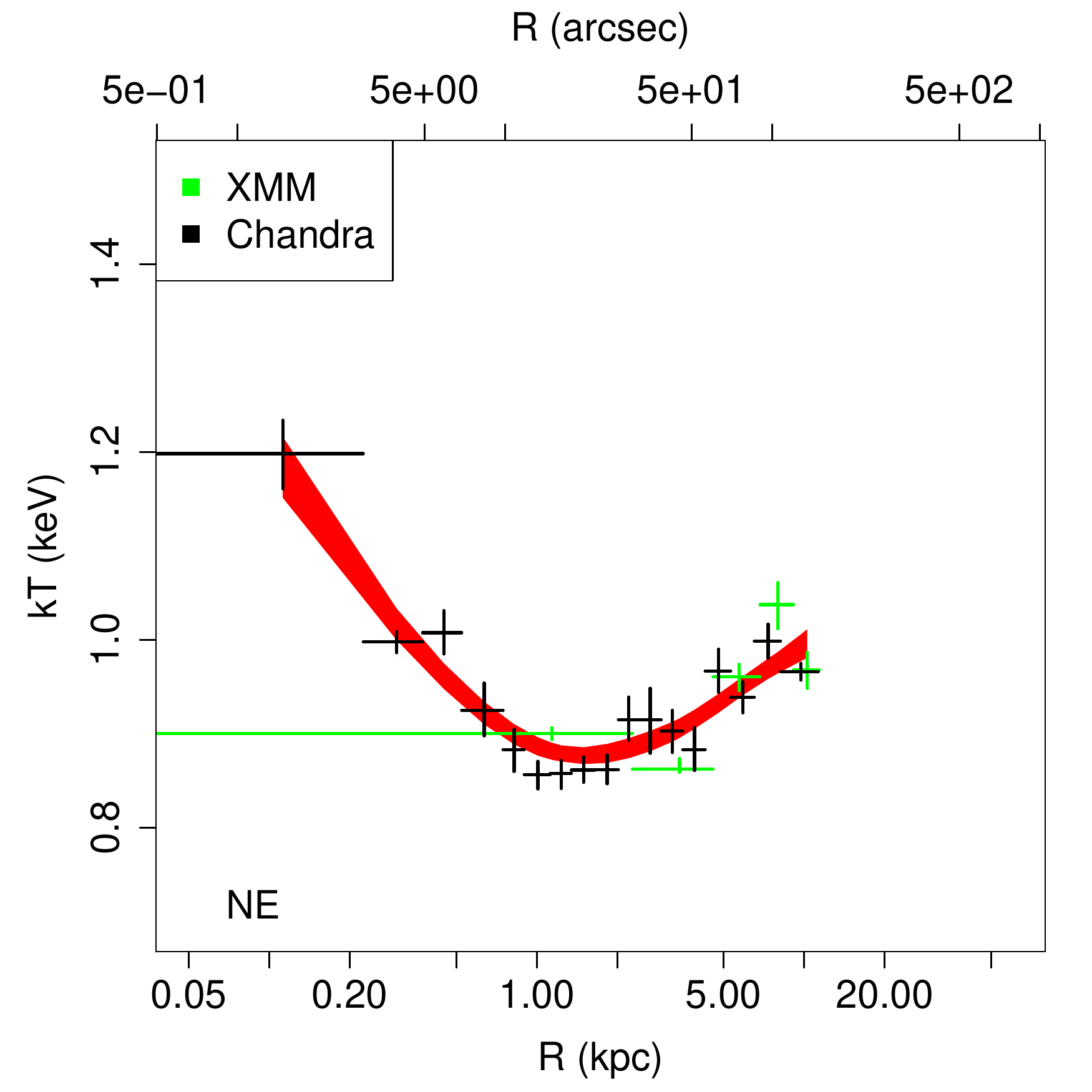}
\includegraphics[scale=0.2]{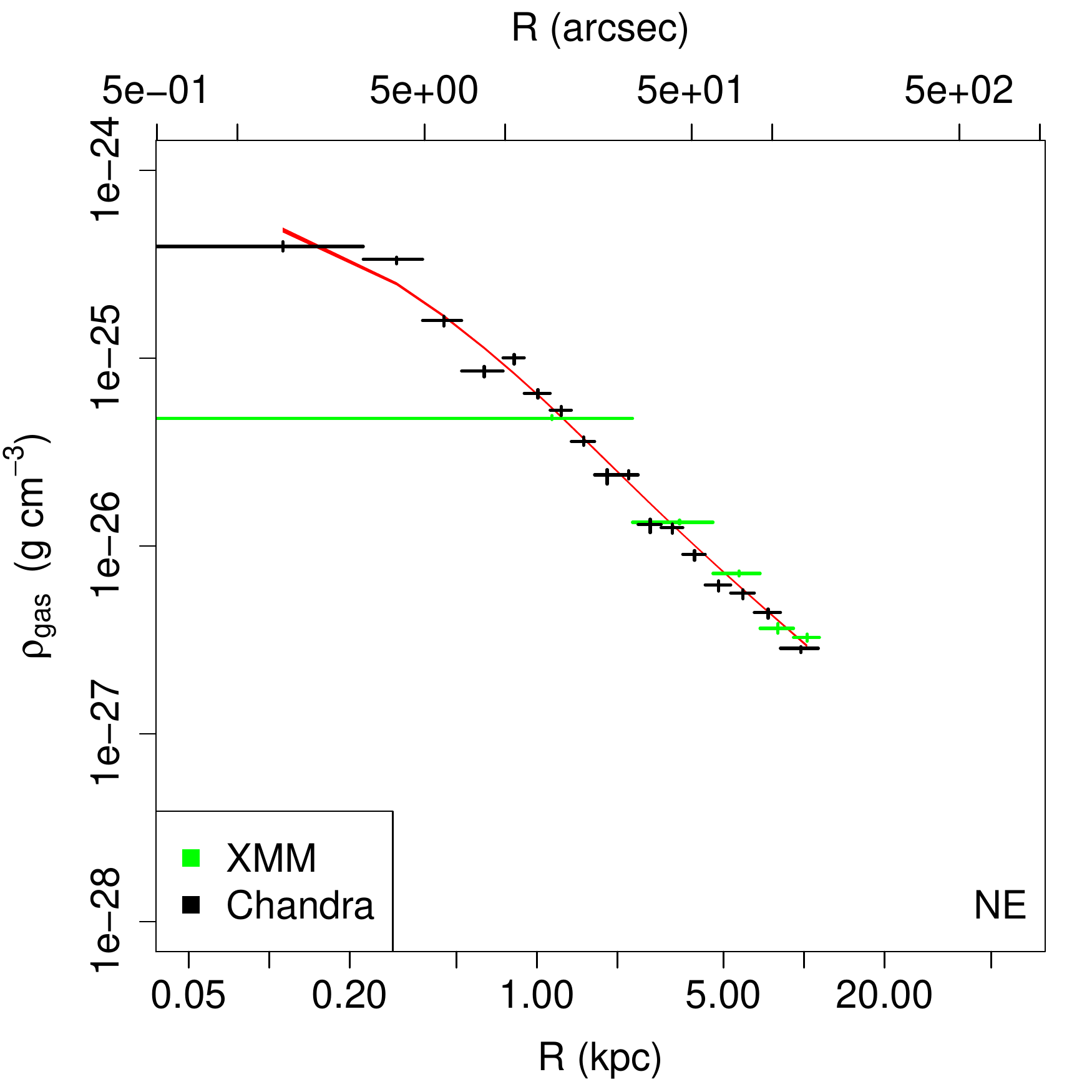}
\includegraphics[scale=0.2]{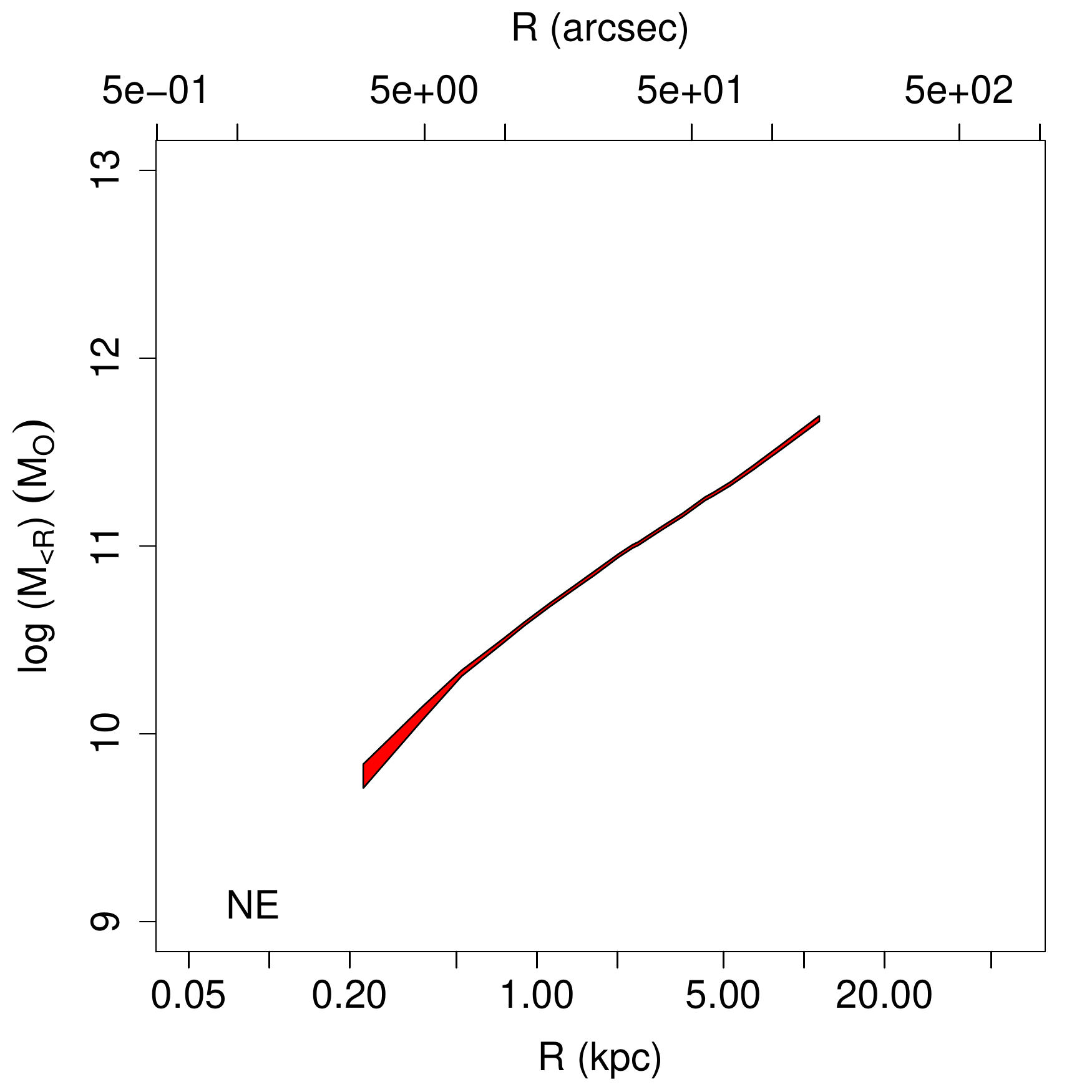}\\
\includegraphics[scale=0.2]{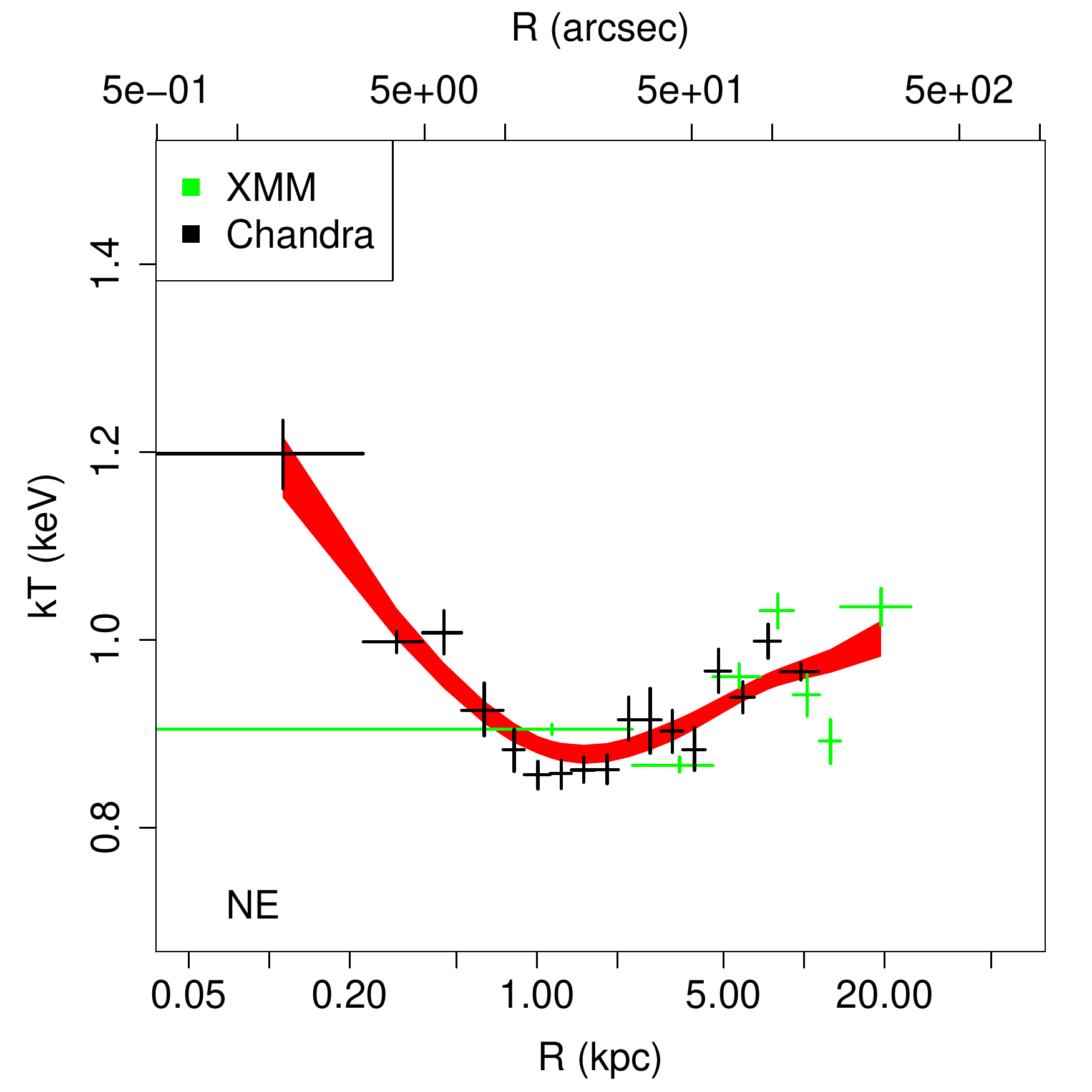}
\includegraphics[scale=0.2]{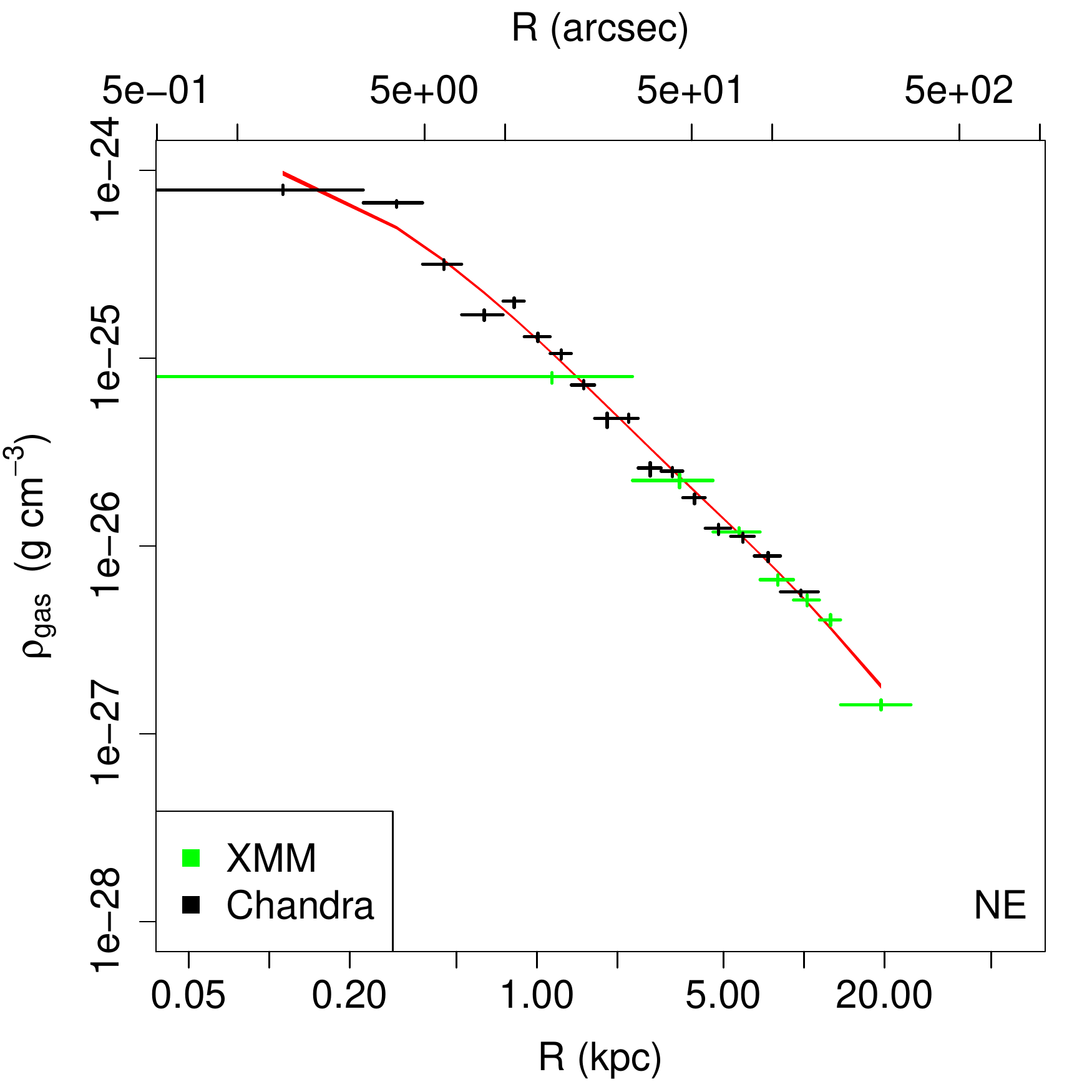}
\includegraphics[scale=0.2]{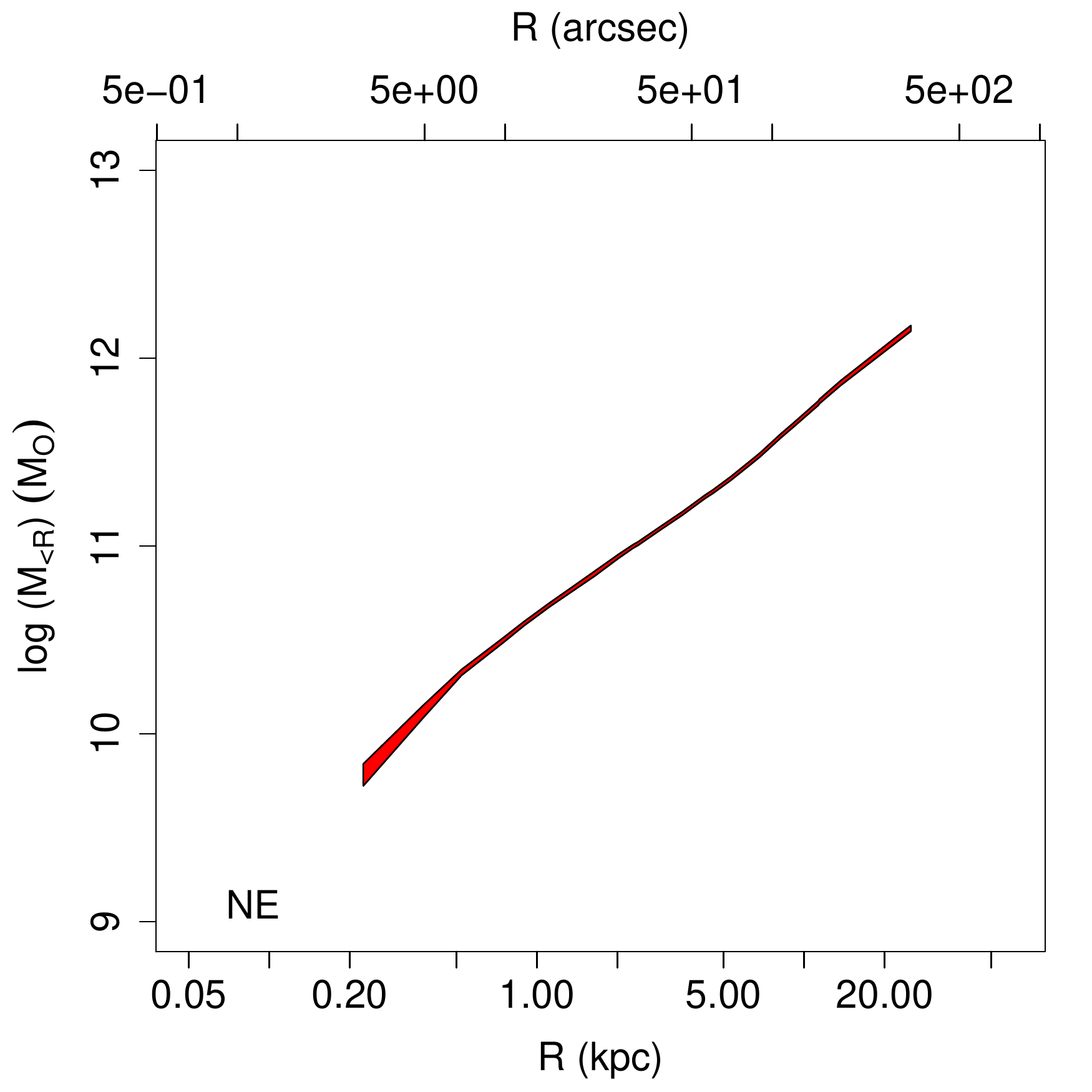}\\
\includegraphics[scale=0.2]{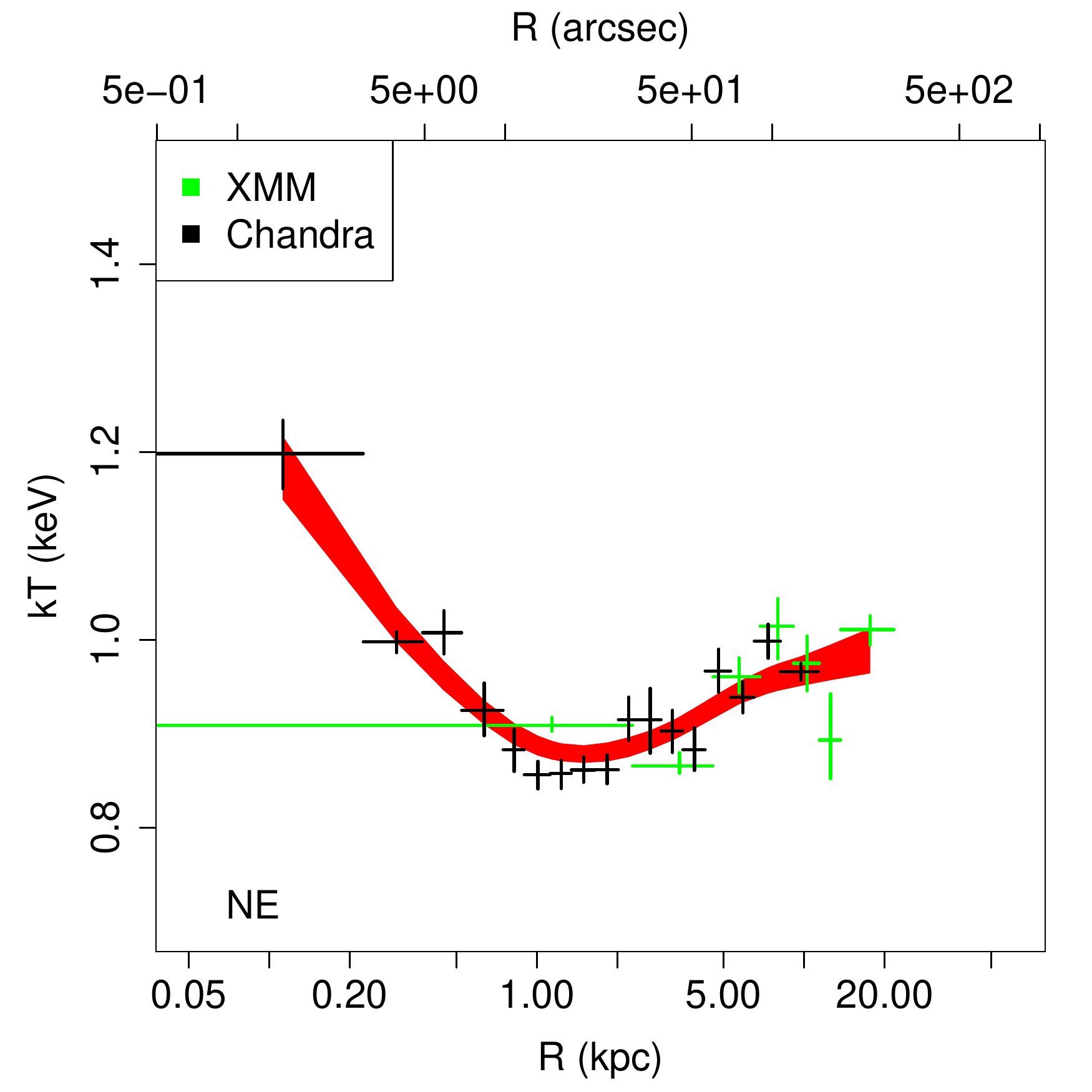}
\includegraphics[scale=0.2]{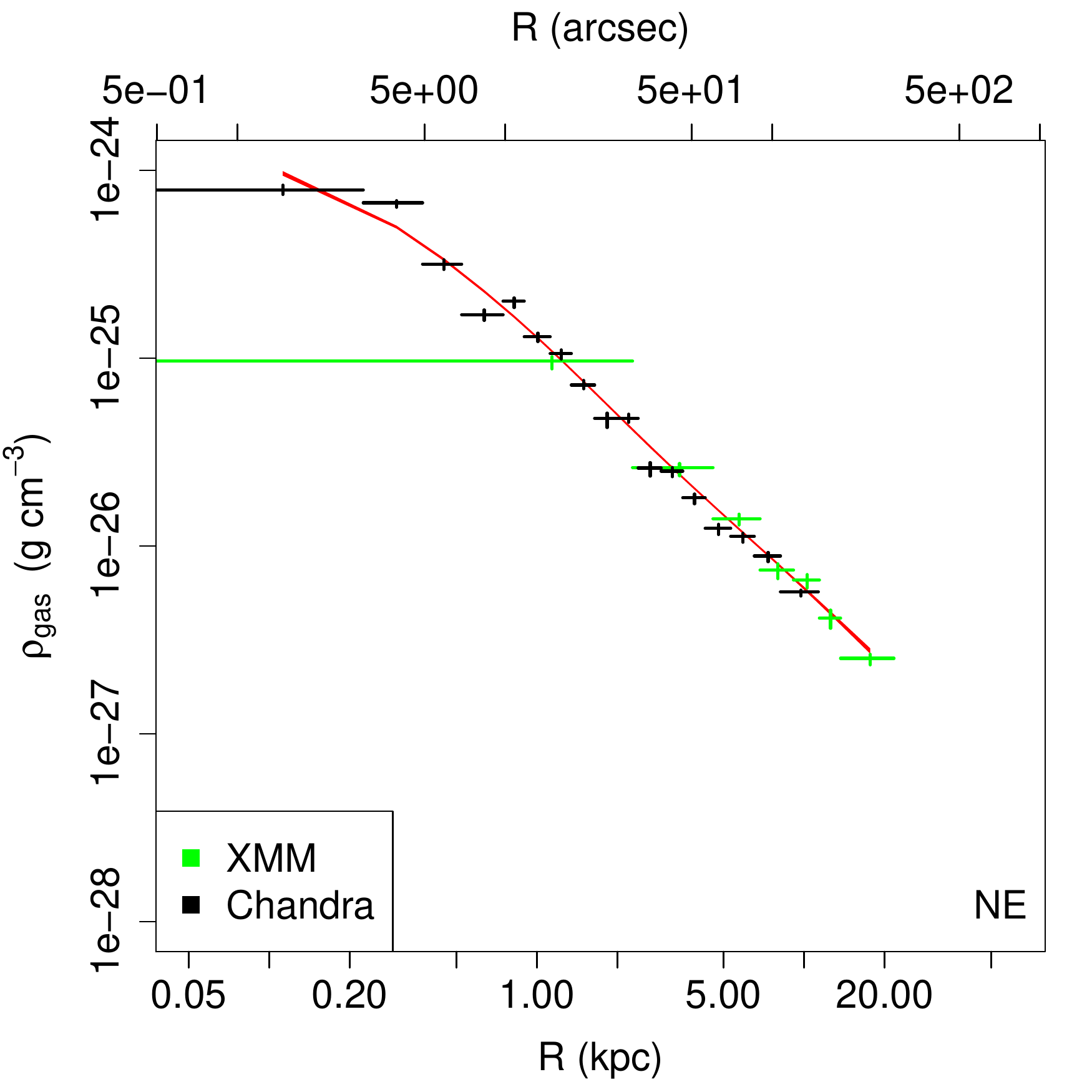}
\includegraphics[scale=0.2]{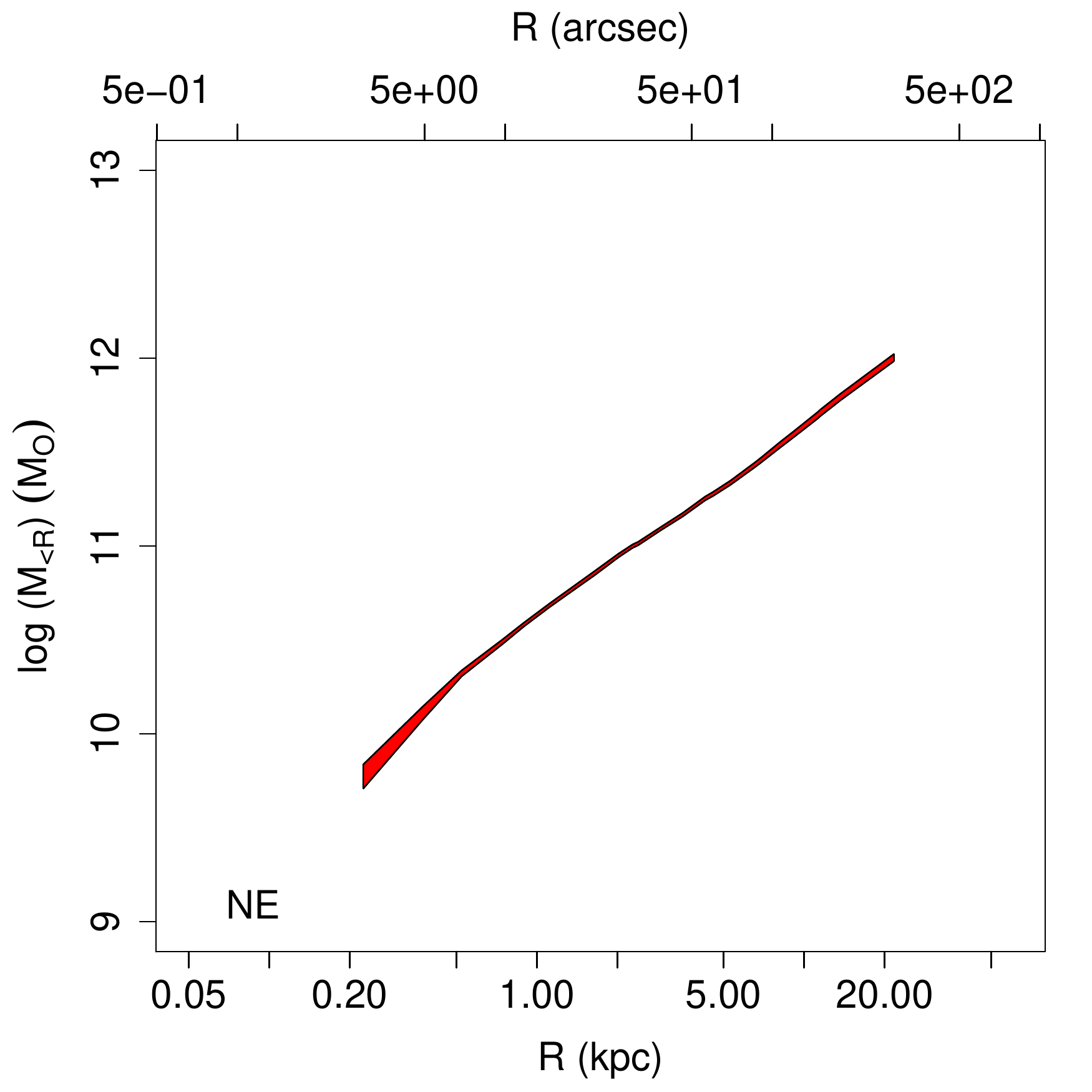}
\caption{{Effect of the different background subtraction procedures on the X-ray gas profiles of NGC 4649 in the NE (0-90) sector. From left to right we show the gas temperature, gas density (with overplotted the best fit spline model with a smoothing parameter of 0.7) and resulting mass profile for the \citeauthor{2005ApJ...629..172N}, \citeauthor{2001A&A...365L..80A}, \citeauthor{2011AAS...21734417S} in the top, middle and bottom rows, respectively. The minimum signal to noise ratio is 30 and 50 for \textit{XMM-Newton} and \textit{Chandra} data, respectively.}}\label{fig:effects_background}
\end{figure}

\newpage

\section{Profiles}\label{app:profiles}

Here we present the complete set of the gas and mass profiles obtained for NGC 4649 and NGC 5846. In Figs. \ref{fig:N4649_bp_chandra_app} and \ref{fig:N5846_bp_chandra_app} we show the surface brightness profiles. In Figs. \ref{fig:N4649_gas_profiles_merged_app}, \ref{fig:N5846_gas_profiles_merged_app}, \ref{fig:N4649_gas_profiles_merged_abund_app} and \ref{fig:N5846_gas_profiles_merged_abund_app} we present the gas and mass profiles for fixed and variable abundances. Finally in Figs. \ref{fig:N4649_mass_fits_app} and \ref{fig:N5846_mass_fits_app} we show the multi-component fits to these mass profiles.

\begin{figure}[!b]
\centering
\includegraphics[scale=0.19]{N4649_chandra_surface_brightness_0_360_0_0.pdf}
\includegraphics[scale=0.19]{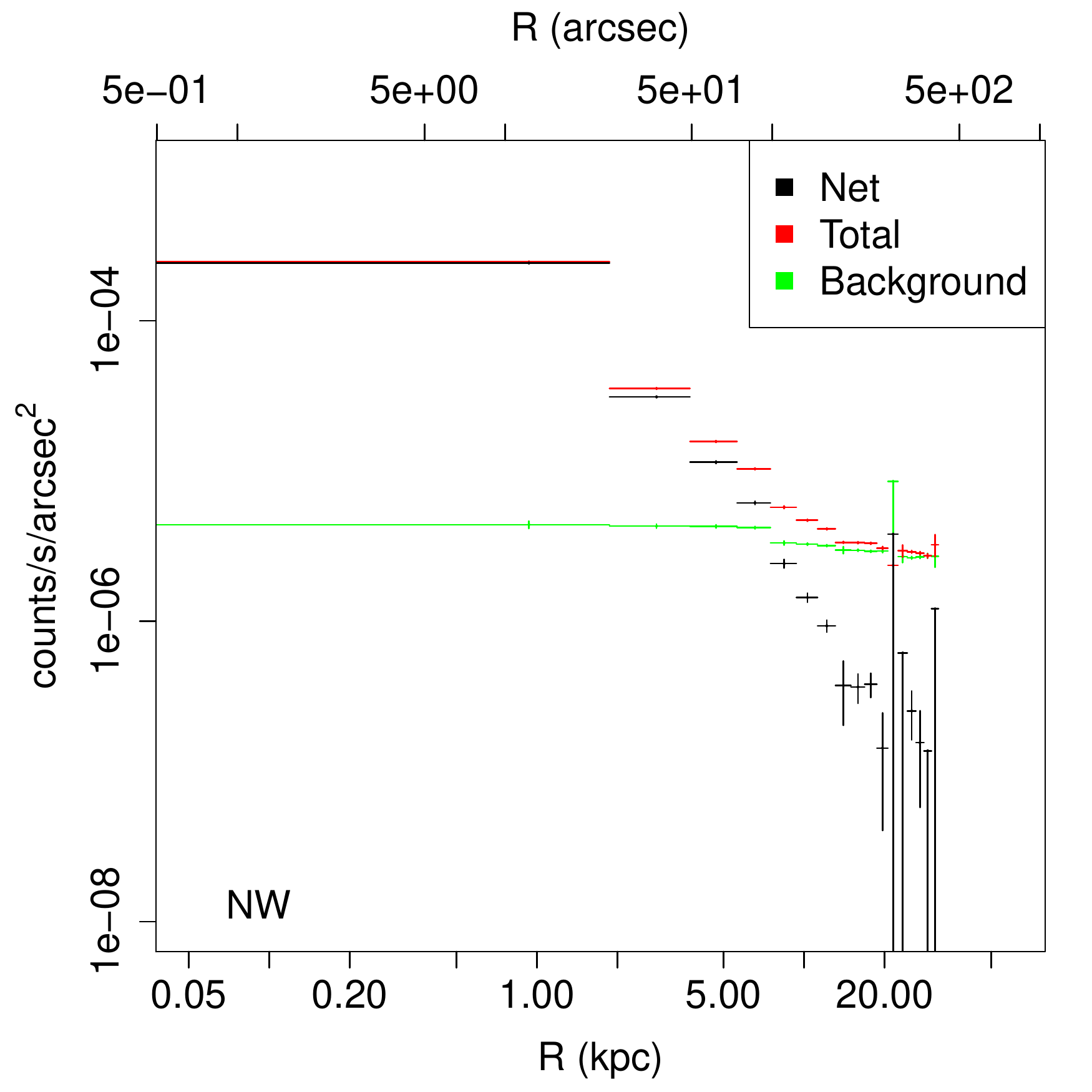}
\includegraphics[scale=0.19]{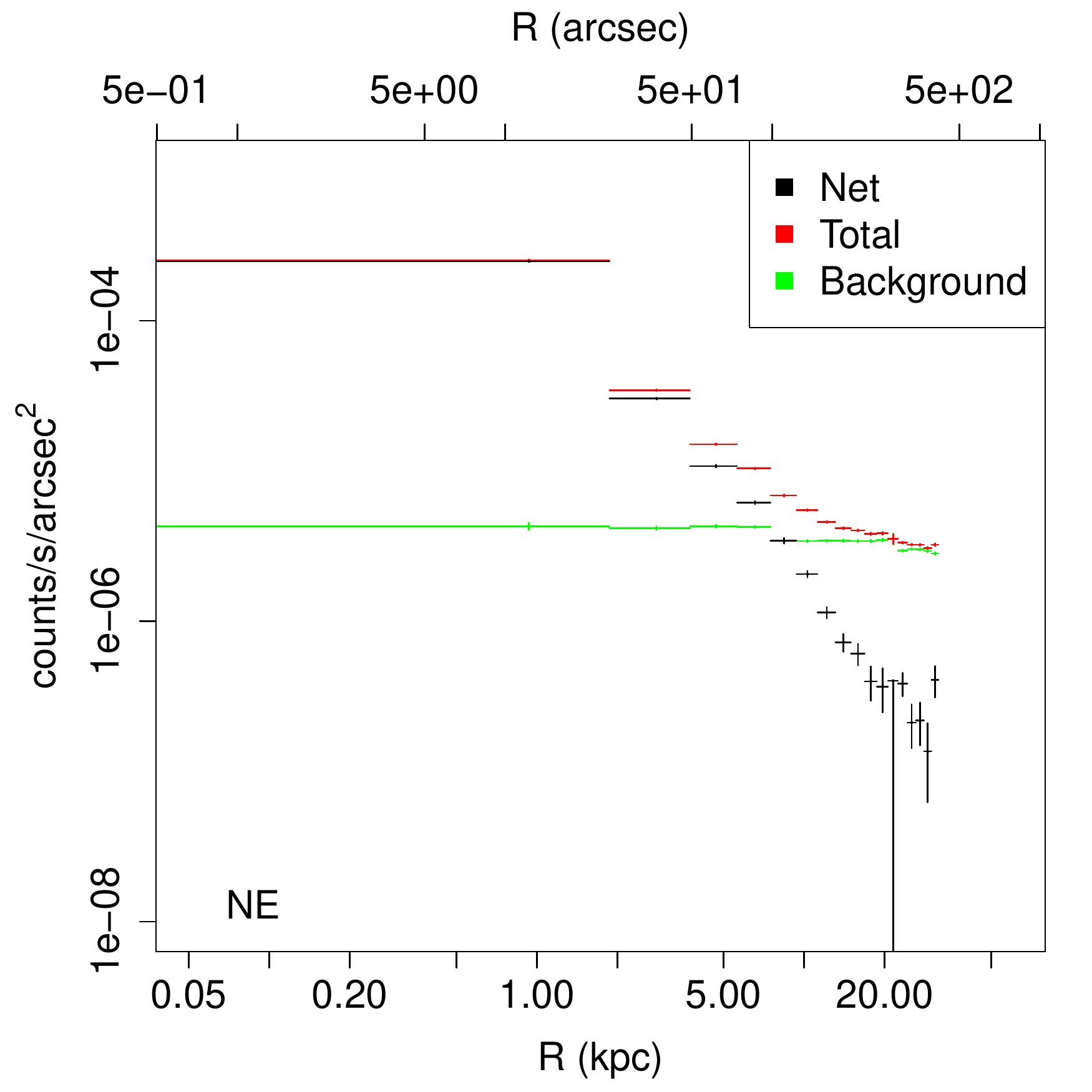}
\includegraphics[scale=0.19]{N4649_chandra_surface_brightness_180_270_0_0.pdf}
\includegraphics[scale=0.19]{N4649_chandra_surface_brightness_270_360_0_0.pdf}\\
\includegraphics[scale=0.19]{N4649_surface_brightness_0_360_0_0.pdf}
\includegraphics[scale=0.19]{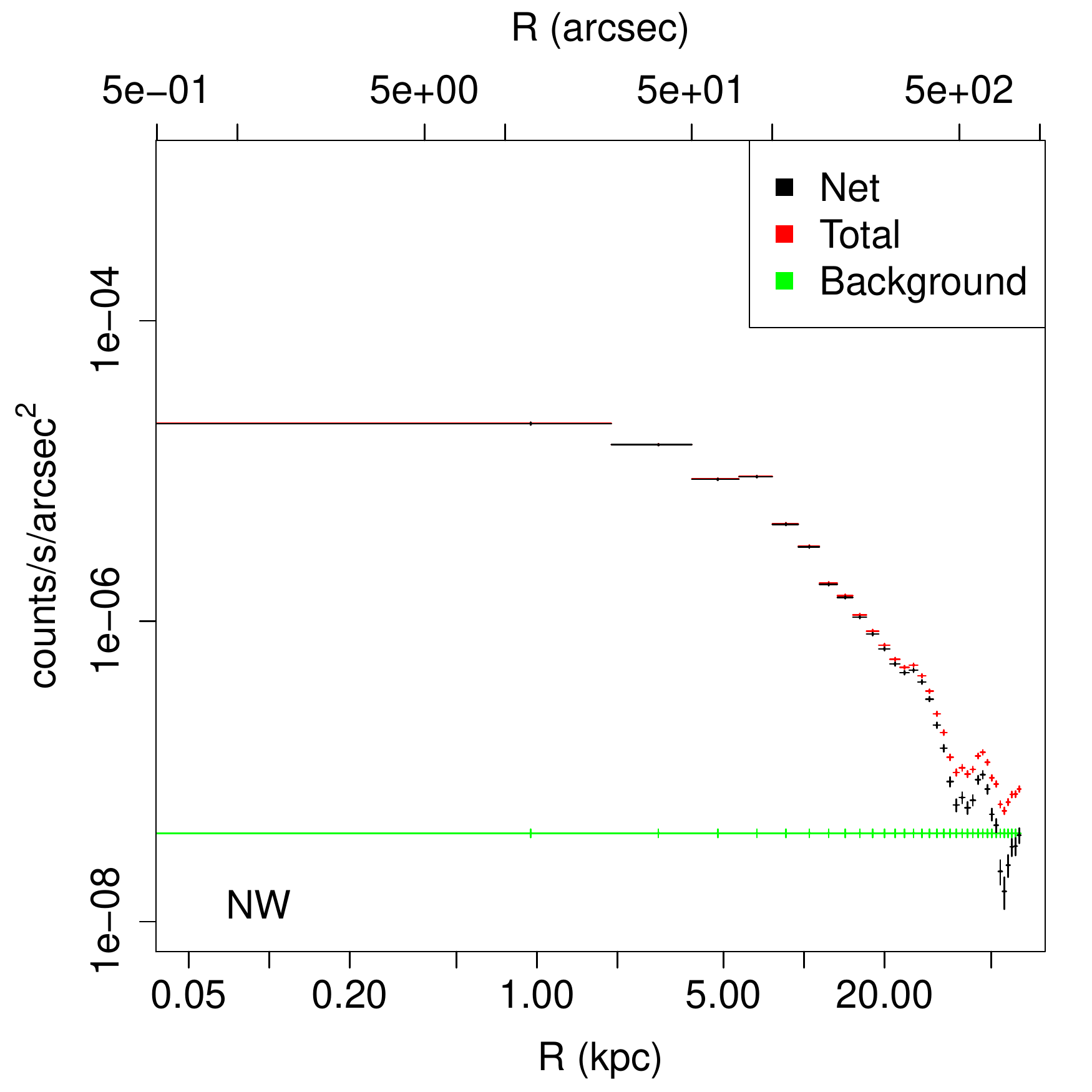}
\includegraphics[scale=0.19]{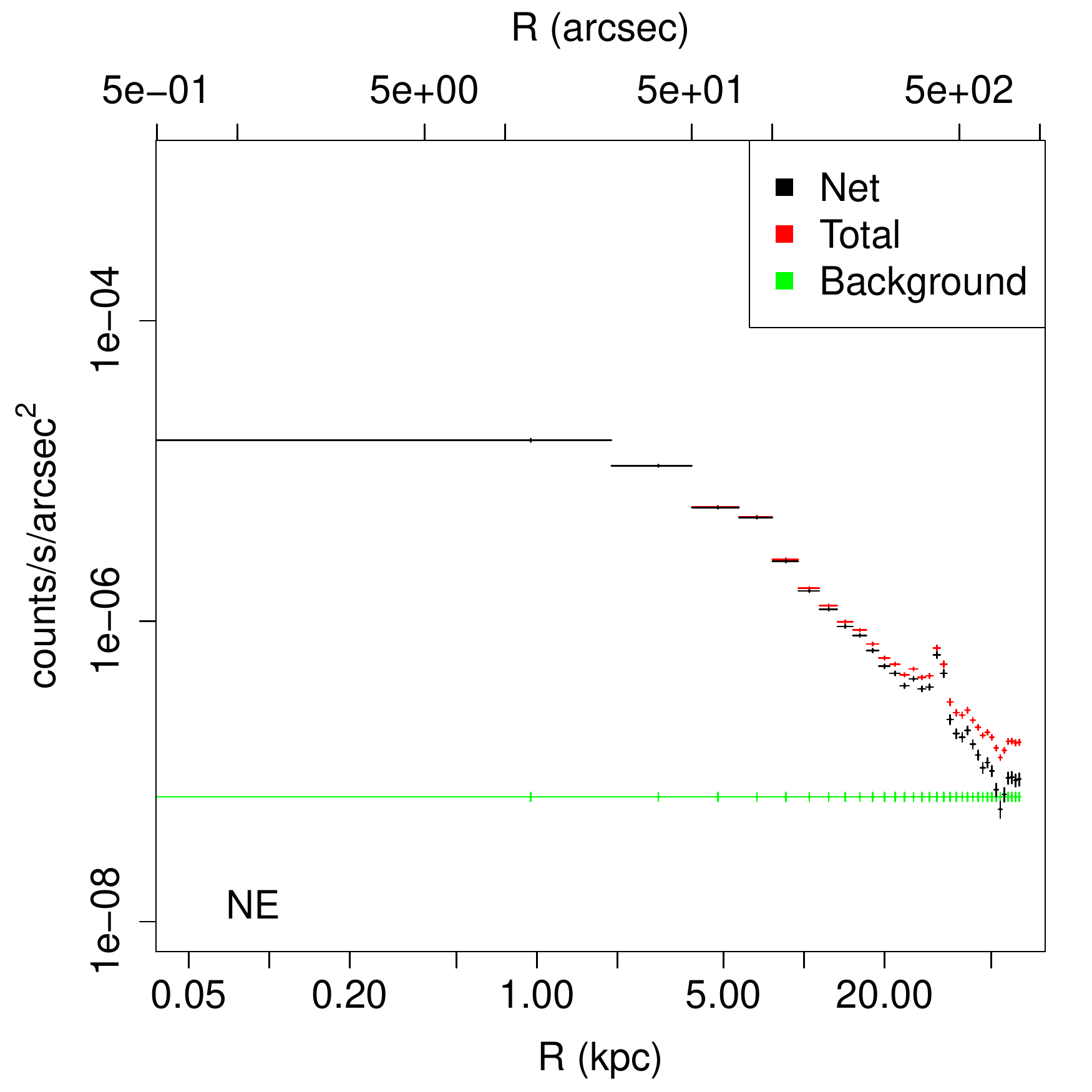}
\includegraphics[scale=0.19]{N4649_surface_brightness_180_270_0_0.pdf}
\includegraphics[scale=0.19]{N4649_surface_brightness_270_360_0_0.pdf}
\caption{Brightness profiles for NGC 4649 in the \(0.3-10\) keV band in the different sectors shown in Fig. \ref{fig:N4649_mos}, from left to right full (0-360), NW (270-360), NE (0-90), SE (90-180), SW (180-270), respectively. In particular we show brightness profiles for \textit{Chandra} ACIS data {in the top row}, for the \textit{XMM}-MOS data obtained from the reduction procedure proposed by \citet{2005ApJ...629..172N} in the bottom row. The annuli width is \(\sim 25''\), 50 pixels for \textit{Chandra} ACIS data and 500 pixels for \textit{XMM}-MOS data. Red, black and green points represent total, net, and background brightness profiles, respectively.}\label{fig:N4649_bp_chandra_app}
\end{figure}

\begin{figure}
\centering
\includegraphics[scale=0.19]{{N4649_temp_profile_merged_0_360_0_0_fit_0.7}.pdf}
\includegraphics[scale=0.19]{{N4649_nh_profile_merged_0_360_0_0_fit_0.7}.pdf}
\includegraphics[scale=0.19]{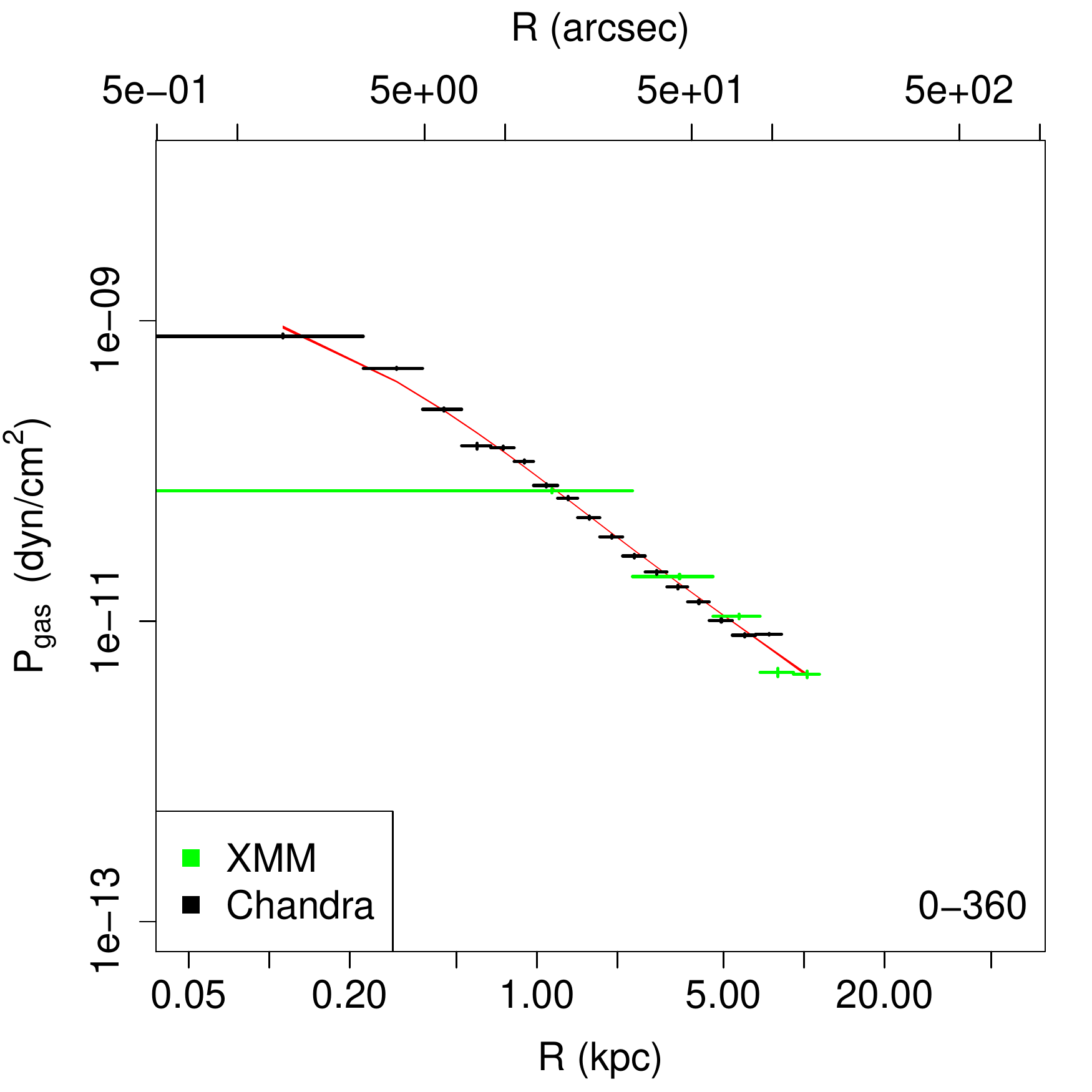}
\includegraphics[scale=0.19]{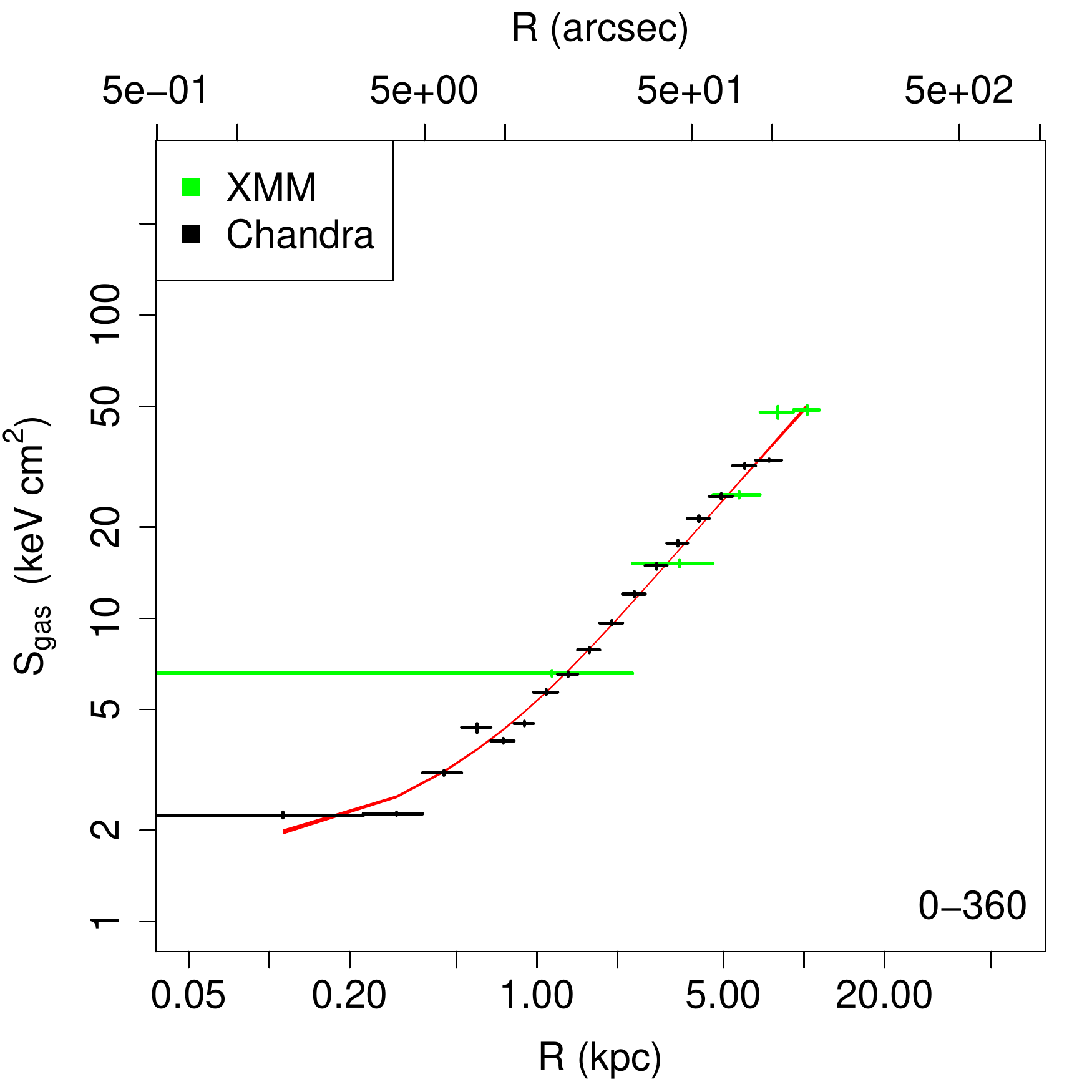}
\includegraphics[scale=0.19]{{N4649_mass_profile_comparison_0_360_0_0_0.7}.pdf}\\
\includegraphics[scale=0.19]{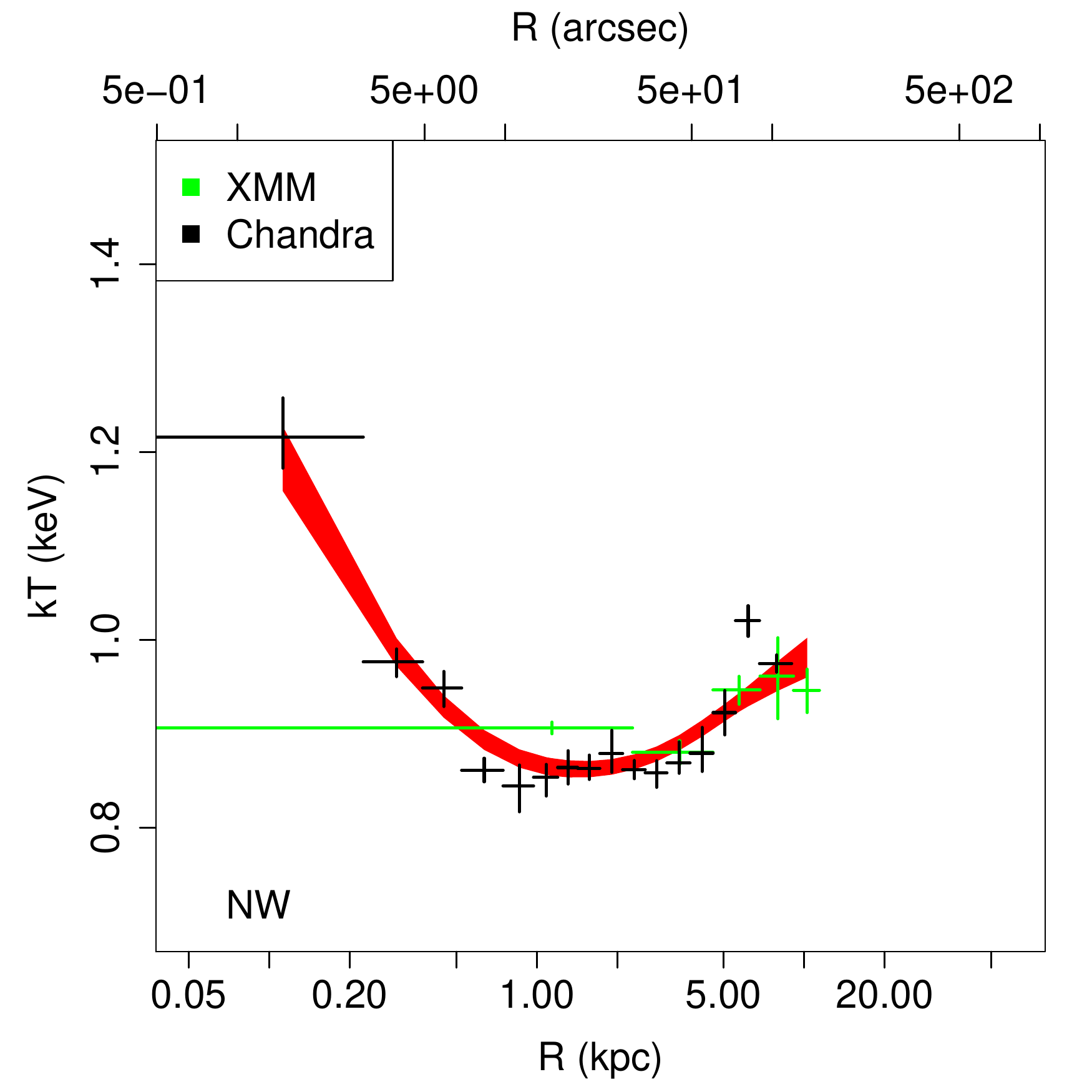}
\includegraphics[scale=0.19]{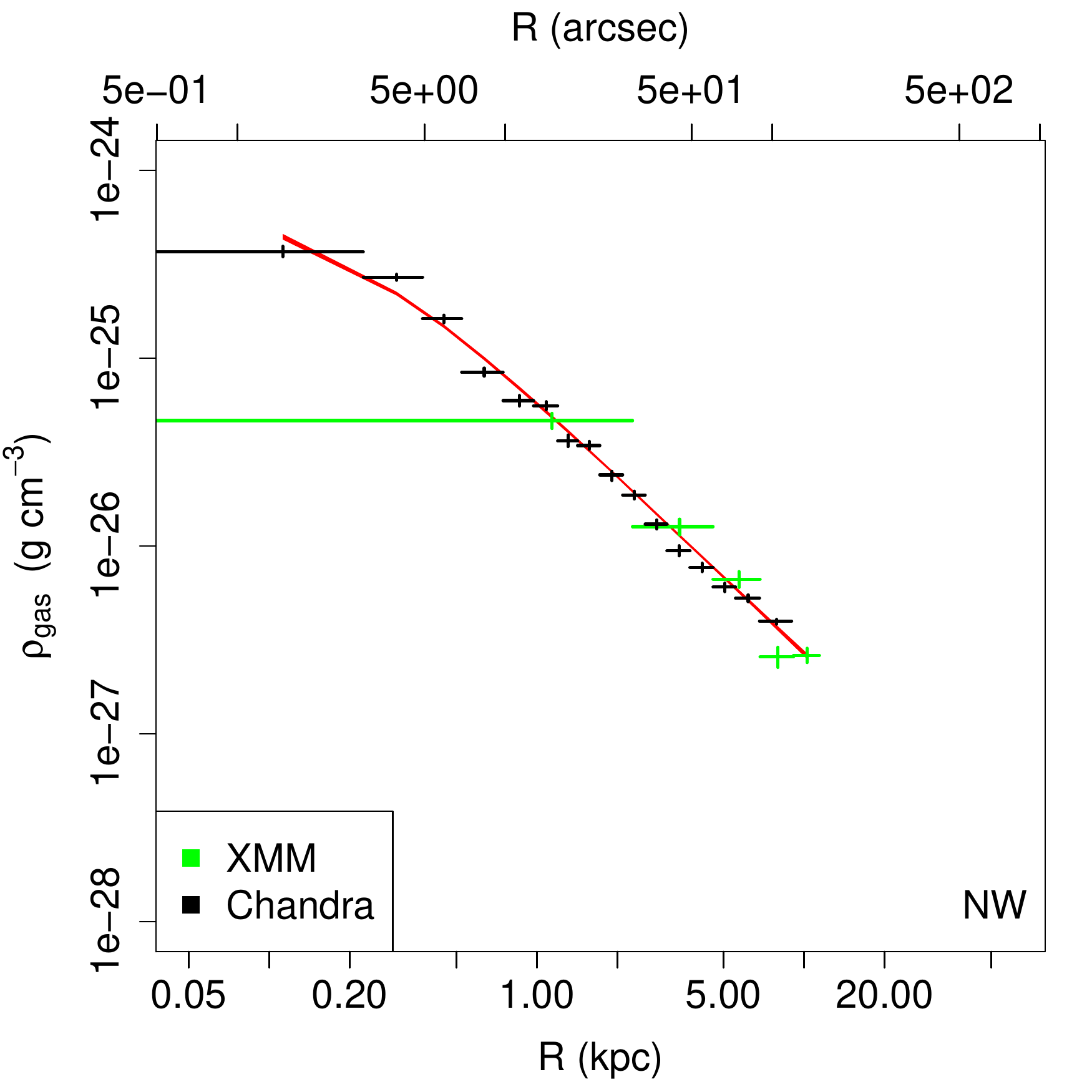}
\includegraphics[scale=0.19]{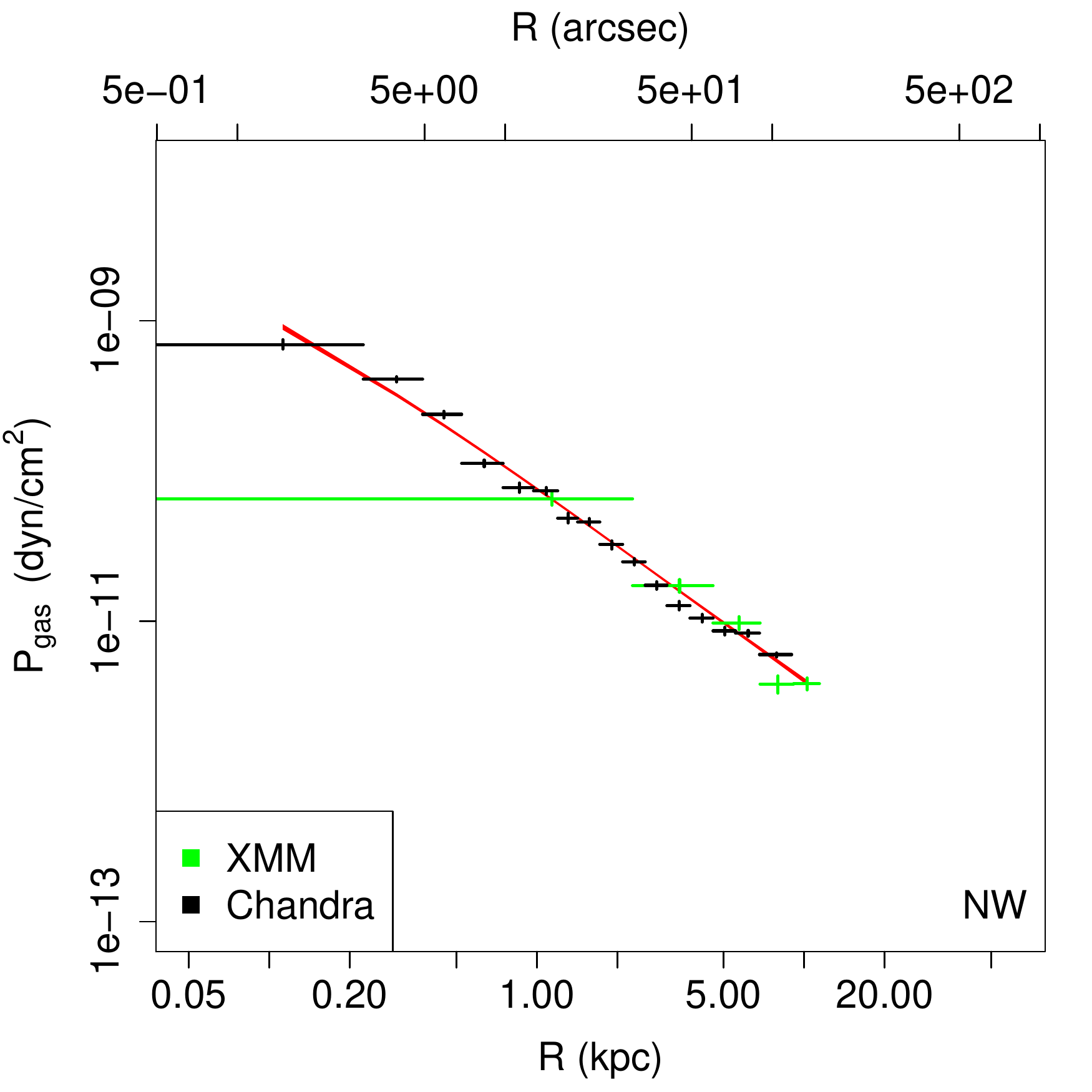}
\includegraphics[scale=0.19]{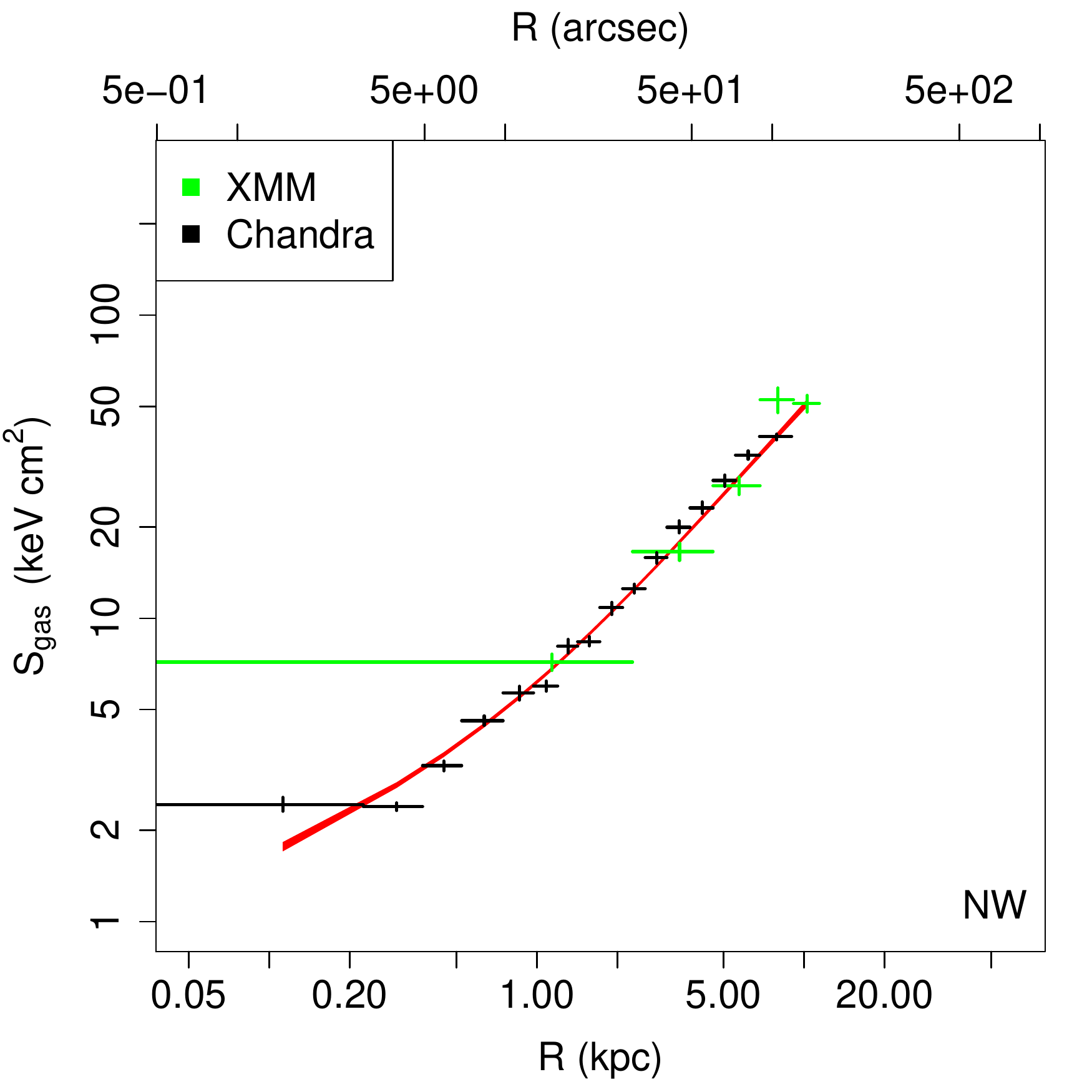}
\includegraphics[scale=0.19]{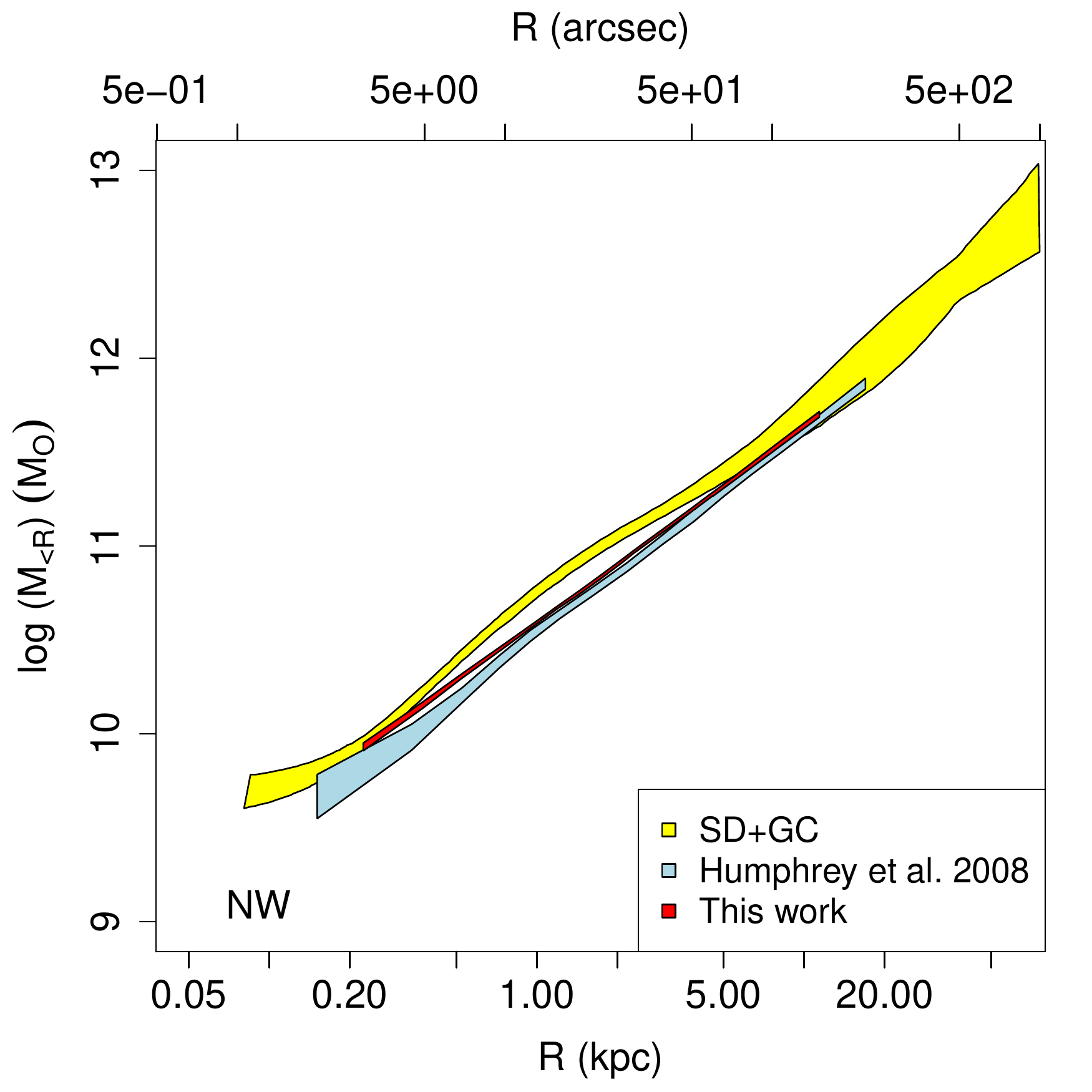}\\
\includegraphics[scale=0.19]{{N4649_temp_profile_merged_90_180_0_0_fit_0.7}.pdf}
\includegraphics[scale=0.19]{{N4649_nh_profile_merged_90_180_0_0_fit_0.7}.pdf}
\includegraphics[scale=0.19]{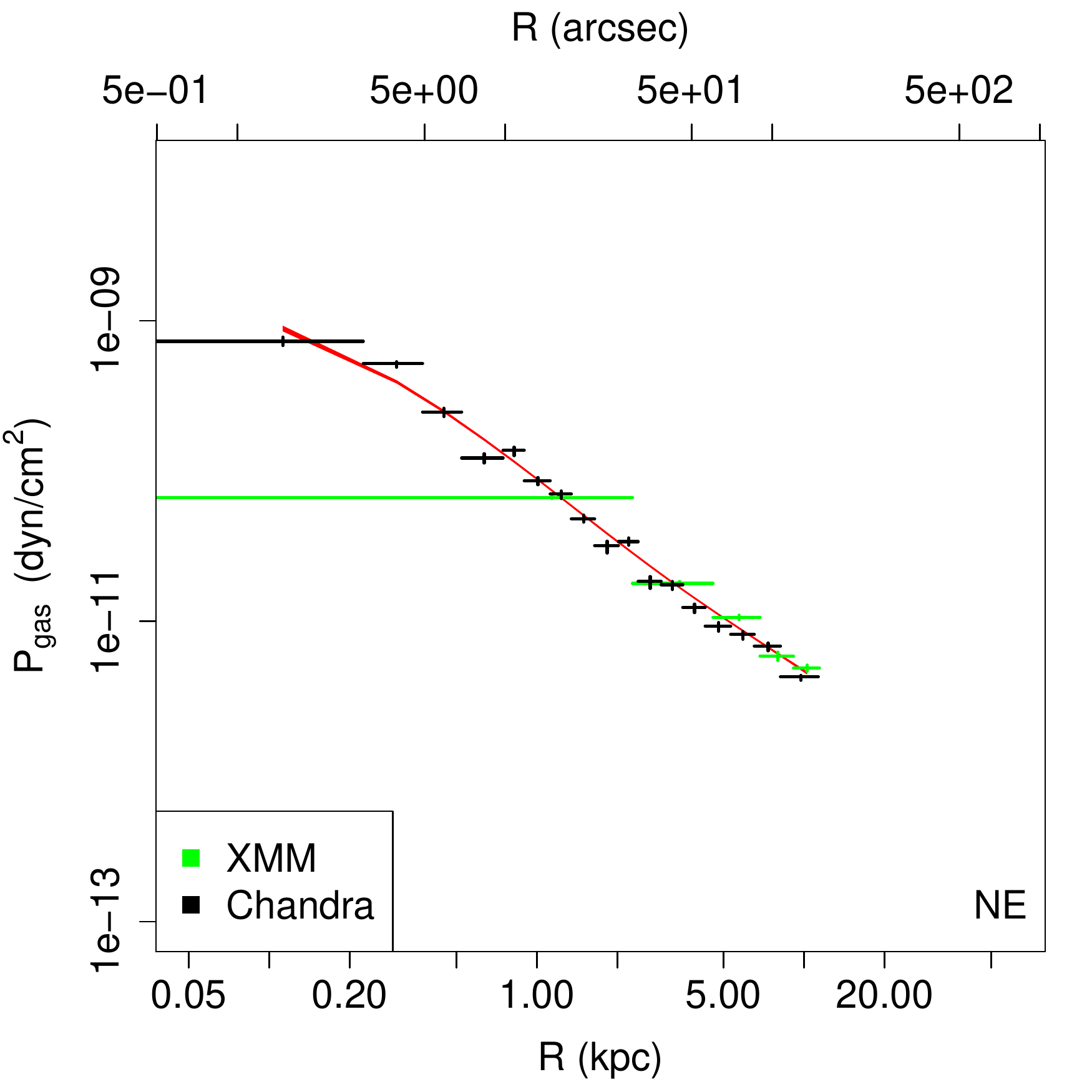}
\includegraphics[scale=0.19]{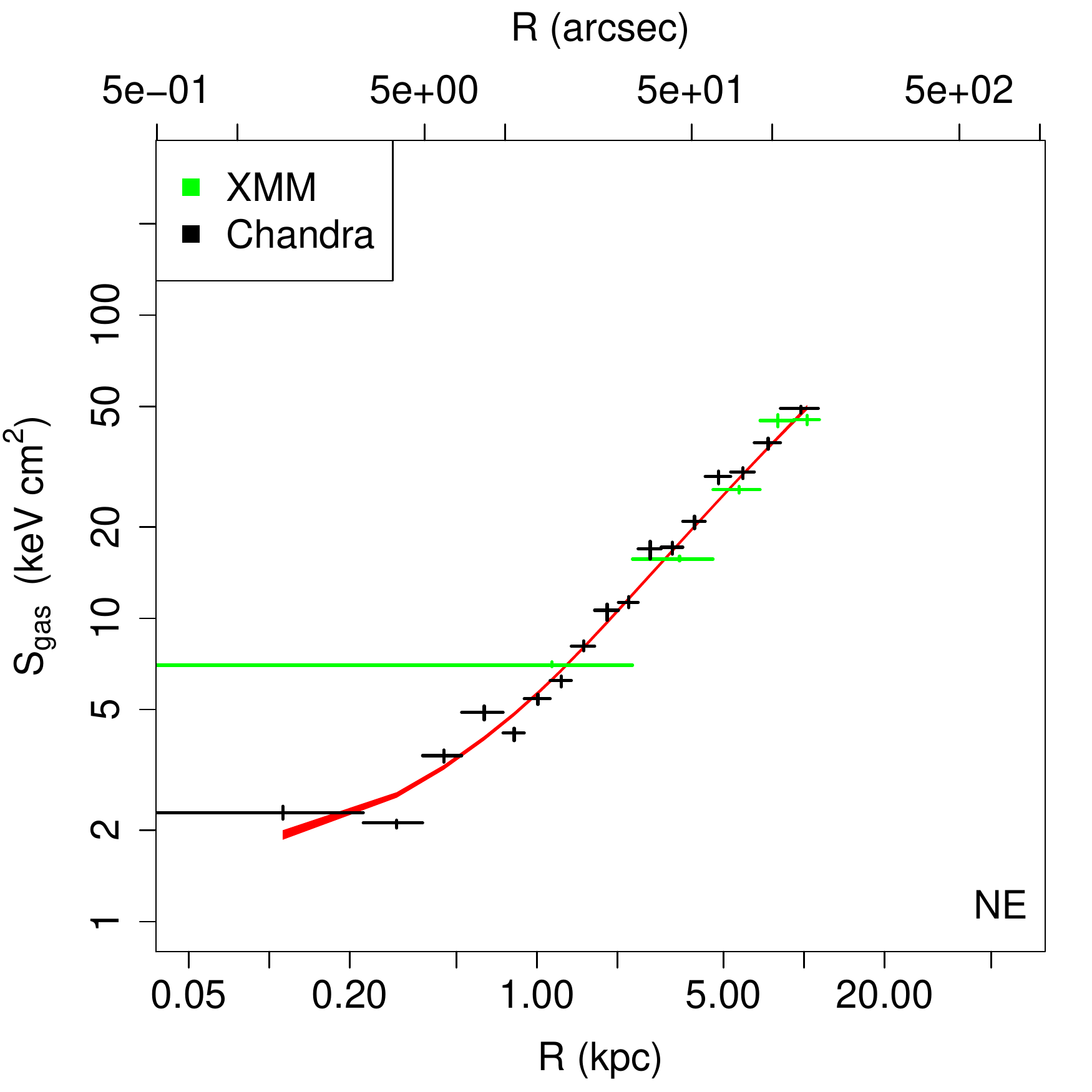}
\includegraphics[scale=0.19]{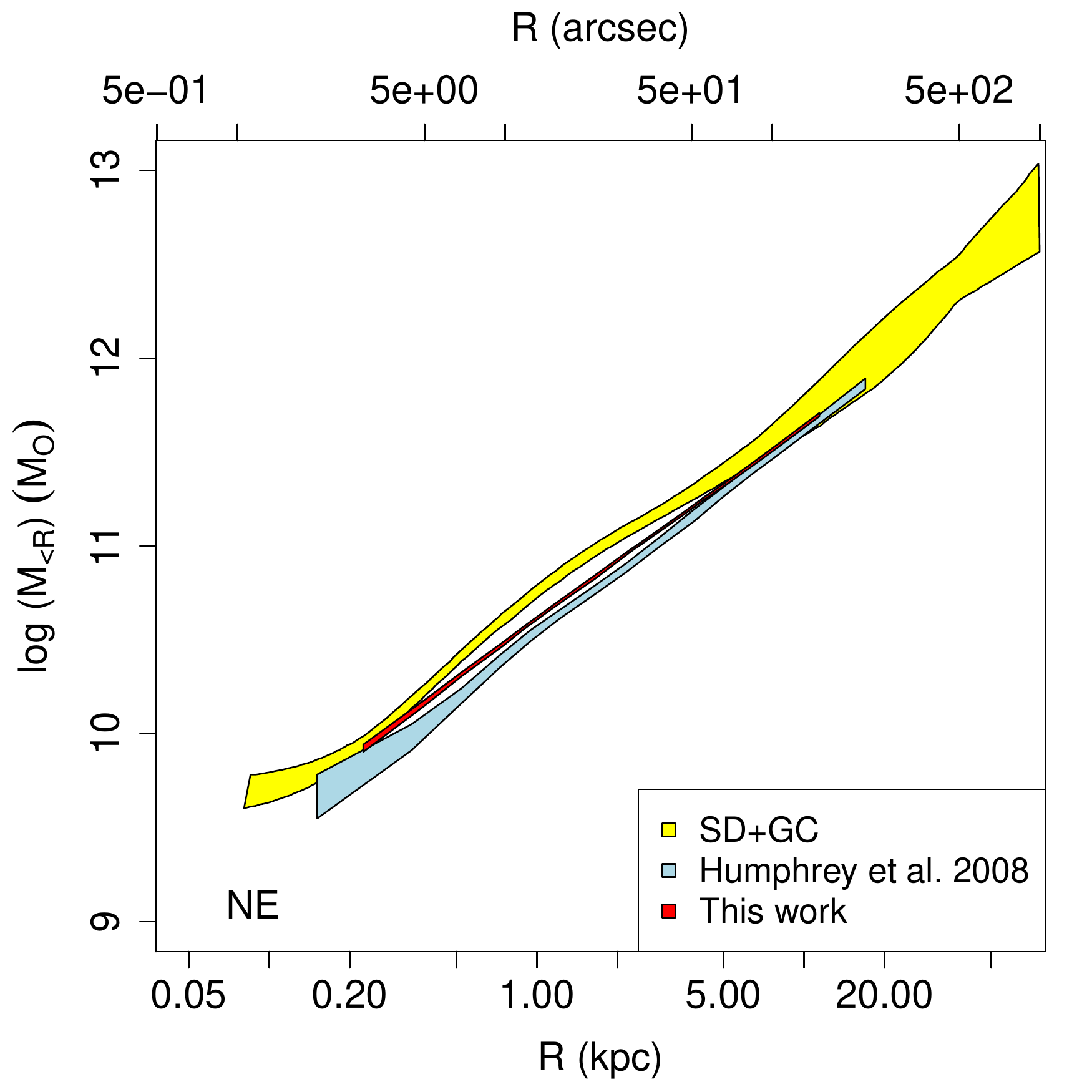}\\
\includegraphics[scale=0.19]{{N4649_temp_profile_merged_180_270_0_0_fit_0.8}.pdf}
\includegraphics[scale=0.19]{{N4649_nh_profile_merged_180_270_0_0_fit_0.8}.pdf}
\includegraphics[scale=0.19]{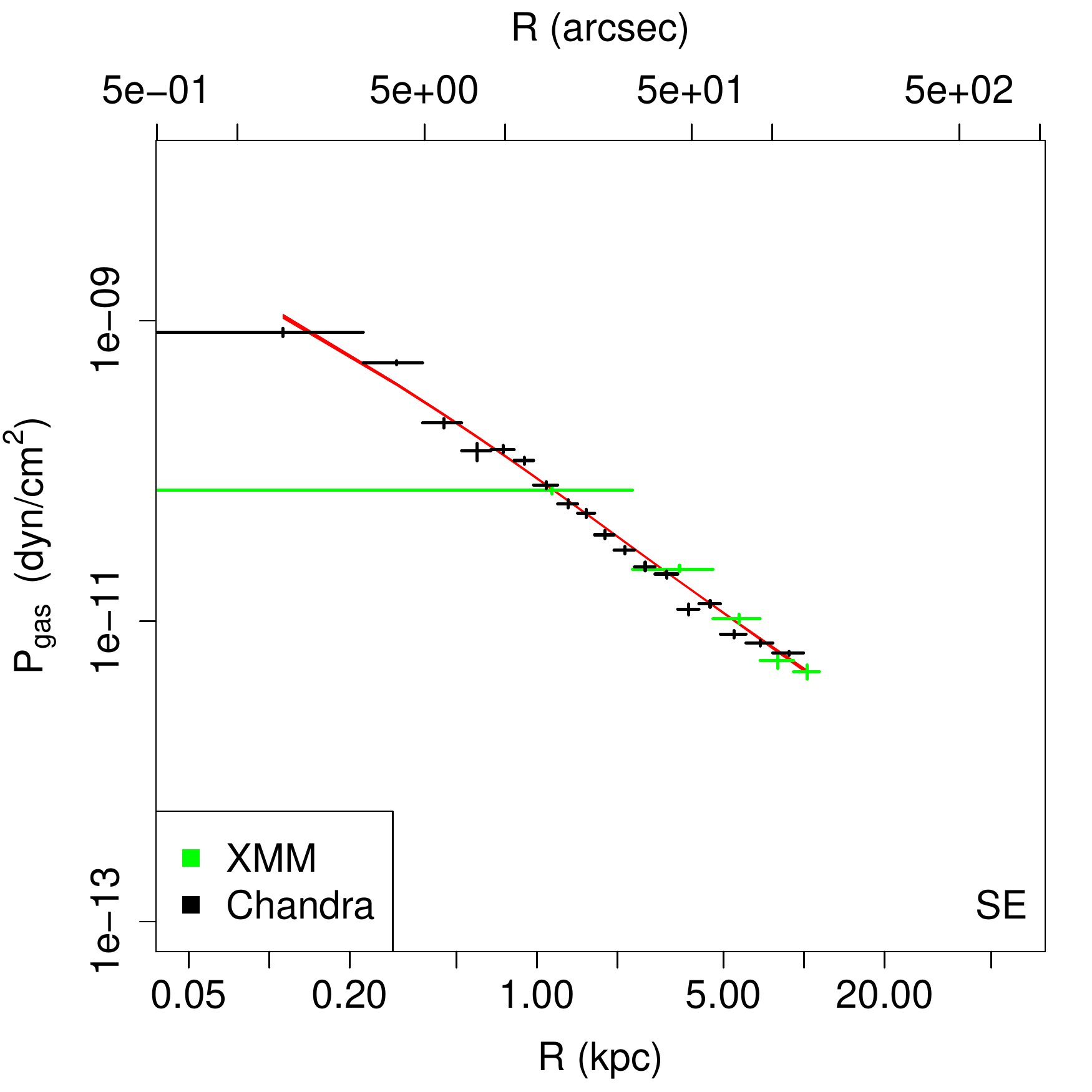}
\includegraphics[scale=0.19]{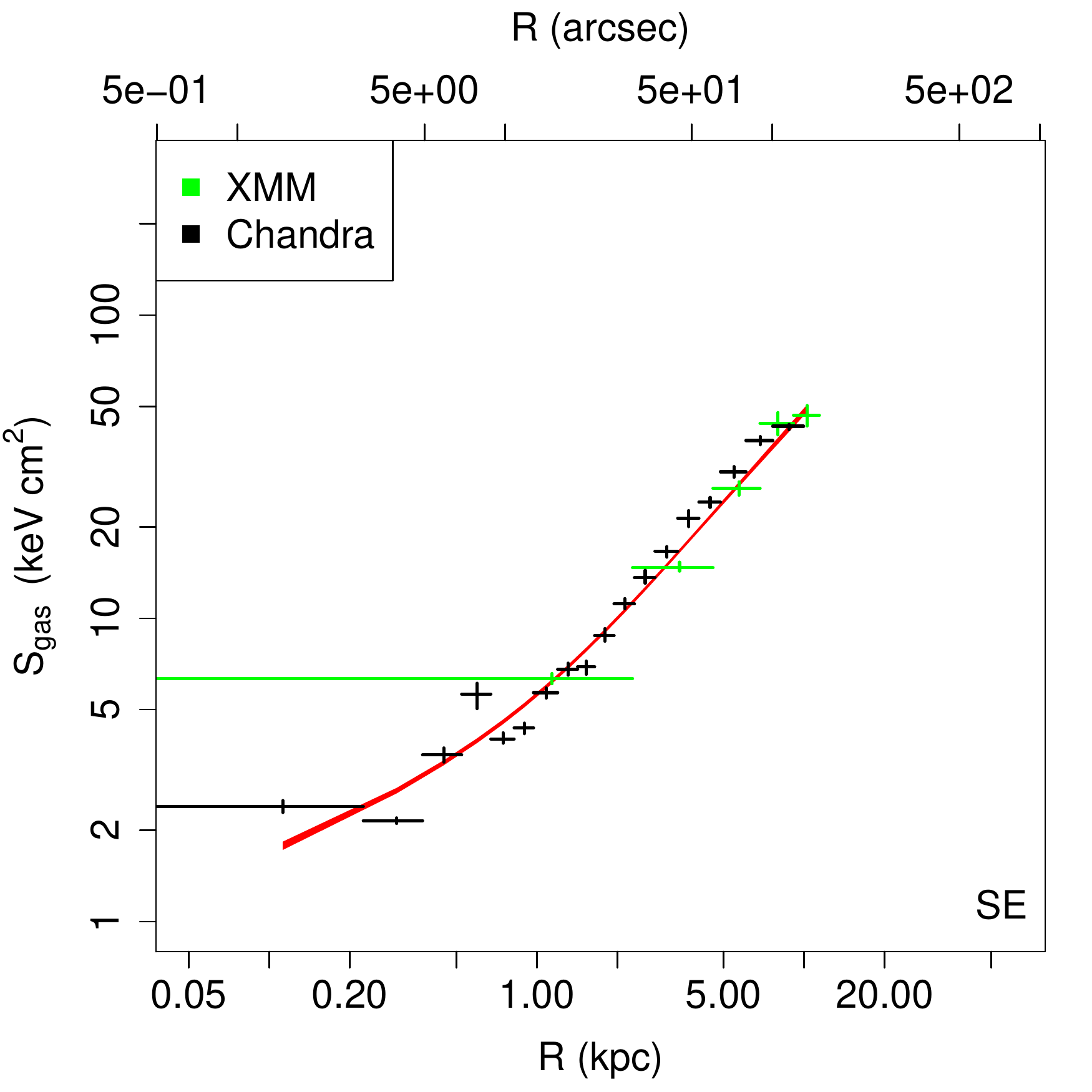}
\includegraphics[scale=0.19]{{N4649_mass_profile_comparison_180_270_0_0_0.8}.pdf}\\
\includegraphics[scale=0.19]{{N4649_temp_profile_merged_270_360_0_0_fit_0.8}.pdf}
\includegraphics[scale=0.19]{{N4649_nh_profile_merged_270_360_0_0_fit_0.8}.pdf}
\includegraphics[scale=0.19]{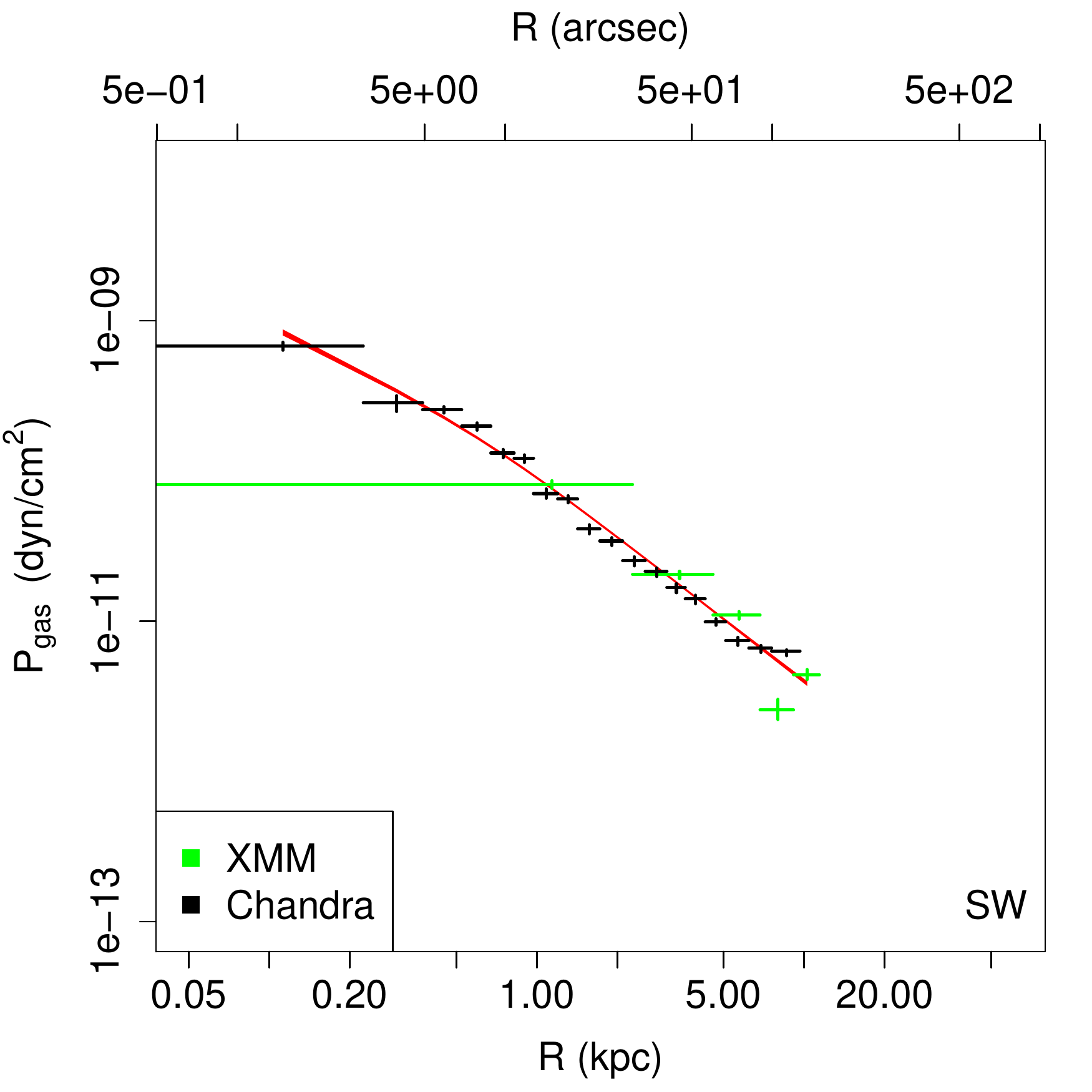}
\includegraphics[scale=0.19]{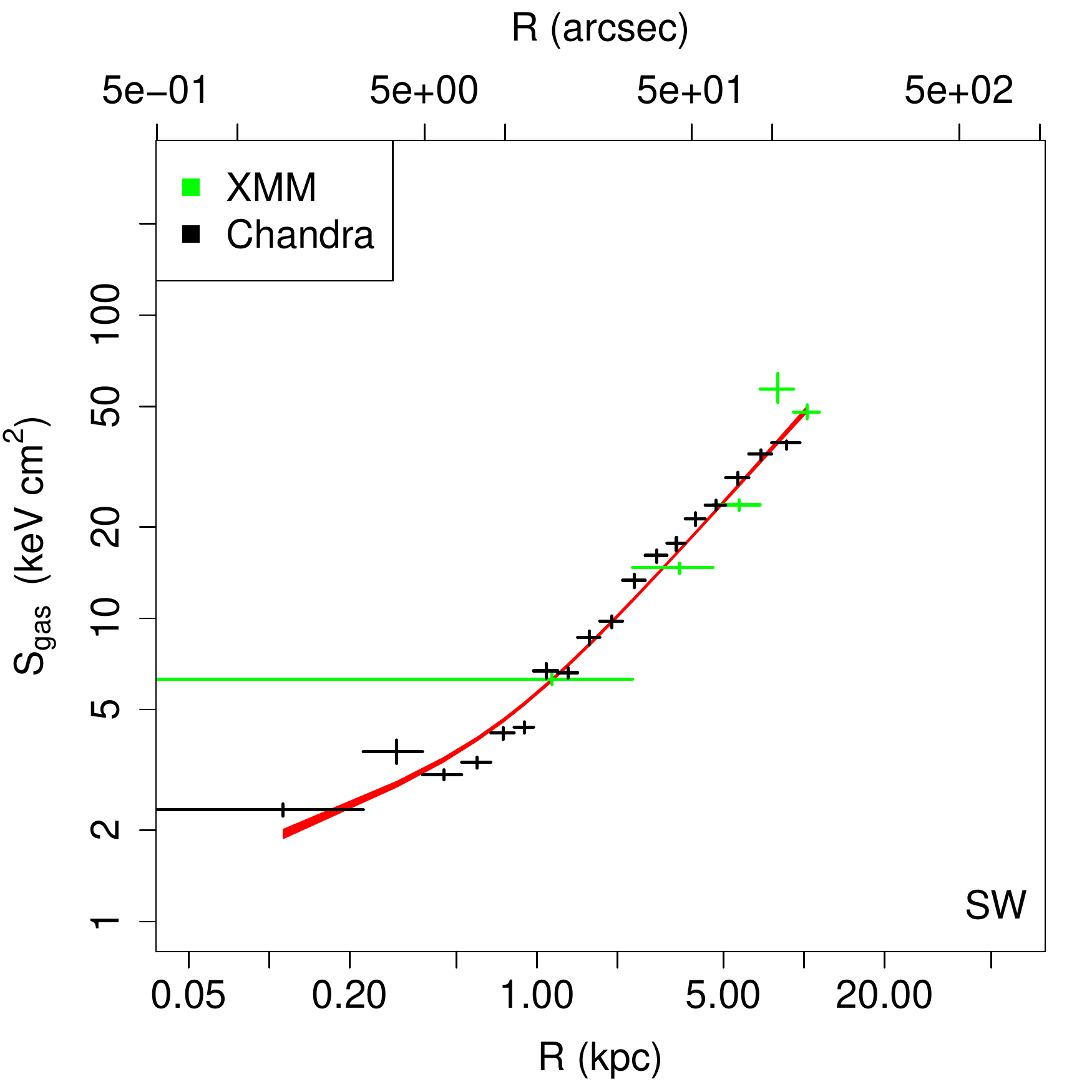}
\includegraphics[scale=0.19]{{N4649_mass_profile_comparison_270_360_0_0_0.8}.pdf}
\caption{Gas profiles obtained in NGC 4649 with the reduction procedure proposed by \citet{2005ApJ...629..172N}. From top to bottom we show the profiles obtained in the full (0-360), NW (270-360), NE (0-90), SE (90-180), SW (180-270) sector, respectively. In each row we show from left to right the gas temperature, density, pressure and entropy, respectively. Spectra extracted in the annuli are then simultaneously fitted (separately for \textit{XMM} and \textit{Chandra} data) with the fixed abundance model, and de-projected using \textsc{projct} model. The annuli width is chosen to reach a signal to noise ratio of 30 for \textit{XMM}-MOS data (represented in red) and of 50 for \textit{Chandra} ACIS data (represented in black) with the exception of the full (0-360) sector for which we chose a signal to noise ratio of 50 for \textit{XMM} data and 100 for \textit{Chandra} data. Best fits of a smooth cubic spline are presented in red, with smoothing parameter from top to bottom of 0.8, 0.8, 0.7, 0.8 and 0.8. In the rightmost panels we present in red the total mass profiles obtained by mean of Eq. \ref{eq:hee} from the best fits to gas temperature and density profiles. In the same panels we show in yellow the optical mass profile obtained from SD and GC reported by \citep{2010ApJ...711..484S}, and in light blue the X-ray mass profiles obtained by \citet{2008ApJ...683..161H}.}\label{fig:N4649_gas_profiles_merged_app}
\end{figure}

\begin{figure}
\centering
\includegraphics[scale=0.16]{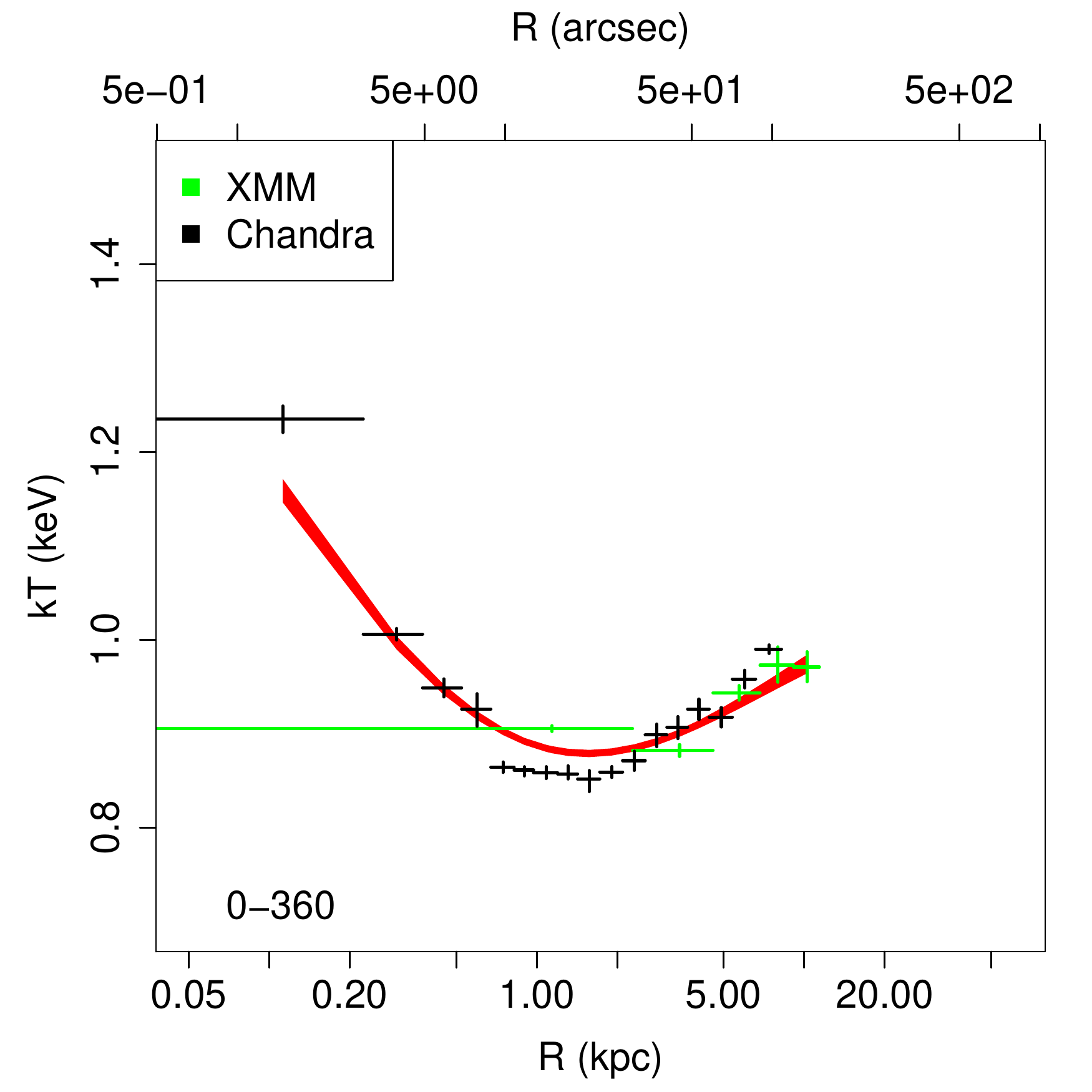}
\includegraphics[scale=0.16]{{N4649_nh_profile_merged_0_360_0_0_fit_0.8_abund}.pdf}
\includegraphics[scale=0.16]{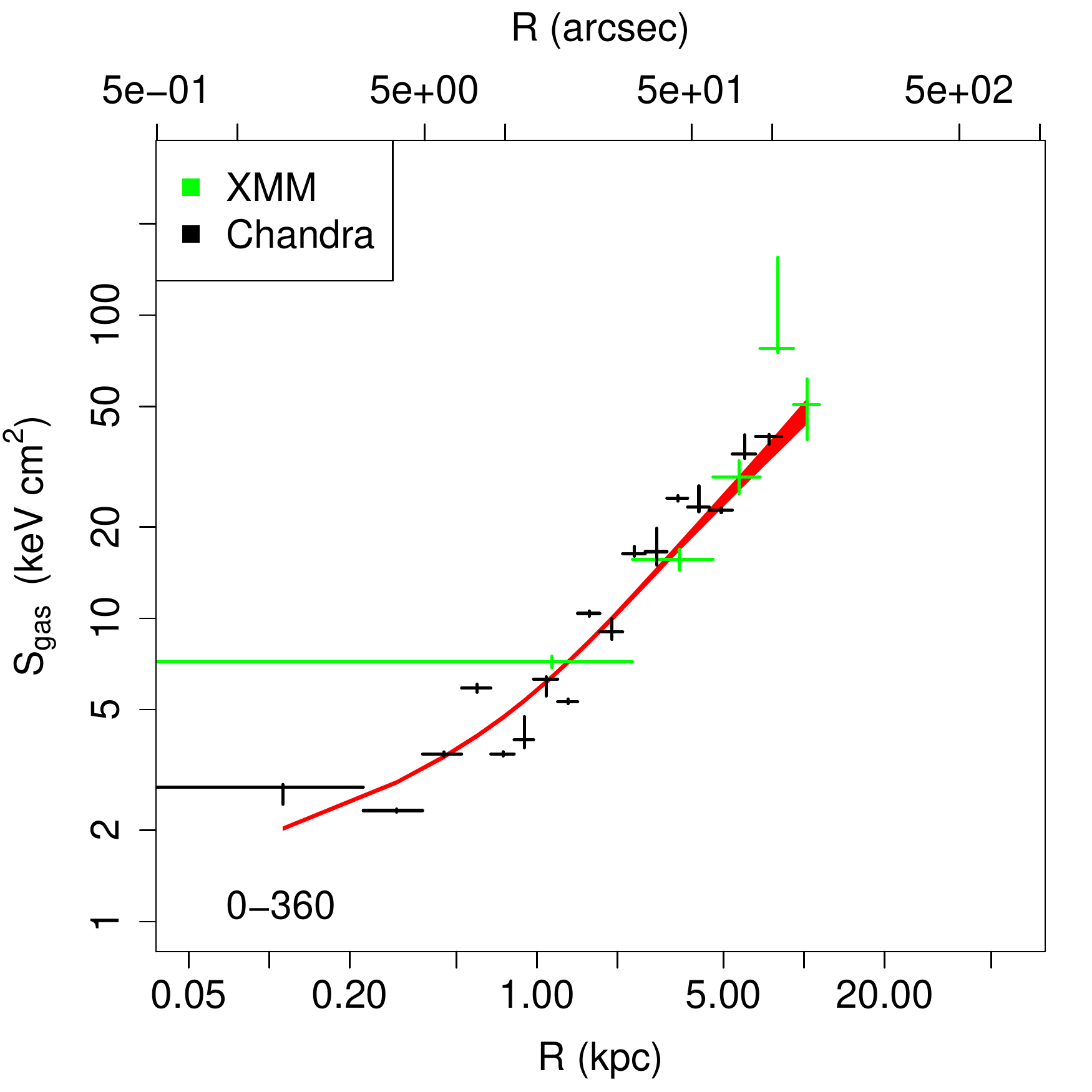}
\includegraphics[scale=0.16]{{N4649_entropy_profile_merged_0_360_0_0_fit_0.8_abund}.pdf}
\includegraphics[scale=0.16]{{N4649_mass_profile_comparison_0_360_0_0_0.8_abund}.pdf}
\includegraphics[scale=0.16]{{N4649_abund_profile_merged_0_360_0_0}.pdf}\\
\includegraphics[scale=0.16]{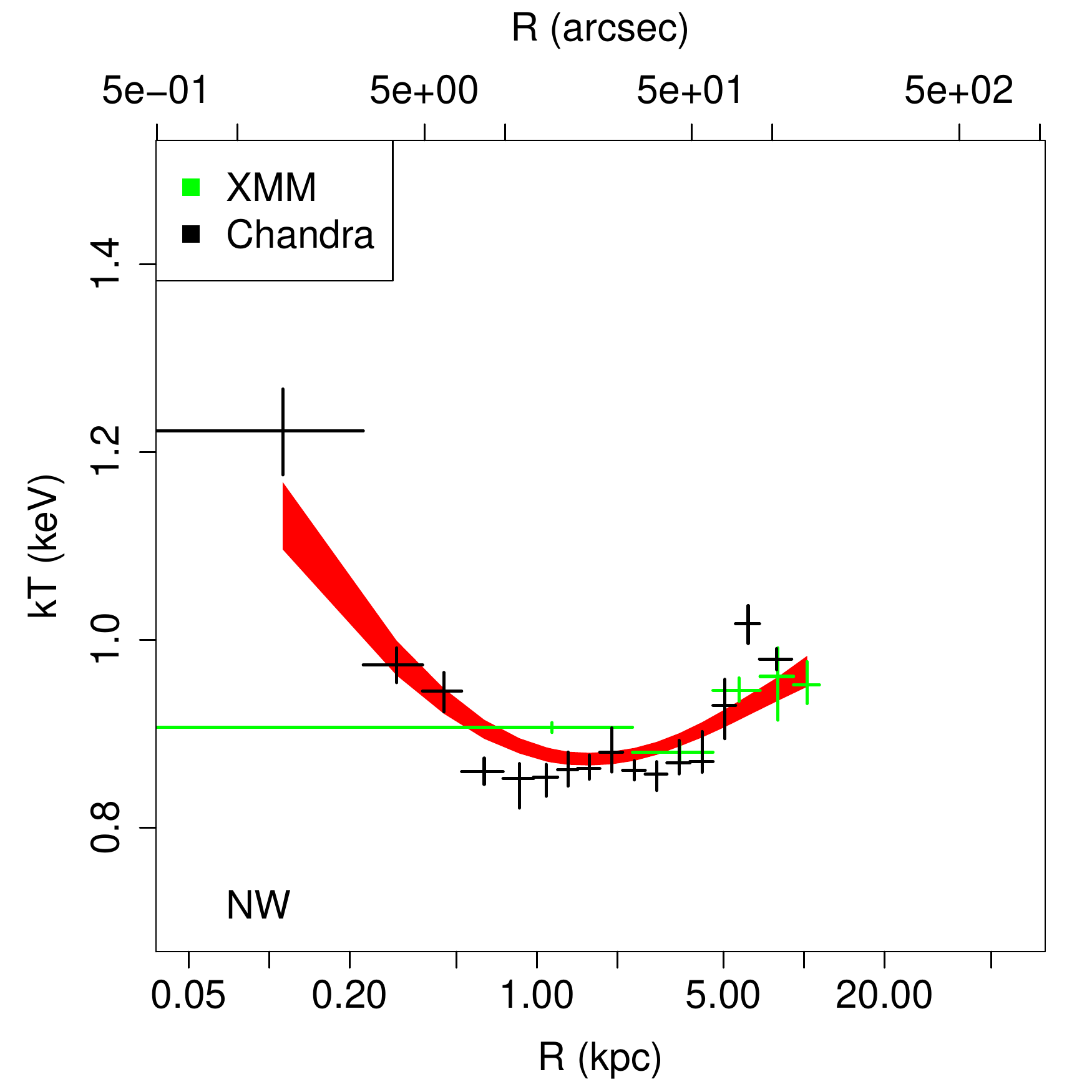}
\includegraphics[scale=0.16]{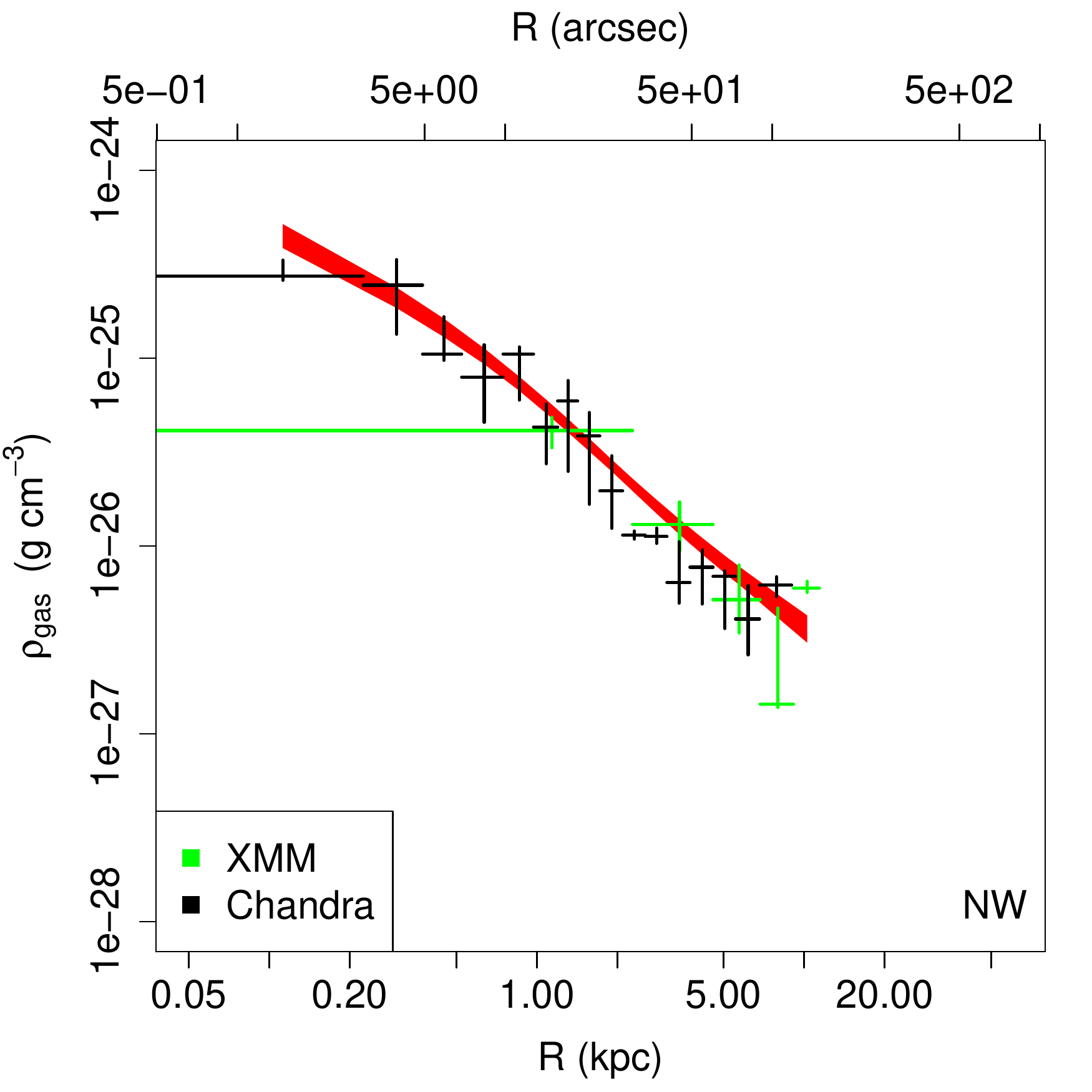}
\includegraphics[scale=0.16]{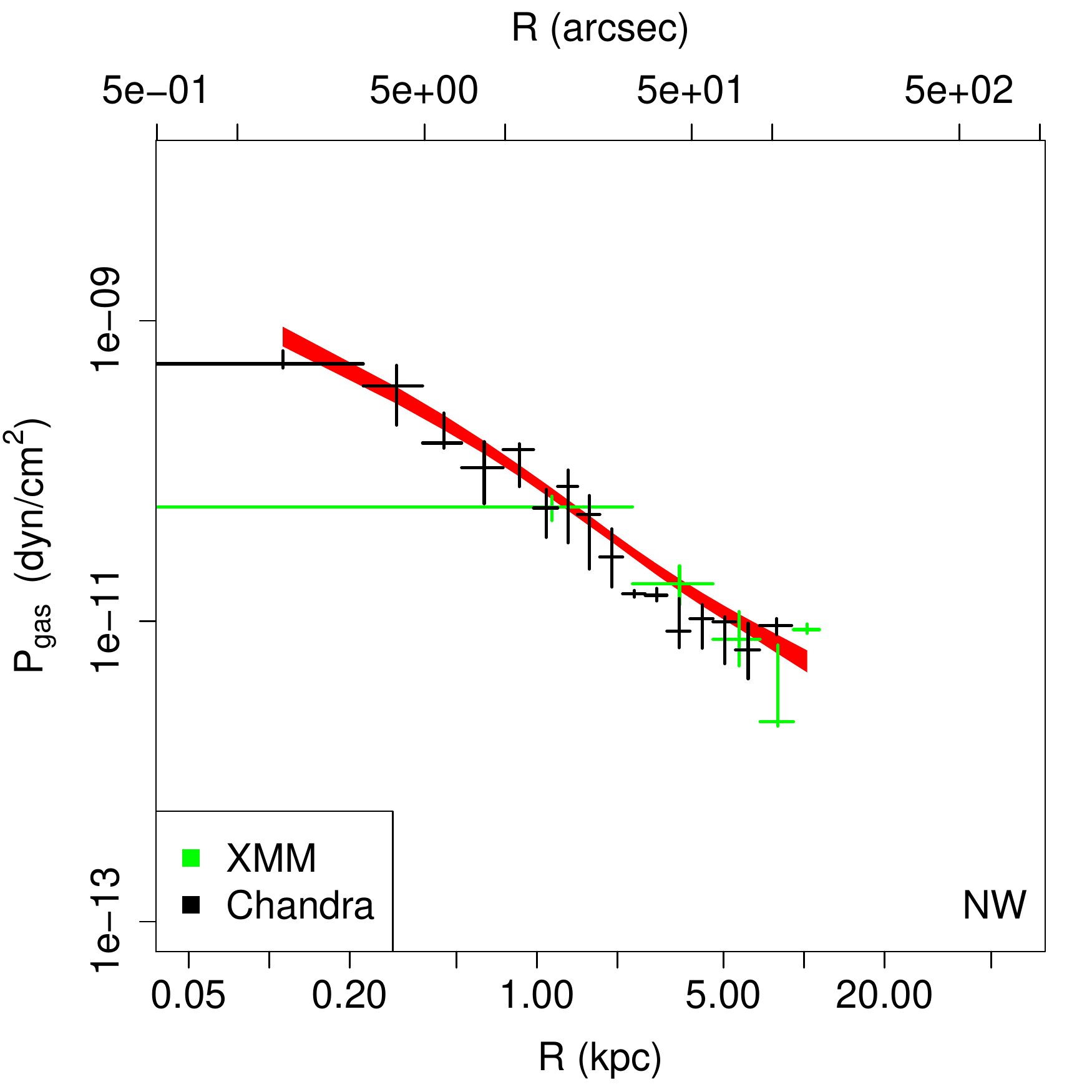}
\includegraphics[scale=0.16]{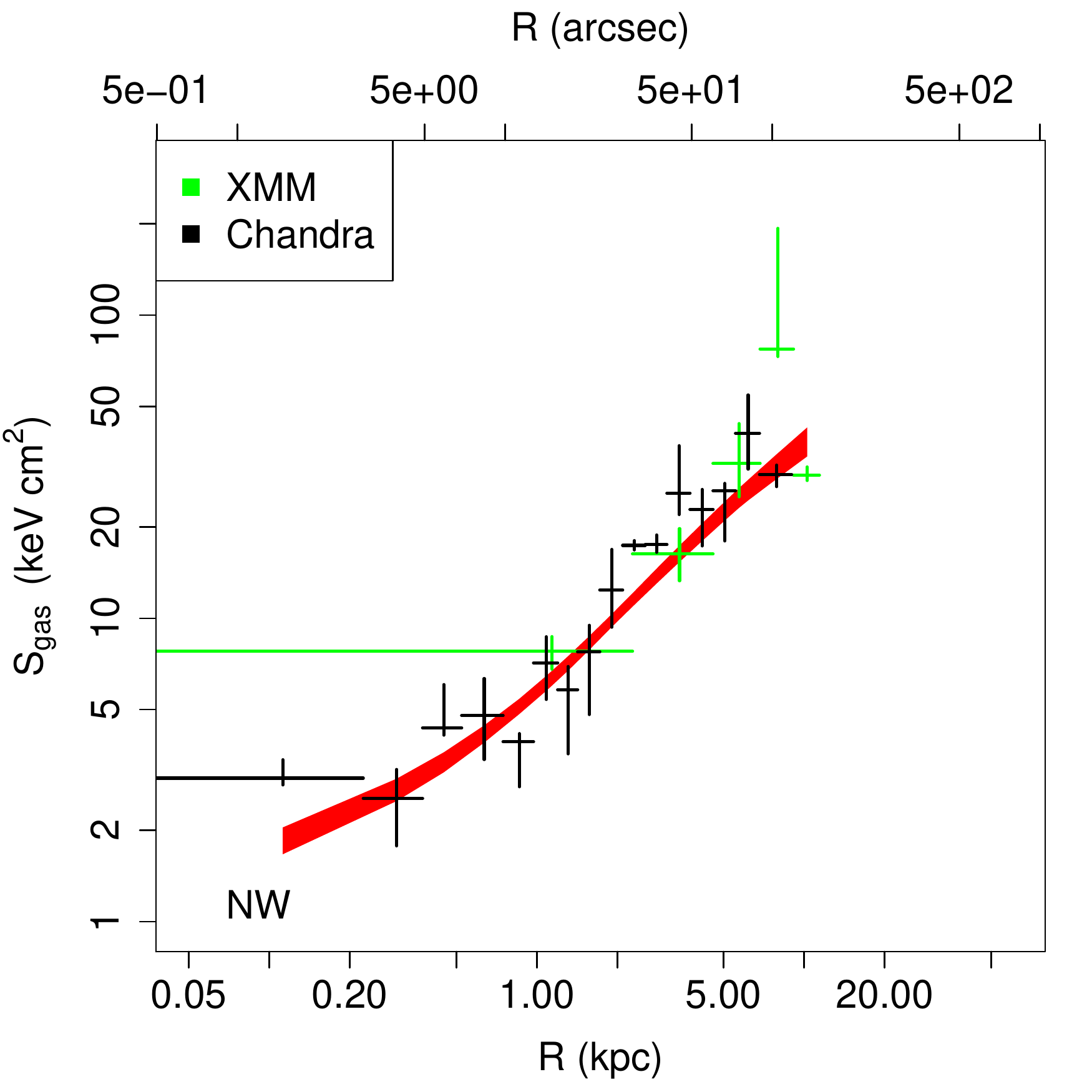}
\includegraphics[scale=0.16]{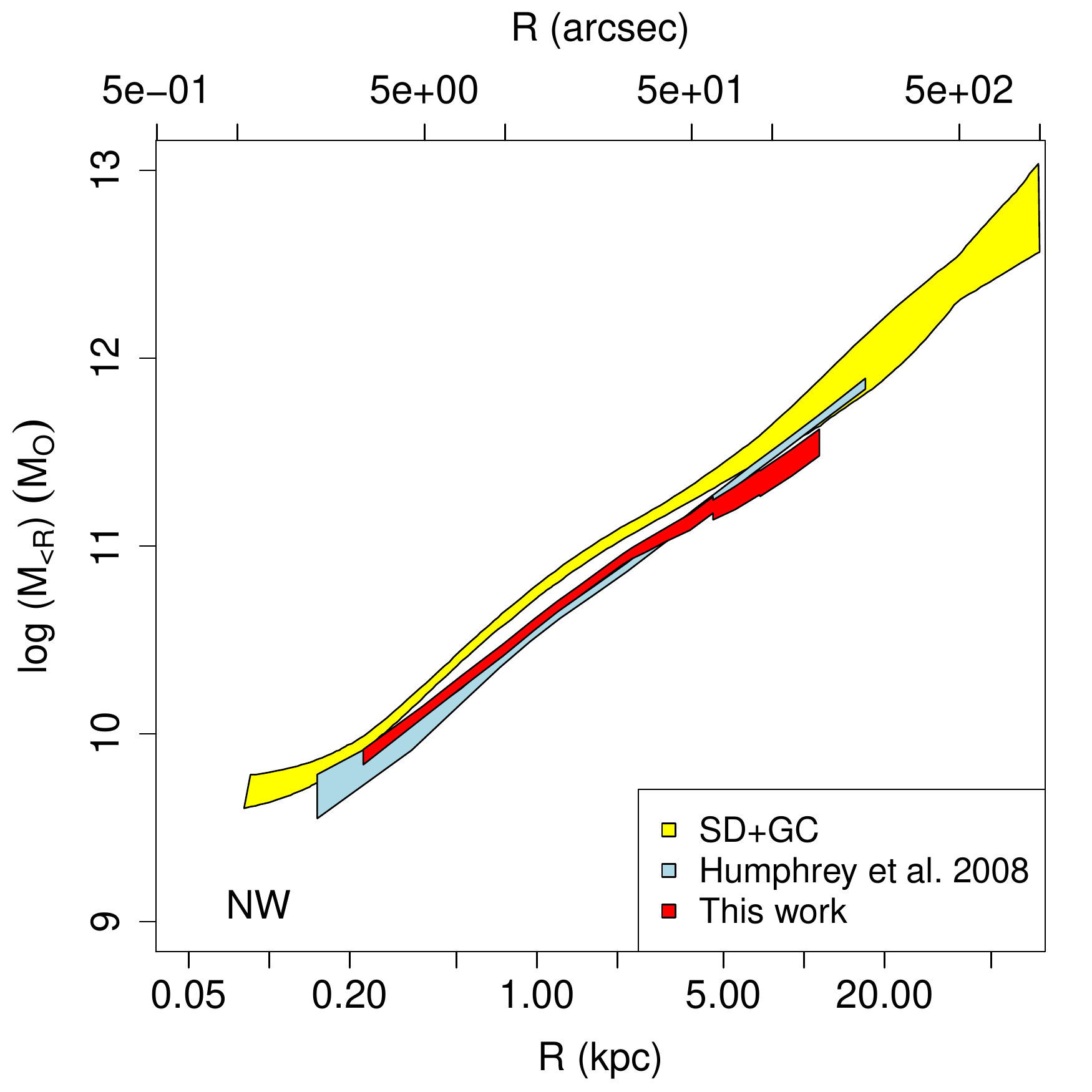}
\includegraphics[scale=0.16]{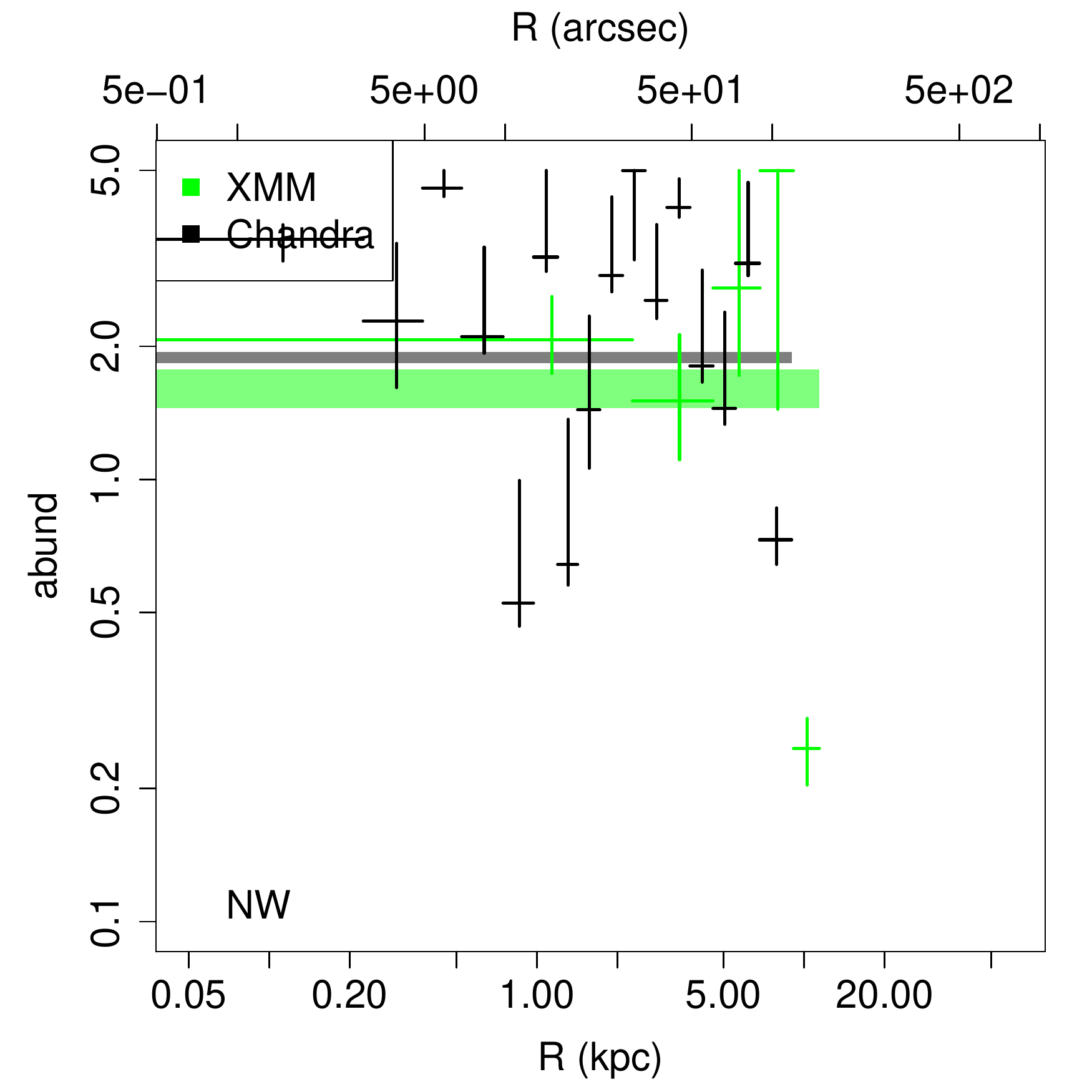}\\
\includegraphics[scale=0.16]{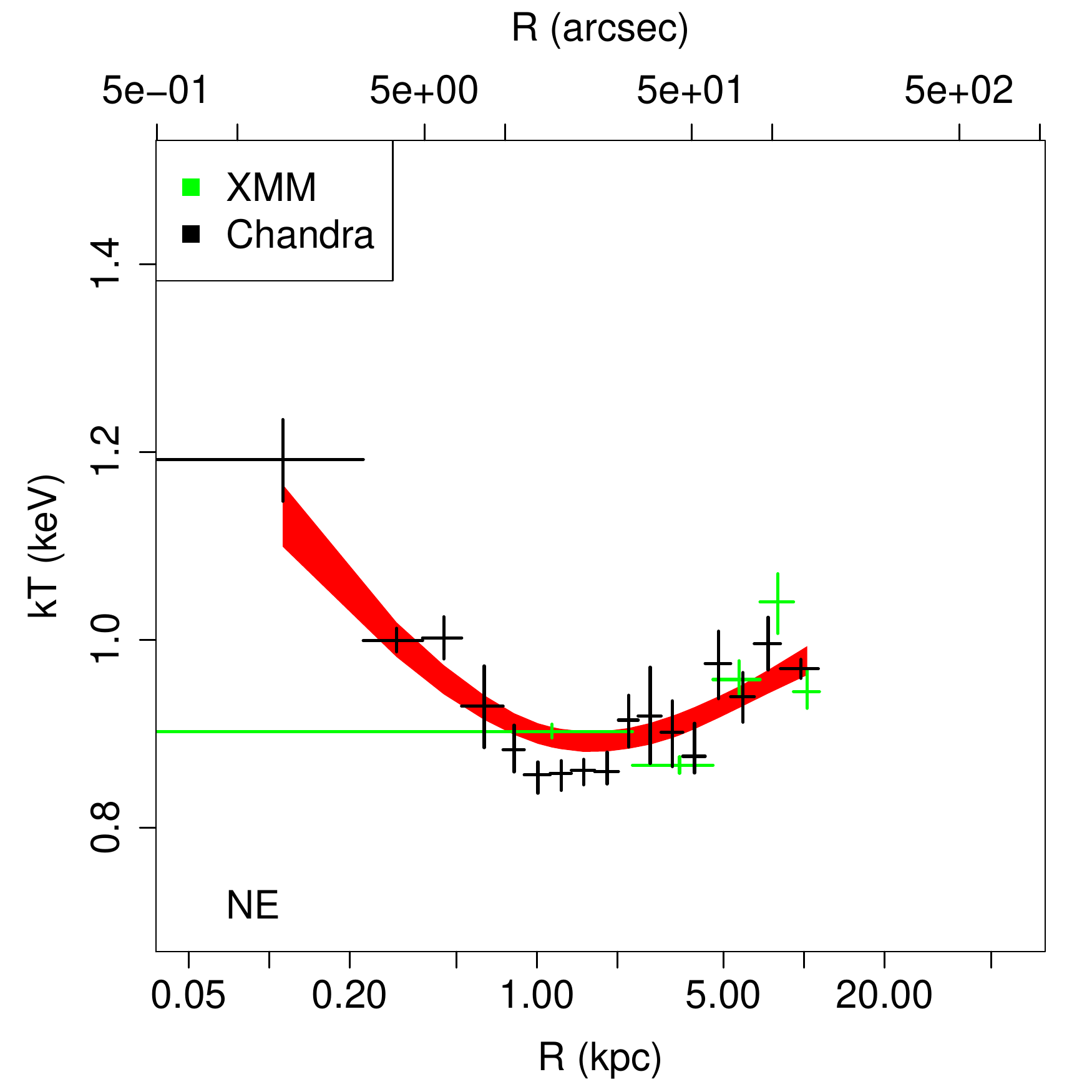}
\includegraphics[scale=0.16]{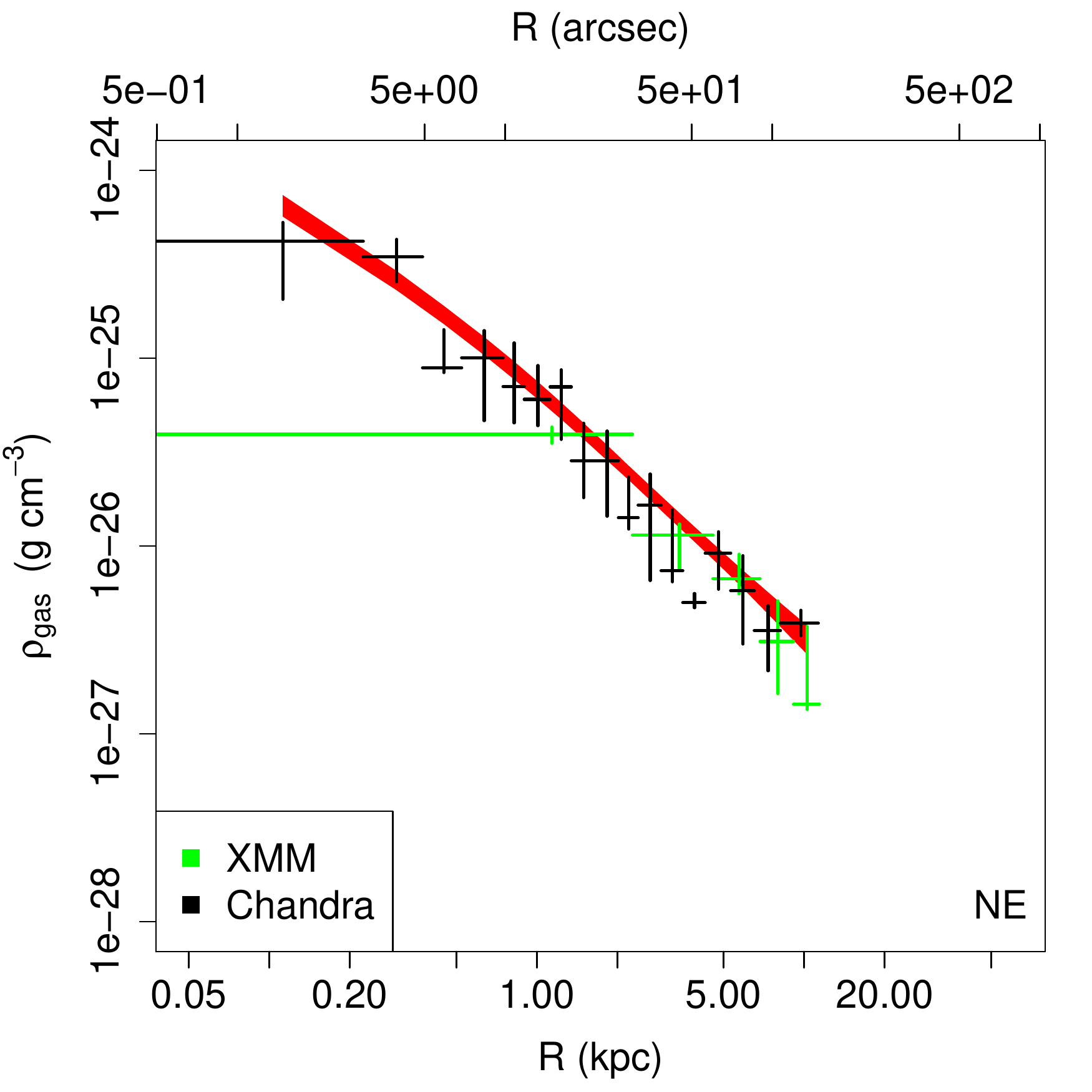}
\includegraphics[scale=0.16]{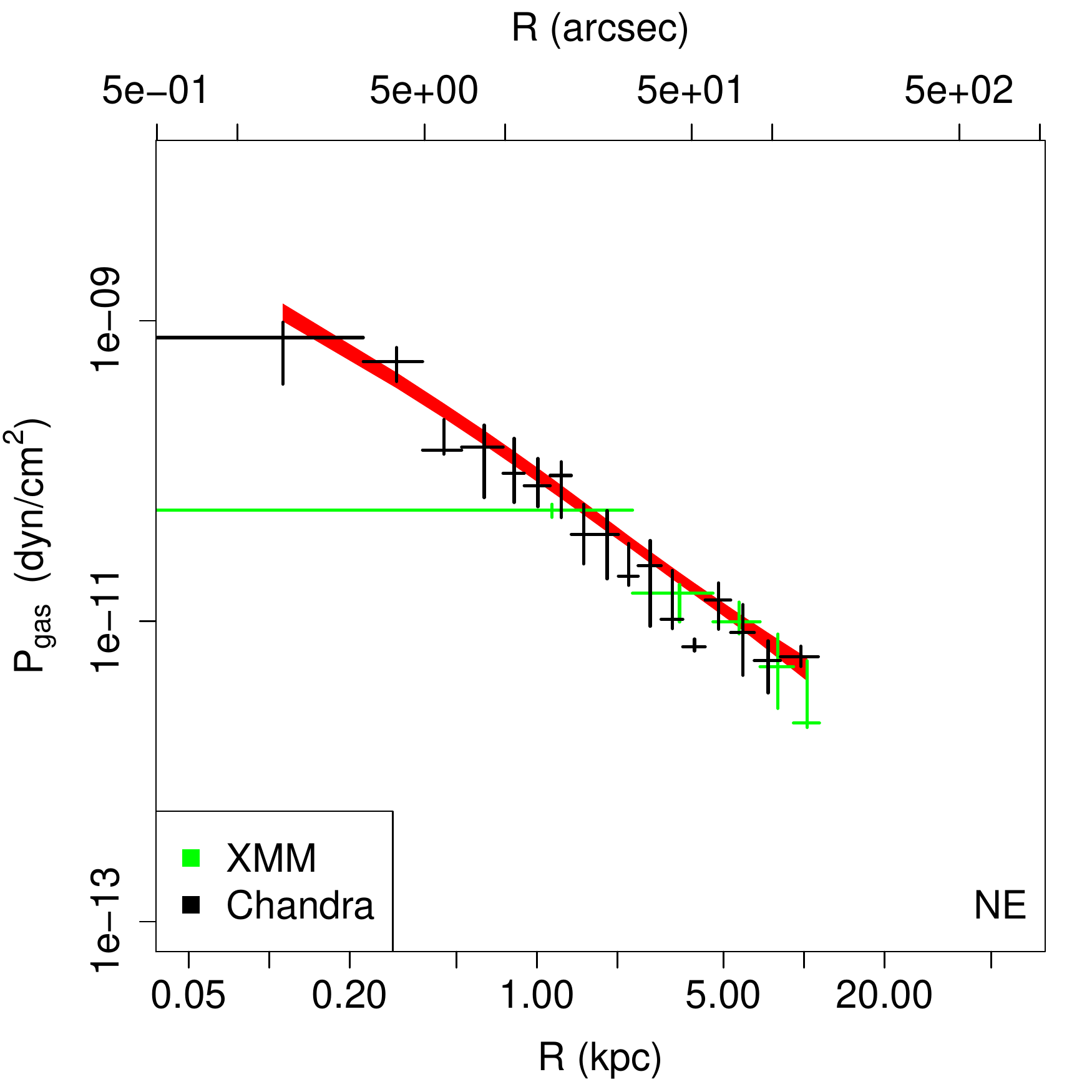}
\includegraphics[scale=0.16]{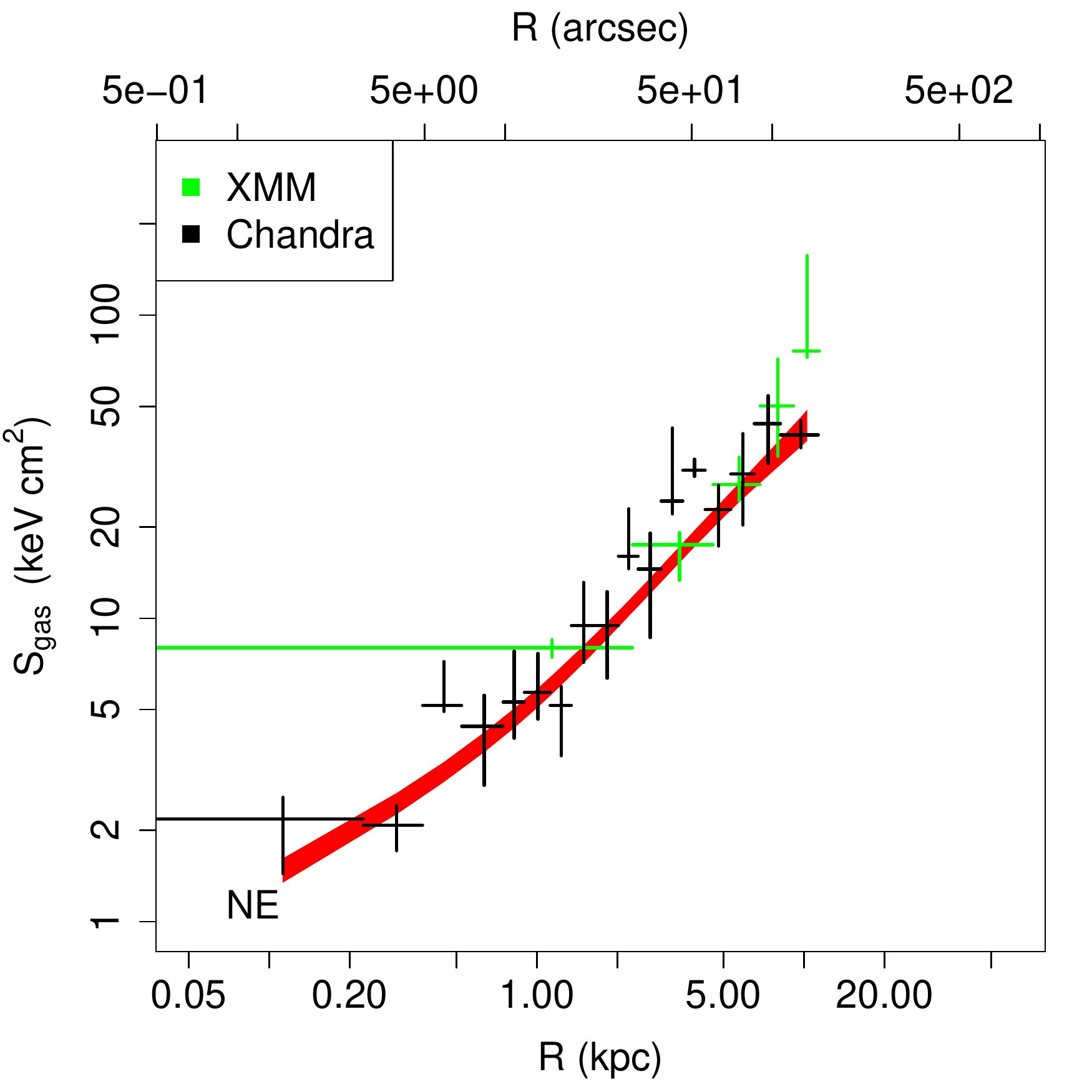}
\includegraphics[scale=0.16]{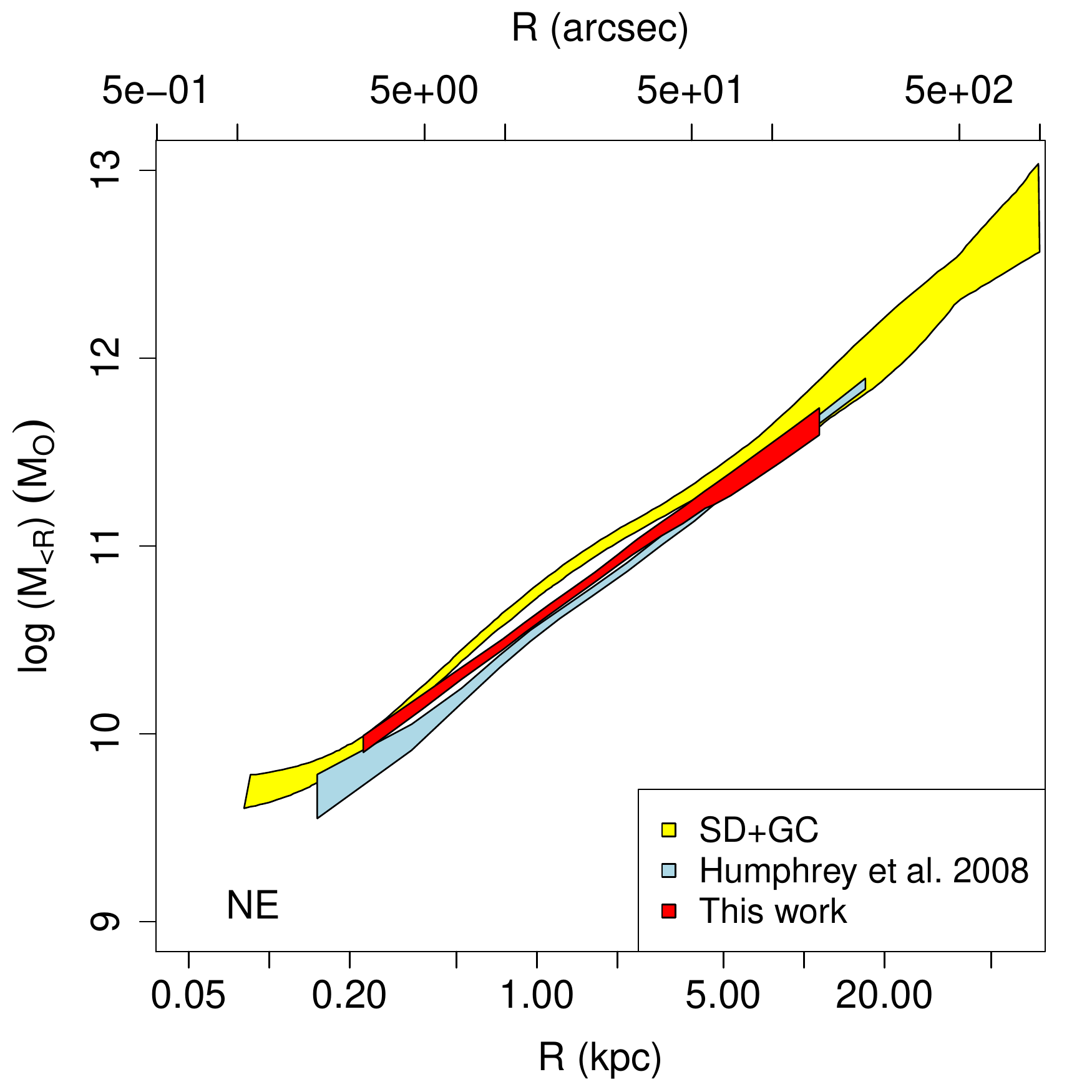}
\includegraphics[scale=0.16]{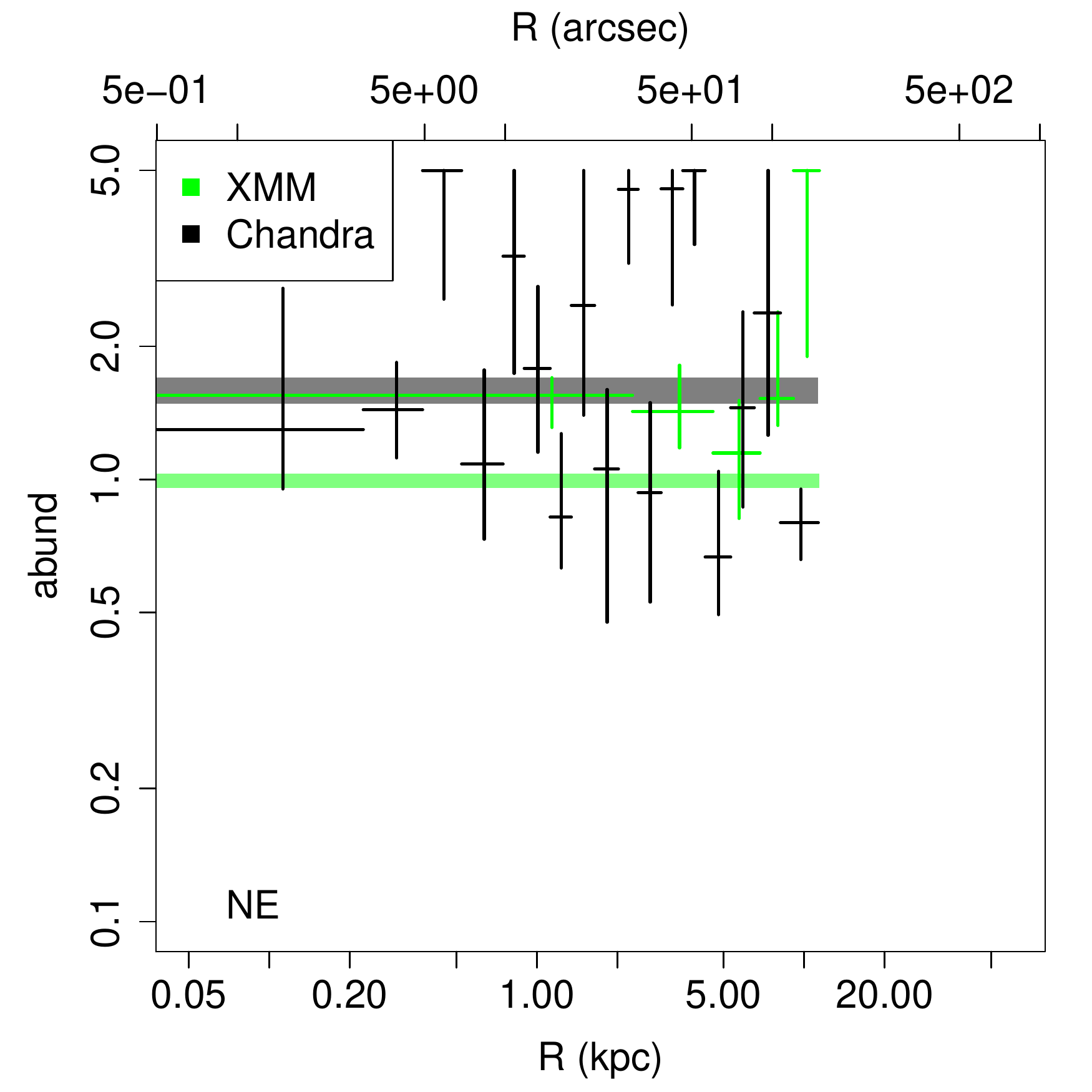}\\
\includegraphics[scale=0.16]{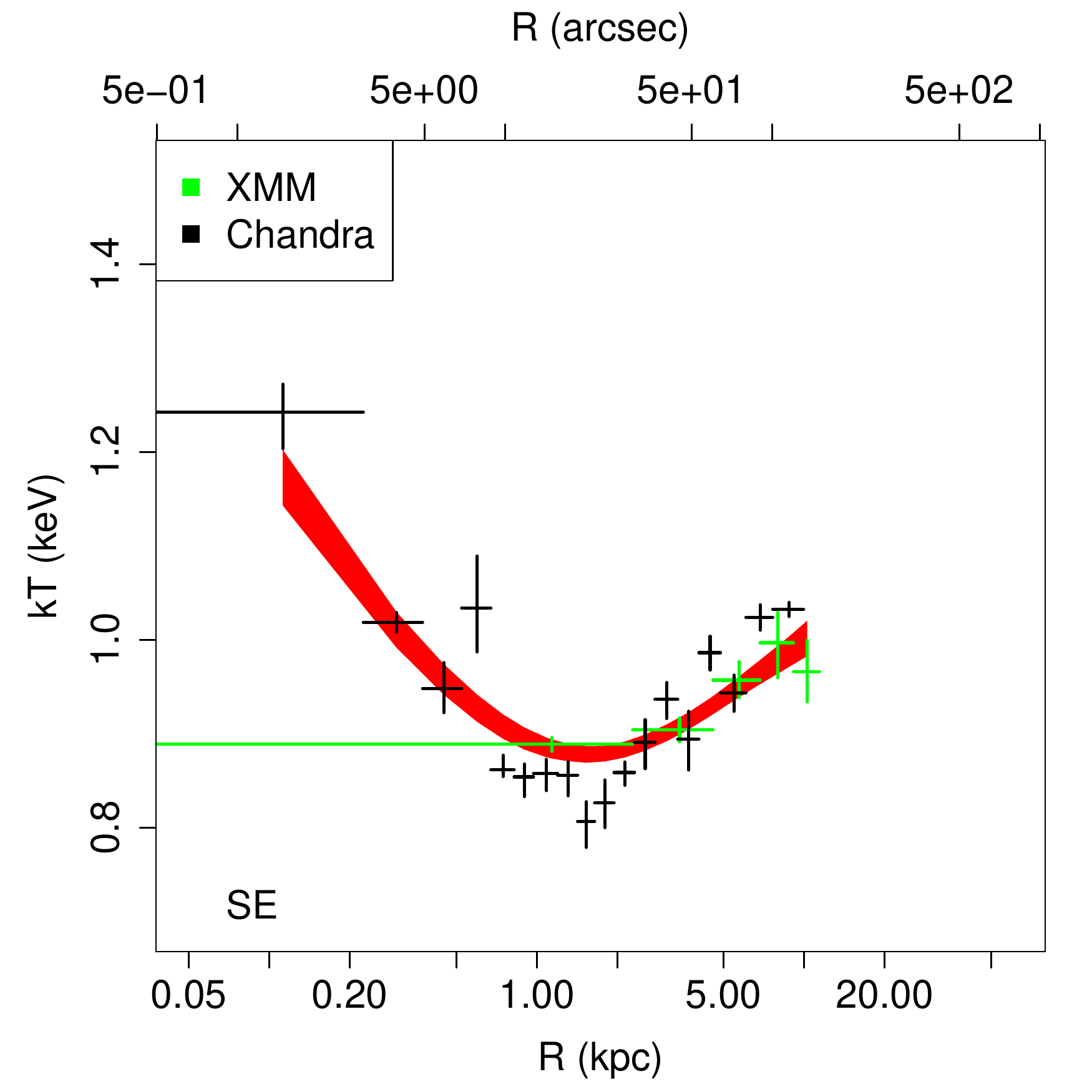}
\includegraphics[scale=0.16]{{N4649_nh_profile_merged_180_270_0_0_fit_0.8_abund}.pdf}
\includegraphics[scale=0.16]{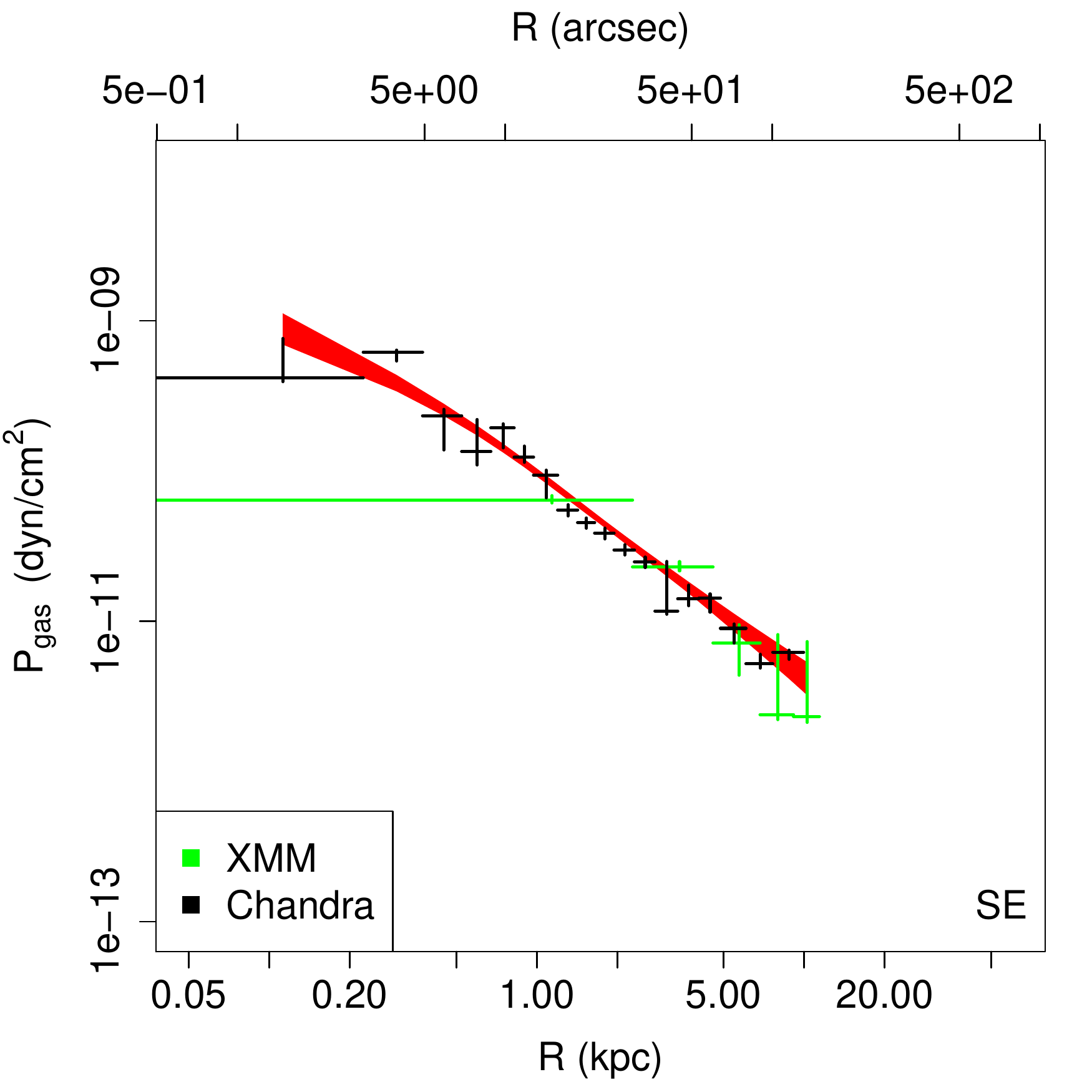}
\includegraphics[scale=0.16]{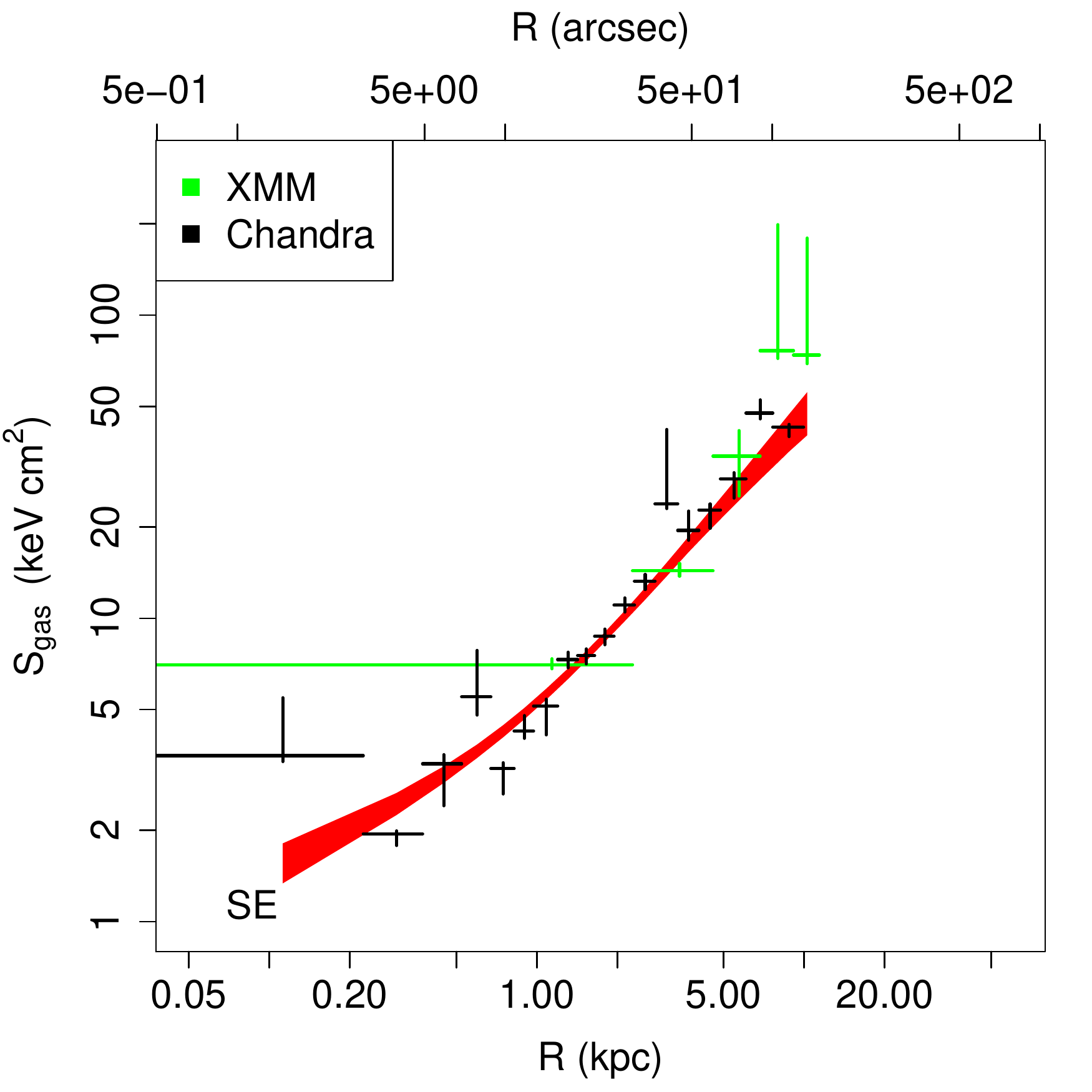}
\includegraphics[scale=0.16]{{N4649_mass_profile_comparison_180_270_0_0_0.8_abund}.pdf}
\includegraphics[scale=0.16]{{N4649_abund_profile_merged_180_270_0_0}.pdf}\\
\includegraphics[scale=0.16]{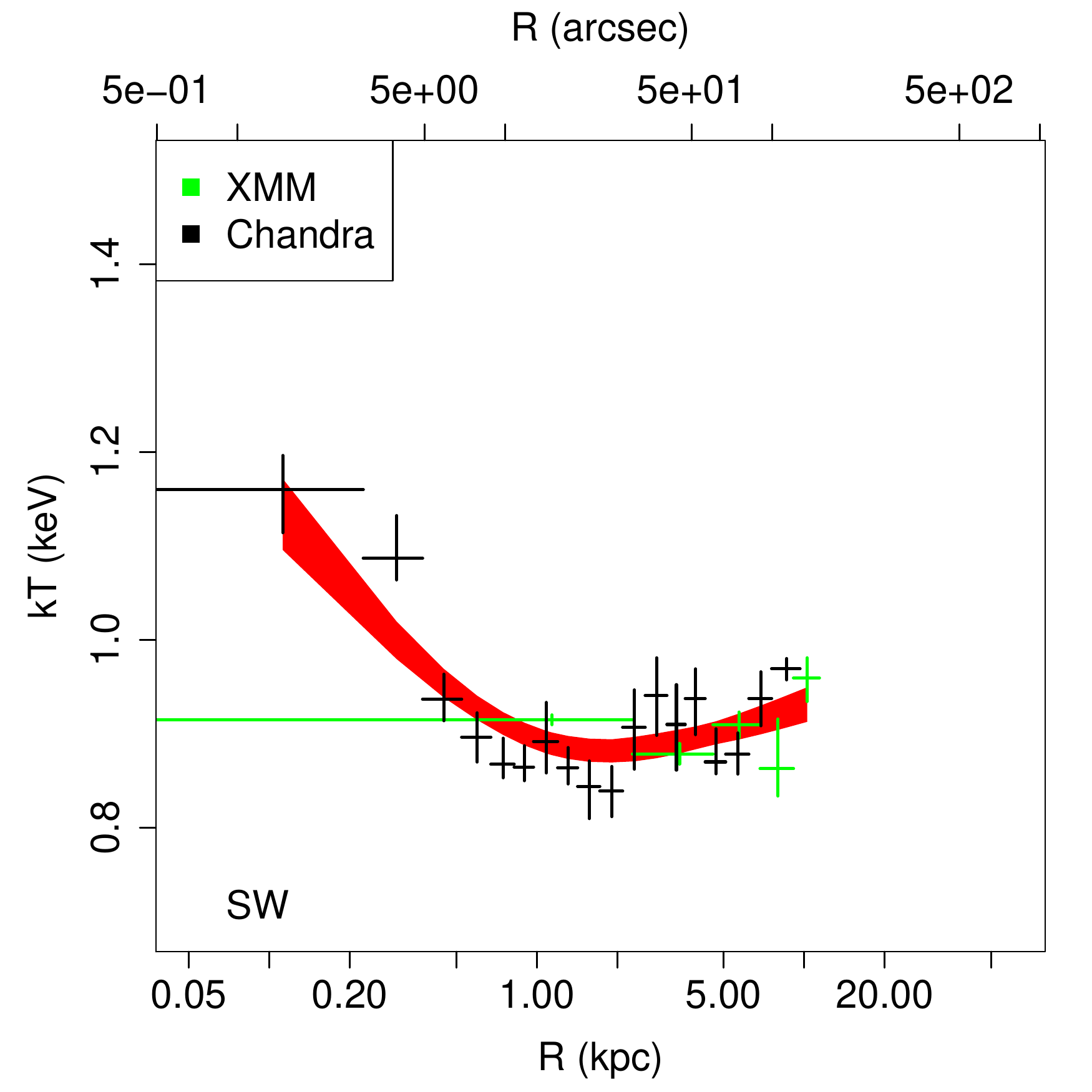}
\includegraphics[scale=0.16]{{N4649_nh_profile_merged_270_360_0_0_fit_0.8_abund}.pdf}
\includegraphics[scale=0.16]{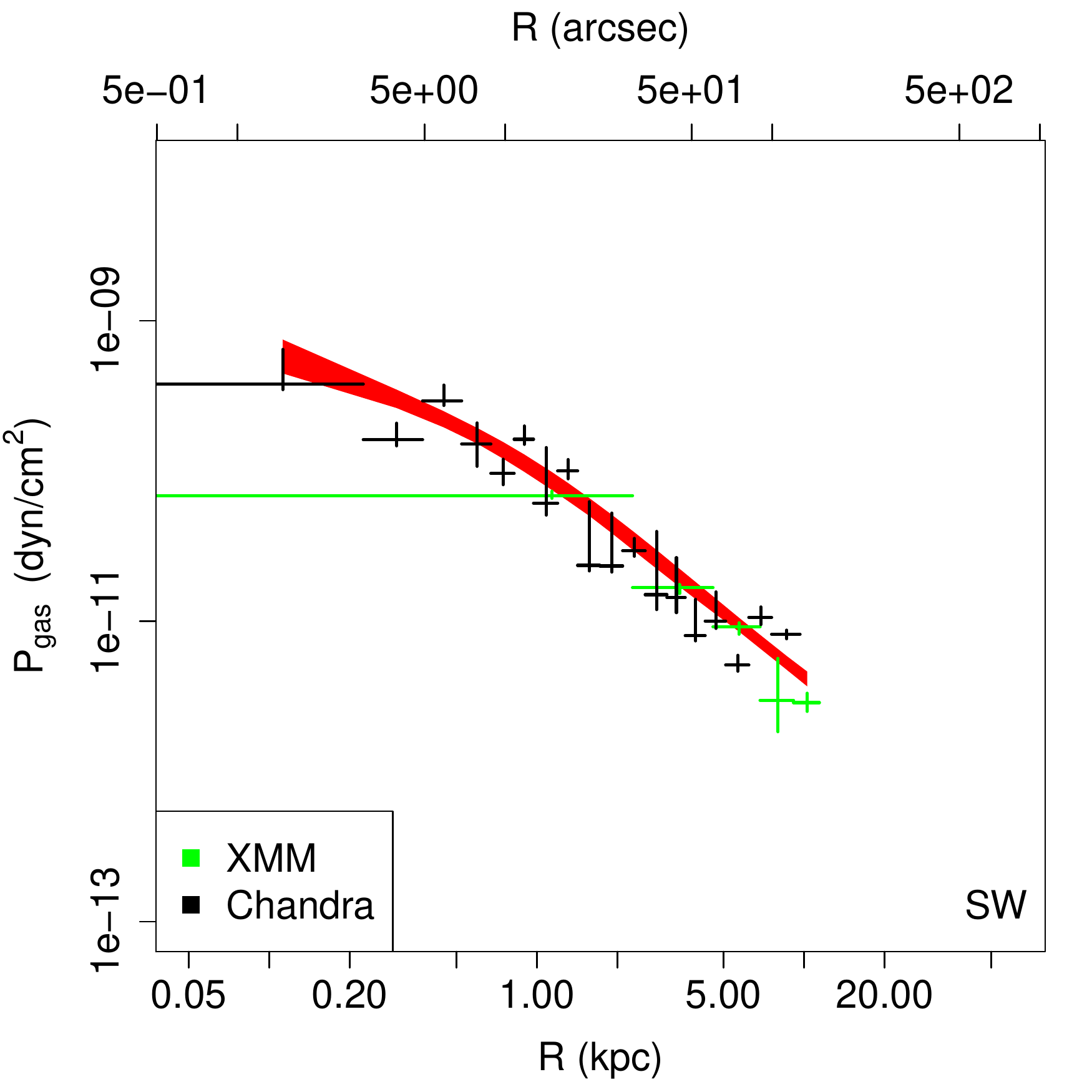}
\includegraphics[scale=0.16]{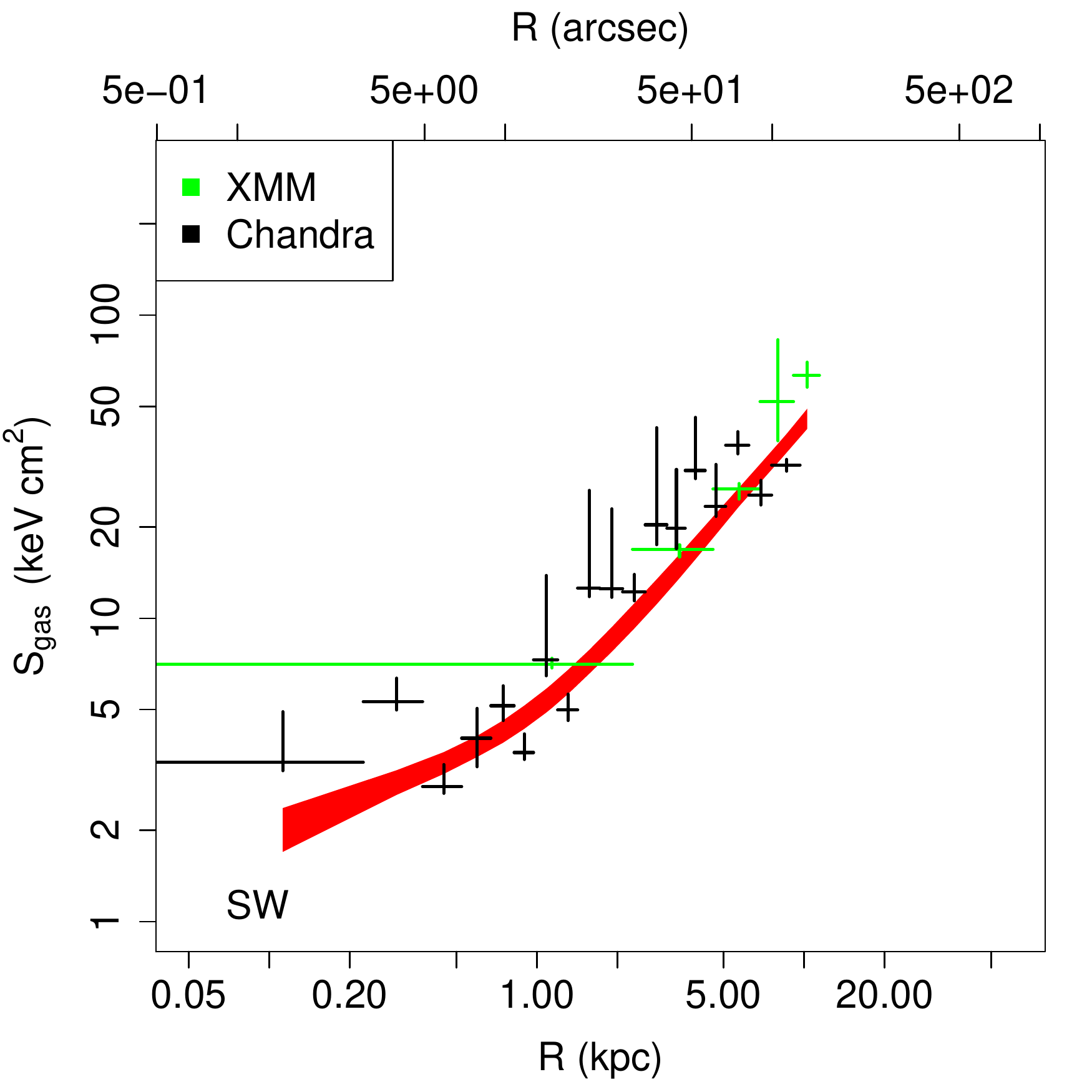}
\includegraphics[scale=0.16]{{N4649_mass_profile_comparison_270_360_0_0_0.8_abund}.pdf}
\includegraphics[scale=0.16]{{N4649_abund_profile_merged_270_360_0_0}.pdf}
\caption{Same as Figure \ref{fig:N4649_gas_profiles_merged_app} but with the free abundance model. In addition, on the rightmost panel of each row we show the element abundances profiles for \textit{XMM}-MOS data (in green) and for \textit{Chandra} ACIS data (represented in black). In the same panels we overplot with green and black rectangles the values of the element abundances obtained with the fixed abundance model for \textit{XMM}  and \textit{Chandra} data, respectively.}\label{fig:N4649_gas_profiles_merged_abund_app}
\end{figure}

\begin{sidewaysfigure}
\centering
\includegraphics[scale=0.26]{{N4649_mass_fit_0_360_0_0_0.8}.pdf}
\includegraphics[scale=0.26]{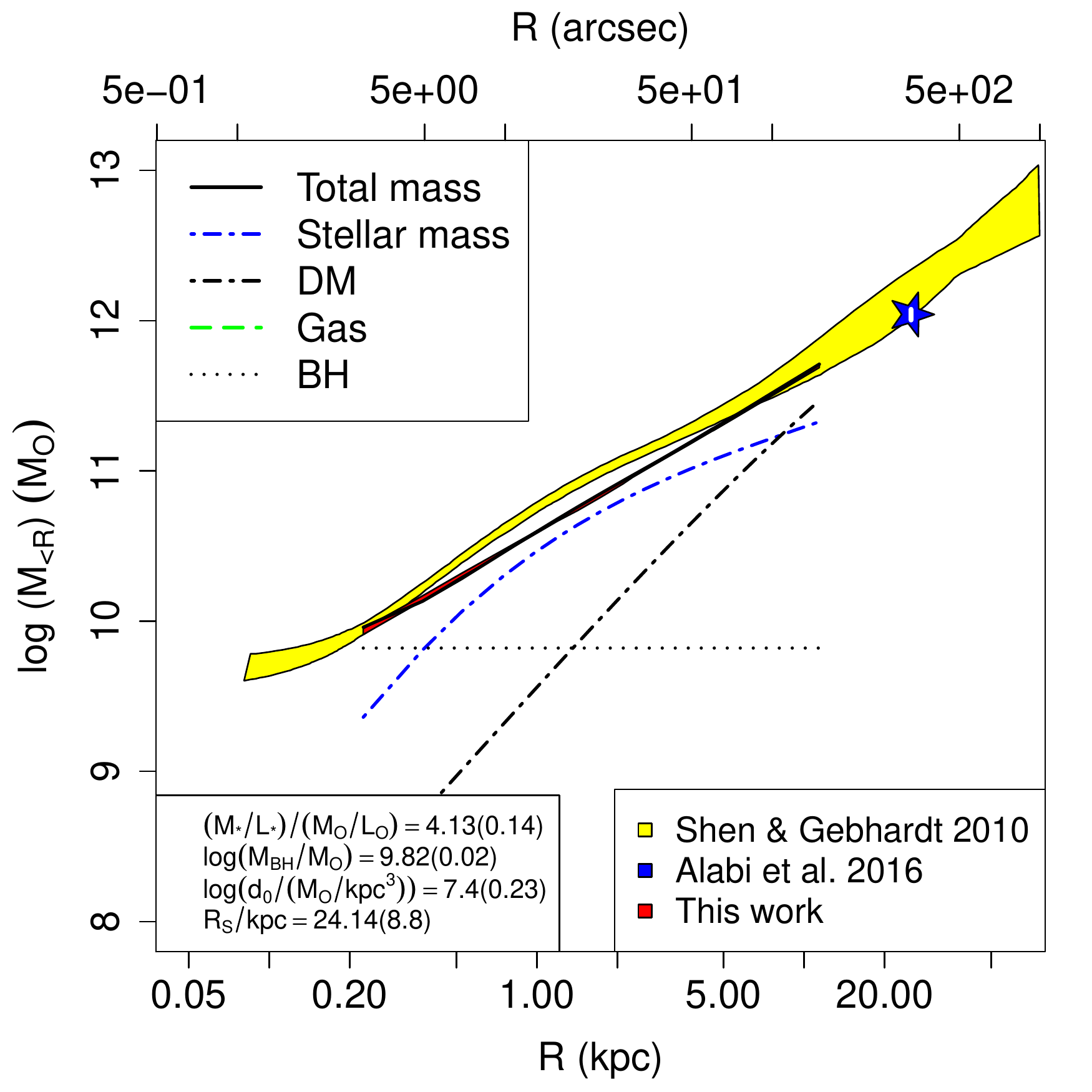}
\includegraphics[scale=0.26]{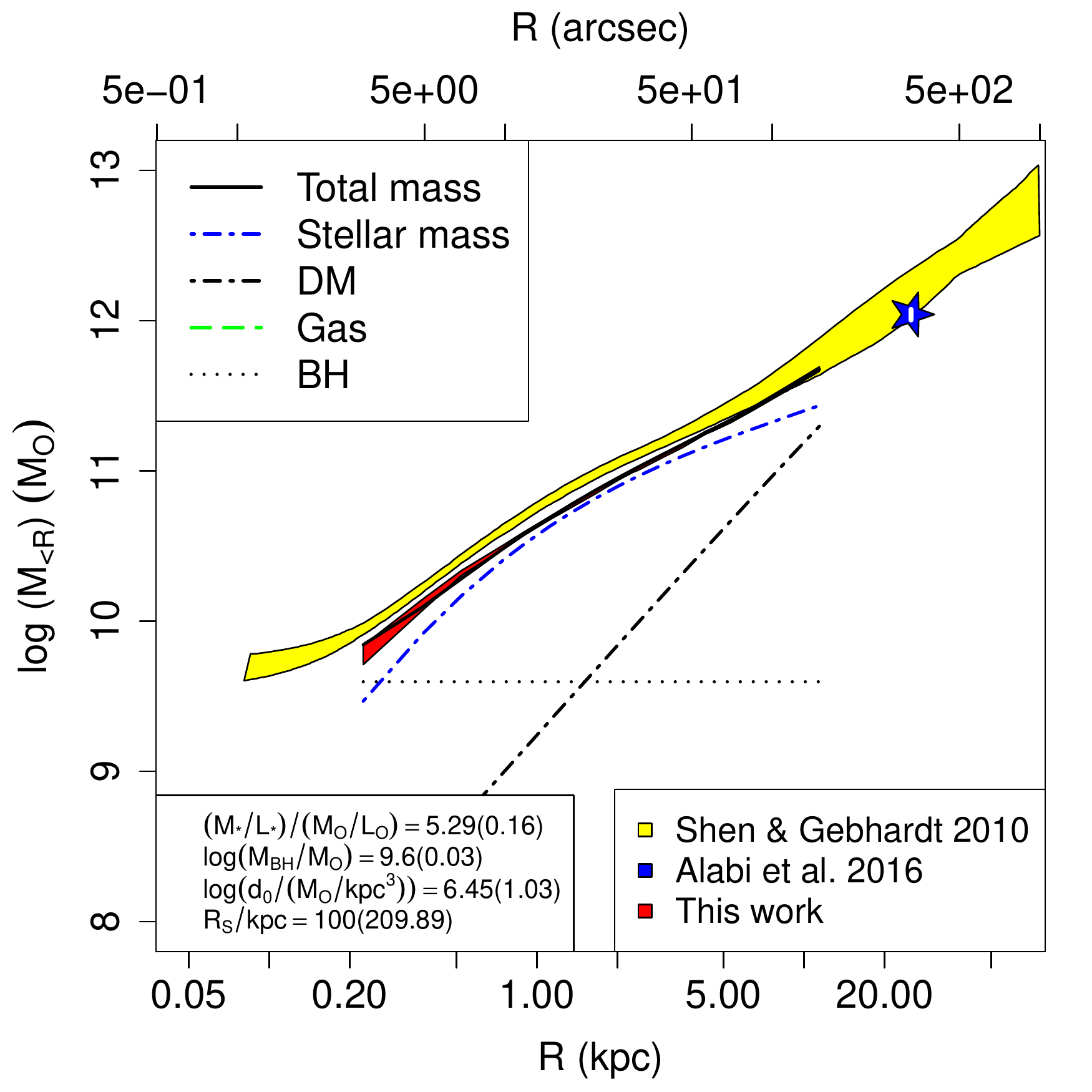}
\includegraphics[scale=0.26]{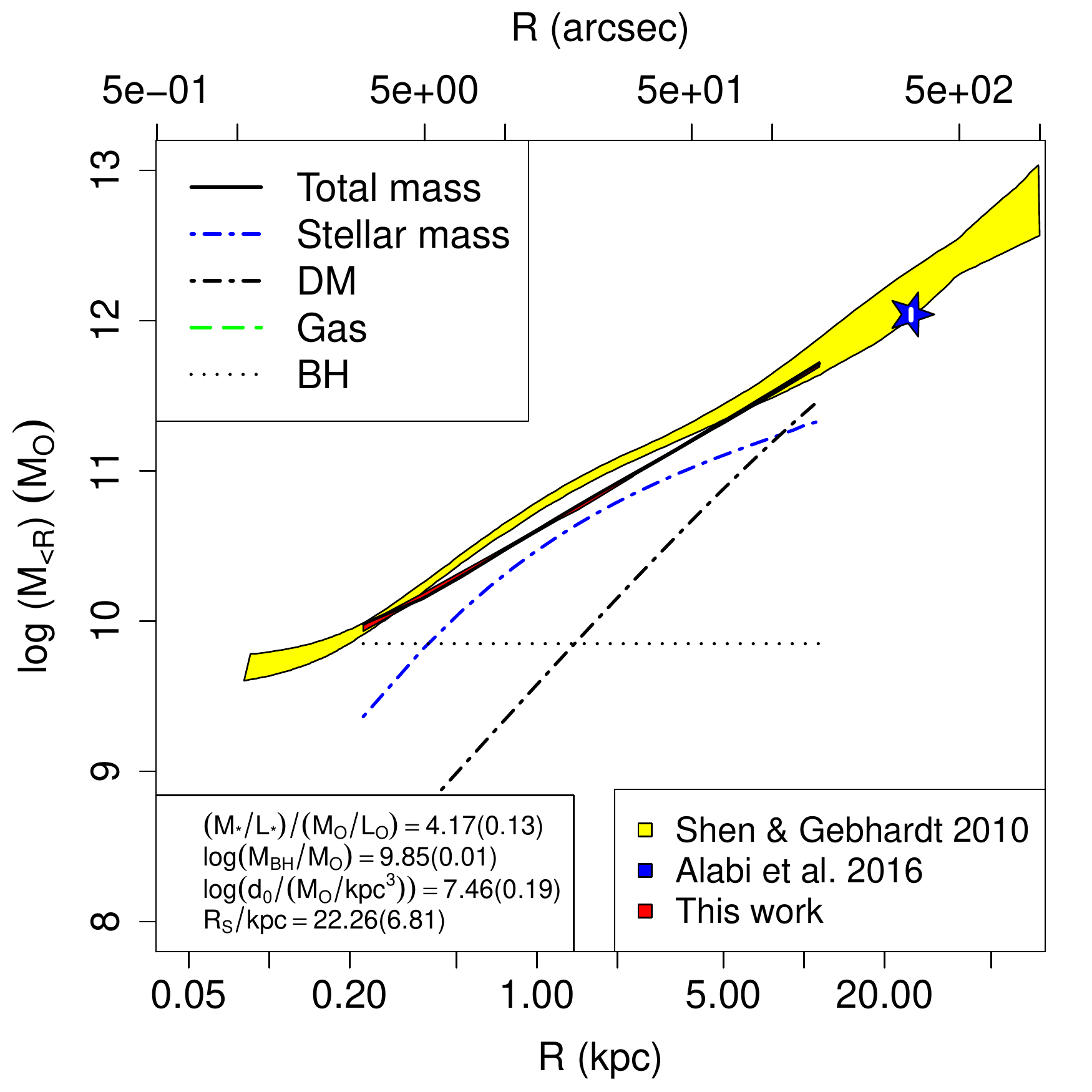}
\includegraphics[scale=0.26]{{N4649_mass_fit_270_360_0_0_0.8}.pdf}\\
\includegraphics[scale=0.26]{{N4649_mass_fit_0_360_0_0_0.8_abund}.pdf}
\includegraphics[scale=0.26]{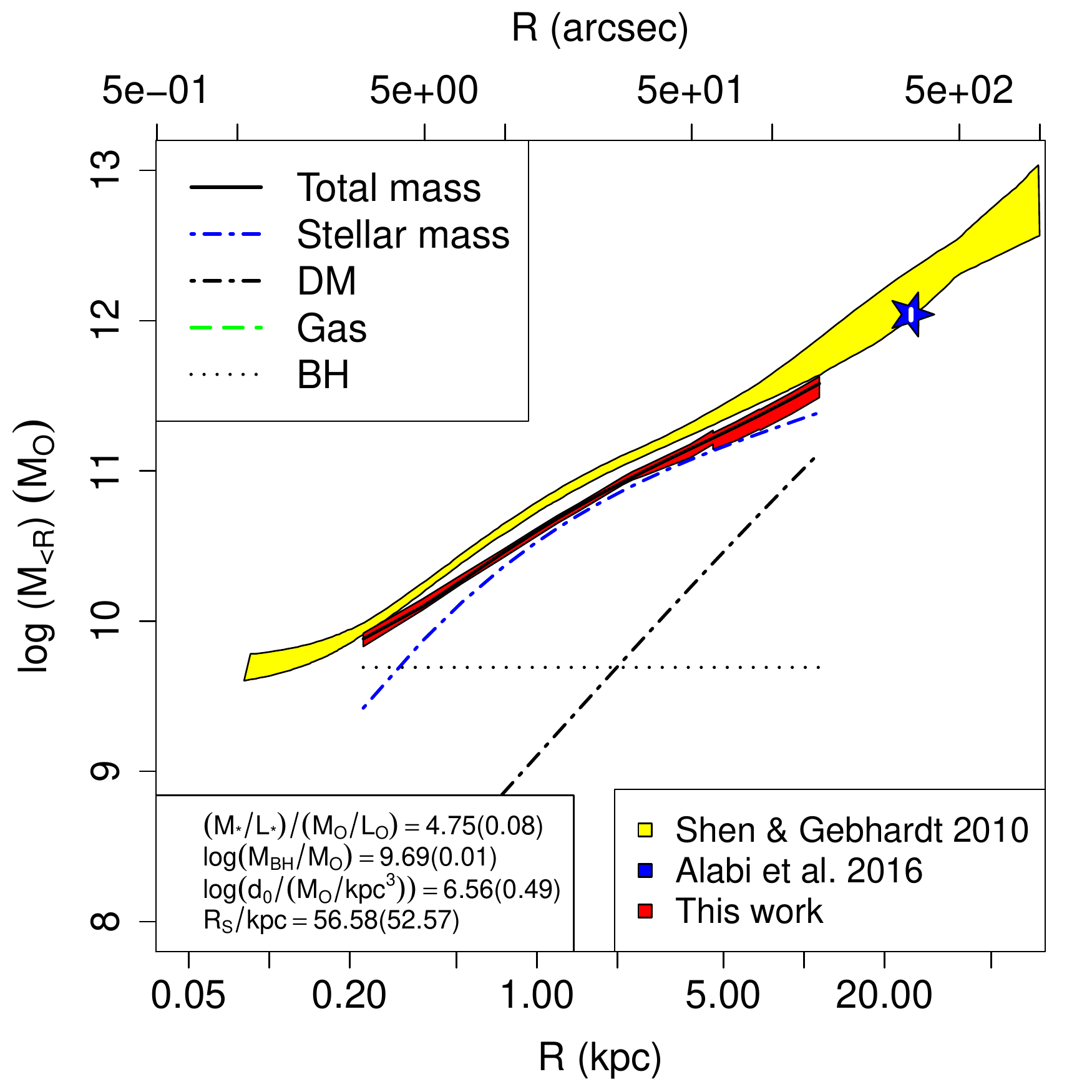}
\includegraphics[scale=0.26]{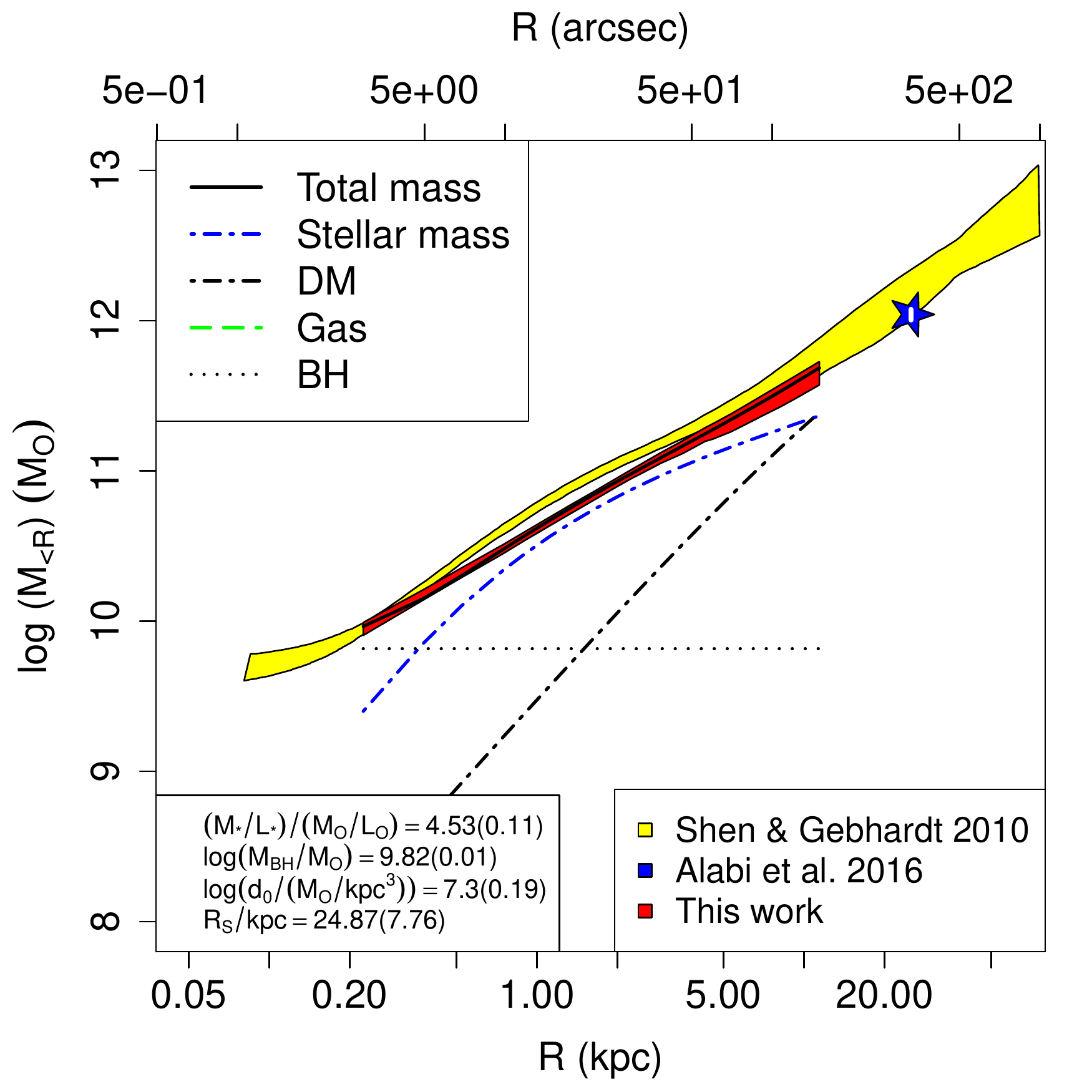}
\includegraphics[scale=0.26]{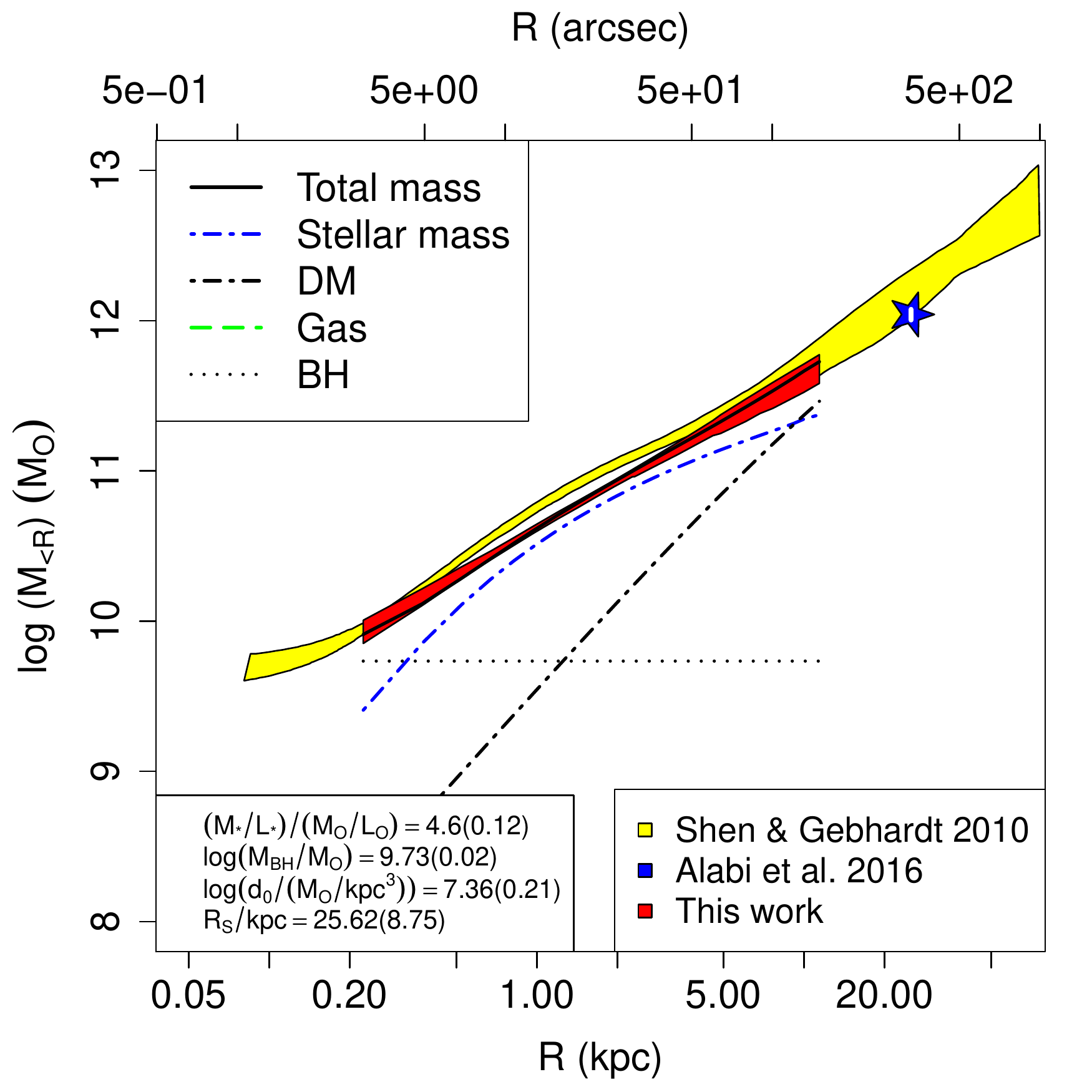}
\includegraphics[scale=0.26]{{N4649_mass_fit_270_360_0_0_0.8_abund}.pdf}
\caption{Fit to mass profiles of NGC 4649 shown in Fig. \ref{fig:N4649_gas_profiles_merged_app} (top row) and in Fig. \ref{fig:N4649_gas_profiles_merged_abund_app} (bottom row), from left to right in the full (0-360), NW (270-360), NE (0-90), SE (90-180), SW (180-270) sector, respectively. The mass profile form the {HE} equation is presented in red, and the best fit contributions of the various mass components (gas mass, stellar mass, black hole and NFW DM profile) are presented with different colors as reported in the legend. The best fit parameters are reported in lower left box. In yellow we show the optical mass profile obtained from SD and GC reported by \citep{2010ApJ...711..484S}, while the blue star represents the measurement by \citet{2016MNRAS.460.3838A} with the corresponding uncertainty shown as a white vertical line side the star itself.}\label{fig:N4649_mass_fits_app}
\end{sidewaysfigure}

\begin{figure}
\centering
\includegraphics[scale=0.19]{N5846_chandra_surface_brightness_0_360_0_0.pdf}
\includegraphics[scale=0.19]{N5846_chandra_surface_brightness_120_180_0_0.pdf}
\includegraphics[scale=0.19]{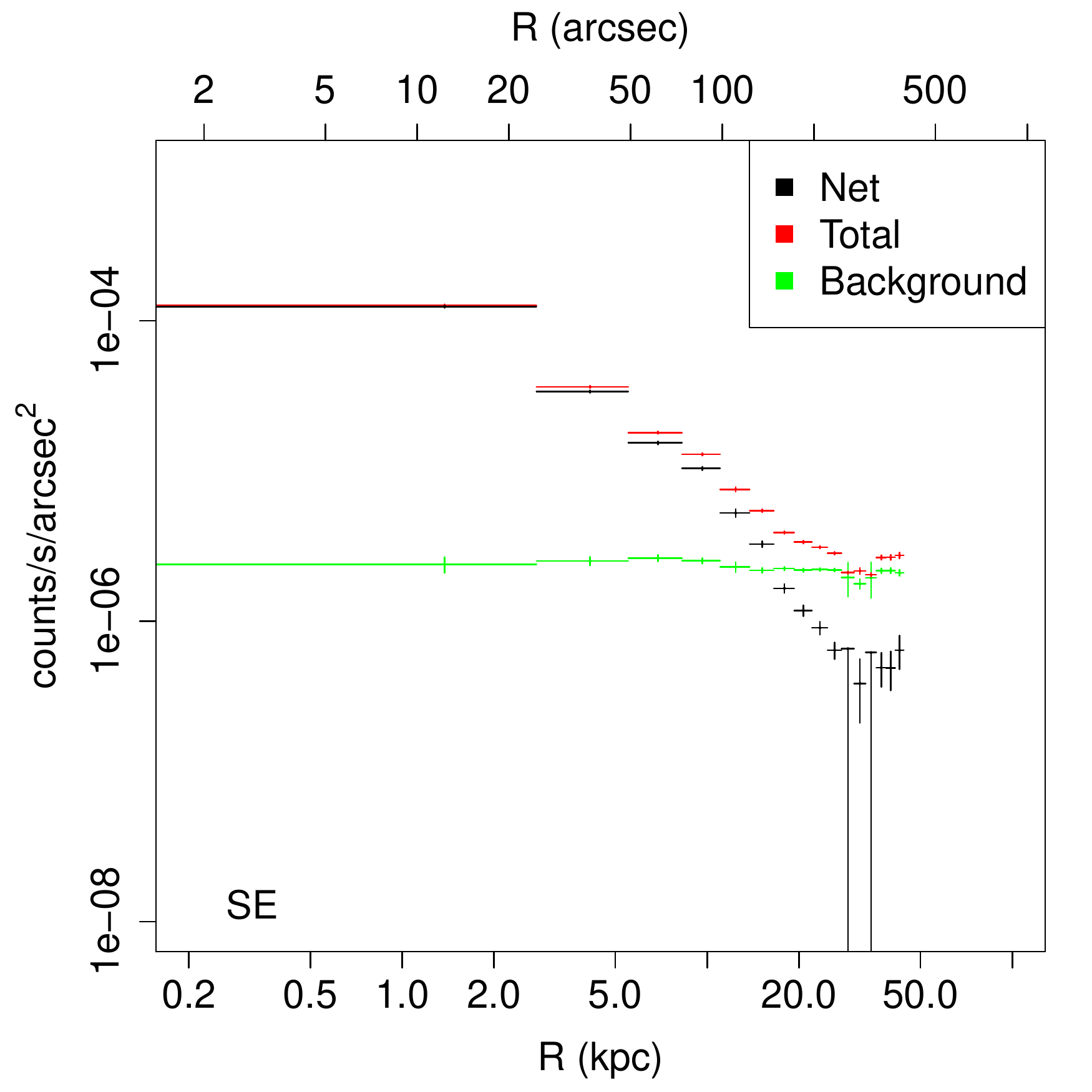}
\includegraphics[scale=0.19]{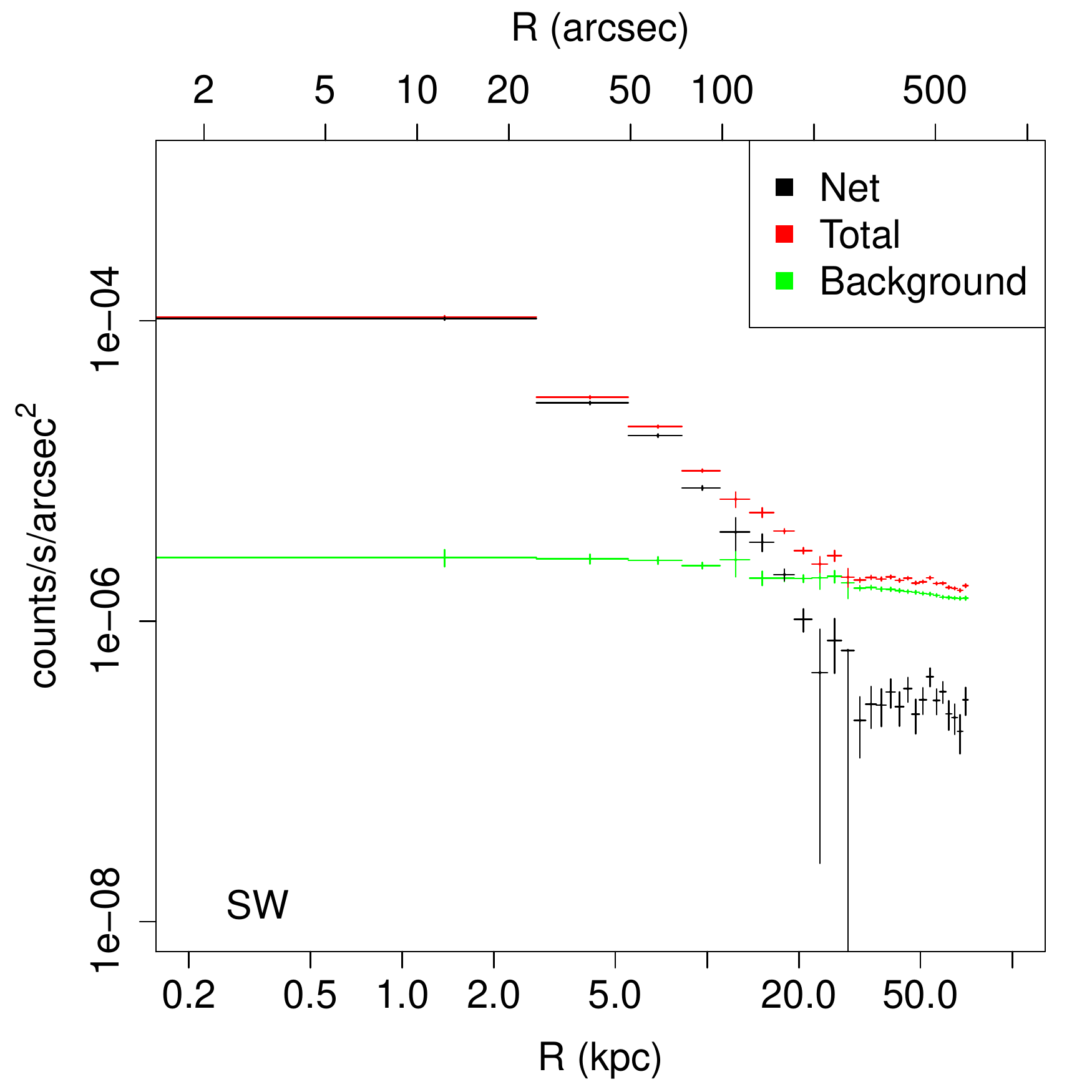}
\includegraphics[scale=0.19]{N5846_chandra_surface_brightness_340_480_0_0.pdf}\\
\includegraphics[scale=0.19]{N5846_surface_brightness_0_360_0_0.pdf}
\includegraphics[scale=0.19]{N5846_surface_brightness_120_180_0_0.pdf}
\includegraphics[scale=0.19]{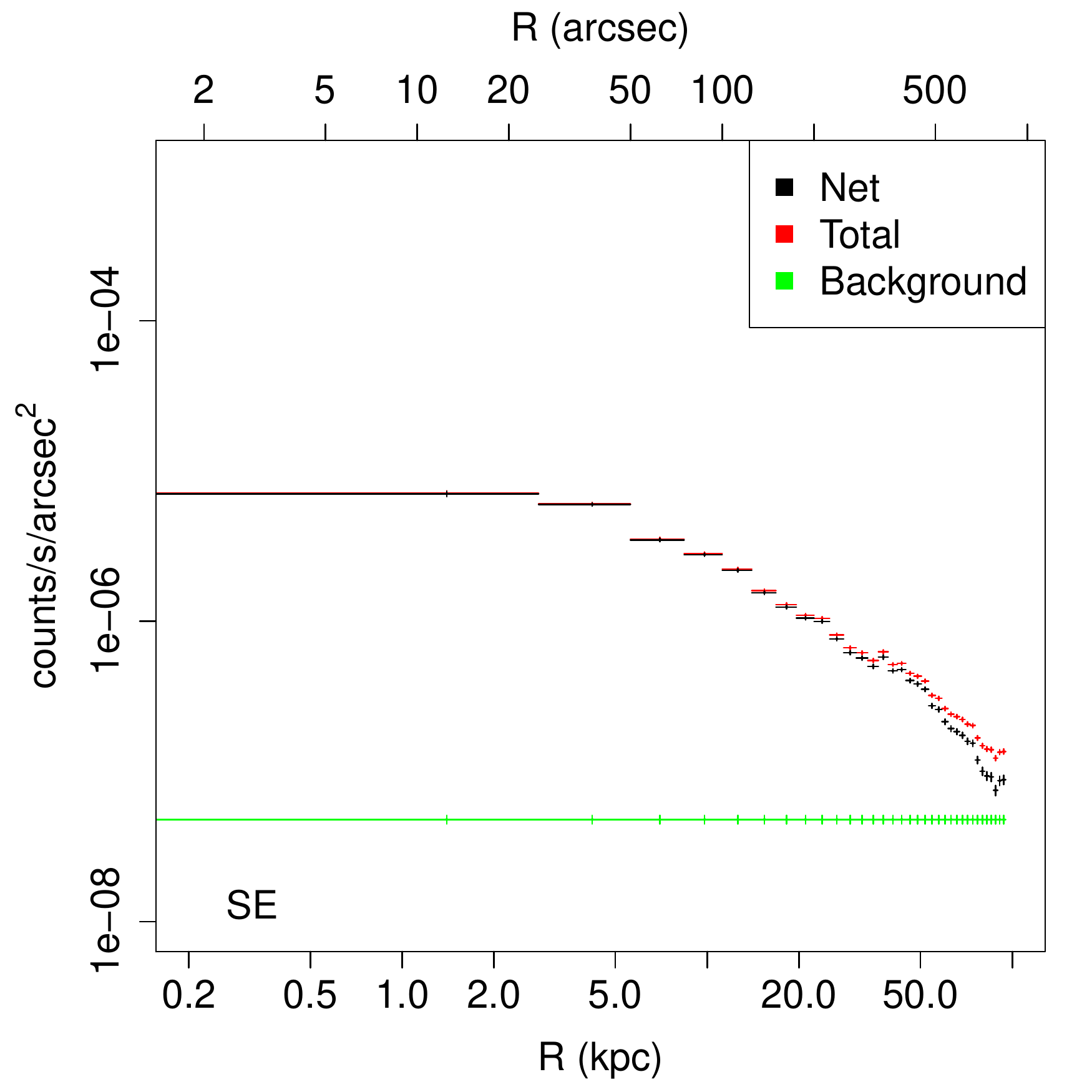}
\includegraphics[scale=0.19]{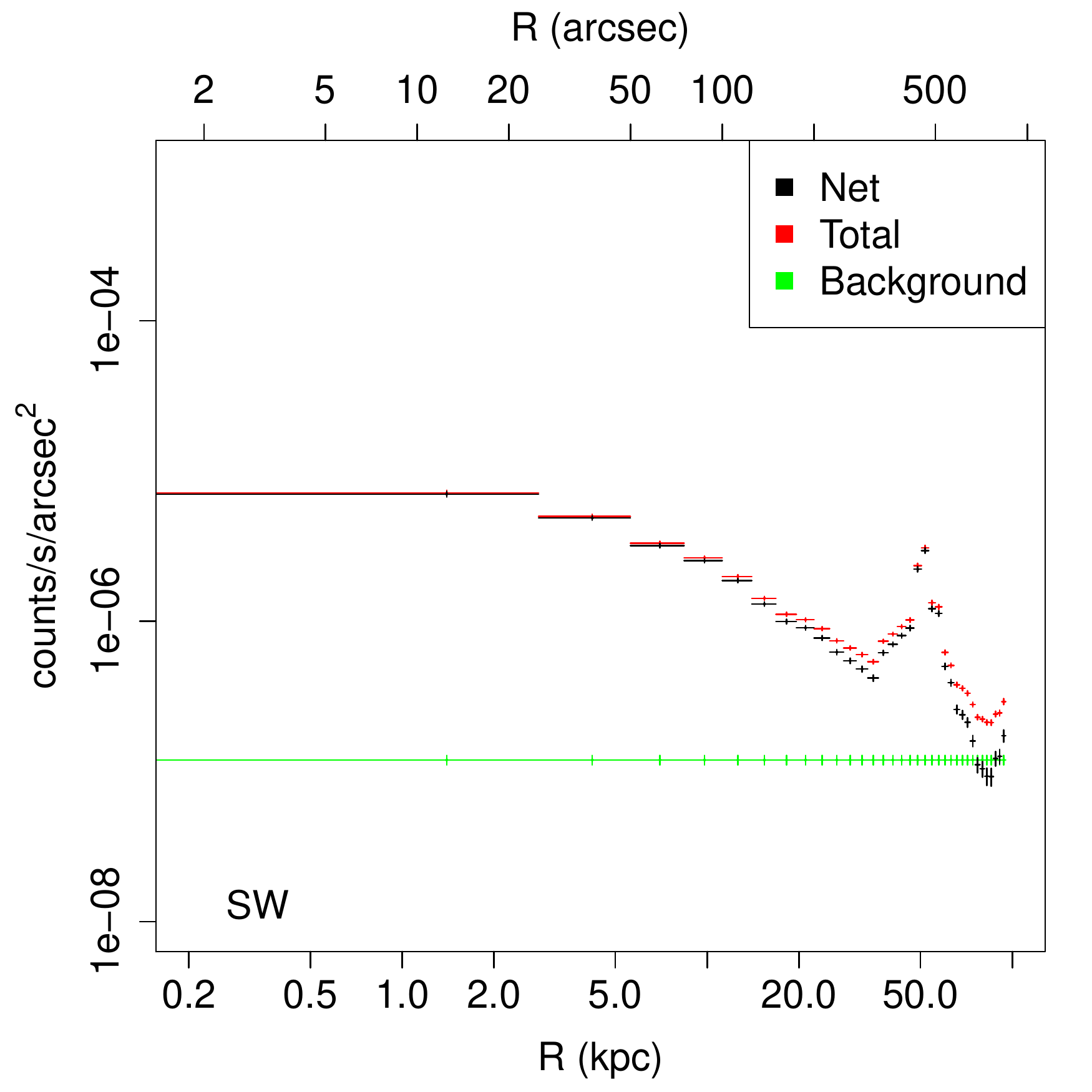}
\includegraphics[scale=0.19]{N5846_surface_brightness_340_480_0_0.pdf}
\caption{Brightness profiles for NGC 5846 in the \(0.3-10\) keV band in the different sectors shown in Fig. \ref{fig:N5846_mos}, from left to right full (0-360), NE (30-90), SE (90-180), SW (180-250) and NW (250-30), respectively. In particular we show brightness profiles for \textit{Chandra} ACIS data {in the top row} and for the \textit{XMM}-MOS data obtained from the reduction procedure proposed by \citet{2005ApJ...629..172N} in the bottom row. The annuli width is \(\sim 25''\), 50 pixels for \textit{Chandra} ACIS data and 500 pixels for \textit{XMM}-MOS data. Red, black and green points represent total, net, and background brightness profiles, respectively.}\label{fig:N5846_bp_chandra_app}
\end{figure}

\begin{figure}
\centering
\includegraphics[scale=0.19]{{N5846_temp_profile_merged_0_360_0_0_fit_0.6}.pdf}
\includegraphics[scale=0.19]{{N5846_nh_profile_merged_0_360_0_0_fit_0.6}.pdf}
\includegraphics[scale=0.19]{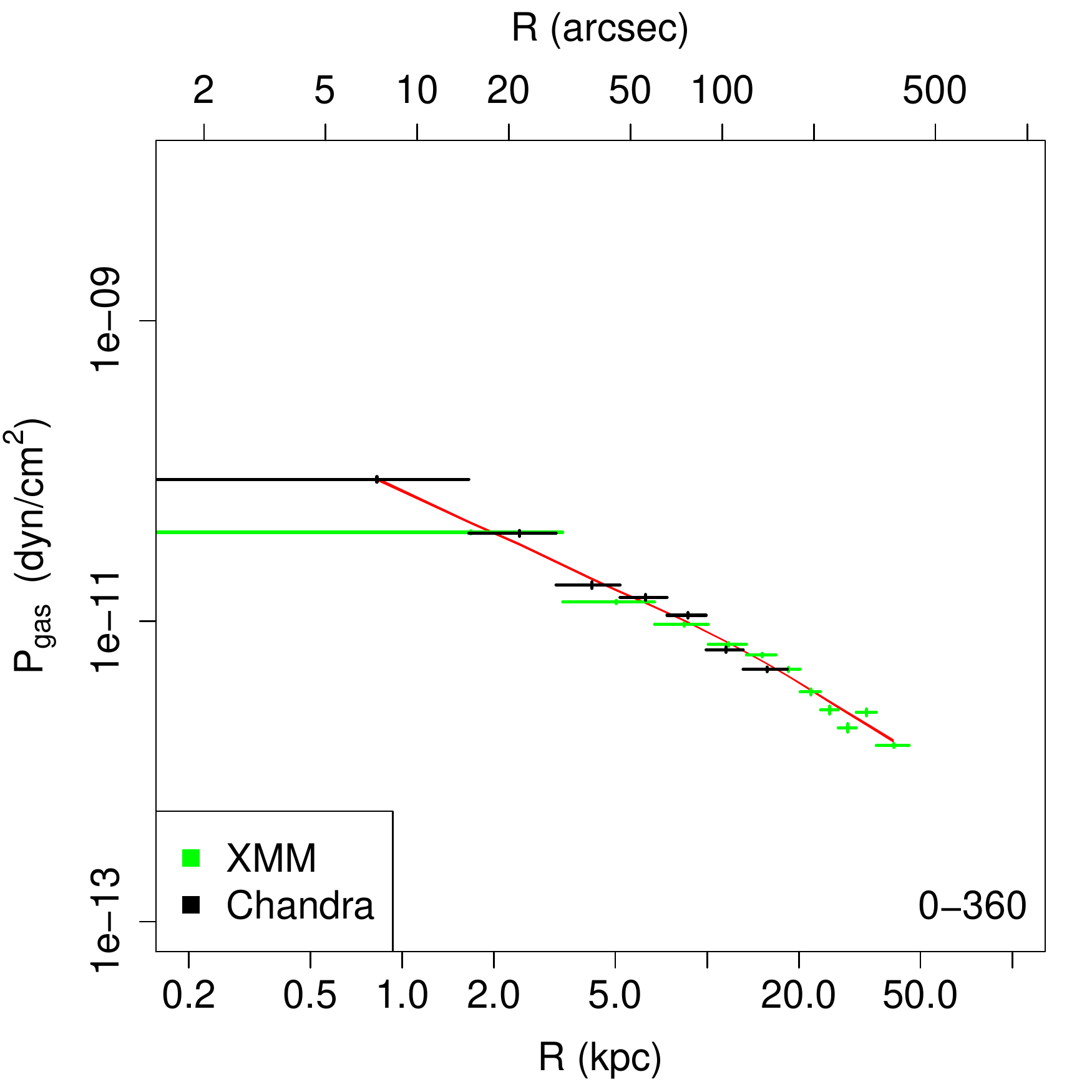}
\includegraphics[scale=0.19]{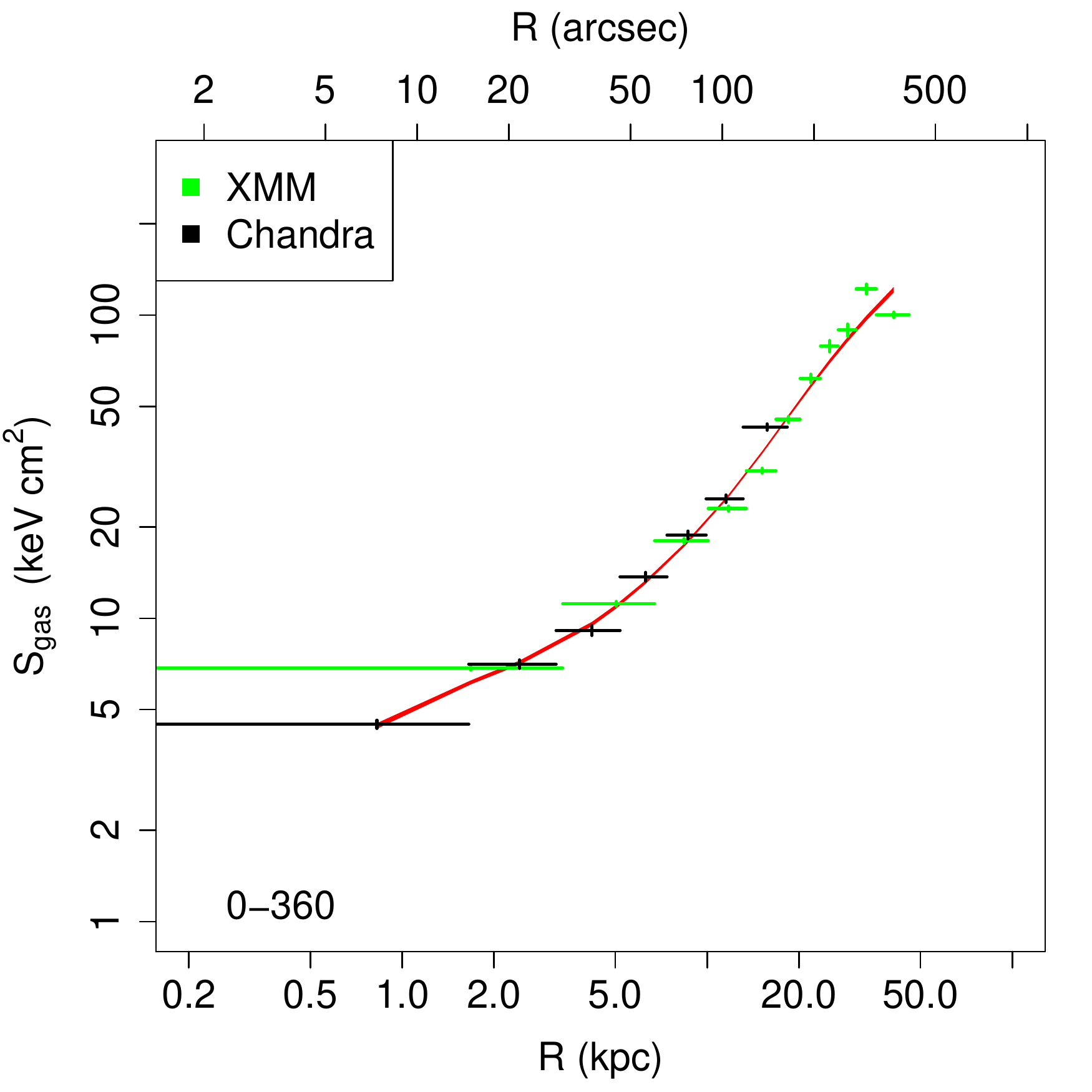}
\includegraphics[scale=0.19]{{N5846_mass_profile_comparison_0_360_0_0_0.6}.pdf}\\
\includegraphics[scale=0.19]{{N5846_temp_profile_merged_120_180_0_0_fit_0.6}.pdf}
\includegraphics[scale=0.19]{{N5846_nh_profile_merged_120_180_0_0_fit_0.6}.pdf}
\includegraphics[scale=0.19]{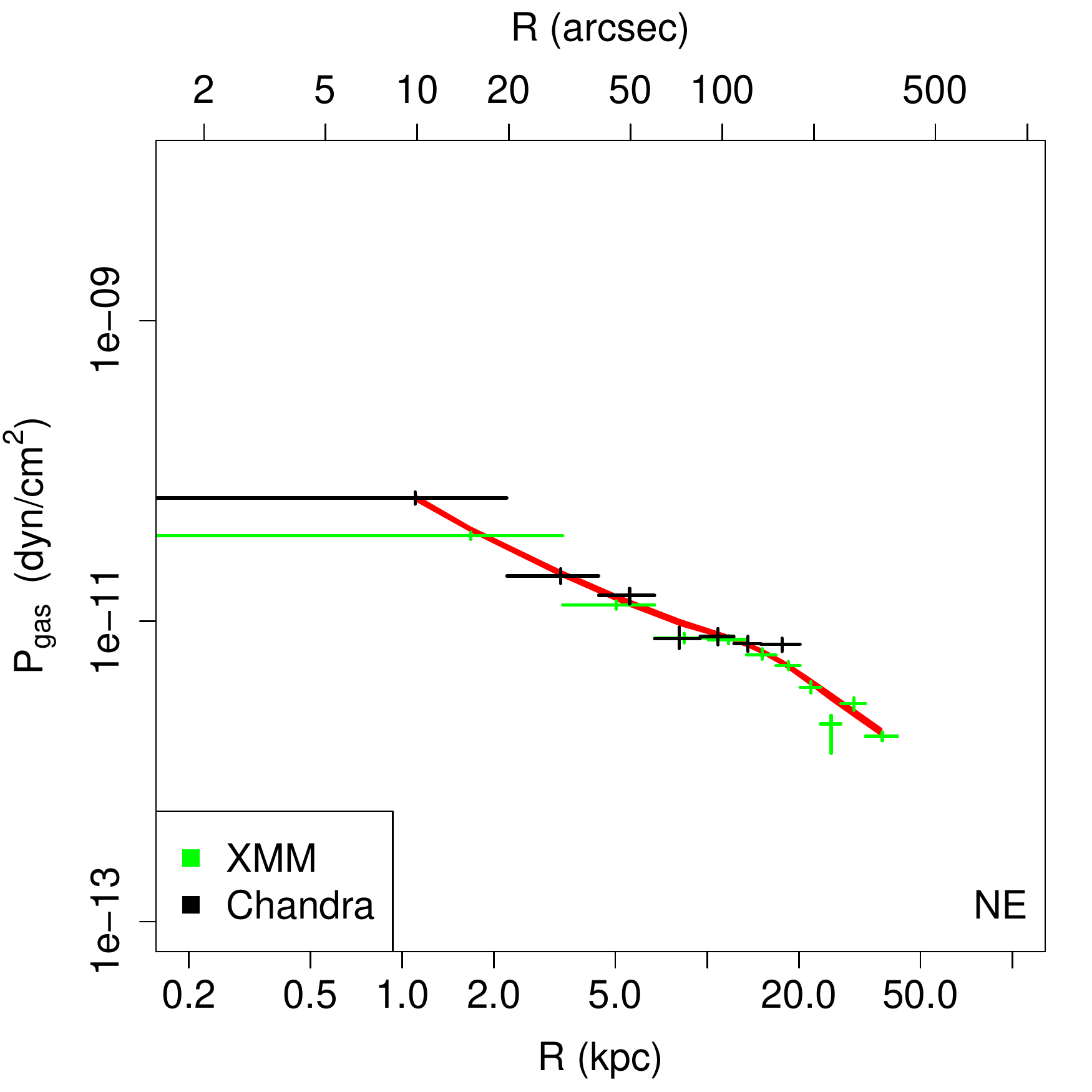}
\includegraphics[scale=0.19]{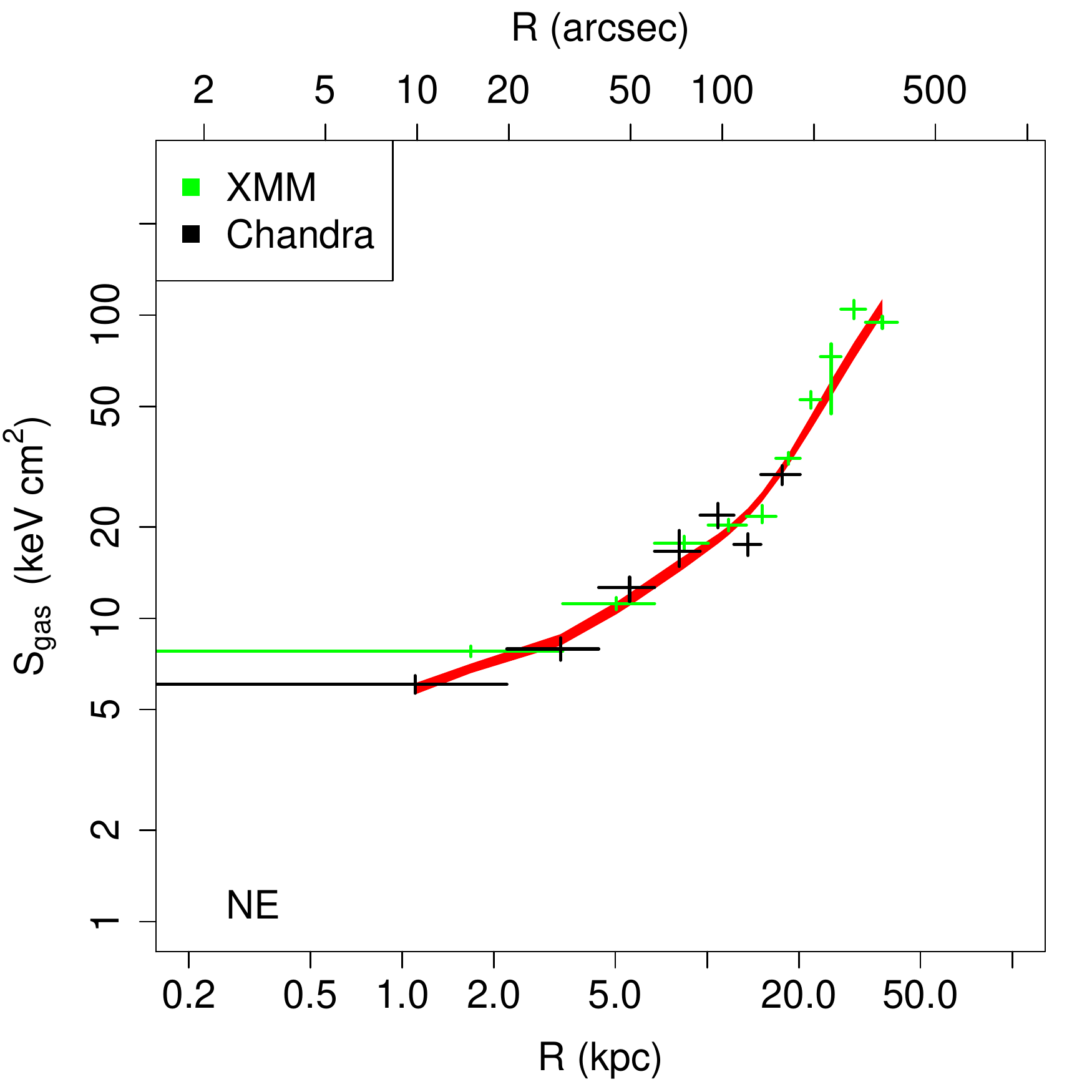}
\includegraphics[scale=0.19]{{N5846_mass_profile_comparison_120_180_0_0_0.6}.pdf}\\
\includegraphics[scale=0.19]{{N5846_temp_profile_merged_180_270_0_0_fit_0.7}.pdf}
\includegraphics[scale=0.19]{{N5846_nh_profile_merged_180_270_0_0_fit_0.7}.pdf}
\includegraphics[scale=0.19]{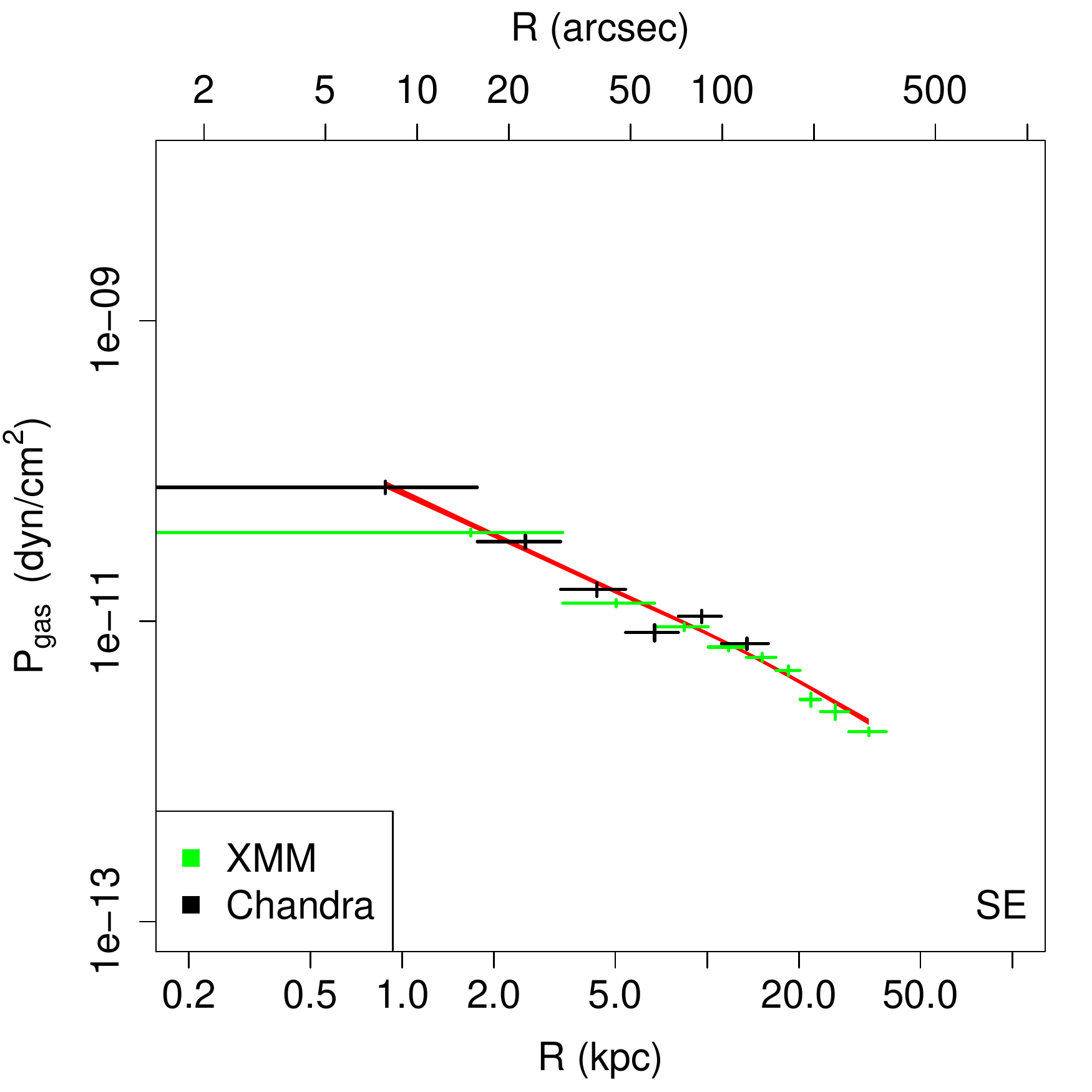}
\includegraphics[scale=0.19]{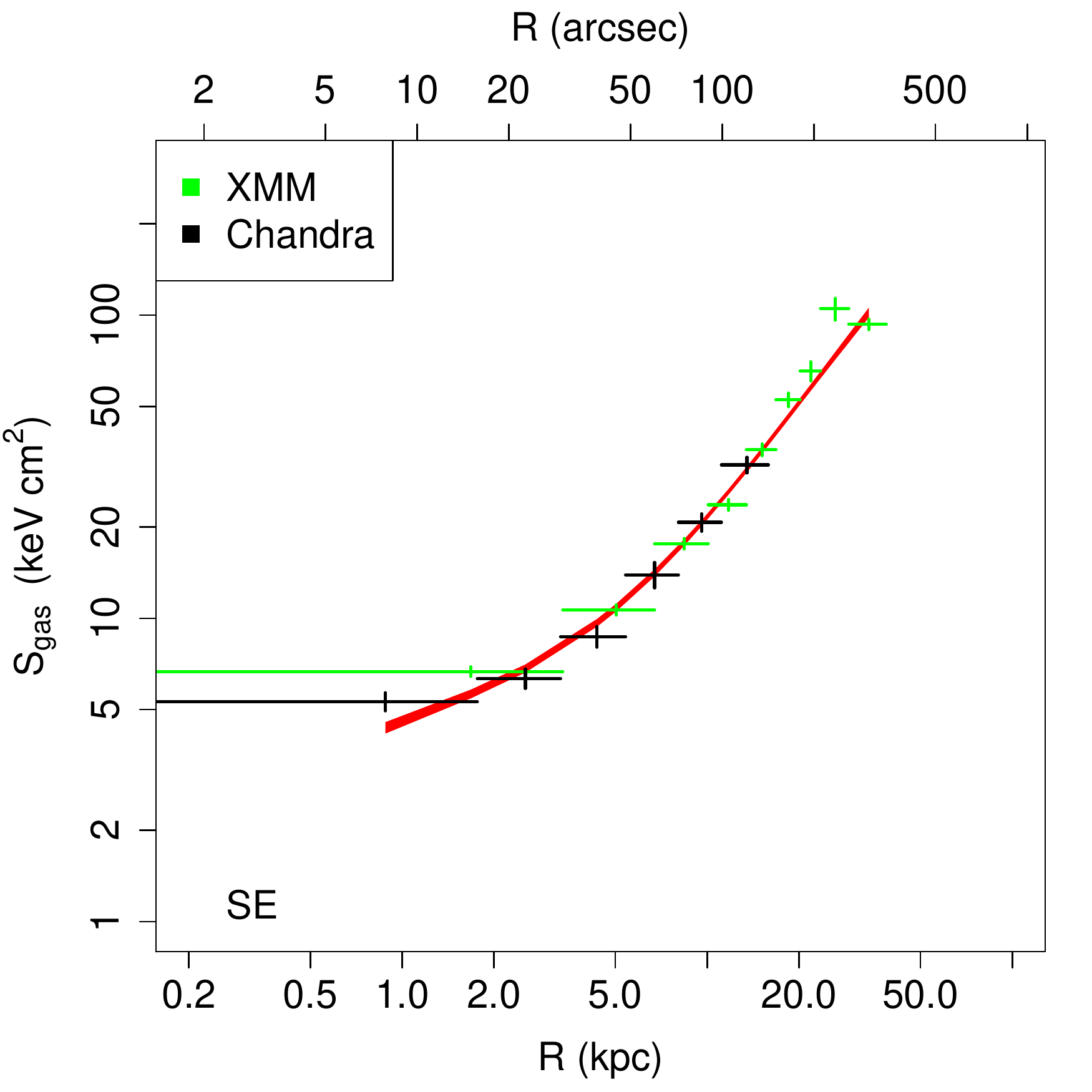}
\includegraphics[scale=0.19]{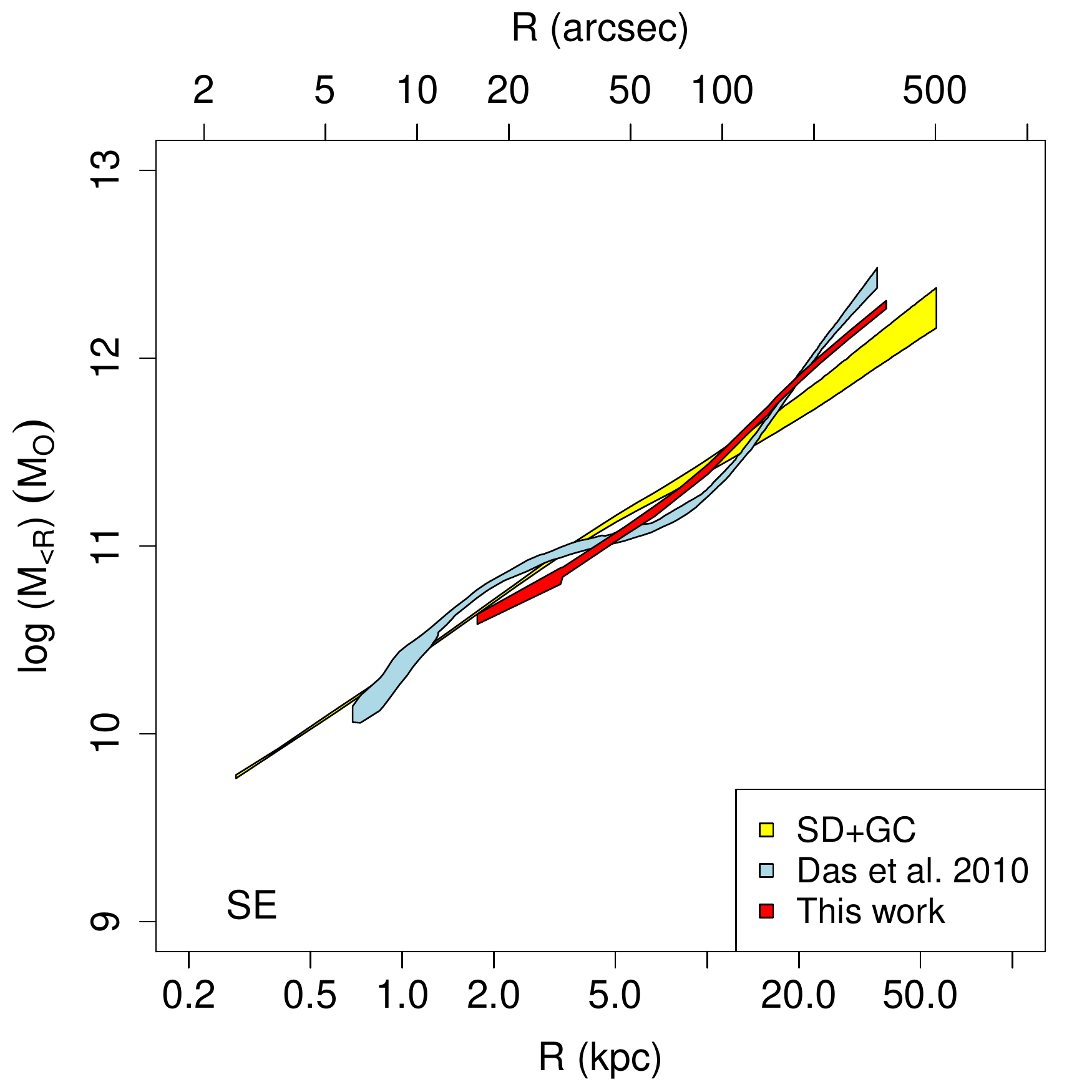}\\
\includegraphics[scale=0.19]{{N5846_temp_profile_merged_270_340_0_0_fit_0.7}.pdf}
\includegraphics[scale=0.19]{{N5846_nh_profile_merged_270_340_0_0_fit_0.7}.pdf}
\includegraphics[scale=0.19]{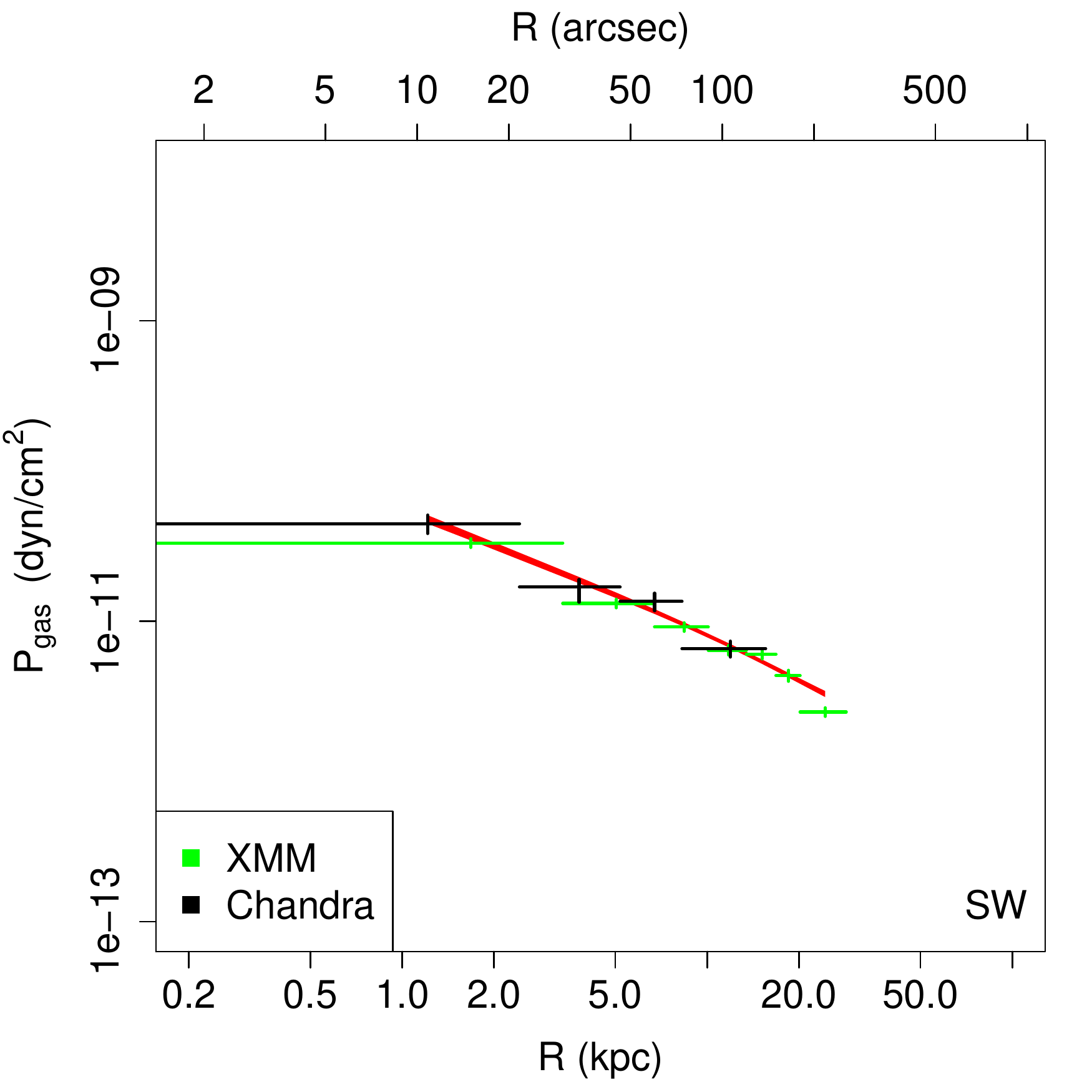}
\includegraphics[scale=0.19]{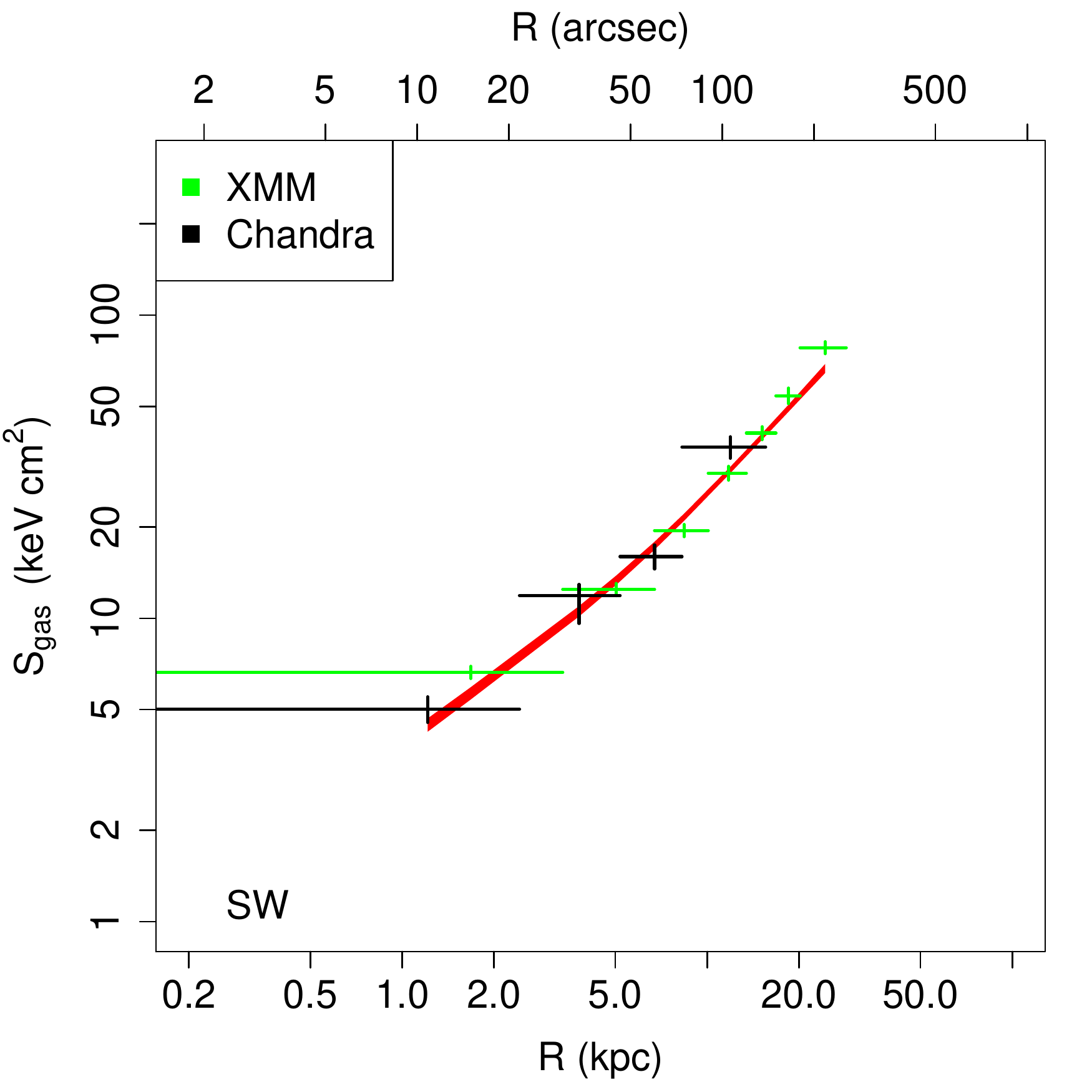}
\includegraphics[scale=0.19]{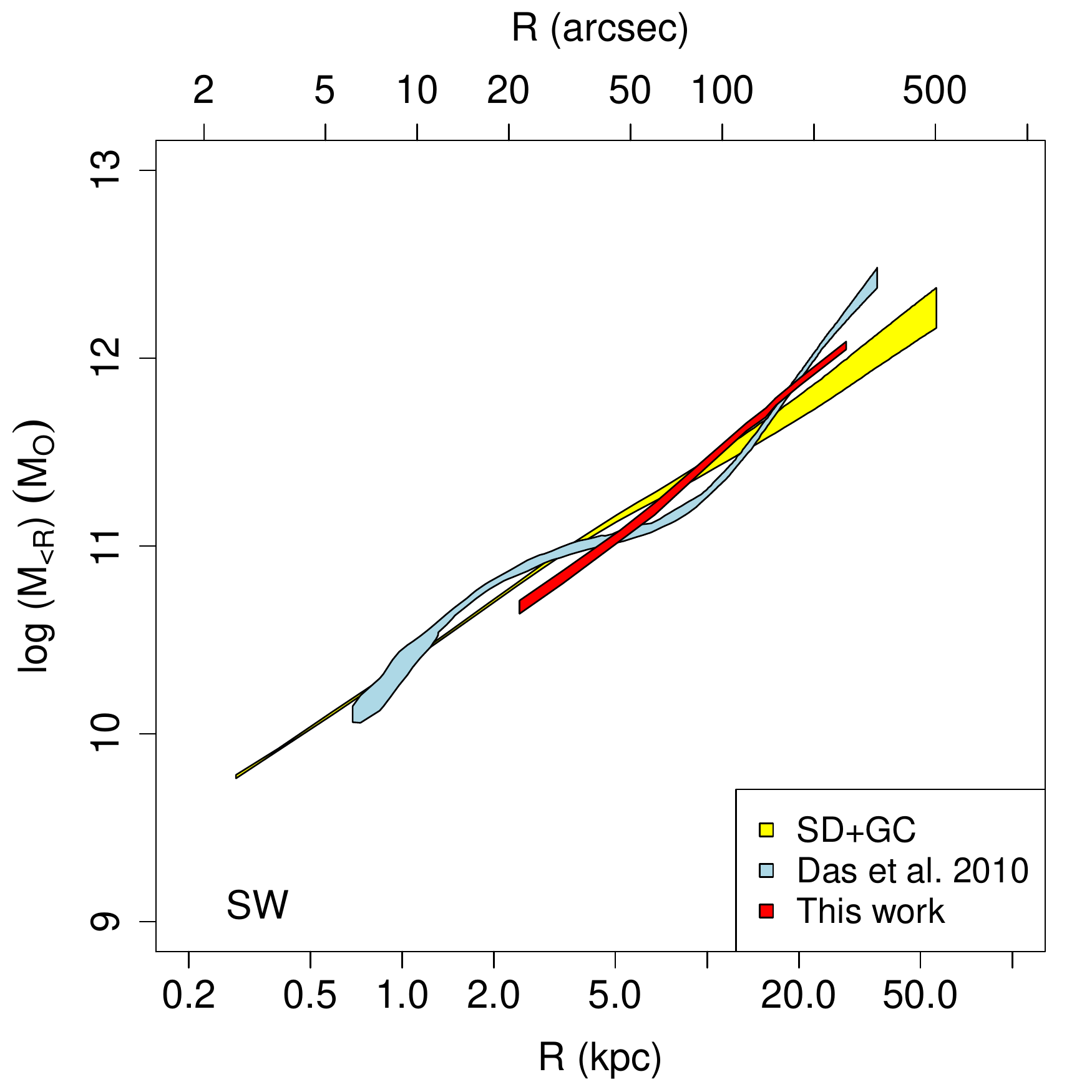}\\
\includegraphics[scale=0.19]{{N5846_temp_profile_merged_340_480_0_0_fit_0.7}.pdf}
\includegraphics[scale=0.19]{{N5846_nh_profile_merged_340_480_0_0_fit_0.7}.pdf}
\includegraphics[scale=0.19]{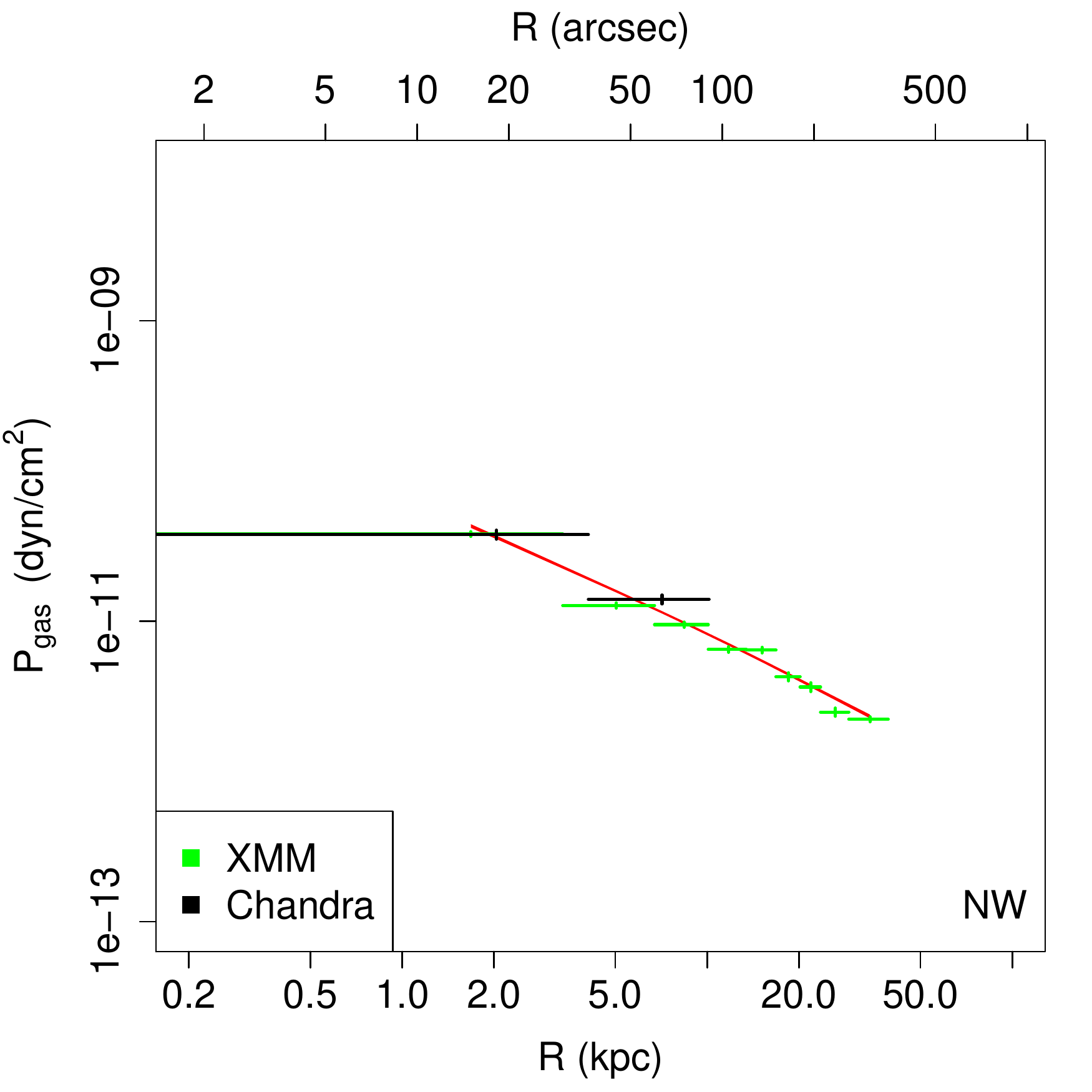}
\includegraphics[scale=0.19]{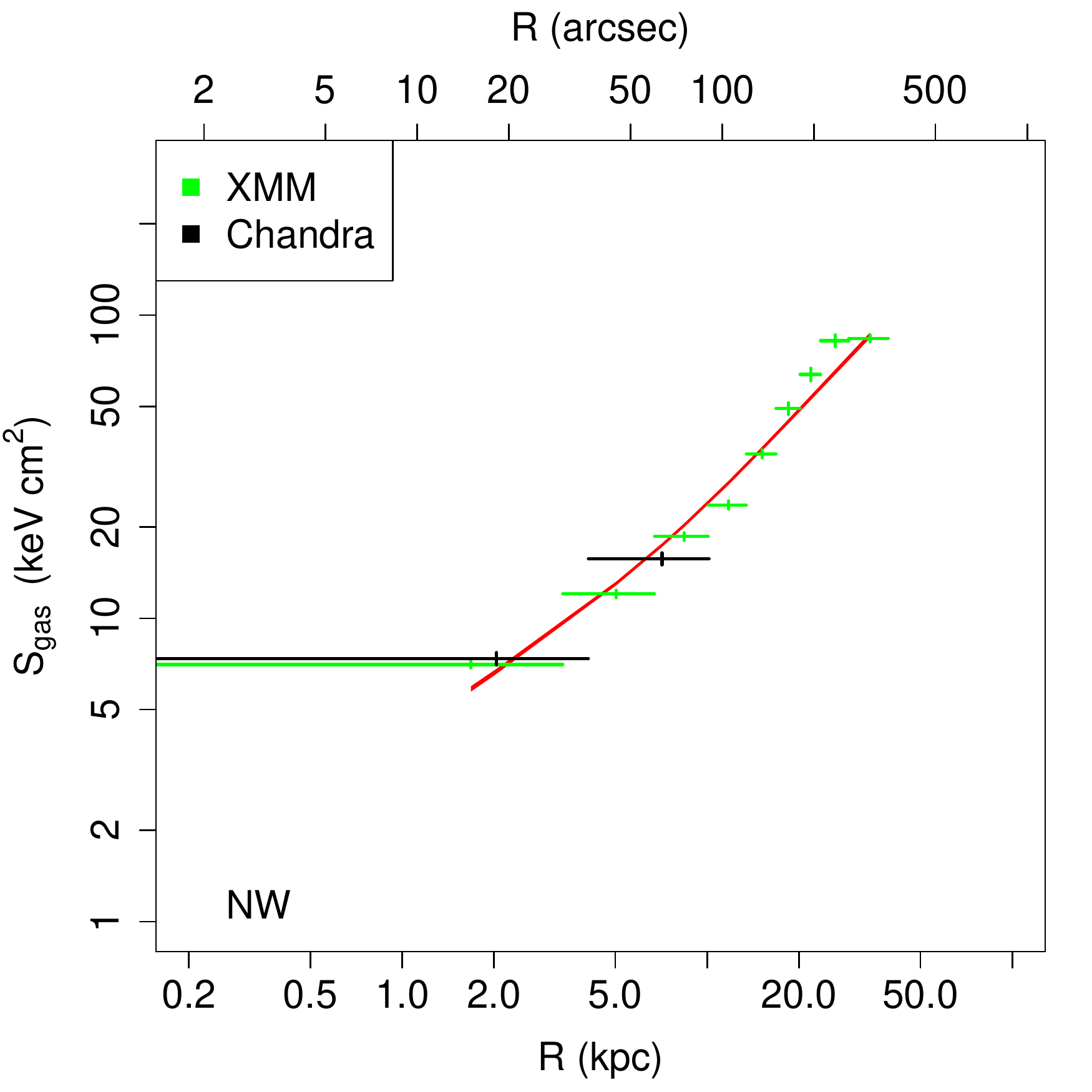}
\includegraphics[scale=0.19]{{N5846_mass_profile_comparison_340_480_0_0_0.7}.pdf}
\caption{Gas profiles obtained in NGC 5846 with the reduction procedure proposed by \citet{2005ApJ...629..172N}. From top to bottom we show the profiles obtained in the full (0-360), NE (30-90), SE (90-180), SW (180-250) and NW (250-30) sector, respectively. In each row we show from left to right the gas temperature, density, pressure and entropy, respectively. Spectra extracted in the annuli are then simultaneously fitted (separately for \textit{XMM} and \textit{Chandra} data) with the fixed abundance model, and de-projected using \textsc{projct} model. The annuli width is chosen to reach a signal to noise ratio of 30 for \textit{XMM}-MOS data (represented in red) and of 50 for \textit{Chandra} ACIS data (represented in black), with the exception of the full (0-360) sector for which we chose a signal to noise ratio of 50 for \textit{XMM} and 100 for \textit{Chandra} data and of the SW (250-30) sector for which we use a signal to noise ratio of 50 for \textit{XMM} and 50 for \textit{Chandra} data. Best fits of a smooth cubic spline are presented in red, with smoothing parameter from top to bottom of 0.6, 0.6, 0.7, 0.7 and 0.7. In the rightmost panels in red we present the total mass profiles obtained by mean of Eq. \ref{eq:hee} from the best fits to gas temperature and density profiles. In the same panels we show in yellow the optical mass profile obtained from SD and GC reported by \citep{2014MNRAS.439..659N}, and in light blue the X-ray mass profiles obtained by \citet{2010MNRAS.409.1362D}.}\label{fig:N5846_gas_profiles_merged_app}
\end{figure}

\begin{figure}
\centering
\includegraphics[scale=0.16]{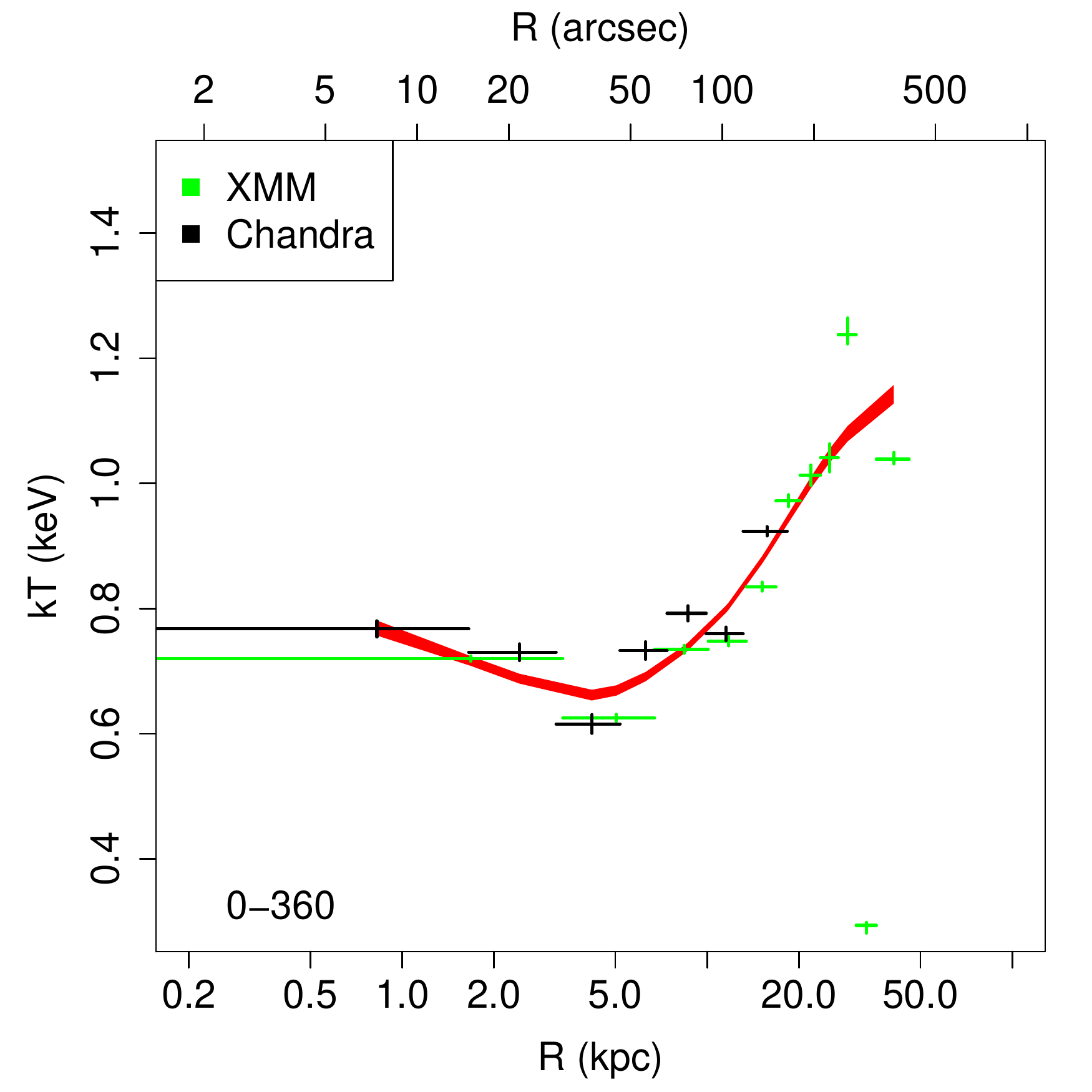}
\includegraphics[scale=0.16]{{N5846_nh_profile_merged_0_360_0_0_fit_0.6_abund}.pdf}
\includegraphics[scale=0.16]{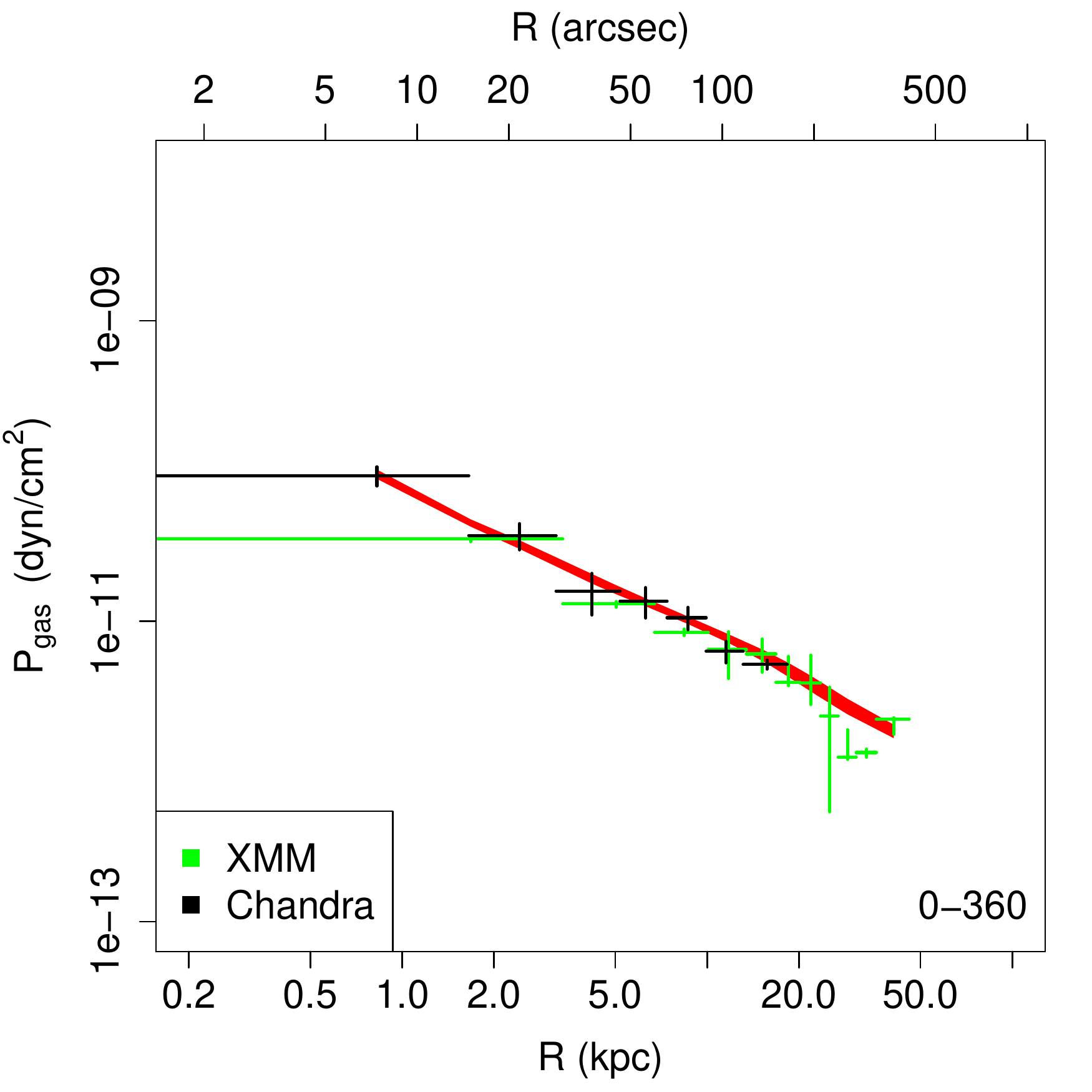}
\includegraphics[scale=0.16]{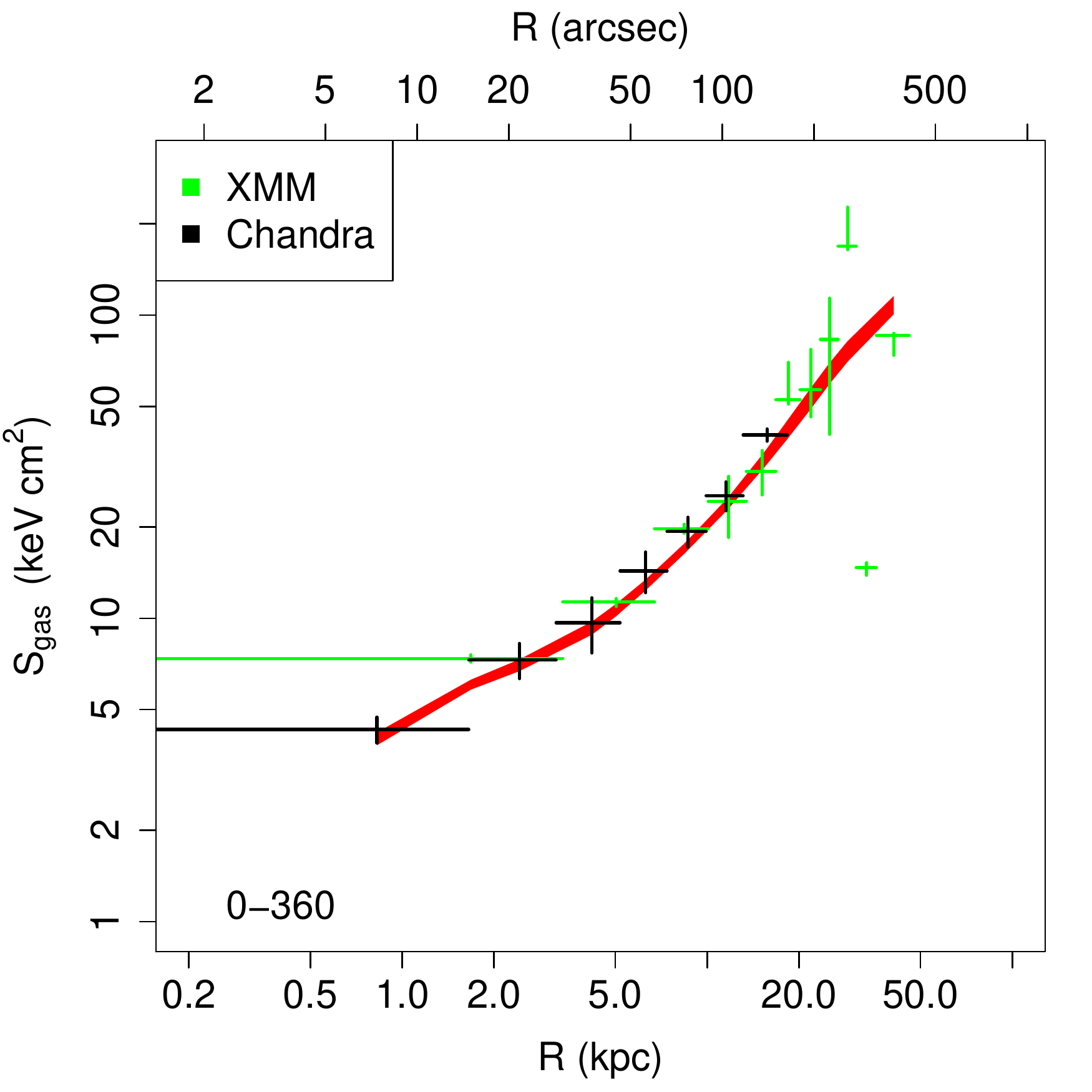}
\includegraphics[scale=0.16]{{N5846_mass_profile_comparison_0_360_0_0_0.6_abund}.pdf}
\includegraphics[scale=0.16]{{N5846_abund_profile_merged_0_360_0_0}.pdf}\\
\includegraphics[scale=0.16]{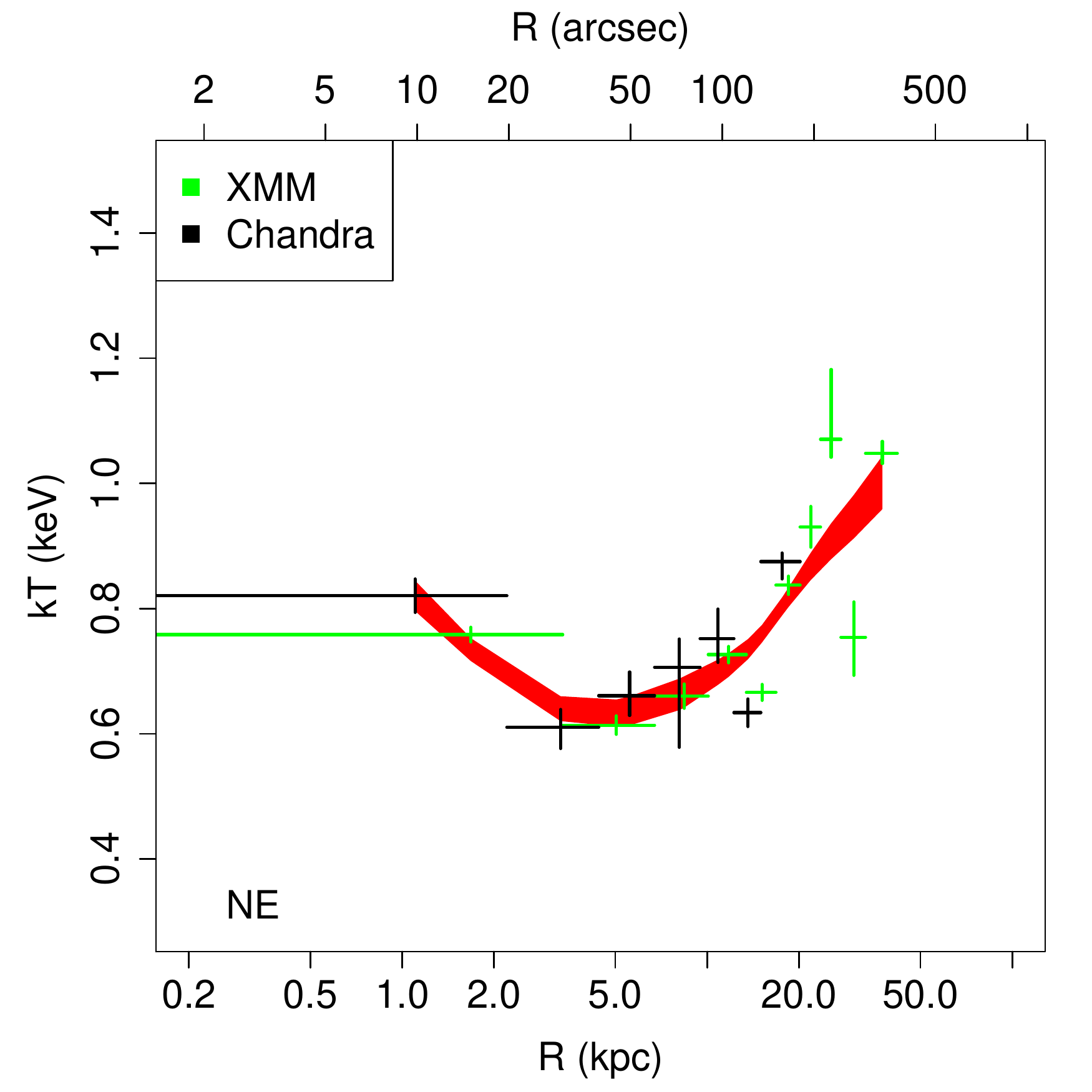}
\includegraphics[scale=0.16]{{N5846_nh_profile_merged_120_180_0_0_fit_0.6_abund}.pdf}
\includegraphics[scale=0.16]{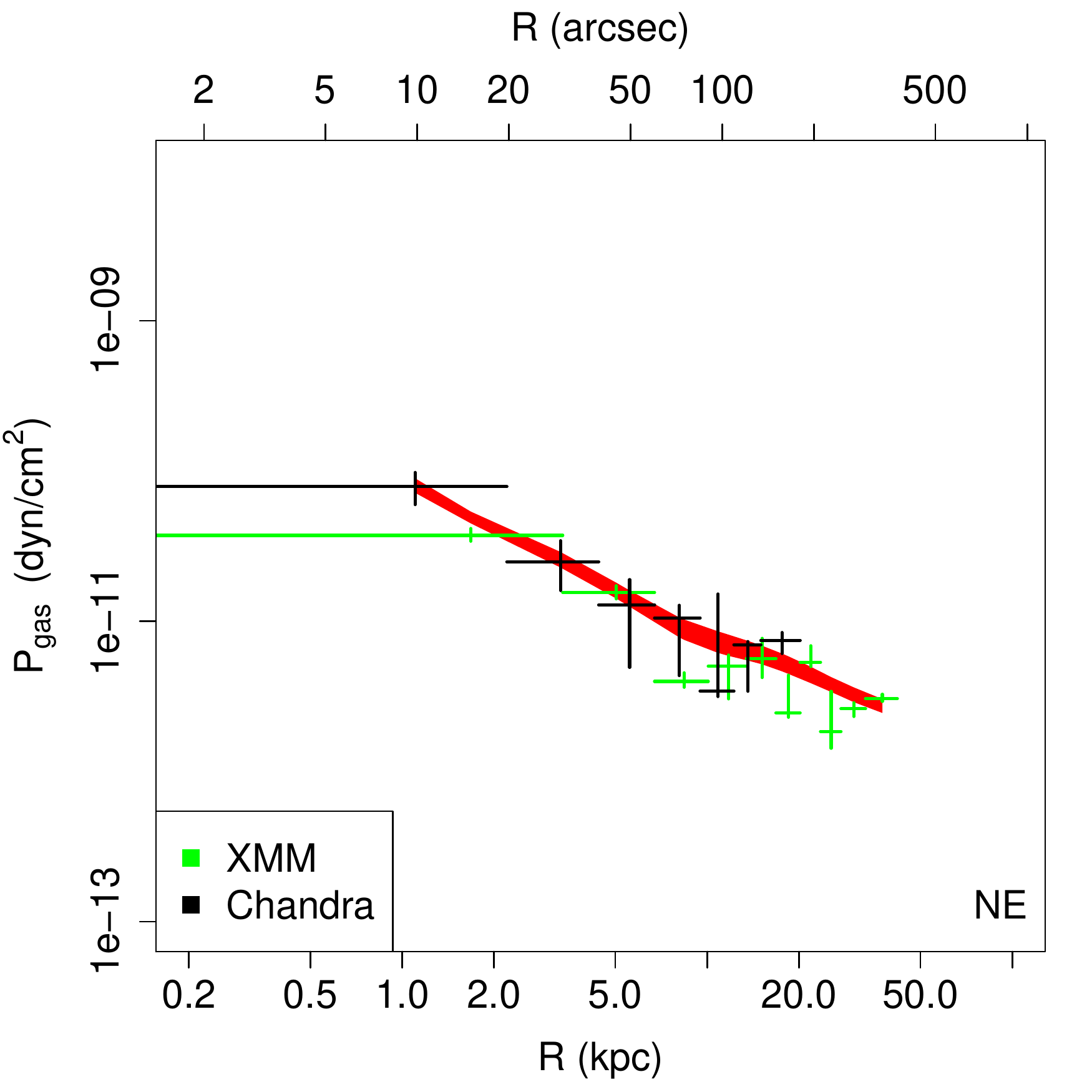}
\includegraphics[scale=0.16]{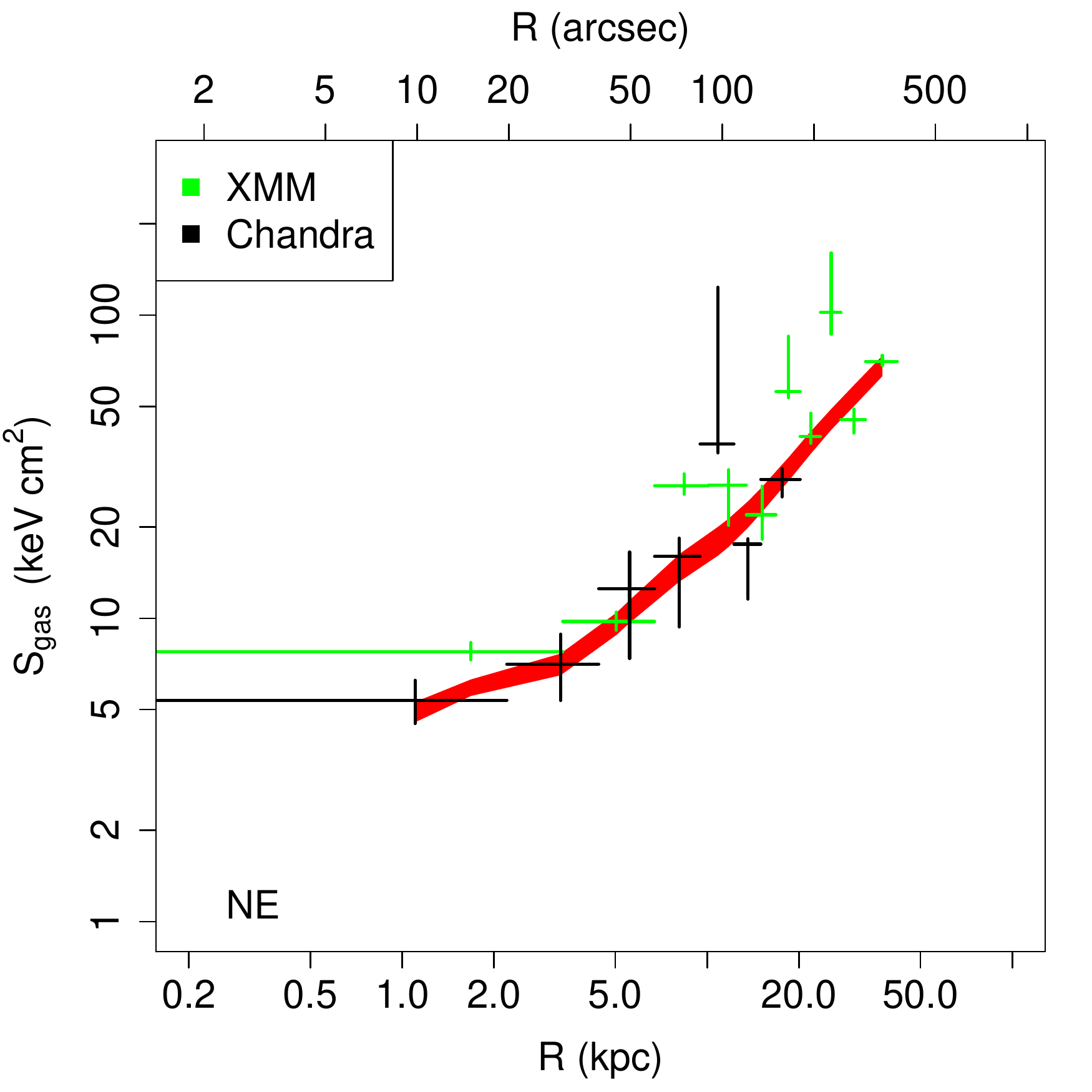}
\includegraphics[scale=0.16]{{N5846_mass_profile_comparison_120_180_0_0_0.6_abund}.pdf}
\includegraphics[scale=0.16]{{N5846_abund_profile_merged_120_180_0_0}.pdf}\\
\includegraphics[scale=0.16]{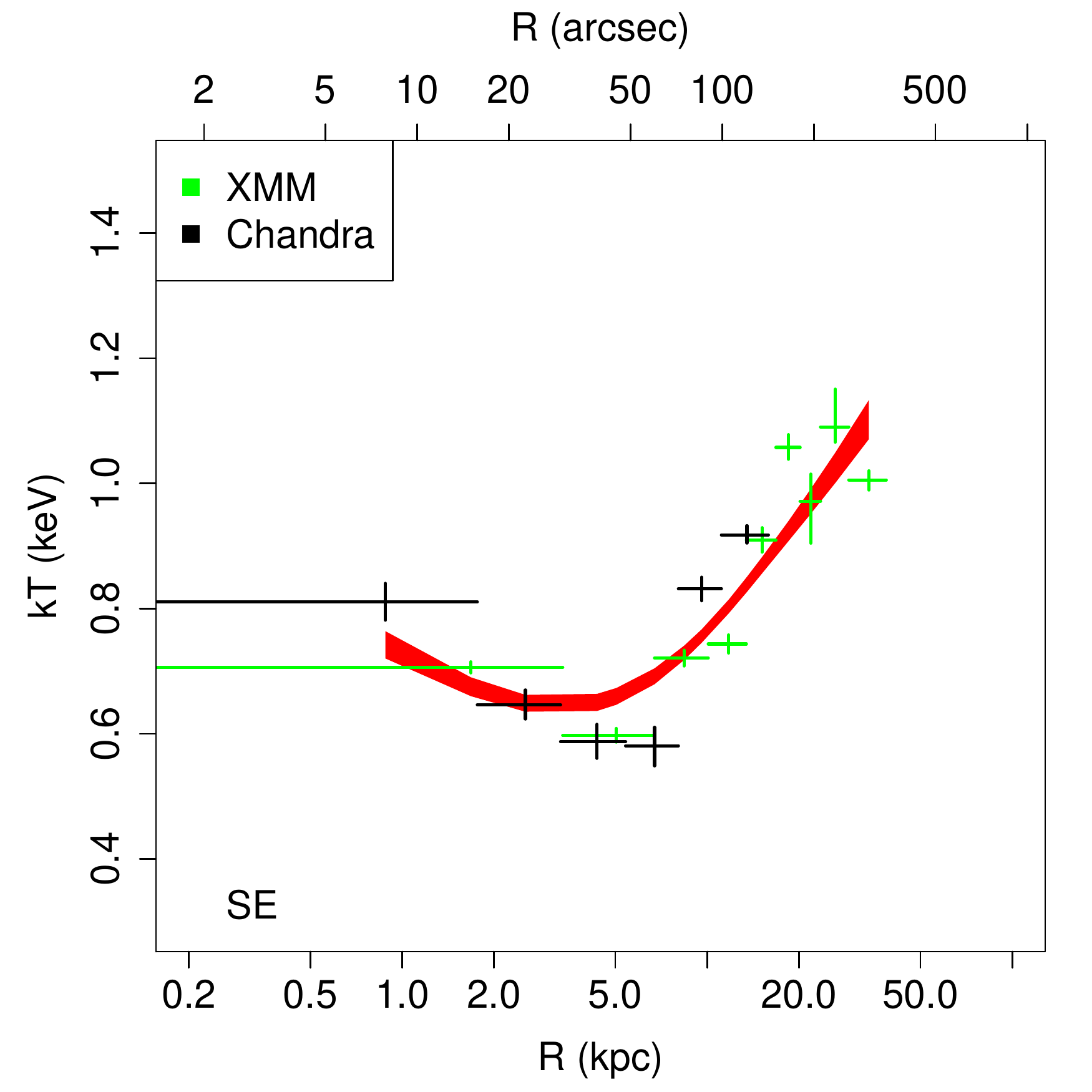}
\includegraphics[scale=0.16]{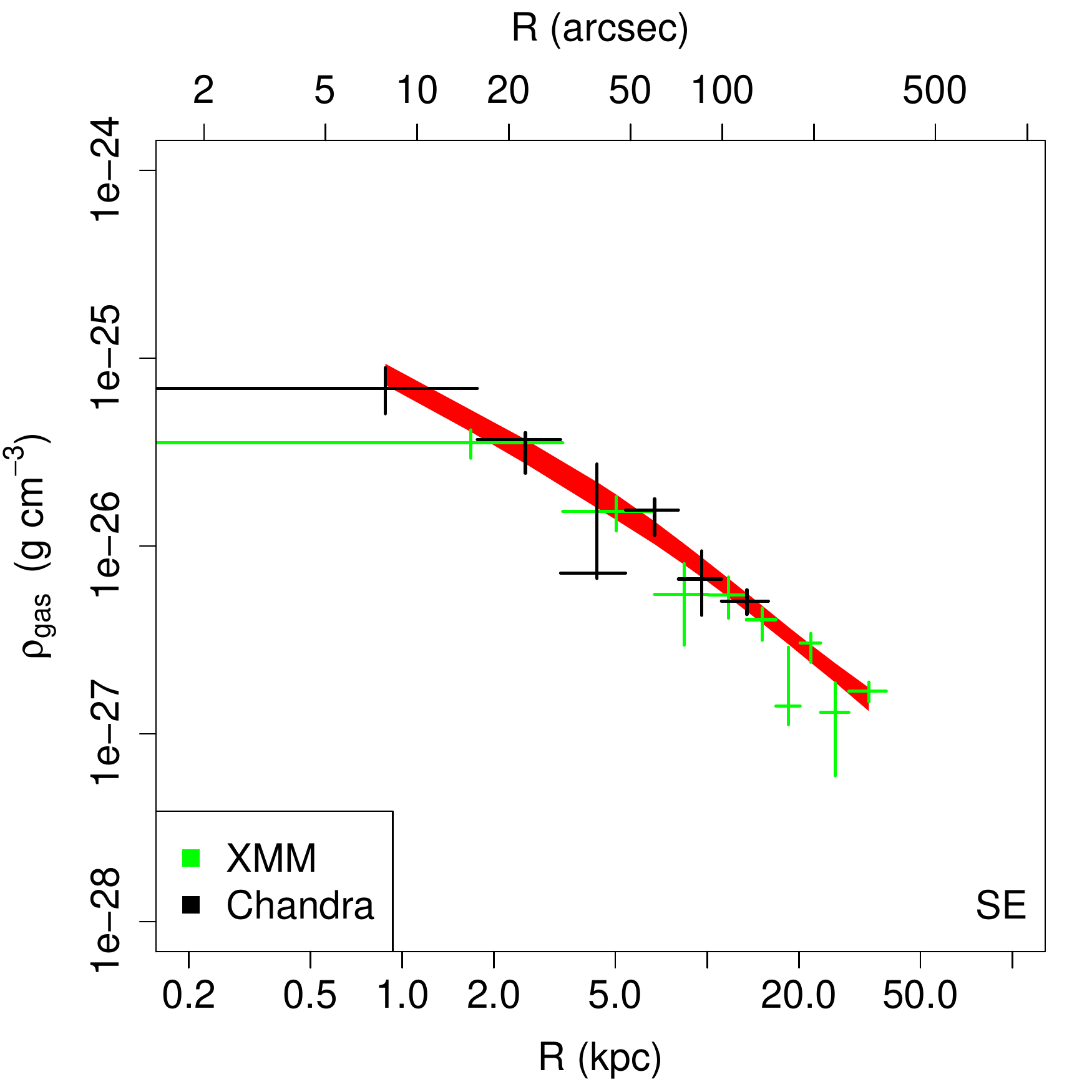}
\includegraphics[scale=0.16]{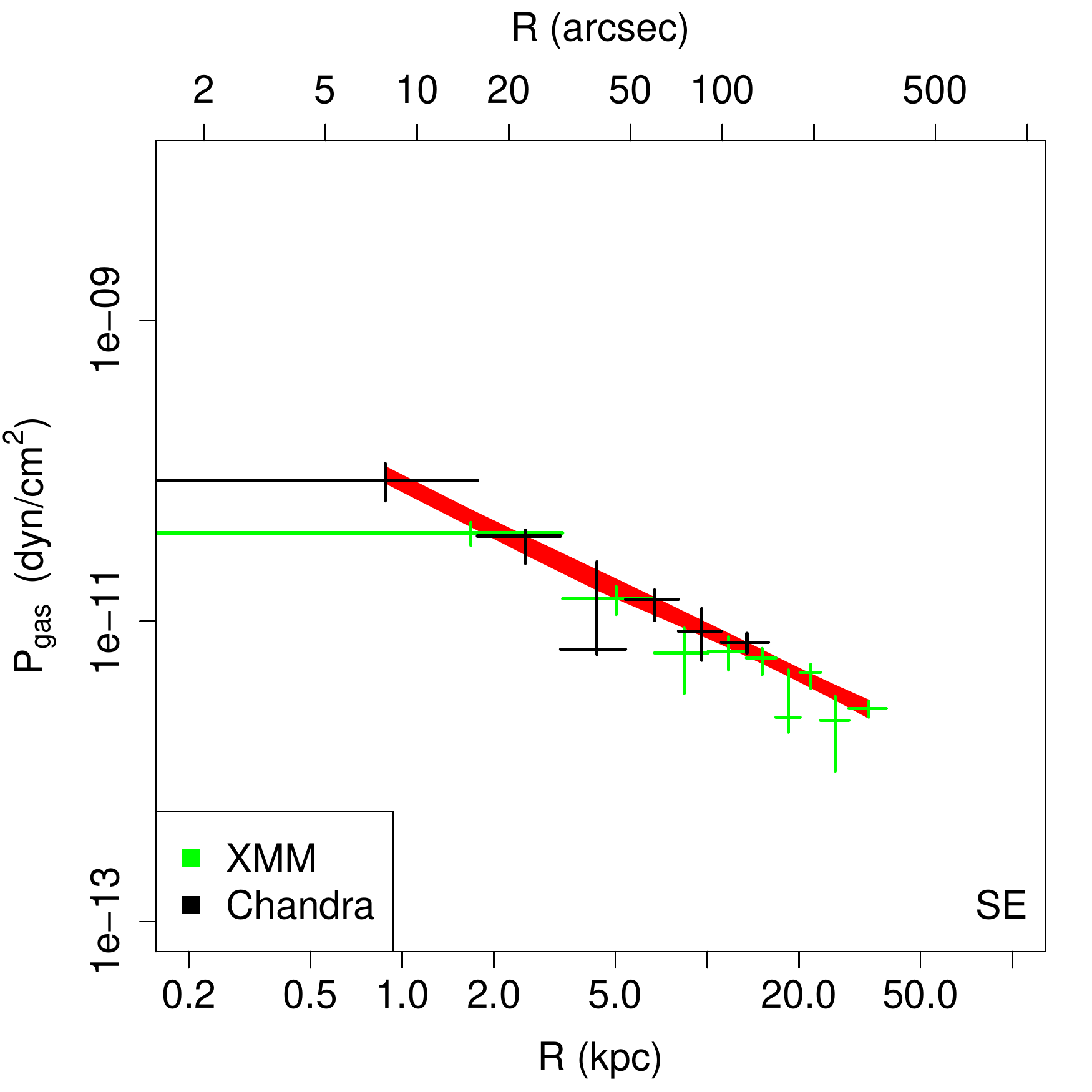}
\includegraphics[scale=0.16]{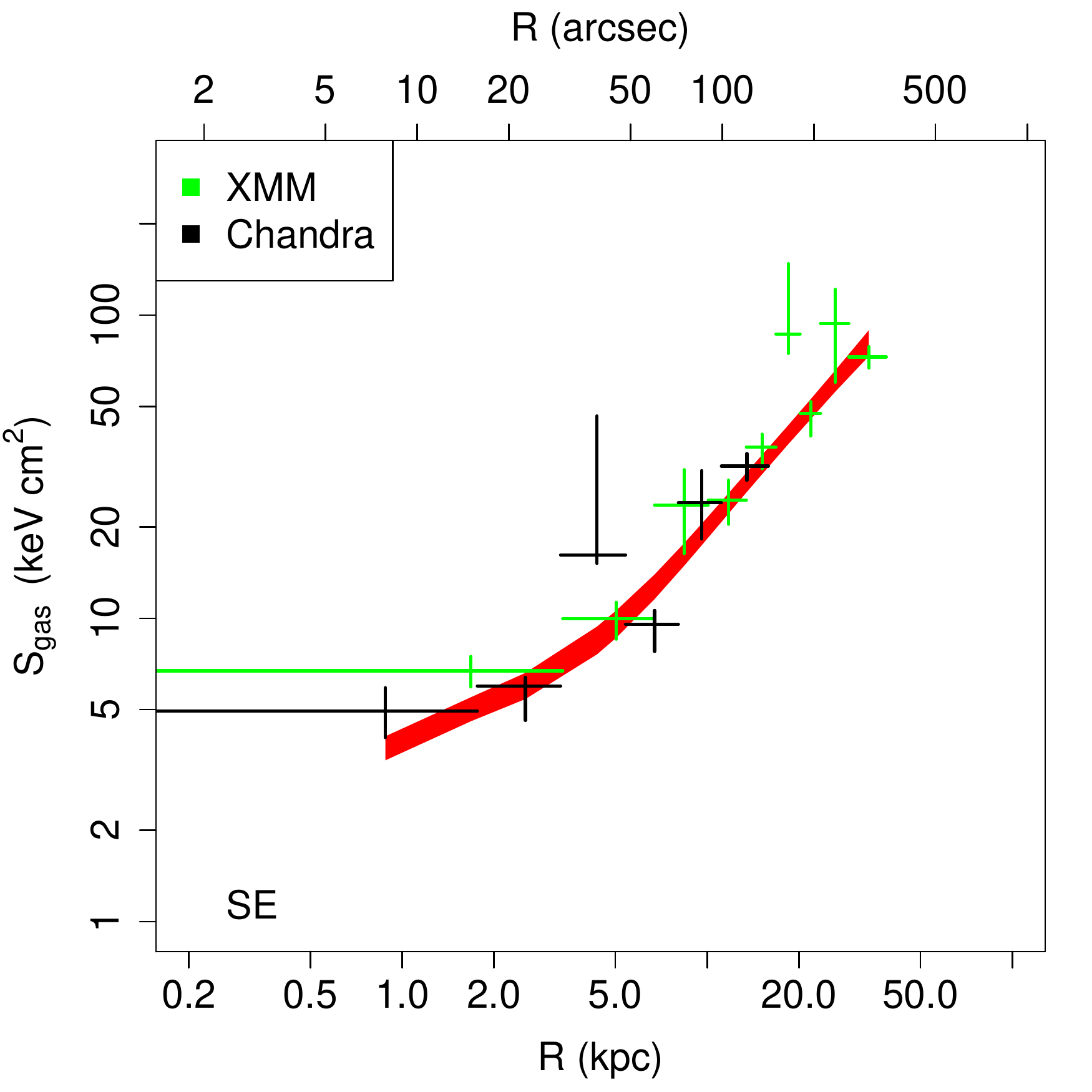}
\includegraphics[scale=0.16]{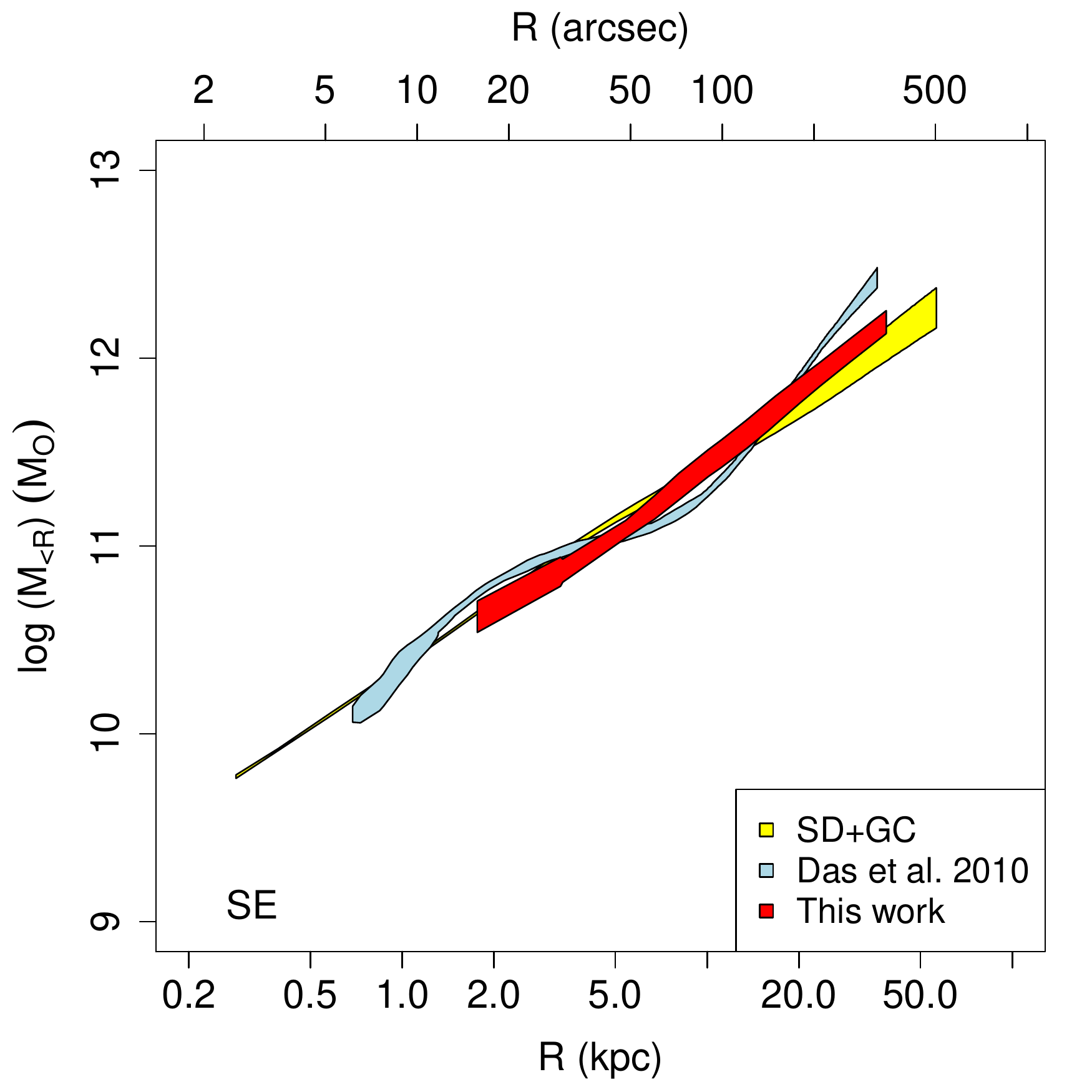}
\includegraphics[scale=0.16]{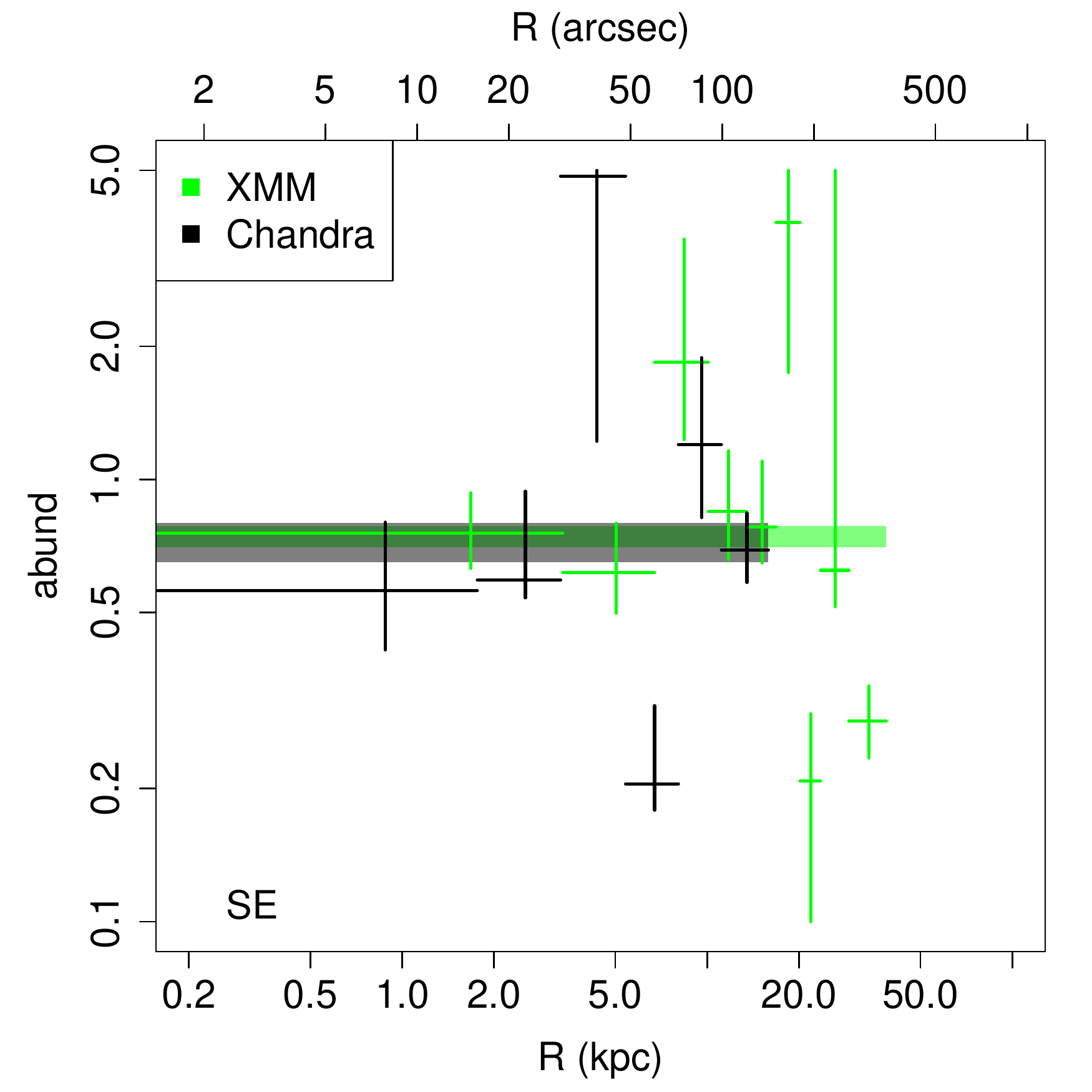}\\
\includegraphics[scale=0.16]{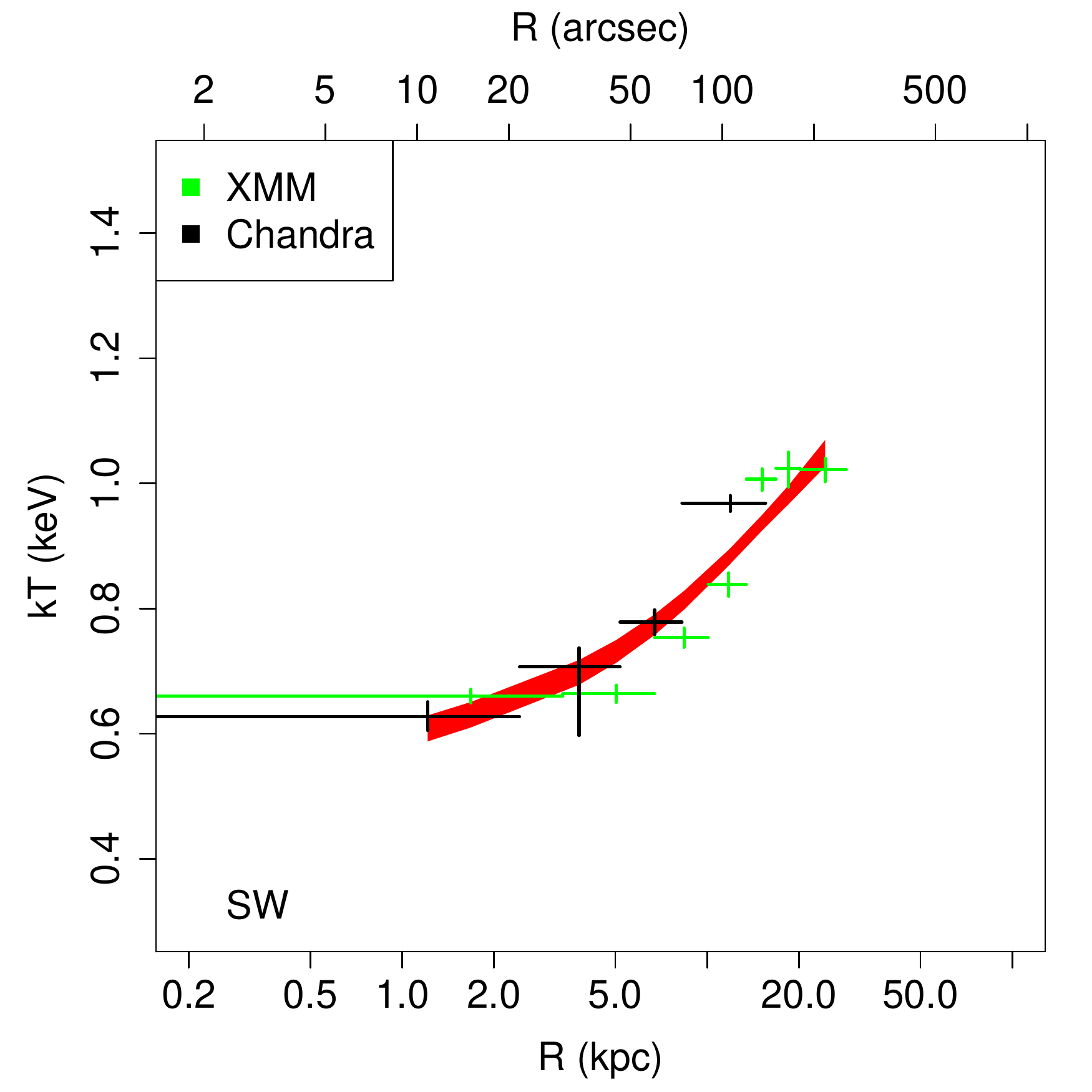}
\includegraphics[scale=0.16]{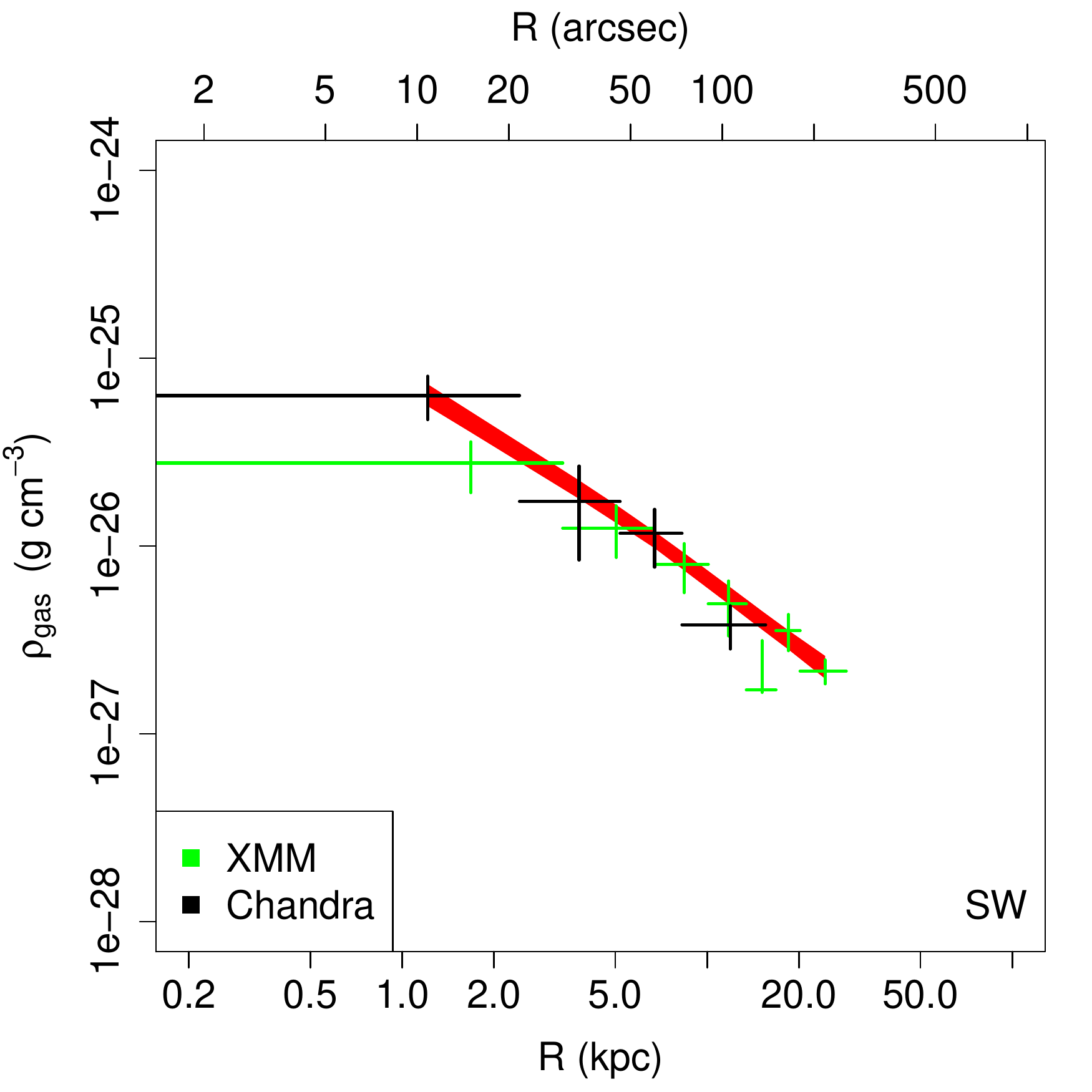}
\includegraphics[scale=0.16]{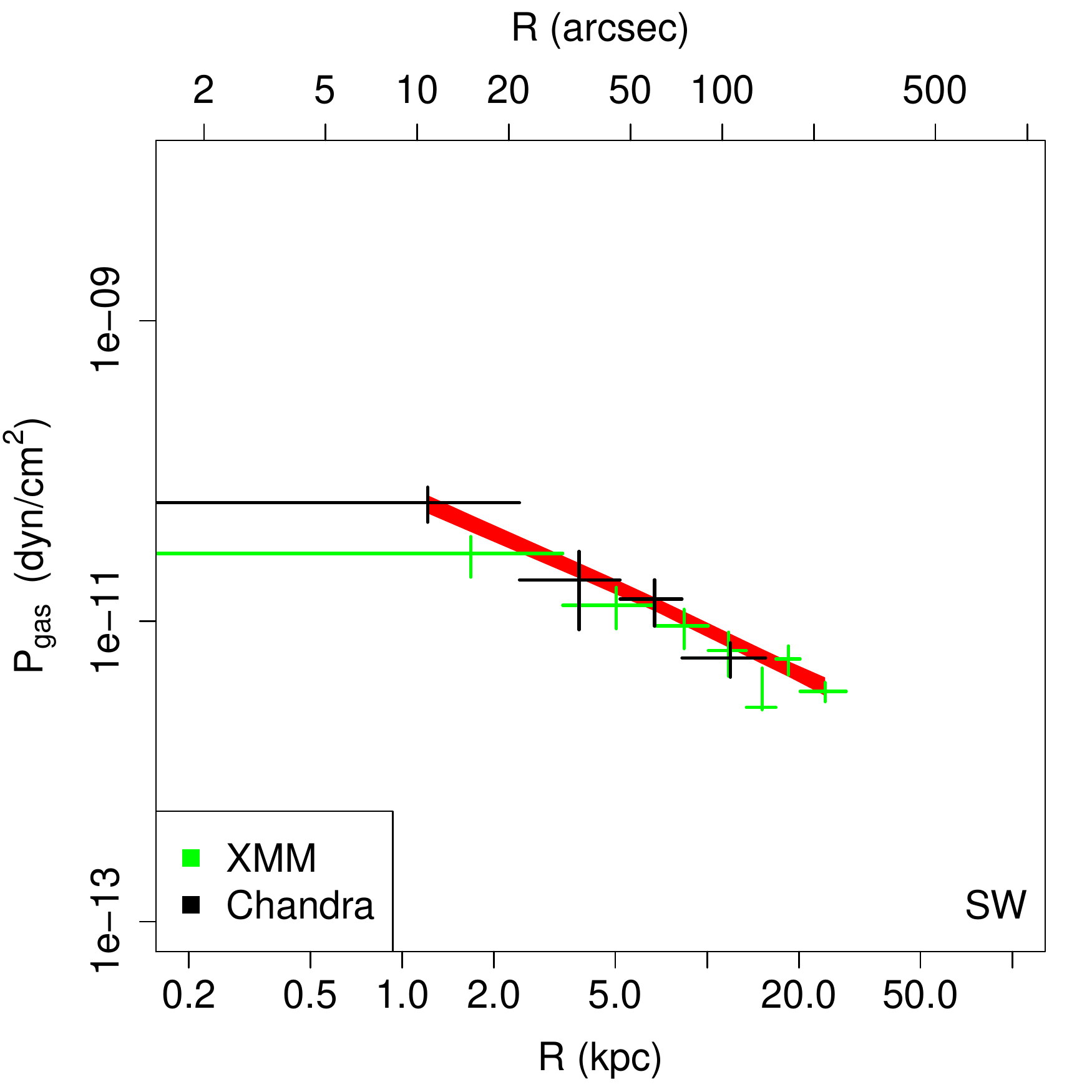}
\includegraphics[scale=0.16]{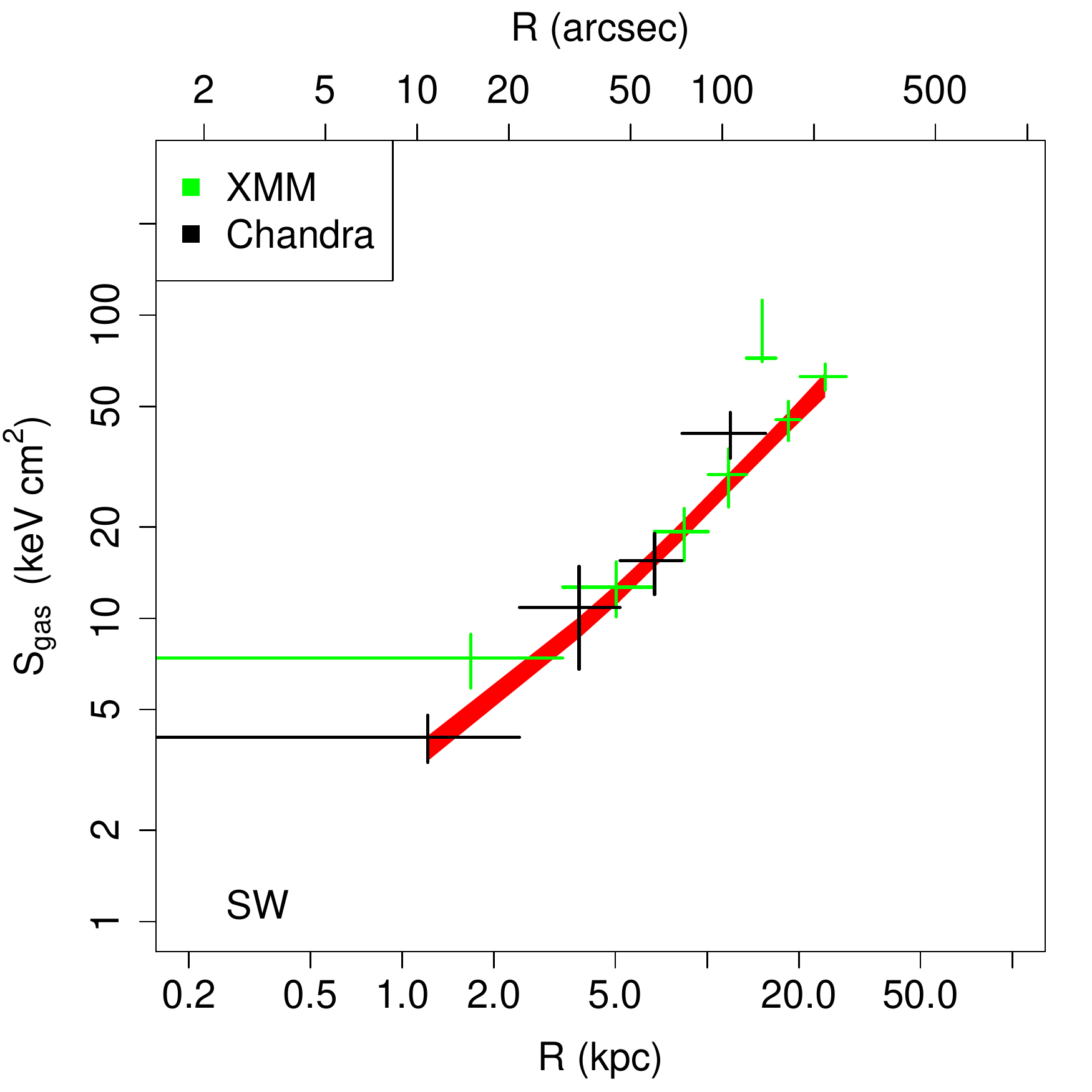}
\includegraphics[scale=0.16]{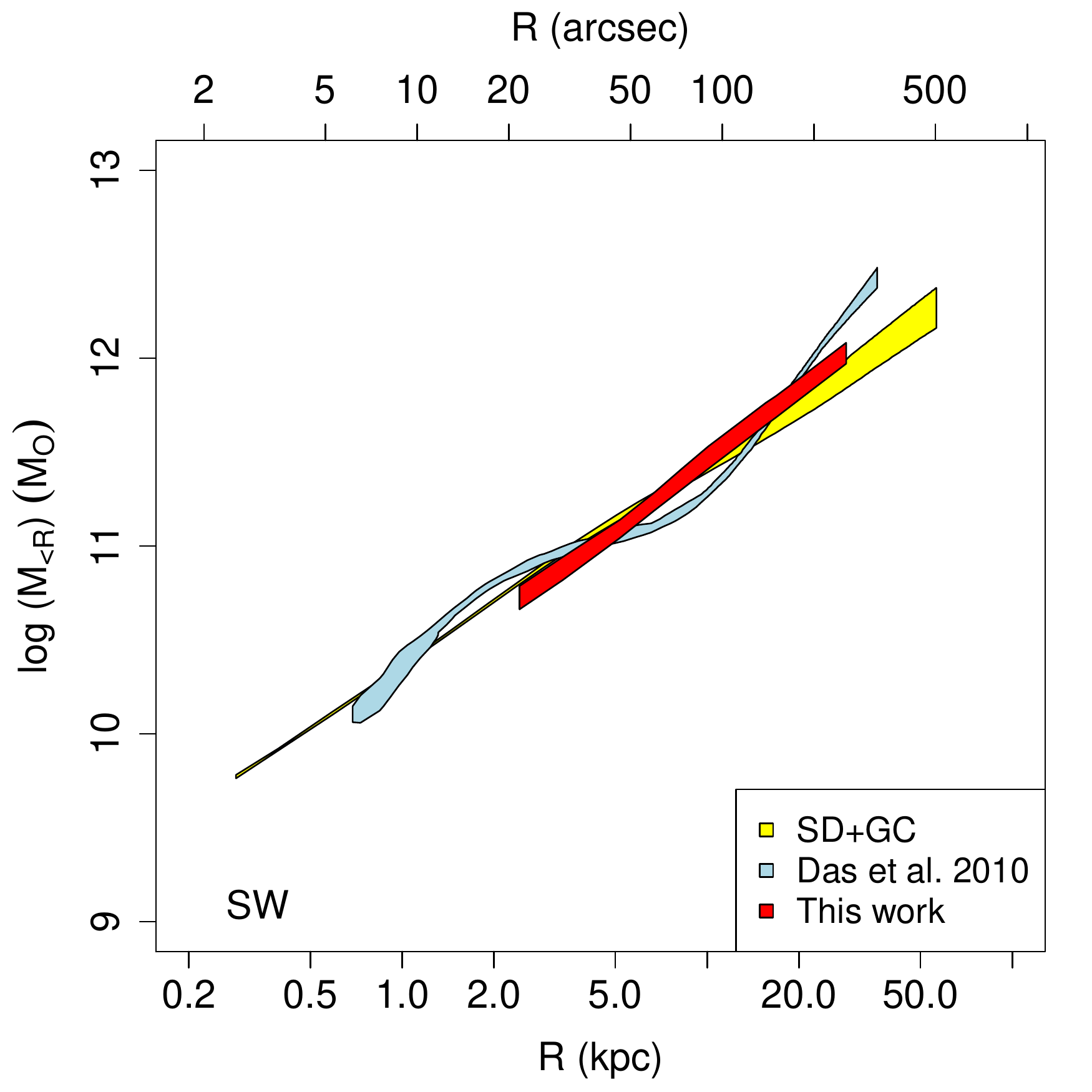}
\includegraphics[scale=0.16]{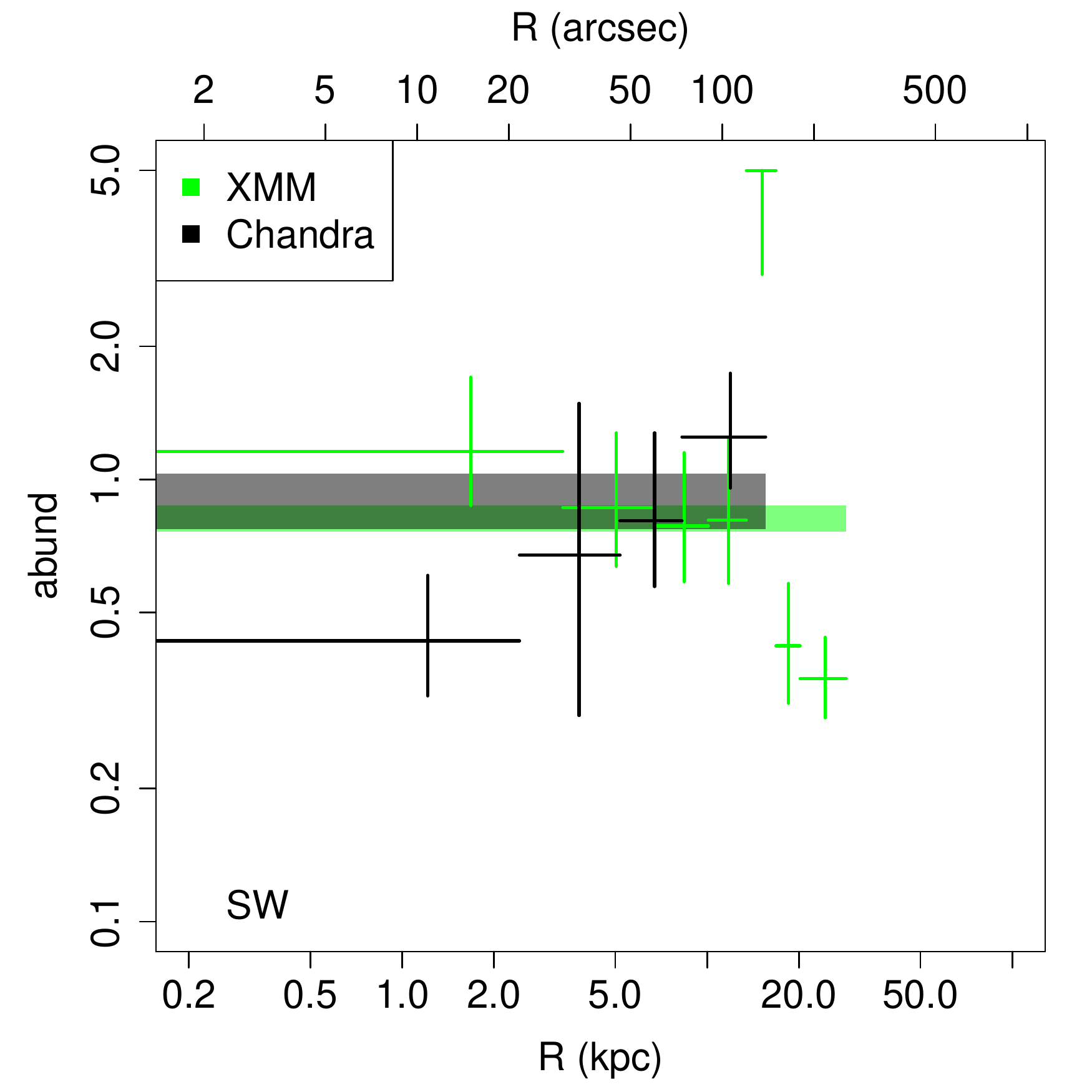}\\
\includegraphics[scale=0.16]{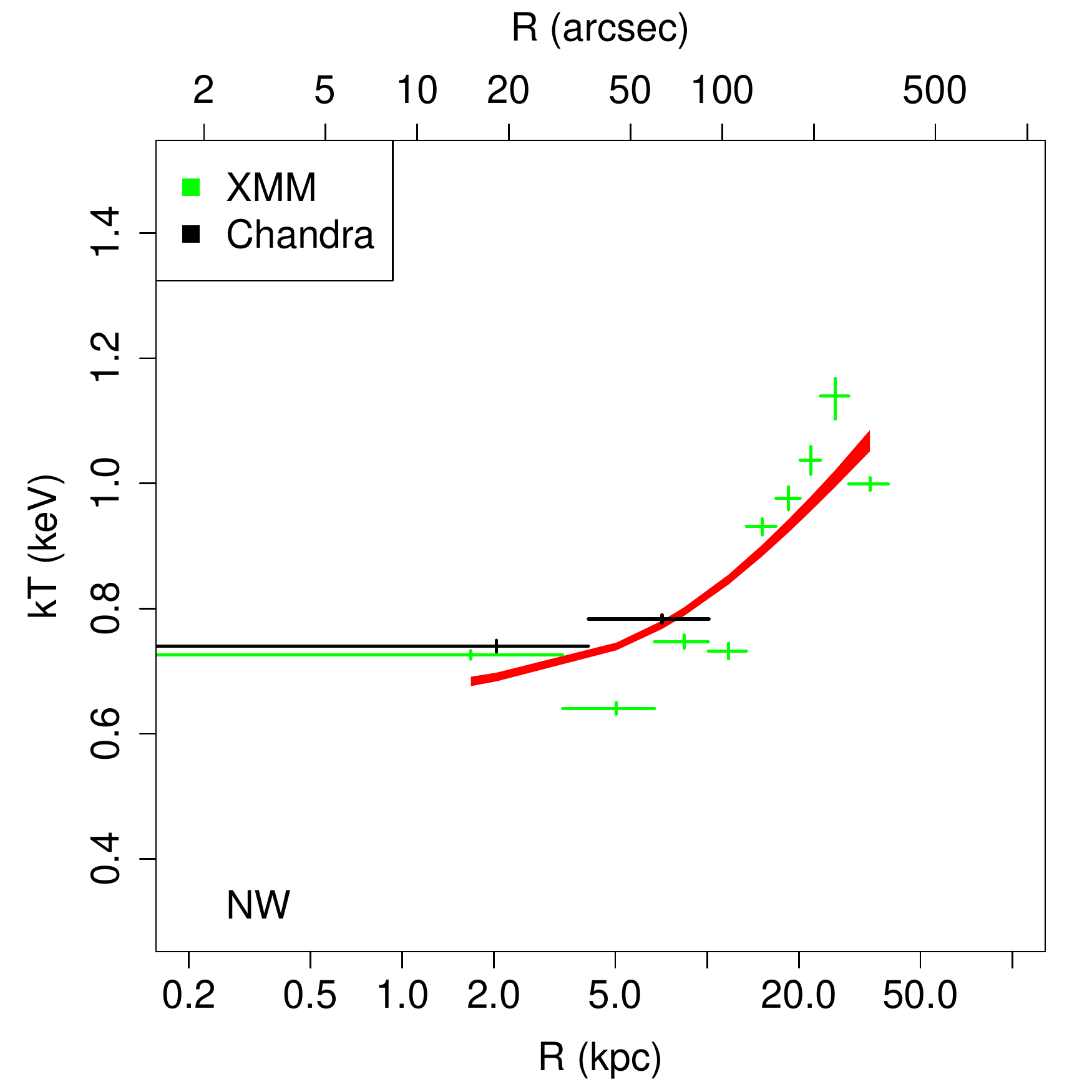}
\includegraphics[scale=0.16]{{N5846_nh_profile_merged_340_480_0_0_fit_0.7_abund}.pdf}
\includegraphics[scale=0.16]{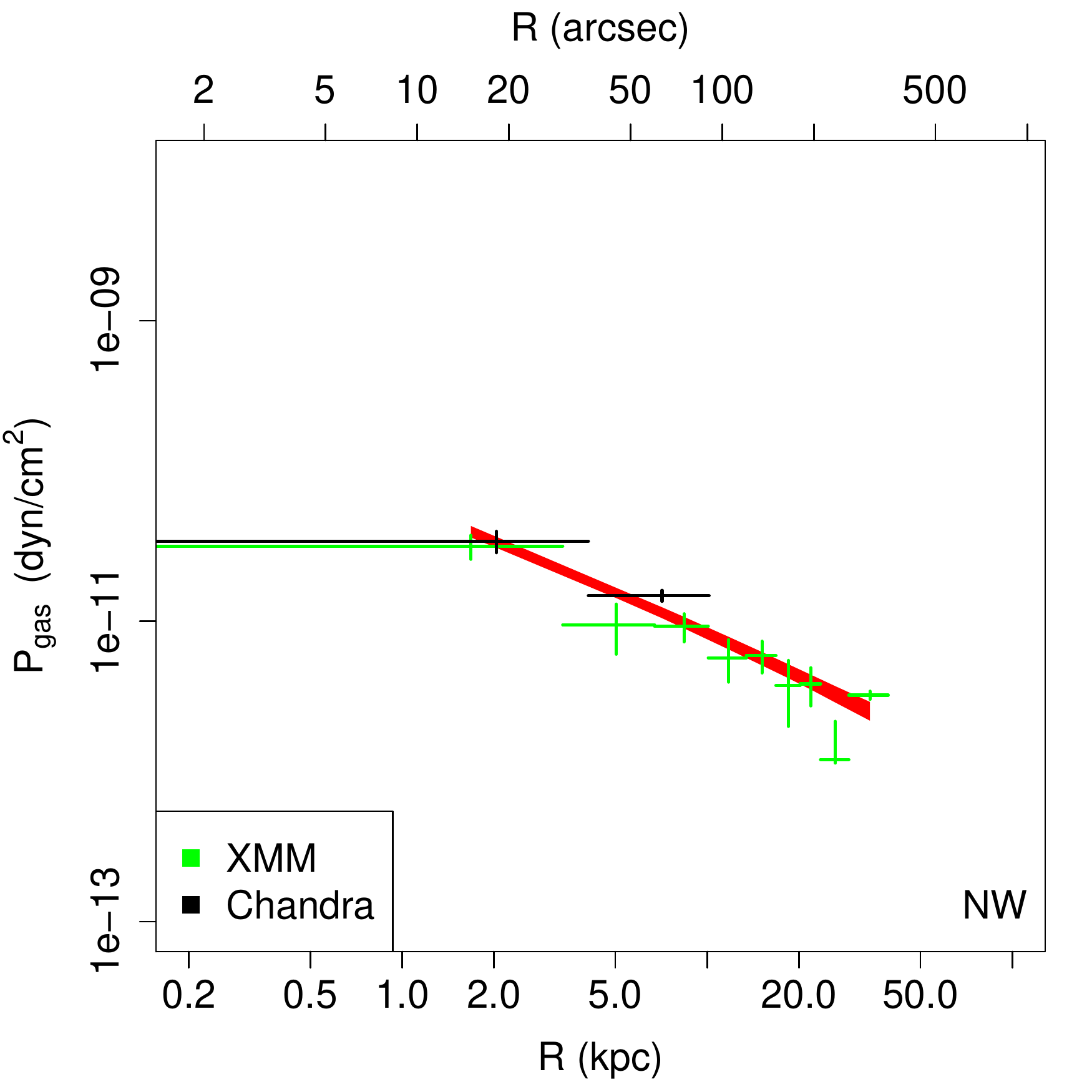}
\includegraphics[scale=0.16]{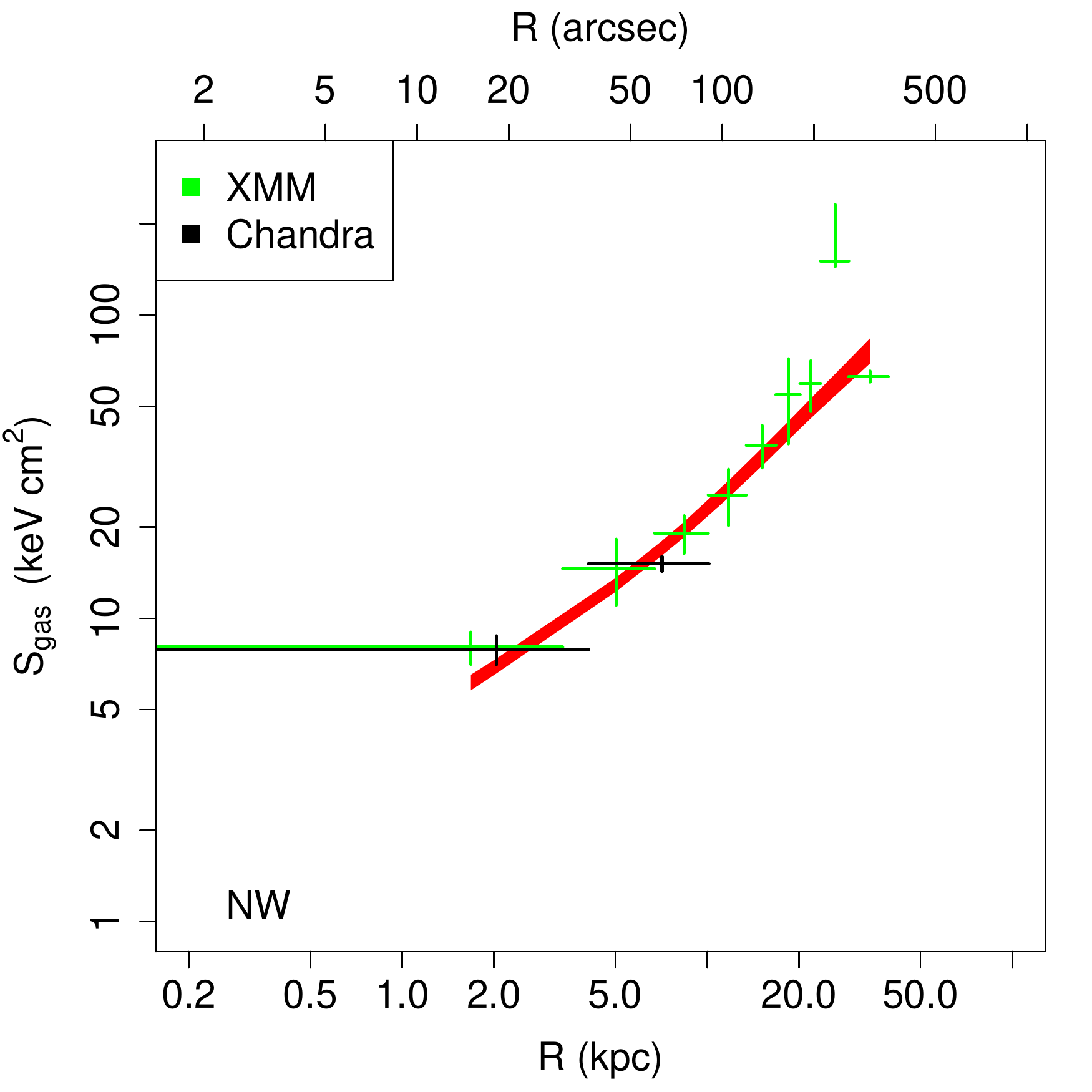}
\includegraphics[scale=0.16]{{N5846_mass_profile_comparison_340_480_0_0_0.7_abund}.pdf}
\includegraphics[scale=0.16]{{N5846_abund_profile_merged_340_480_0_0}.pdf}
\caption{Same as Figure \ref{fig:N5846_gas_profiles_merged_app} but with the free abundance model. In addition, on the rightmost panel of each row we show the element abundances profiles for \textit{XMM}-MOS data (in green) and for \textit{Chandra} ACIS data (represented in black). In the same panels we overplot with green and black rectangles the values of the element abundances obtained with the fixed abundance model for \textit{XMM}  and \textit{Chandra} data, respectively.}\label{fig:N5846_gas_profiles_merged_abund_app}
\end{figure}

\begin{sidewaysfigure}
\centering
\includegraphics[scale=0.26]{{N5846_mass_fit_0_360_0_0_0.6}.pdf}
\includegraphics[scale=0.26]{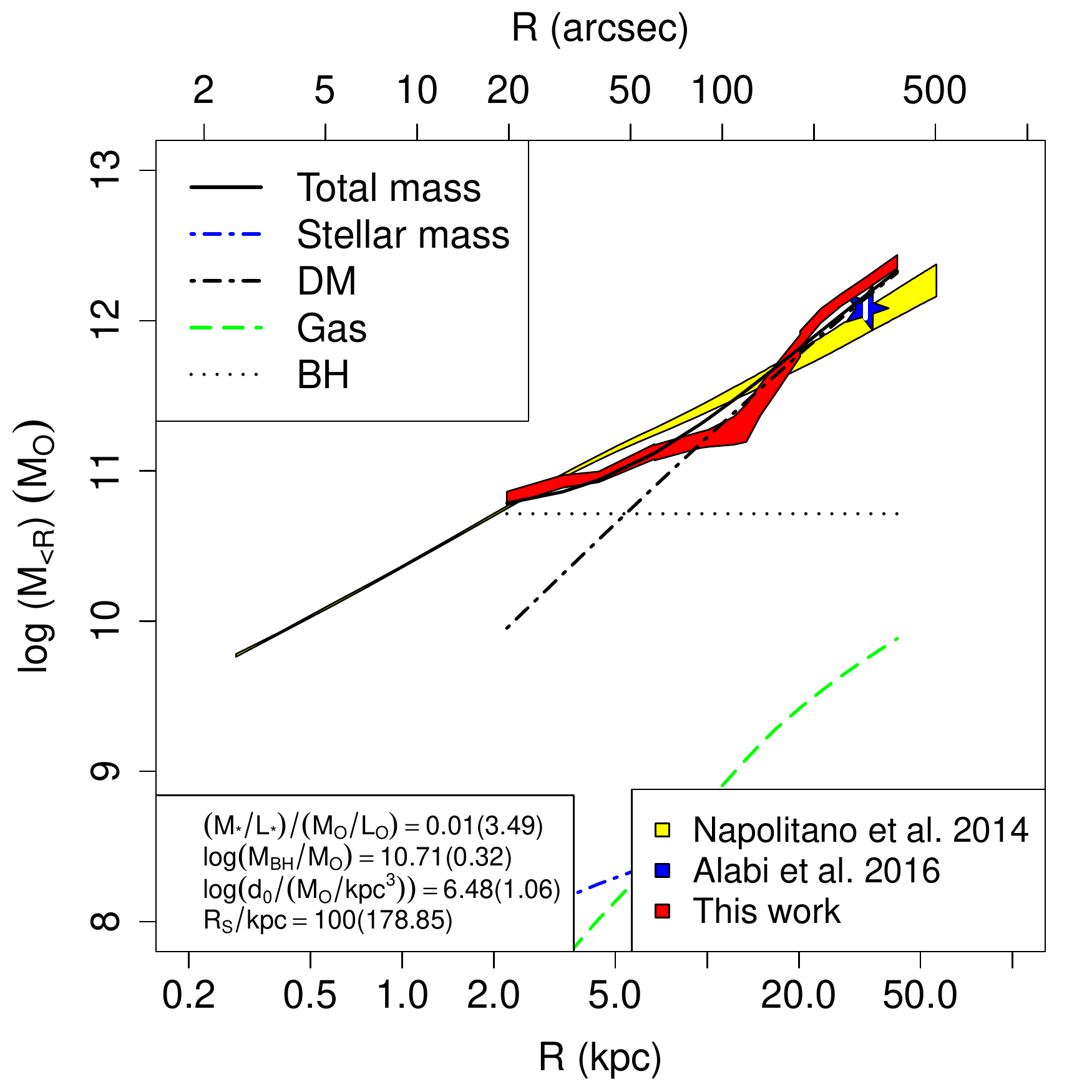}
\includegraphics[scale=0.26]{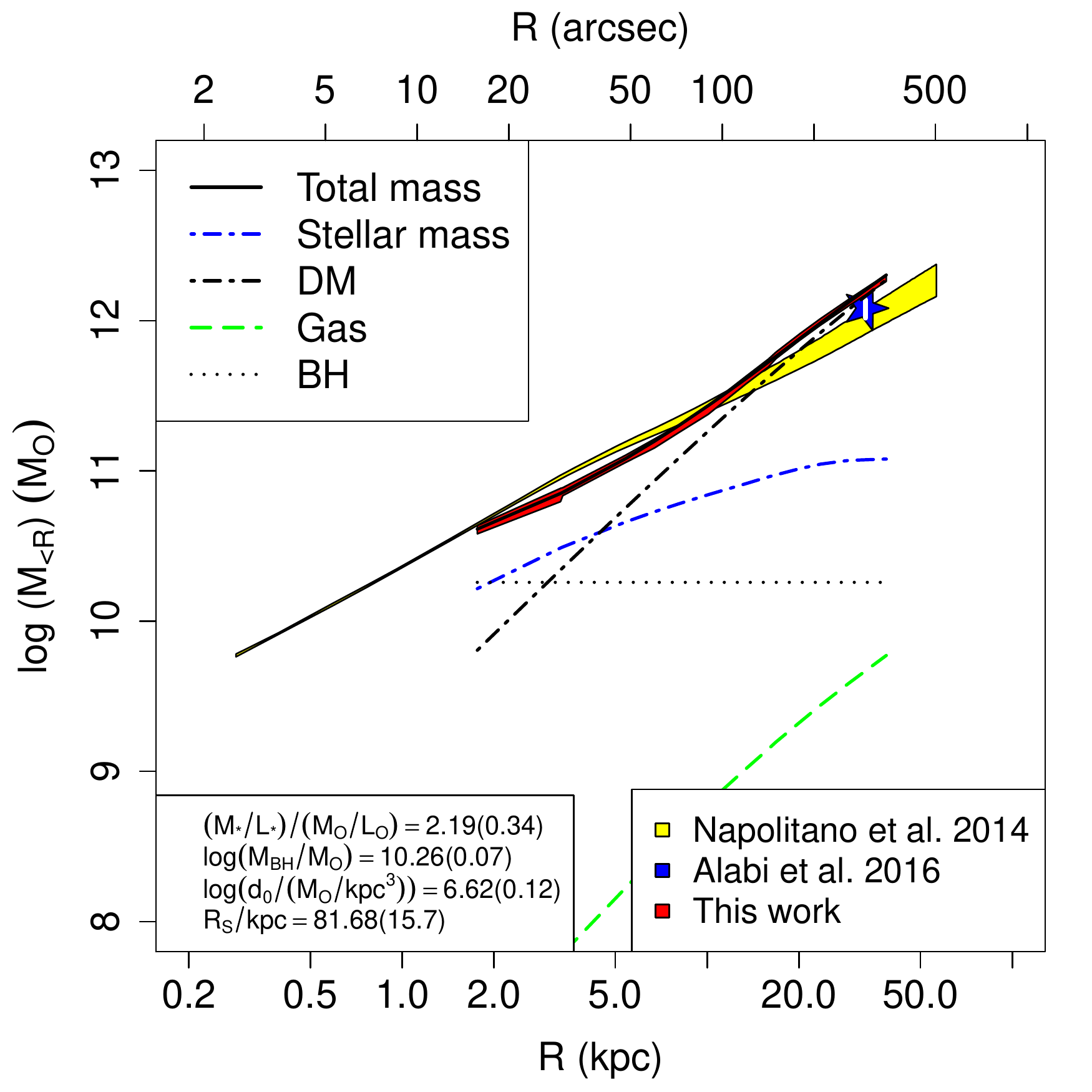}
\includegraphics[scale=0.26]{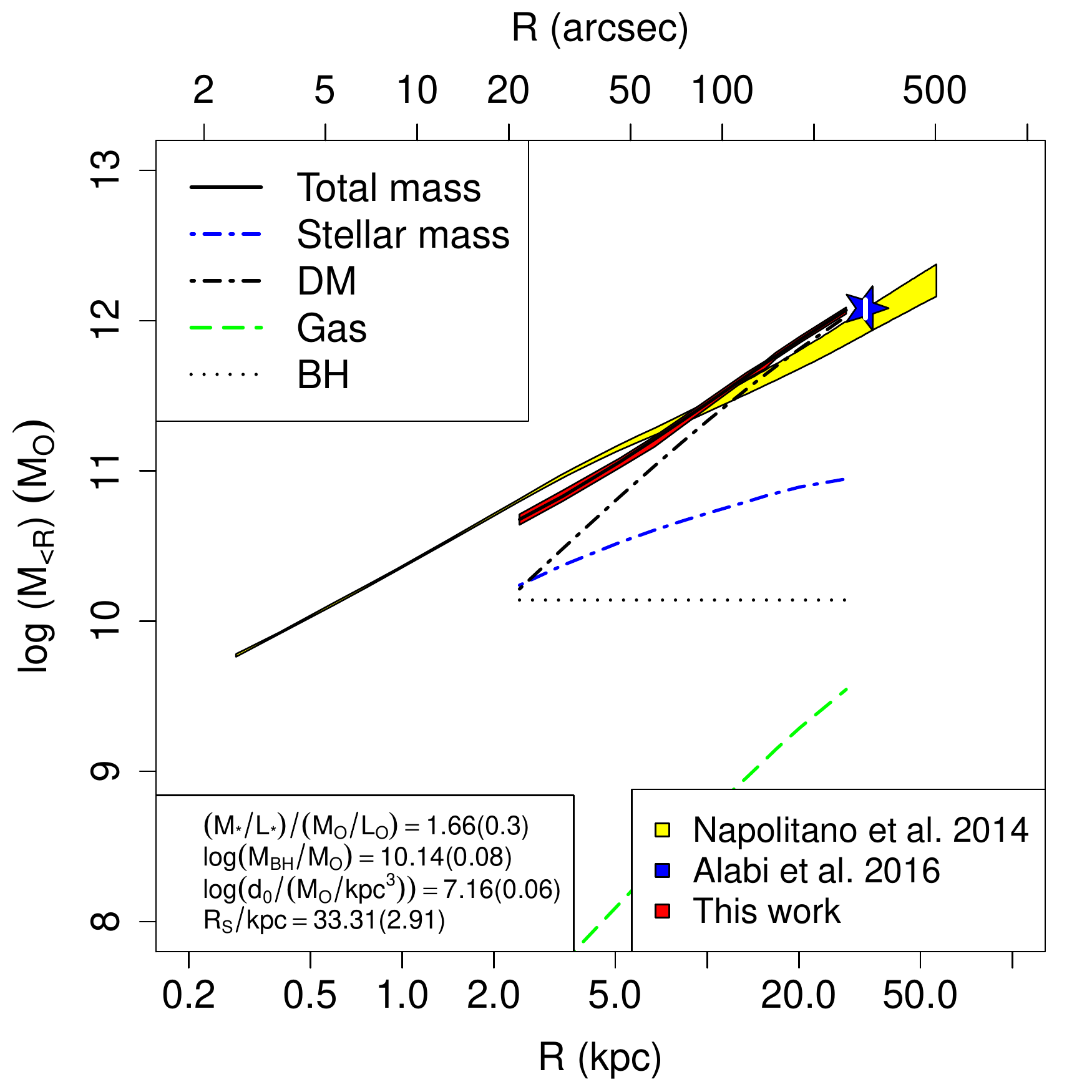}
\includegraphics[scale=0.26]{{N5846_mass_fit_340_480_0_0_0.7}.pdf}\\
\includegraphics[scale=0.26]{{N5846_mass_fit_0_360_0_0_0.6_abund}.pdf}
\includegraphics[scale=0.26]{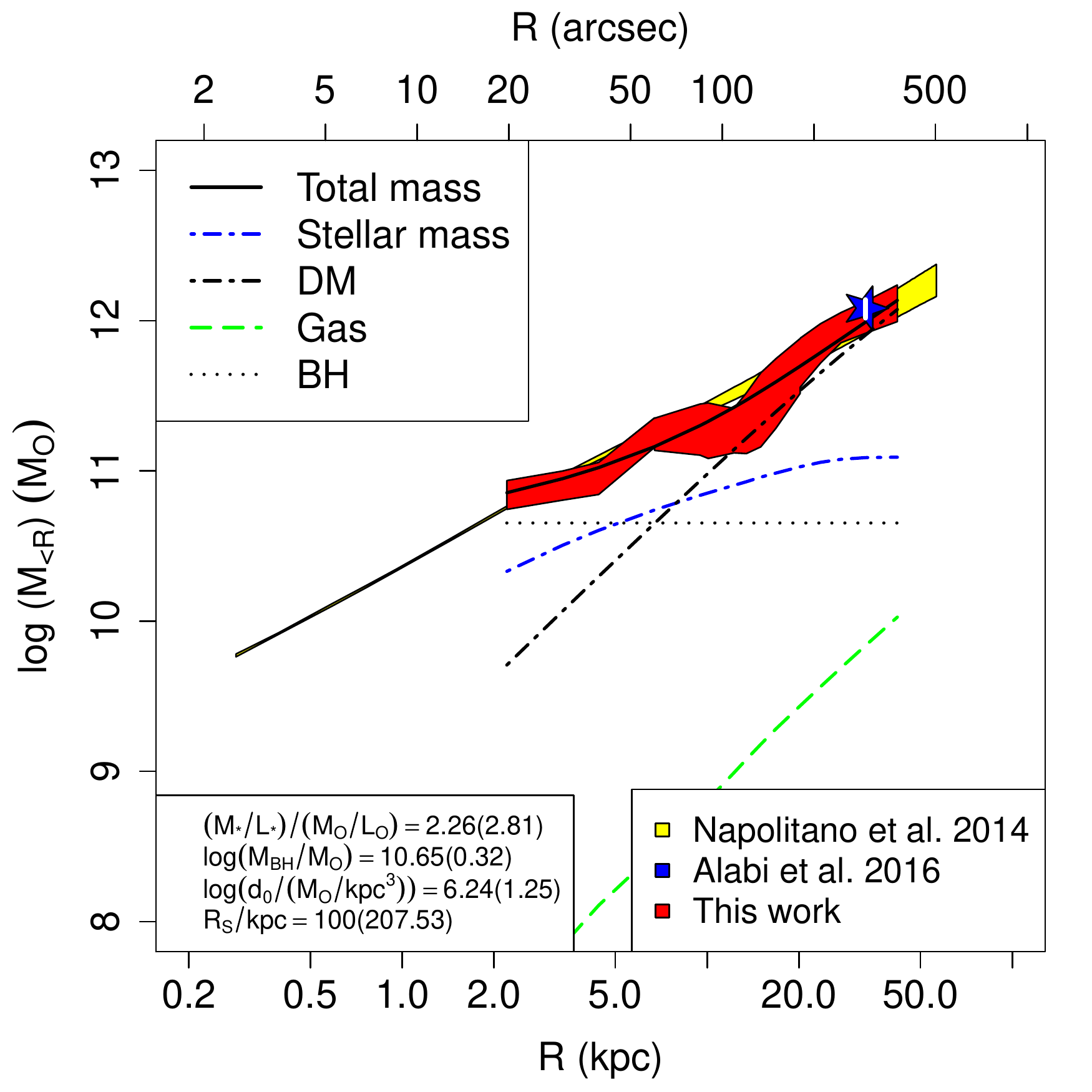}
\includegraphics[scale=0.26]{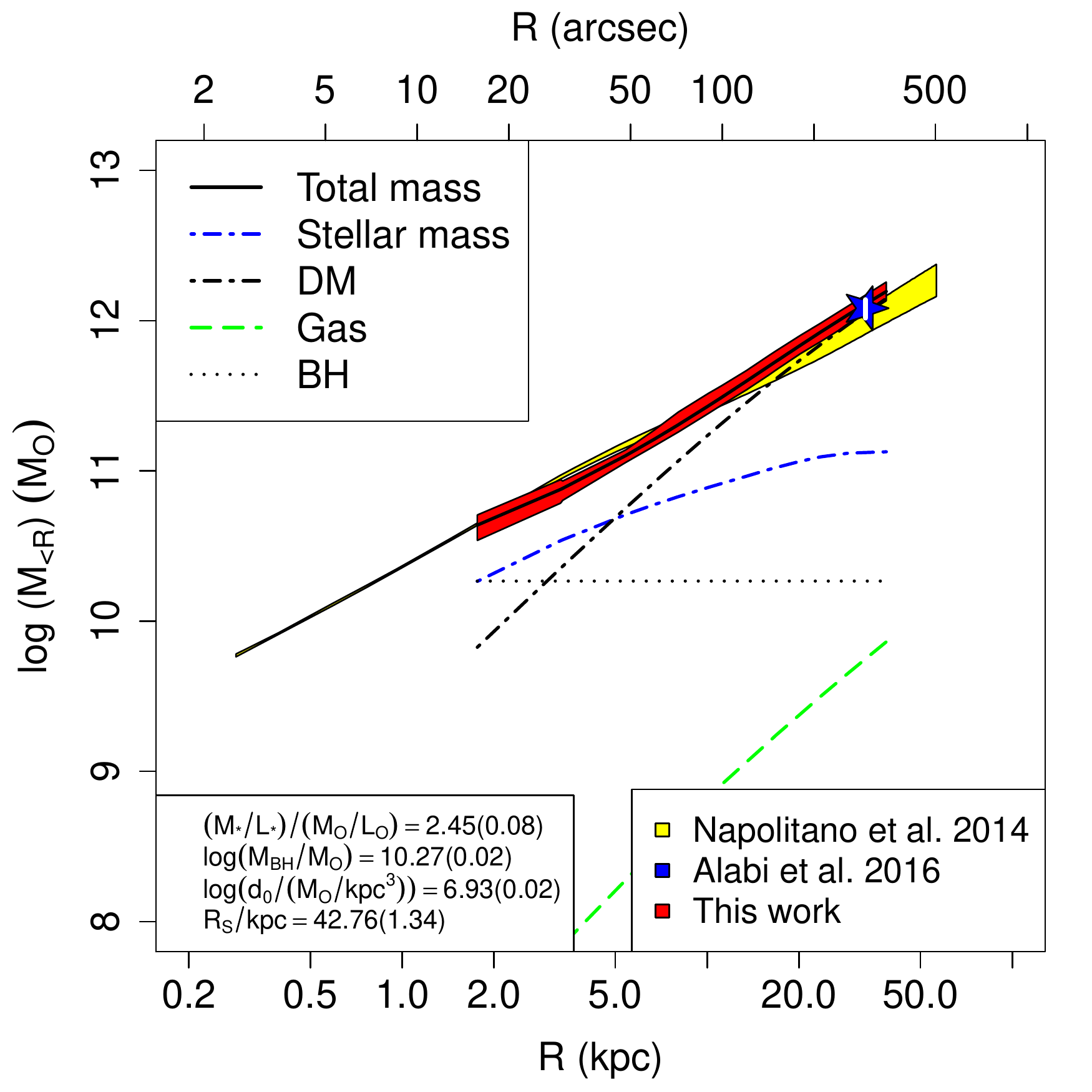}
\includegraphics[scale=0.26]{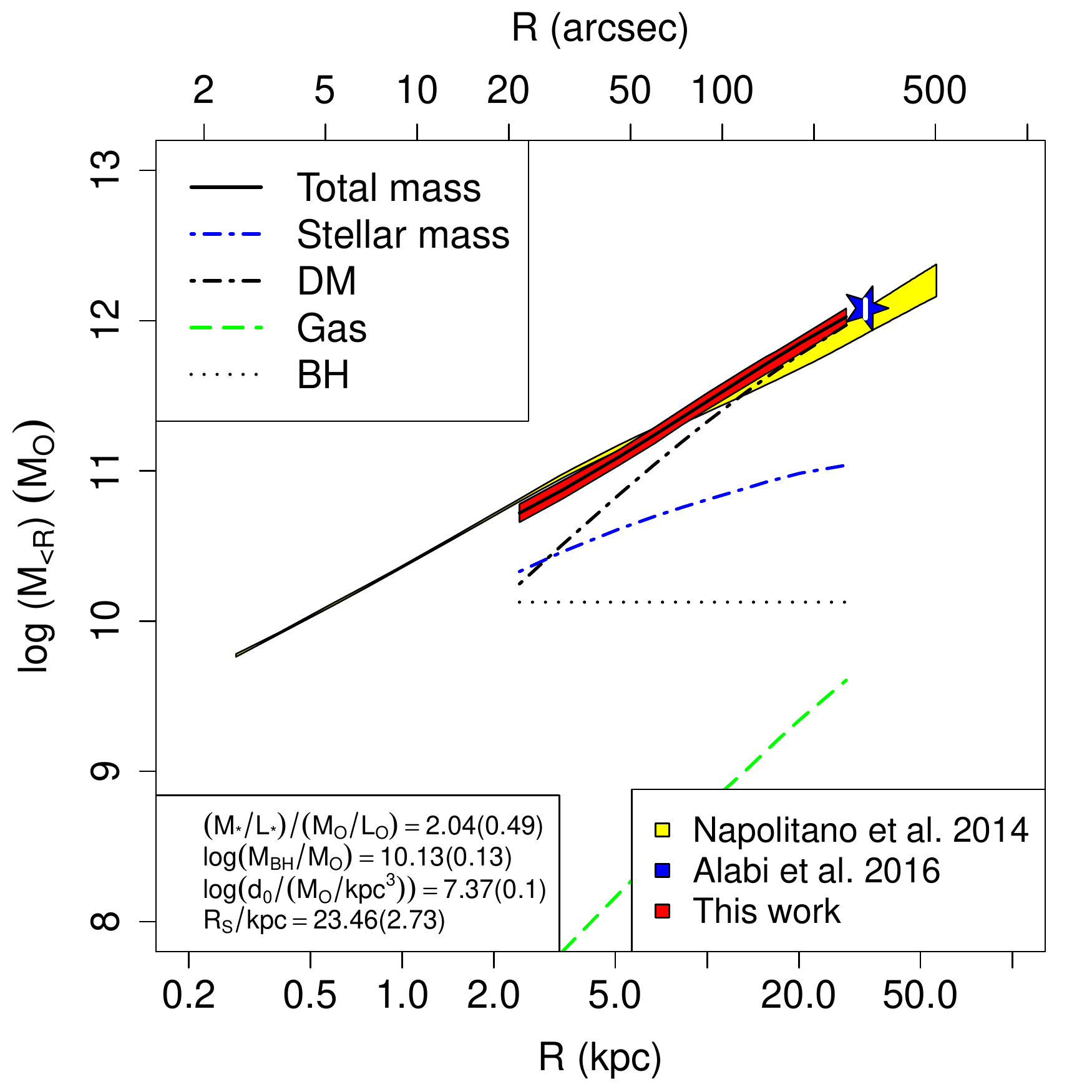}
\includegraphics[scale=0.26]{{N5846_mass_fit_340_480_0_0_0.7_abund}.pdf}
\caption{Fit to mass profiles of NGC 5846 shown in Fig. \ref{fig:N5846_gas_profiles_merged_app} (top row) and in Fig. \ref{fig:N5846_gas_profiles_merged_abund_app} (bottom row), from left to right in the full (0-360), SE (120-180) and NW (250-30) sector, respectively. The mass profile form the {HE} equation is presented in red, and the best fit contributions of the various mass components (gas mass, stellar mass, black hole and NFW dark matter profile) are presented with different colors as reported in the legend. The best fit parameters are reported in lower left box. In yellow we show the optical mass profile obtained from SD and GC reported by \citep{2014MNRAS.439..659N}, while the blue star represents the measurement by \citet{2016MNRAS.460.3838A} with the corresponding uncertainty shown as a white vertical line side the star itself.}\label{fig:N5846_mass_fits_app}
\end{sidewaysfigure}

\end{appendix}
}

\end{document}